\documentclass[a4paper,12pt]{scrartcl}
\DeclareOldFontCommand{\rm}{\normalfont\rmfamily}{\mathrm}
\usepackage[utf8]{inputenc}				%Umlautdarstellung
\usepackage{graphicx}						%Grafikeinbindung
\usepackage{fancyhdr}						%Kopfzeilengestaltung
\usepackage{units}							%Einheitendarstellung
\usepackage{cite}								%für Literaturverzeichnis nötig
\usepackage{hyperref}						%Verlinktes Inhaltsverzeichnis fuer die PDF Variante
\usepackage{psfrag}
\usepackage{amsmath}
\usepackage{amssymb} %for the symbol of the set of complex numbers, C
\usepackage{bbold} %for the symbol of the identity matrix with 1
\usepackage{psfrag}
\usepackage{color}
\usepackage{longtable}
\usepackage{chngcntr}
\usepackage{latexsym} %for the command \Box
\usepackage{xcolor,cancel}
\newcommand\hcancel[2][black]{\setbox0=\hbox{$#2$}%
	\rlap{\raisebox{.5\ht0}{\textcolor{#1}{\rule{\wd0}{0.5pt}}}}#2}
\usepackage{physics} %for braket

\numberwithin{equation}{section}
\numberwithin{table}{section}
\definecolor{lblue}{RGB}{225,232,242}

\usepackage{tikz}
\usepackage{subcaption}

\begin{document}
\parindent 0pt
\parskip 1.1ex

\newcommand{\X}{\overrightarrow{X}}
\newcommand{\Y}{\overrightarrow{Y}}
\newcommand{\nv}{\overrightarrow{n}}
\newcommand{\Phic}{\underline{p}}
\newcommand{\vc}{\underline{v}}
\newcommand{\kc}{\underline{k}}
\newcommand{\Fc}{\underline{F}}
\newcommand{\rc}{\underline{r}}
\newcommand{\Green}{\underline{G}}
\newcommand{\parta}[1]{\frac{\partial}{\partial #1}}

\pagenumbering{roman}

\begin{titlepage}

\begin{figure}[t]
	\centering
		\includegraphics[width=0.25\textwidth]{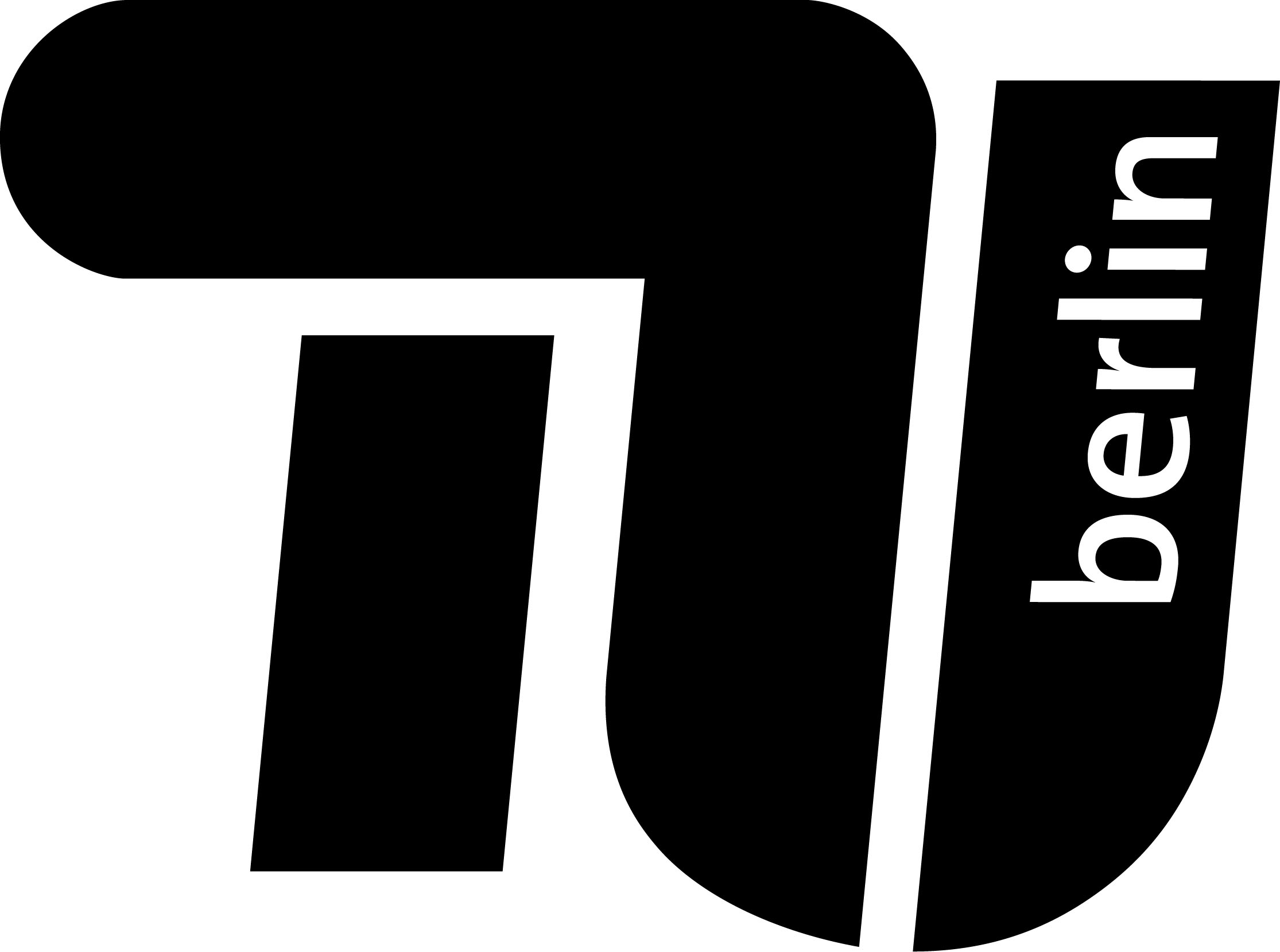}
\end{figure}

\vspace*{0,5cm}

\begin{center}
	\large{\textbf{Technische Universität Berlin}\\
	Fakultät V - Verkehrs- und Maschinensysteme\\
	Institut für Strömungsmechanik und Technische Akustik\\
	Fachgebiet Technische Akustik\\
	Sekr. TA 7  -  Einsteinufer 25  -  10587 Berlin}\\
\end{center}

\vspace*{1cm}

%Hier werden Vorlesungsreihe und Titel des Labors eingefügt
\begin{center}
	\Large{\textbf{Acoustic analogies with general relativity, \\quantum fields, and thermodynamics\\
	}}
\end{center}
\begin{flushleft}
	\vspace*{\fill}
	\large Drasko Masovic, TU Berlin, 2018 (last update: \today)
\end{flushleft}

\end{titlepage}

%Definition der Kopfzeile; Bitte Thema des Labors einfügen
\pagestyle{fancy}
\addtolength{\headheight}{20pt}
\lhead{TU Berlin}
\chead{\scriptsize{Institut für Strömungsmechanik und Technische Akustik \\
				Fachgebiet Technische Akustik}}
\rhead{}
%\rhead{wag. '85, weit. '02}

\vspace*{\fill}
The author owes a gratitude to Prof. Ennes Sarradj from the Technical University of Berlin for his valuable help and support on many occasions during the work.

All constructive comments on the ideas presented in the text are welcome and should be sent to the author (drasko.masovic@tu-berlin.de).

\newpage

%Hier können die einzelnen Kapitel eingefügt werden
\tableofcontents
\pagenumbering{arabic}
\section{Preface}\label{ch:preface}
\setcounter{footnote}{0}
%\addcontentsline{toc}{section}{Preface}

\begin{figure}[h]
	\centering
	\begin{subfigure}{.5\textwidth}
		\centering
		\includegraphics[width=1\linewidth]{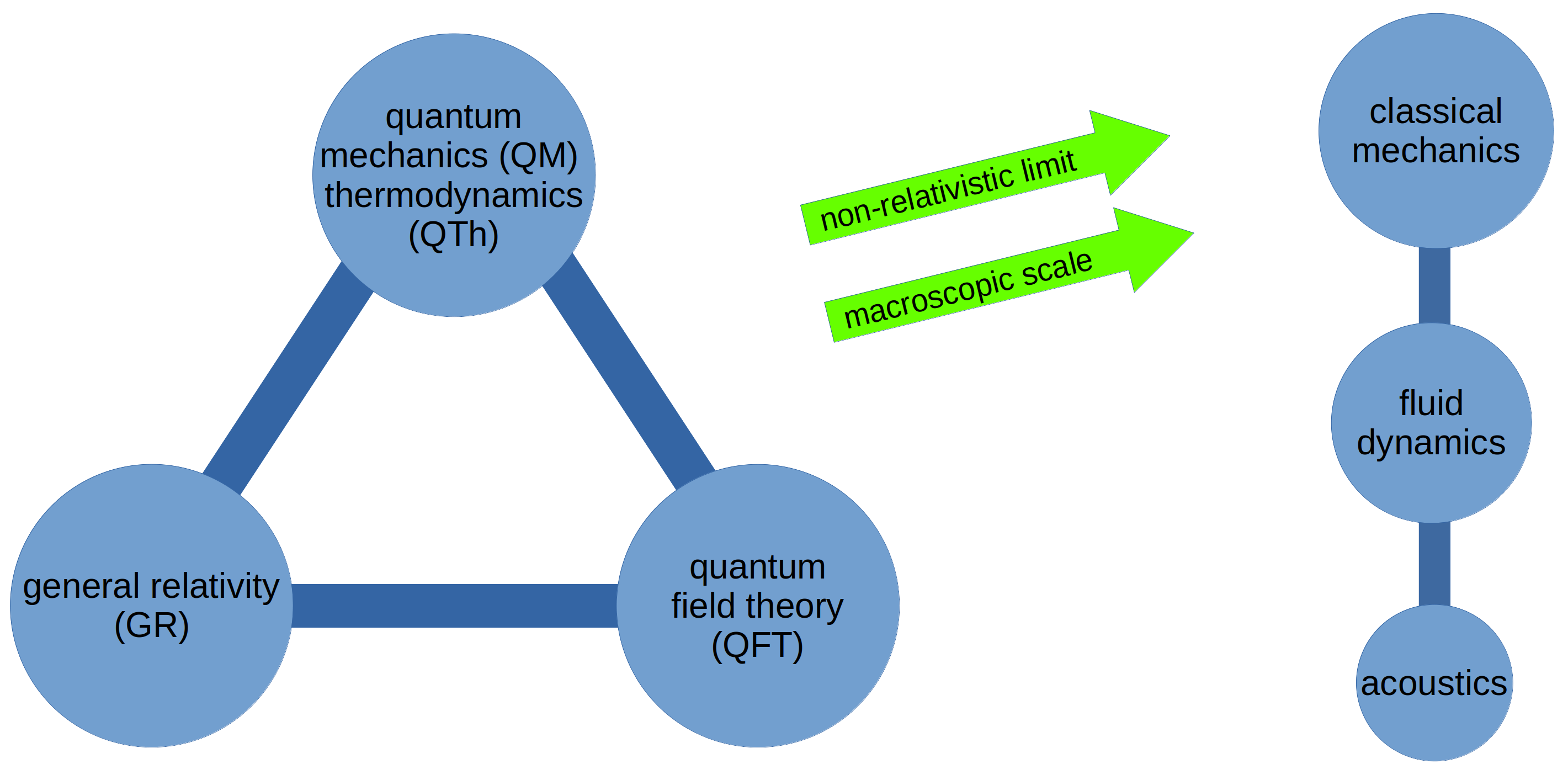}
		%\caption{A subfigure}
		\label{fig:physics_acoustics_usual}
	\end{subfigure}%
	\begin{subfigure}{.5\textwidth}
		\centering
		\includegraphics[width=1\linewidth]{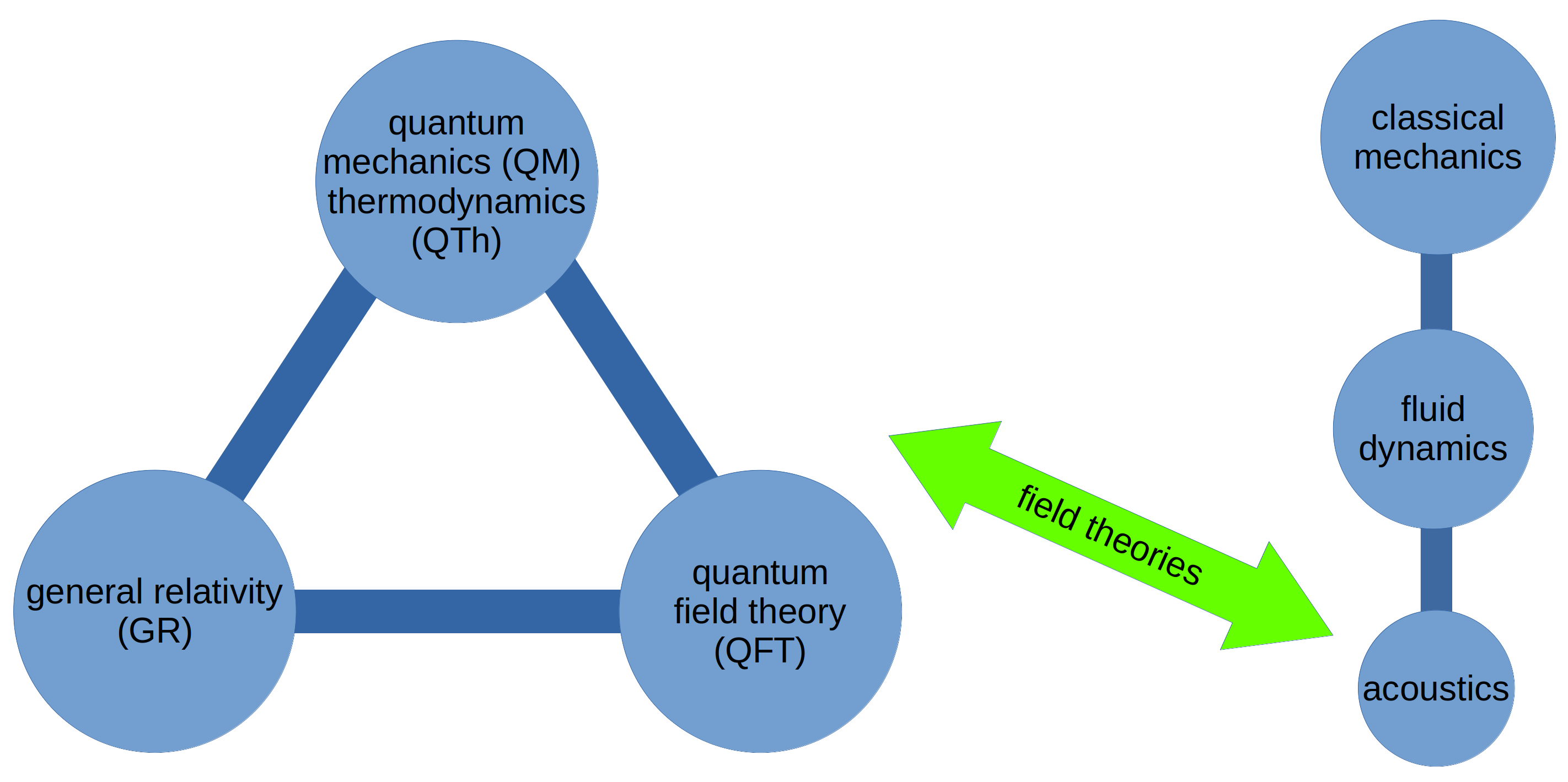}
		%\caption{A subfigure}
		\label{fig:physics_acoustics_analogies}
	\end{subfigure}
	\caption{Modern physics and classical acoustics: (left) usual relation and (right) relation in terms of acoustic analogies.}
	\label{fig:physics_and_acoustics}
\end{figure}

Modern physics is based on three major theories -- general relativity, quantum field theory, and thermodynamics, in particular quantum thermodynamics. Classical acoustics in fluids is usually regarded and studied as a part of classical mechanics, more precisely fluid dynamics (Fig.~\ref{fig:physics_and_acoustics} left). This assumes velocities much below the speed of light and length scales much above the molecular scale. However, like acoustics, all three modern theories are field theories, whether classical (gravitation) or quantum (quantum field theory and quantum thermodynamics). This fact opens possibilities for establishing direct analogies between acoustics and other theories (Fig.~\ref{fig:physics_and_acoustics} right) and they are in the focus of the work presented in the following.

Unlike the mentioned analogies, which have been considered in literature rather sporadically and unsystematically, the analogy between acoustics and classical electromagnetism as another field theory has been studied and exploited by many authors, bringing a large benefit to theoretical acoustics. These results will not be repeated here, but can be found elsewhere across the acoustic literature. However, we shall address the analogy with the more modern covariant (as well as quantum) vector theory of electromagnetism, as the most natural link between the analogue scalar theory of acoustics and the second-order tensor theory of general relativity. The analogy with general relativity will be treated first and in more detail, since it allows an elegant introduction of Lorentz invariance in classical acoustics. The emphasis is again not on the already established analogue models of gravity, based on the concept of acoustic spacetime, but on possible extensions and new views on the nature of sound waves and their generation in the analogue spacetime, which could complement the existing theory of wave propagation in the (curved) acoustic spacetime.

Classical acoustic and quantum fields appear to be even more remote and possible analogies have been investigated and exploited even less. Nevertheless, two facts allow bridging the apparent gap. First, quantum particles are described essentially in terms of continuous state functions (fields and eigenmodes). Second, continuity of sound fields is also necessarily interrupted at the microscopic length scale of molecular motion, which thus forces discretization of the theory. Importantly, the formal analogy with quantum mechanics provides not only a relation with quantum fields, but quantum thermodynamics, as well. These in combination with the analogy with general relativity allow an analogue unified theory to be defined, in which the fundamentals of classical acoustics in fluids are expressed using the formalisms of modern field theories. This should further help narrowing the gap between the apparently different physics.

As already suggested, the following text does not provide a detailed description of already established acoustic analogies, but only the necessary basics. It rather focuses on possible extensions of the analogies with general relativity (sections~\ref{ch:motivation}, \ref{ch:sound_wave_generation_in_spacetime}, and \ref{ch:dark_energy}), covariant electromagnetism (section~\ref{ch:analogy_with_EMG}), and quantum fields and thermodynamics (section~\ref{ch:acoustic_Lagrangians}). The analogies with classical fields are briefly summarized in section~\ref{ch:unified_analogy}, for a better overview before addressing the quantum theories and the unified theory in section~\ref{ch:unified_theory}.

The proposed extensions are still hypothetical and require proper validation and evaluation. The first scarce (and overhasty?) reactions are not very encouraging. Most of them can be divided in three categories:
\begin{itemize}
	\item recognition of the research potential without support in terms of scientifically valid arguments
	\item rejection of any deeper relationship between classical acoustics and the listed Lorentz-invariant and quantum theories\\
	Typical arguments used against the proposed analogies are that classical acoustics
	\begin{itemize}
		\item is Newtonian (although the Lorentz-invariant acoustic spacetime with the reference speed of sound has already been introduced, accepted, and successfully used both by the relativistic community for the models of analogue gravity and the acoustic community for modelling sound propagation in non-uniform media),
		\item is a scalar field theory (although conceptually very similar and well-established analogy of gravitoelectromagnetism relates no less remote vector fields of classical electromagnetism and the second-order tensor fields of general relativity),
		\item does not share the dynamics and the role of mass with the theory of gravitation (although this is irrelevant for the entire kinematics of fields as well as incompressible flows),
		\item is a continuous macroscopic theory free of quantum phenomena (even though quantum fields are also essentially described by continuous functions and many quantum particles are ``macroscopic" with respect to the Plank scale).
	\end{itemize}
	\item general acceptance of the analogies, but disapproval of their scientific value or novelty, because they merely provide an acoustic re-interpretation of the existing results of other theories\\ However, this neglects the fact that similar formal analogies (such as with classical electromagnetism mentioned above) have shown very fruitful for theoretical acoustics and that similar idealistic requirements for novelty would have excluded many important crossdisciplinary findings. It also neglects a potential didactic significance of bringing acoustics and other field theories closer than they currently are in science.
\end{itemize}

Challenging such conceptually very diverse, inconsistent, and arguably superficial criticism, the aim of this text is to introduce and promote the extended analogies and bring solid arguments for their further consideration. Potential benefits for the acoustic theory are large -- new formalisms and paradigms for modelling sound generation and propagation phenomena, new approaches and mathematical tools for solving acoustic problems, broader and more versatile acoustic education, and a more prominent role of acoustics in modern science. The text should therefore motivate primarily theoretical acousticians to embrace modern (both classical and quantum) field theories more vigorously and profit from the always fertile ground of physical analogies. Even when developed for the use in other areas, the acoustic analogies cannot be fully evaluated and appreciated without an active involvement of the acoustic community. This applies even more taking into account the diversity of modern acoustics (particularly aeroacoustics), which remains unknown to most of the theoretical physicists.

Being rather formal, the investigated analogies do not imply new physically meaningful findings in other field theories. Still, this does not contradict a reevaluation of the shared phenomena or even the establishment of a unified theory using (less abstract) acoustic terminology. The author's belief is that acoustics in fluids is best approached as essentially the theory of scalar fields and monopole sources, which are thus complementary to and side-by-side with the vector fields and dipole sources of electromagnetism and second-order tensor fields and quadrupole sources of gravitation. As such, it is didactically well-suited as a prototype-theory strongly rooted in everyday life for more complex and abstract field theories. References to hypothetical, heuristic, or ``dummy'' scalar fields, which are common in literature on theoretical physics, can thus be avoided. May it also be said that the acoustic theory can be as rich as other (both classical and quantum) field theories, including dipole and quadrupole sources of waves, discretization, and thermodynamic treatment, and involve even more sophisticated boundary conditions and interactions (for example, with turbulence) or transverse waves (in solid media, which will not be studied here). This is by no means recognized or appreciated enough by theoretical physicists. Unfortunately, their general interest for and knowledge of highly non-trivial and constantly developing acoustic theory does not appear to be much greater than the interest for and knowledge of their areas among theoretical acousticians, even after more than a century of co-existence.

\section{Sound waves in acoustic spacetime}\label{ch:motivation}

The first observations of geometric similarities between special relativity and propagation of sound waves in fluids date back to W. Gordon~\cite{Gordon1923}. However, the extended analogy between acoustics and general relativity is most commonly attributed to W.~G.~Unruh~\cite{Unruh1981}, who used it primarily for studying Hawking radiation by means of much better understood acoustics in transonic flows\footnote{Although usually associated with quantum effects, Hawking radiation is practically completely described with classical mechanics and geometry of spacetime~\cite{Unruh2014}.}. Since then the analogy has been further developed and promoted by M.~Visser and C.~Barcel\'{o} in their studies of analogue models of general relativity~\cite{Barcelo2004} or simply analogue gravity~\cite{Visser2011}. In particular, the authors have shown that sound propagation in inhomogeneous flows of the background fluid can be described by differential geometry of a curved acoustic spacetime, while keeping the usual d'Alembert operator on the left-hand side of the wave equation. The only essential difference from the relativistic spacetime is the occurrence of the speed of sound rather than light as the reference speed. The acoustic analogy was used to provide new insights into various gravitational phenomena, such as black holes, with the aid of simpler Newtonian physics of fluid dynamics and acoustics. For example, gravitational ergo-surfaces are treated as surfaces in fluids where Mach number of the background flow reaches one~\cite{Visser1998}.

Described as above, the acoustic analogy with relativity should not be confused with relativistic acoustics. The first is based on the formal similarity of the two mathematical descriptions when the speed of light is replaced by (typically much lower) speed of sound. In contrast to this, the second is concerned with physical scenarios in which the speed of sound (or compressible waves in the fluid) is actually comparable to the speed of light. It is obviously relevant only under certain extreme conditions, such as in early universe, close to black holes and galactic nuclei, in the cores of neutron stars, and similar~\cite{Faccio2013}. In the following we consider relations between acoustics and relativity only in the former sense of an analogy.

Although the acoustic analogy has raised a certain interest in the relativistic community, the methods and results of general relativity have been used very sporadically for illuminating acoustic phenomena. The recent works of Gregory et al.~\cite{Gregory2015a,Gregory2015b} as well as Ewert and Proskurov~\cite{Ewert2020} are rare examples of such applications. Similarly as Unruh and others, the authors treat sound propagation in uniform and non-uniform background flows with the aid of geometric algebra of the four-dimensional acoustic spacetime. Apart from this, the Lorentz transformations of special relativity had been used even earlier for sound propagation in uniform mean flows~\cite{Jones1986,Hirschberg2018}. Still, the physical analogy is generally believed to be limited only to four-dimensional geometrical treatment of sound propagation in various media, due to different governing equations of general relativity (the Einstein field equations) and fluid dynamics~\cite{Barcelo2004,Gregory2015b}. In other words, the analogy is suitable only for certain flow acoustic problems of wave propagation in such treated background acoustic spacetime, but not for general aeroacoustics or even sound waves as a perturbation of the fluid.

In the theory of general relativity, a gravitational wave is a weak perturbation (or curvature) of the background spacetime. A question which naturally appears in the analogue acoustic theory is whether sound waves can also be associated with a weak curvature of the acoustic spacetime. The answer which appears to be given by the relativistic (that is, non-acoustic) and as such accepted by the acoustic community~\cite{Ewert2020} is negative. The main supporting argument is that metrics of the acoustic spacetime capture only kinematics, while gravitational waves, which obey the (linearized) Einstein field equations, depend also on the dynamics of general relativity~\cite{Visser1998,Ewert2020}, which is determined by the mass-involving source term on the right-hand side of the Einstein field equations. However, this argument neglects the fact that sound waves can also be described in terms of purely kinematic quantities (for example, velocity potential), independently of the dynamic source terms. Both types of waves are defined equally well with homogeneous wave equations. Moreover, many important sources of sound are essentially kinematic, for example, incompressible low Mach number turbulent flows. Hence, the limitation to kinematic effects does not a priori exclude an extension of the analogy to the nature of sound waves and their generation. In addition to this, the fact that most of the work on the acoustic models of analogue gravity focuses on the background fluid motion rather than on sound waves calls for a more critical evaluation of the given arguments by the acoustic community\footnote{As an illustrative example and another argument why the analogy should be studied and used more actively by the acoustic community, may it be mentioned that in Ref.~\cite{Visser1998} it is also stated that the difference between a wind gust and sound waves is basically conventional (a matter of frequency of the associated fluid perturbation) and without a deeper physical meaning. The modern aeroacoustic theory, however, clearly distinguishes between sound waves associated with compressible and irrotational motion of fluid particles and incompressible or solenoidal motion, regardless of frequency. It is also stated that sound waves in a background vorticity cannot be represented by a scalar potential, which is actually the defining property of sound waves.}.

As opposed to this, some previous investigations~\cite{Masovic2019,Masovic2020} demonstrated that sound waves do correspond to a weak perturbation of the background acoustic spacetime, that a non-relativistic (that is, involving mass moving at velocities much smaller than the speed of light) quadrupole source of gravitational waves is entirely analogous to the quadrupole source of sound in an incompressible (low Mach number) turbulent flow, and that the linearized Einstein field equations can be used for both types of waves. These results are elaborated here. After presenting the essentials of the state-of-the-art relativistic treatment of the background acoustic spacetime in section~\ref{background_spacetime}, we hypothesize in section~\ref{ch:sound_wave_in_spacetime} that the sound waves represent a small metric perturbation of the background acoustic spacetime. The obvious difference between longitudinal (compressible, scalar) sound waves and transverse (incompressible, second-order tensor) gravitational waves is solely due to the observer (listener or microphone), who (unlike a relativistic observer) does not exist in the acoustic spacetime, but the external Newtonian space and time, and thus violates the Lorentz invariance. Information on the sound waves is nevertheless contained in the metric perturbation. The generation of perturbation governed by the linearized Einstein field equations will be considered in section~\ref{ch:sound_wave_generation_in_spacetime}.

\subsection{Background spacetime}\label{background_spacetime}

Acoustic spacetime is a four-dimensional Lorentzian manifold analogous to the relativistic spacetime, with the constant speed of light in vacuum replaced by the constant speed of sound in the background fluid. This definition already sets certain limits for applications of the analogy, namely the acoustic problems for which a constant speed of sound can be reasonably defined (for example, in a quiescent fluid or in an inhomogeneous fluid at infinity) and it represents the maximum speed of motion of the fluid particles (as in subsonic flows). The background fluid thus acts as the vacuum and its particles constitute the fabric of space. Such a concept of analogue acoustic spacetime has been successfully used for studying sound propagation in uniform flows (flat spacetime)~\cite{Gregory2015a} and non-uniform flows (curved spacetime)~\cite{Gregory2015b,Ewert2020}. This section summarizes briefly the main concepts and results of these studies as a basis for extensions of the theory in the rest of the text.

The central quantity for describing geometry of a spacetime is metric tensor, $\boldsymbol g$ (bold symbol indicates a second-order tensor written independent of a frame of reference). By definition, pseudo-Riemannian (including Lorentzian) manifolds considered here are differentiable manifolds supplied at each point with a metric. Thereby, the metric gives a certain shape to the manifold, which would otherwise be merely a set of disconnected points. More particularly, the metric tensor determines the scalar product of two vectors, for example, $\vec A$ and $\vec B$. In a specific frame of reference the two vectors are given by their components $A^\alpha$ and $B^\alpha$ and their scalar product equals
\begin{equation}\label{scalar_product}
\vec{A}\cdot\vec{B} = A^\alpha B^\beta g_{\alpha\beta},
\end{equation}
where $g_{\alpha\beta}$ are the components of $\boldsymbol g$ in the same frame. As usual in general relativity, we use Greek letters to denote four-dimensional coordinates ($\alpha, \beta = 0,1,2,3$, where 0 corresponds to the time coordinate and the remaining are spatial coordinates) and Latin letters for the spatial components only (for example, $i = 1,2,3$). Therefore, $x^0 = c_0 t$, where $c_0$ is the constant reference speed of light or sound\footnote{In this section we will allow the speed of sound $c$ to vary over space, but $c_0$ remains constant, for example the value of $c$ at infinity, far from the inhomogeneities of the fluid. Moreover, it will be constant in the rest of the text, since we assume a simple quiescent fluid.} and $x^1$ to $x^3$ are spatial components, for example, $x^1 = x$, $x^2 = y$, and $x^3 = z$ in Cartesian coordinates. Equation~(\ref{scalar_product}) also assumes Einstein's summation convention, according to which the summation is performed over each letter which appears in an expression once as a subscript and once as a superscript. The two positions of the letters indicate the covariant and contravariant vector bases, respectively. Since the scalar product commutes, metric tensor is necessarily symmetric ($g_{\alpha\beta} = g_{\beta\alpha}$) with maximum ten independent components in any frame. 

In classical acoustics sound propagation in moving and inhomogeneous media is typically described by replacing the simplest wave operator, the d'Alembertian $\Box$, with a more general (and complex) differential operator on the left-hand side of the wave equation. Implementation of the acoustic spacetime as defined above avoids this by introducing more general, non-flat spacetimes, while keeping the d'Alembert operator as the wave operator. In a general spacetime the d'Alembertian of a scalar $\phi$ equals
\begin{equation}\label{eq:d'Alembertian}
\Box \phi = (g^{\alpha \beta} \phi_{,\beta})_{;\alpha} = \frac{1}{\sqrt{-g}}(\sqrt{-g}g^{\alpha\beta}\phi_{,\beta})_{,\alpha}.
\end{equation}
Comma denotes derivative with respect to the coordinate(s) which follow(s) it (for example, $\phi_{,\alpha} = \partial \phi/\partial x^\alpha$ and $\phi_{,i} = \nabla \phi$ is usual gradient of $\phi$ in a three-dimensional space), while semicolon denotes covariant derivatives in curved manifolds (for a vector $V^\beta$, $V^\beta {}_{;\alpha} = V^\beta {}_{,\alpha} + V^\mu \Gamma^\beta {}_{\mu\alpha}$, where $\Gamma^\beta {}_{\mu\alpha}$ are Christoffel symbols, which will be introduced later in eq.~(\ref{Christoffel_symbols_metric}); in a flat spacetime the covariant derivative reduces to the ordinary derivative, because $\Gamma^\beta{}_{\mu\alpha}$ equals zero). Determinant of the matrix ($g^{\alpha\beta}$) is written shortly as $g = \det(g^{\alpha\beta})$. It is interesting to note that the d'Alembertian in eq.~(\ref{eq:d'Alembertian}) is also the operator of the massless Klein-Gordon equation. The Klein-Gordon equation is the most general Lorentz-invariant equation of motion which a non-interacting scalar field can satisfy~\cite{Schwartz2014}. It is, therefore, natural that the same operator appears in the Lorentz-invariant description of (massless) scalar sound fields in unbounded fluids. This will be used in section~\ref{ch:acoustic_Lagrangians} as the starting point for the quantum field analogy.

Temporal dimension of the spacetime is not merely the fourth spatial dimension coupled by the constant $c_0$. It comes with the sign opposite to the three spatial dimensions. Adopting the mixed signature $[-+++]$, the simplest, flat (also called Minkowski) spacetime has the metric
\begin{equation}\label{eq:metric_tensor_eta}
g^{\alpha\beta} = \eta^{\alpha\beta} =
\begin{bmatrix}
-1 & 0 & 0 & 0\\
0 & 1 & 0 & 0\\
0 & 0 & 1 & 0\\
0 & 0 & 0 & 1
\end{bmatrix}
\end{equation}
in a certain frame of reference. The determinant equals $g=-1$, so the square root in eq.~(\ref{eq:d'Alembertian}) delivers a real number. Of course, the metric tensor can take more complicated forms in other frames of reference, but the frame in which $g^{\alpha\beta}$ reduces to $\eta^{\alpha\beta}$ from eq.~(\ref{eq:metric_tensor_eta}) must exist if the spacetime is Minkowski spacetime. Since it is flat, the covariant derivative becomes the ordinary derivative and in this particular frame the d'Alembertian becomes the classical wave operator:
\begin{equation}\label{eq:d'Alembertian_quiescent_flow}
\Box \phi = (\eta^{\alpha\beta}\phi_{,\beta})_{,\alpha} = \eta^{\alpha\beta} \phi_{,\alpha\beta} = \phi^{,\alpha}{}_{,\alpha} = \left( -\frac{1}{c_0^2} \frac{\partial^2}{\partial t^2} + \nabla^2 \right) \phi,
\end{equation}
since $\eta^{\alpha\beta}$ is constant, the derivative with respect to the temporal coordinate is $_{,0} = 
\partial/\partial x^0 = (1/c_0)\partial/\partial t$, and the Laplacian is $\nabla^2 = \partial^2/\partial (x^1)^2 + \partial^2/\partial (x^2)^2 + \partial^2/\partial (x^3)^2$. As a general rule, multiplication with the spacetime metric $g^{\alpha\beta}$ raises the index $_\alpha$ (or $_\beta$) while multiplication with $g_{\alpha\beta}$ lowers it, so the summation over $\alpha$ applies in the equation above.

An example of a flat spacetime metric in another frame of reference is
\begin{equation}\label{eq:metric_tensor_M}
g^{\alpha\beta} =
\begin{bmatrix}
-1 & 0 & 0 & -M\\
0 & 1 & 0 & 0\\
0 & 0 & 1 & 0\\
-M & 0 & 0 & 1-M^2
\end{bmatrix},
\end{equation}
where $0 \leq M<1$ is a dimensionless constant ($g^{\alpha\beta} = \eta^{\alpha\beta}$ for $M=0$). The determinant is again $g = -1$, which is a property of a flat spacetime. In this frame, however, the d'Alembertian obtains a somewhat more complicated form:
\begin{equation}\label{eq:d'Alembertian_uniform_flow}
\begin{aligned}
\Box \phi &= g^{\alpha\beta} \phi_{,\alpha\beta} = \left( -\frac{1}{c_0^2} \frac{\partial^2}{\partial t^2} + \nabla^2 - \frac{2M}{c_0} \frac{\partial^2}{\partial t \partial x^3} - M^2 \frac{\partial^2}{\partial (x^3)^2} \right) \phi &\\
&= \left( - \frac{1}{c_0^2} \left( \frac{\partial}{\partial t} + M c_0 \frac{\partial}{\partial x^3} \right)^2 + \nabla^2 \right) \phi = \left( - \frac{1}{c_0^2} \frac{D^2}{Dt^2} + \nabla^2 \right) \phi,
\end{aligned}
\end{equation}
with $D/Dt = \partial/\partial t + M c_0 \partial/\partial x^3$. Here we recognize the differential operator of the convected wave equation. It describes sound propagation in a constant subsonic flow, that is, a background medium moving uniformly in the direction of the $x^3$-axis, with the subsonic Mach number $M$. Since the frame of reference can always be chosen such that the flow is in the direction of the $x^3$-axis, we can conclude that the d'Alembertian in eq.~(\ref{eq:d'Alembertian}) can describe sound propagation in both a quiescent fluid and a uniform mean flow, when supplied with appropriate metric tensor components. In the context of analogue gravity the metrics such as the one in eq.~(\ref{eq:metric_tensor_M}), which typically occur in acoustic problems, are sometimes called acoustic metrics \cite{Visser1998}.

In both cases considered so far the background acoustic spacetime was flat. Different operators of the wave equations were obtained solely due to different frames of reference and, consequently, different forms of the metric tensors, but the Lorentz-invariant operator, the d'Alembertian, remained unchanged. The last example of an acoustic metric describing sound propagation in a background flow which we consider is
\begin{equation}\label{eq:metric_tensor_Pierce}
g^{\alpha\beta} = -\left( \frac{\rho c_0^5}{\rho_0 c^5} \right)^{1/3}
\begin{bmatrix}
1 & u^1/c_0 & u^2/c_0 & u^3/c_0\\
u^1/c_0 & (u^1/c_0)^2 - (c/c_0)^2 & u^1 u^2/c_0^2 & u^1 u^3/c_0^2\\
u^2/c_0 & u^1 u^2/c_0^2 & (u^2/c_0)^2 - (c/c_0)^2 & u^2 u^3/c_0^2\\
u^3/c_0 & u^1 u^3/c_0^2 & u^2 u^3/c_0^2 & (u^3/c_0)^2 - (c/c_0)^2
\end{bmatrix}.
\end{equation}
In flow acoustics $\rho$ and $c$ denote density and speed of sound in a non-uniform unsteady background fluid, respectively. Since these are not constant, they can in general differ from the reference values $\rho_0$ and $c_0$, which are usually taken at infinity (and often denoted accordingly as $\rho_\infty$ and $c_\infty$). In addition to this, the vector $\vec u = [u^1, u^2, u^3]$ is velocity vector of the moving medium. The determinant of this matrix is $g = -(\rho c_0^{1/2})^{4/3}/(\rho_0 c^{1/2})^{4/3}$, which is negative (as required by the square root in eq.~(\ref{eq:d'Alembertian})), but generally different from -1, which implies a curved spacetime. The spacetime becomes flat when $\rho = \rho_0$ and $c = c_0$, even if $\vec u \neq 0$ (if $\vec u = 0$, then $g^{\alpha\beta} = \eta^{\alpha\beta}$). Using again eq.~(\ref{eq:d'Alembertian}) as well as the mass conservation low, $\partial \rho/\partial t + \nabla \cdot (\rho \vec u) = 0$, one can show that the metric in eq.~(\ref{eq:metric_tensor_Pierce}) gives the wave operator of the Pierce equation~\cite{Pierce1990}:
\begin{equation}\label{eq:d'Alembertian_Pierce}
\Box \phi = \frac{D}{Dt} \left( \frac{1}{c^2} \frac{D \phi}{D t} \right) - \frac{1}{\rho} \nabla \cdot (\rho \nabla \phi),
\end{equation}
where $D/Dt = \partial /\partial t + \vec u \cdot \nabla$, multiplied by the irrelevant factor $-(\rho c)^{1/3}/(\rho_0 c_0)^{1/3}$. This operator describes sound propagation in irrotational low Mach number inhomogeneous unsteady flows, with the characteristic length and time scales of the inhomogeneities much larger than those of the acoustic perturbation (that is, the wave equation is accurate up to the first order in the derivatives of the background flow quantities). This result suggests that the effects of a non-uniform background flow on sound propagation, such as refraction, can be captured by a curved acoustic spacetime metric~\cite{Gregory2015b}. In analogue gravity a metric very similar to the one in eq.~(\ref{eq:metric_tensor_Pierce}) is demonstrated to describe horizon of a black hole~\cite{Unruh2014} and even more general metrics for background flows which are not irrotational are discussed in Ref.~\cite{Visser2004}.

Different metrics of the flat or curved acoustic spacetime expressed in suitable reference frames are obviously capable of capturing effects of uniform or non-uniform background flows on sound propagation, such as convection and refraction. This has been noted by many authors who used the analogy for the treatment of both flow acoustic and relativistic problems. However, the use has been restricted to the mentioned wave propagation effects (for example, calculation of sound propagation paths), due to apparently different physics (governing equations) of acoustics and general relativity, nature of the two types of waves (compressible longitudinal sound waves described with a scalar and transverse gravitational waves described with a second-order tensor), and mechanisms of their generation (the role of mass and dynamics in the two theories). Nevertheless, it seems very unlikely that a possible weak perturbation of the background acoustic spacetime cannot be associated by any means with propagating sound waves, analogously to gravitational waves in the relativistic spacetime, at least under certain special conditions. If so, the questions of interest are: which special conditions must be satisfied, do the same governing equations apply, and how can the two different types of waves be explained?

In the following we argue that most of the apparent differences can be circumvented if scalar sound waves are described primarily with kinematic quantities, particularly the velocity potential, and the sources of sound are purely kinematic aeroacoustic sources in unbounded fluids, such as in incompressible low Mach number turbulent flows. Thereby we extend the analogy to include both nature of waves in the background spacetime and their generation, by using the linearized Einstein field equations as the governing equations for acoustic problems. It turns out that, unlike gravitational waves, sound waves do not have a geometrically invariant meaning. An acoustically relevant result is given by a single component of the weak metric perturbation in a specific (Newtonian) frame of reference. This is, however, dictated solely by the acoustic observer existing in the Newtonian space and time, which are external to the Lorentz-invariant acoustic spacetime, and not by different nature of the spacetime perturbation. For simplicity, we will assume a weakly perturbed, essentially flat background spacetime with the Minkowski metric tensor from eq.~(\ref{eq:metric_tensor_eta}). This corresponds to a quiescent fluid with constant values of $\rho_0$ and $c_0$. After providing the basic description of sound waves in the relativistic framework in the rest of this section, the mechanisms of their generation will be considered in section~\ref{ch:sound_wave_generation_in_spacetime}.

\subsection{Sound waves}\label{ch:sound_wave_in_spacetime}

Next we show how longitudinal sound waves in fluids can be related to a weak metric perturbation of the background acoustic spacetime. The perturbation is expressed using the linearized theory of general relativity, which is otherwise used for description of transverse gravitational waves propagating with the speed of light. It should be immediately pointed out that this by no means implies existence of longitudinal gravitational waves. As already indicated, longitudinality of sound waves is an artefact of the acoustic observer existing in an external ``spacetime'' and violating the Lorentz invariance. Thus, a preferred frame exists, which will turn out to be the Newtonian frame, and it is only in this particular gauge that a single component of the metric perturbation takes the role of a classical acoustic scalar.

Even though both gravitational and sound waves are described by linearized theories, it is instructive to start with the non-linear Einstein field equations, the governing equations of general relativity. They relate curvature of spacetime, expressed by the metric tensor $\boldsymbol g$, with its source, the stress-energy tensor $\boldsymbol T$ (with the unit kg/(m\,s$^2$))\cite{Schutz2017}:
\begin{equation}\label{Einstein_field_equations}
	\boldsymbol G + \Lambda \boldsymbol g = \frac{k G}{c_0^4} \boldsymbol T,
\end{equation}
where $\boldsymbol G$ is Einstein tensor (which depends only on the metric tensor and its derivatives up to the second order) and $\Lambda \approx 1.1 \cdot 10^{-52} \text{\,m}^{-2}$, dimensionless $k$ (not to be confused later with the wave number), and $G \approx 6.67 \cdot 10^{-11}$\,m$^3$/(kg\,s$^2$) are constants. $\Lambda$ takes a very small value and is called the cosmological constant. If we are not interested in steady or, more precisely, very slowly varying, large-scale solutions\footnote{As in acoustics, large time scale implies low frequency and long length scale. This and the place of the cosmological constant in the acoustic analogy will be discussed separately in section~\ref{ch:dark_energy}. Here the cosmological constant can be neglected, as in the theory of gravitational waves. In other words, we assume an effectively unbounded spacetime and the theory does not cover the unsteady solutions for the background metric, that is, the minimum frequency is finite.}, we can adopt the usual $\Lambda = 0$. $G$ is known as the gravitational constant. Equation~(\ref{Einstein_field_equations}) is written in a frame-independent form. Local conservation of mass-energy and momentum provide additional constraints, which in a general curved spacetime and for a specific frame of reference read
\begin{equation}\label{conservation_laws_Einstein_tensor}
	G^{\alpha\beta}{}_{;\beta} = T^{\alpha\beta}{}_{;\beta} = 0.
\end{equation}

Next we suppose that the only weak disturbance of otherwise flat background spacetime is due to the propagating waves. Thereby we imply a simple quiescent background fluid with constant density $\rho_0$ and speed of sound $c_0$. In a suitable frame the metric tensor can be written as a sum of the Minkowski metric $\eta^{\alpha\beta}$ from eq.~(\ref{eq:metric_tensor_eta}) and a weak component $h^{\alpha\beta}$,
\begin{equation}\label{metric_tensor_small_perturbation}
	g^{\alpha\beta} = \eta^{\alpha\beta} + h^{\alpha\beta},
\end{equation}
with $|h^{\alpha\beta}| \ll 1$. Within the first-order approximation of the linearized theory it can be shown~\cite{Schutz2017} that there always exists
\begin{equation}\label{metric_tensor_small_perturbation_Lorenz_gauge}
	\bar h^{\alpha\beta} = h^{\alpha\beta} - \frac{1}{2} \eta^{\alpha\beta} h^\nu{}_\nu,
\end{equation}
such that
\begin{equation}\label{Einstein_tensor_linear_approximation}
G^{\alpha\beta} = -\frac{1}{2} \Box \bar{h}^{\alpha\beta}
\end{equation}
and the condition
\begin{equation}\label{Lorenz_gauge}
\bar{h}^{\alpha\beta}{}_{,\beta} = 0
\end{equation}
holds. The term $h^\nu{}_\nu$ represents the trace of $h^{\alpha\beta}$ and from eq.~(\ref{metric_tensor_small_perturbation_Lorenz_gauge}) it is clear that $|\bar{h}^{\alpha\beta}| \sim |h^{\alpha\beta}| \ll~1$.

Equation~(\ref{Lorenz_gauge}) is the Lorenz gauge condition. It defines a class of frames in which the metric perturbation $\bar h^{\alpha\beta}$ is expressed. If the condition is not satisfied directly by $\bar h^{\alpha\beta}$ from eq.~(\ref{metric_tensor_small_perturbation_Lorenz_gauge}) for $h^{\alpha\beta}$ given in a certain frame, a small change of coordinates (gauging) 
\begin{equation}\label{gauging}
	x^\alpha \rightarrow x^\alpha + \xi^\alpha
\end{equation}
can always be applied, transforming the metric as
\begin{equation}\label{metric_tensor_small_perturbation_gauging}
	h^{\alpha\beta} \rightarrow h^{\alpha\beta} - \xi^{\alpha,\beta} - \xi^{\beta,\alpha},
\end{equation}
such that
\begin{equation}\label{Lorenz_gauge_from_any_gauge_vector}
\Box \xi^\alpha = \xi^{\alpha,\beta}{}_{,\beta} =  \left( h^{\alpha\beta} - \frac{1}{2} \eta^{\alpha\beta} h^\nu{}_\nu \right)_{,\beta} \neq 0
\end{equation}
and therefore
\begin{equation}\label{Lorenz_gauge_from_any_gauge}
	\bar{h}^{\alpha\beta} = h^{\alpha\beta} - \frac{1}{2} \eta^{\alpha\beta} h^\nu{}_\nu - \xi^{\alpha,\beta} - \xi^{\beta,\alpha} + \eta^{\alpha\beta} \xi^\nu{}_{,\nu}
\end{equation}
does satisfy it. Note that
\begin{equation}
	\bar h^\alpha{}_\alpha = \eta_{\alpha\beta} \bar h^{\alpha\beta} = \eta_{\alpha\beta} \left( h^{\alpha\beta} - \frac{1}{2} \eta^{\alpha\beta} h^\nu{}_\nu \right) = h^\alpha{}_\alpha - \frac{1}{2} \eta^\alpha{}_\alpha h^\nu{}_\nu = h^\alpha{}_\alpha - 2 h^\alpha{}_\alpha = -h^\alpha{}_\alpha,
\end{equation}
so $h^{\alpha\beta}$ and $\bar{h}^{\alpha\beta}$ are mutually trace reverse and eq.~(\ref{metric_tensor_small_perturbation_Lorenz_gauge}) can be inverted to
\begin{equation}\label{metric_tensor_small_perturbation_Lorenz_gauge_inverted}
	h^{\alpha\beta} = \bar h^{\alpha\beta} - \frac{1}{2} \eta^{\alpha\beta} \bar h^\nu{}_\nu,
\end{equation}
which will be used later. Inserting eq.~(\ref{Einstein_tensor_linear_approximation}) into eq.~(\ref{Einstein_field_equations}) with $\Lambda = 0$ gives the linearized Einstein field equations:
\begin{equation}\label{Einstein_field_equations_linear}
	\Box \bar{h}^{\alpha\beta} = -\frac{2kG}{c_0^4} T^{\alpha\beta}.
\end{equation}

The stress-energy tensor in the source term will be studied in more detail in section~\ref{ch:sound_wave_generation_in_spacetime}. Here it is of interest to note that for non-relativistic sources involving motion at velocities much smaller than $c_0$ the dominant component is $T^{00} = \rho c_0^2$ (see eq.~(\ref{eq:stress-energy_tensor_perfect_fluid_components})). For such sources the linearized Einstein field equations reduce to
\begin{equation}\label{Einstein_field_equations_linear_non-relativistic_source}
\Box \bar{h}^{00} = -\frac{2kG}{c_0^2} \rho,
\end{equation}
which is a scalar wave equation typical for acoustics in fluids. In the incompressible acoustic near field of the source of mass $\rho$, that is, for $\omega r/c_0 \ll 1$, where $r$ is distance from the source, the d'Alembertian reduces to the Laplacian:
\begin{equation}\label{Einstein_field_equations_linear_non-relativistic_source_near_field}
\nabla^2 \bar{h}^{00} = -\frac{2kG}{c_0^2} \rho.
\end{equation}
On the other hand this is also the equation of Newtonian gravity with the gravitational potential $-c_0^2 \bar h^{00}/4$ (and $k=8\pi$, as will be shown). However, the acoustic theory assumes an unsteady source of mass and the linearized theory discussed here (with $\Lambda = 0$) does not apply to steady quantities. This brings several important indications. First, the second-order tensor theory of linearized general relativity reduces to the scalar acoustic theory for non-relativistic sources. Second, the acoustic near field corresponds to Newtonian gravity\footnote{Or Coulomb's electrostatics, if mass is replaced with charge (see eq.~(\ref{Maxwell_currents_Lorenz_gauge_non_relativistic_near_field})). The analogy with the vector theory of covariant electromagnetism is the subject of section~\ref{ch:analogy_with_EMG}.}, if we let the unsteady mass vary slowly enough with regard to the distance of the observer, with frequency $\omega \ll c_0/r$. However, the Newtonian gravitational potential (or $\bar h^{00}$) can be defined also for the far field, as in eq.~(\ref{Einstein_field_equations_linear_non-relativistic_source}). Third, the Lorentz invariance of the relativistic theory is violated in the near field. Fourth, the mass does have the same role in gravitation and the acoustic analogy in the non-relativistic regime\footnote{In section~\ref{ch:pulsating_sphere}, which deals with monopole sources, we will learn that this is due to the incompressibility of the fluid in a compact region of an acoustic source. Therefore, the analogy does not have to capture compressible fluid dynamics in order to include acoustic sources. In section~\ref{ch:aeroacoustic_sound_generation} we will also see that every non-relativistic source is compact.}. 

In this section we consider only wave propagation and therefore set the source term on the right-hand side of the wave equation to zero. Since the background spacetime is flat and the higher-order terms can be neglected, eq.~(\ref{Einstein_field_equations_linear}) is wave equation with the classical d'Alembert operator from eq.~(\ref{eq:d'Alembertian_quiescent_flow}), that is
\begin{equation}\label{Einstein_field_equations_linear_no_source}
	\Box \bar{h}^{\alpha\beta} = \left( -\frac{1}{c_0^2} \frac{\partial^2}{\partial t^2} + \nabla^2 \right) \bar{h}^{\alpha\beta} = 0.
\end{equation}
The simplest solution is a plane wave, the real part of
\begin{equation}\label{plane_wave}
	\bar{h}^{\alpha\beta} = A^{\alpha\beta} e^{jk_\nu x^\nu},
\end{equation}
where components of the polarization tensor $A^{\alpha\beta}$ are complex constants and the four-vector $k^\alpha$ is null vector in the flat Minkowski spacetime: $k_\alpha k^\alpha = \eta_{\alpha\beta} k^\beta k^\alpha = 0$. For example, if the plane wave propagates in the direction of the $x^3$-axis, $k^\alpha = [\omega/c_0, 0, 0, \omega/c_0]$, $k_\alpha = \eta_{\alpha\beta} k^\beta = [-\omega/c_0, 0, 0, \omega/c_0]$, and we obtain the usual exponent $-j \omega (t - z/c_0)$ after replacing $x^0$ with $c_0 t$ and $x^3$ with the Cartesian coordinate $z$.

The gauge condition in eq.~(\ref{Lorenz_gauge}) provides additional constraints
\begin{equation}\label{plane_wave_Lorenz_gauge_condition}
	k^\beta A_{\alpha\beta} = 0,
\end{equation}
following from the equality $\bar{h}^{\alpha\beta}{}_{,\nu} = jk_\nu \bar{h}^{\alpha\beta} = jk_\nu A^{\alpha\beta} e^{jk_\mu x^\mu}$. Since it involves four equations, the condition reduces the number of unknown components $A^{\alpha\beta}$ from ten (due to the symmetry of the metric tensor) to six. This number can be decreased even further. We note that eq.~(\ref{Lorenz_gauge_from_any_gauge_vector}) suggests that another small change of coordinates, achieved by adding a vector $\zeta^\mu$ to $\xi^\mu$ from eq.~(\ref{gauging}), leaves equations~(\ref{metric_tensor_small_perturbation}), (\ref{Einstein_tensor_linear_approximation}), (\ref{Lorenz_gauge}), and thus eq.~(\ref{Einstein_field_equations_linear}) unchanged if
\begin{equation}\label{gauging_freedom}
\Box \zeta^\mu = 0.
\end{equation}
This gauge invariance provides further freedom in selecting a specific gauge within the class of Lorenz gauges, with additional four constraints and reduction of the number of unknowns.

The standard treatment of plane gravitational waves takes the solution of eq.~(\ref{gauging_freedom})
\begin{equation}\label{gauging_vector}
	\zeta_\alpha = B_\alpha e^{jk_\mu x^\mu},
\end{equation}
in which~\cite{Scott2016}
\begin{equation}\label{gauging_vector_amplitude}
	B_\alpha = \frac{1}{U^\nu k_\nu} \left( -k_\alpha U^\beta B_\beta - jU^\beta A_{\alpha\beta} + j\frac{1}{2} A_\beta{}^\beta U_\alpha \right)
\end{equation}
and
\begin{equation}\label{gauging_vector_amplitude_2}
	U^\beta B_\beta = -\frac{j}{2U^\nu k_\nu} \left( U^\beta U^\alpha A_{\beta\alpha} + \frac{1}{2} A_\mu {}^\mu \right).
\end{equation}
Here $\vec{U}$ denotes a dimensionless (normalized with $c_0$) four-velocity vector. In a particle's momentarily co-moving reference frame it is a constant timelike unit vector (time basis vector), $U^\alpha = \delta^\alpha{}_0$, where $\delta^\alpha{}_\beta$ is Kronecker delta. Therefore, $\vec{U} \cdot \vec{U} = -1$. Even if the background spacetime is not flat, one can always perform a background Lorentz transformation such that $U^\beta = \delta^\beta{}_0$ in the specific frame. Furthermore, the spatial axes of the frame can be pointed such that $k^\alpha = [\omega/c_0, 0, 0, \omega/c_0]$, so the results derived here are completely general. In the special case of a slowly moving (non-relativistic) particle, the three-dimensional velocity satisfies $|\vec v| \ll c_0$ and the spatial components of the four-velocity vector are simply $v^j/c_0$. 

After the additional gauging with $\zeta_\alpha$, equations~(\ref{metric_tensor_small_perturbation_Lorenz_gauge}) and (\ref{Lorenz_gauge_from_any_gauge}) give
\begin{equation}\label{switch_to_TT_gauge}
	\bar{h}^{TT}_{\alpha\beta} = \bar{h}_{\alpha\beta} - \zeta_{\alpha,\beta} - \zeta_{\beta,\alpha} + \eta_{\alpha\beta} \zeta^\nu{}_{,\nu}.
\end{equation}
Components of the metric perturbation are thus given for a specific Lorenz gauge called transverse-traceless (TT) gauge, in which $A^{TT\alpha}{}_\alpha = 0$ (zero trace) and $A^{TT}_{\alpha\beta} U^\beta = A^{TT}_{\alpha\beta} \delta^\beta{}_0 = 0$. From the last equality it follows that $A^{TT}_{\alpha 0} = A^{TT}_{0 \alpha} = 0$ and from eq.~(\ref{plane_wave_Lorenz_gauge_condition}) $k_3 A^{TT\alpha 3} = \omega A^{TT\alpha 3} /c_0 = 0$ for $k^\alpha = [\omega/c_0, 0, 0, \omega/c_0]$, which explains why the gauge is ``transverse''. Hence, in the transverse-traceless gauge the polarisation tensor equals
\begin{equation}\label{eq:wave_amplitude_Lorenz_gauge_TT}
	A^{TT}_{\alpha\beta} =
	\begin{bmatrix}
		0 & 0 & 0 & 0\\
		0 & A^{TT}_{11} & A^{TT}_{12} & 0\\
		0 & A^{TT}_{12} & -A^{TT}_{11} & 0\\
		0 & 0 & 0 & 0
	\end{bmatrix}
\end{equation}
and the plane transverse gravitational waves are determined by two values, $A^{TT}_{11}$ and $A^{TT}_{12}$.

Importantly, the gauging in equations (\ref{Lorenz_gauge_from_any_gauge}) and (\ref{switch_to_TT_gauge}) does not change the order of magnitude of the metric perturbation:
\begin{equation}\label{gauging_order_of_magnitude}
	|h_{\alpha\beta}| \sim |\bar{h}_{\alpha\beta}| \sim |\bar{h}^{TT}_{\alpha\beta}| \ll 1.
\end{equation}
It is the order of magnitude of the initial weak perturbation from eq.~(\ref{metric_tensor_small_perturbation}) and the derived solution in the transverse-traceless gauge is thus naturally associated with gravitational waves in the background spacetime. The waves are transverse, with two polarizations represented by the two components $A^{TT}_{11}$ and $A^{TT}_{12}$. The components are analogous to the electric and magnetic components of transverse electromagnetic waves and in contrast to a single component which determines longitudinal sound waves.

The next question of interest is how a metric perturbation determines kinematics of particles in a weakly perturbed spacetime. Each free particle in a curved spacetime obeys the geodesic equation
\begin{equation}\label{geodesic_equation}
	\frac{1}{c_0} \frac{d \vec U}{d \tau} = 0,
\end{equation}
where $\tau$ is proper time satisfying the equality $c_0^2 (d \tau)^2 = - (d s)^2$ with $d s$ the interval between two infinitesimally close events in spacetime. For example, in a flat spacetime $(ds)^2 = -c_0^2 (dt)^2 + (dx)^2 + (dy)^2 + (dz)^2$. The geodesic equation states that a free particle follows its world line. In a curved spacetime it can also be written as~\cite{Schutz2017}
\begin{equation}\label{geodesic_equation_curved_spacetime}
	U^\alpha {}_{;\beta} U^\beta = \frac{1}{c_0}\frac{d U^\alpha}{d\tau} + \Gamma^\alpha {}_{\mu \nu} U^\mu U^\nu = 0.
\end{equation}
The Christoffel symbols $\Gamma^\alpha {}_{\mu \nu}$ depend on the metric tensor and its first derivatives:
\begin{equation}\label{Christoffel_symbols_metric}
	\Gamma^\alpha {}_{\mu \nu} = \frac{1}{2} g^{\alpha\beta} (g_{\beta\mu,\nu} + g_{\beta\nu,\mu} - g_{\mu\nu,\beta}).
\end{equation}
They obviously vanish in a flat spacetime with constant metric and covariant derivative reduces to an ordinary derivative, as already mentioned in the context of eq.~(\ref{eq:d'Alembertian}).

If the particle is moving slowly (with the speed much smaller than $c_0$) in an essentially flat spacetime of the linearized theory, its four-acceleration (divided with $c_0$) due to a weak metric perturbation $h_{\alpha\beta}$ is within the first-order approximation of eq.~(\ref{geodesic_equation_curved_spacetime})
\begin{equation}\label{particle_acceleration}
	\frac{dU^\alpha}{d\tau} = -c_0 \Gamma^\alpha {}_{00} = -\frac{c_0}{2} \eta^{\alpha\beta} (h_{\beta 0, 0} + h_{0 \beta, 0} - h_{0 0, \beta}),
\end{equation}
with $\vec U = d\vec x/d(c_0 \tau)$. In the analogue linearized acoustic theory fluid particles also move non-relativistically, since the particle velocity due to a sound wave is always much smaller than the speed of sound. For such a particle, $\tau \approx t$ and the three-dimensional acceleration equals
\begin{equation}\label{particle_acceleration_spatial}
	\frac{d^2 x^k}{d t^2} = -\frac{c_0^2}{2} \eta^{kl} (h_{l 0, 0} + h_{0 l, 0} - h_{0 0, l}).
\end{equation}

Particle motion due to a compressible acoustic wave in fluid is irrotational and parallel to the propagation path of the wave. Particle velocity vector associated with acoustic waves can be expressed as gradient of the acoustic scalar potential\footnote{In contrast to this, for example, the vectors of electric and magnetic fields of transverse electromagnetic waves are divergence free and characterized by a vector potential $\vec A$: $\nabla \times \vec A$ (see section~\ref{ch:analogy_with_EMG}).} $\phi$: $\vec{v} = \nabla \phi$. However, the components $h_{l0} = h_{0l}$ may imply transverse motion. Indeed, recalling eq.~(\ref{Einstein_field_equations_linear_non-relativistic_source}), we expect sound waves to be properly described only by the component $h_{00}$ corresponding to the Newtonian gravitational potential. In fact, in the specific Newtonian gauge\footnote{In general relativity the Newtonian gauge is suitable for calculations of relativistic corrections of the Newtonian gravitational potential, which are due to the radiation of gravitational waves (see section~\ref{ch:aeroacoustic_sound_generation}).}, when\footnote{This is not a necessary condition, though. Even if the component $h_{00}$ is not dominant, the gauge can be Newtonian, as long as transverse motion of the particle is negligible. See, for example, equations~(\ref{Maxwell_currents_Lorenz_gauge_split_solution_far_geometric_field_dipole_moment_radiation_j_Taylor_expansion}) and (\ref{Maxwell_currents_Lorenz_gauge_split_solution_far_geometric_field_dipole_moment_radiation_0_Taylor_expansion}) in the electromagnetic analogy. In the relativistic analogy here, the condition will be satisfied in the derivations which follow.} $|h_{l0}| = |h_{0l}| \ll |h_{00}|$,
\begin{equation}\label{particle_acceleration_spatial_Newtonian_gauge}
	\frac{d^2 x^k}{dt^2} = \frac{c_0^2}{2} h_{00}^{,k}.
\end{equation}
Particle motion due to the incoming wave is expressed with a single scalar $h_{00}$, as we expect from the longitudinal acoustic waves in fluids\footnote{Transverse sound waves occur in solids. If they dominate over the longitudinal waves, the components $ h_{0l}$ and $ h_{l0}$ or even the transverse-traceless gauge may still be relevant for acoustics in solids, but this will not be investigated further here.}. In this way we can obtain a measurable acoustic quantity (particle acceleration) from the relativistic analogy. It should be emphasized that the relation holds also in the acoustic far field of the source. The theory is Newtonian not because the near field is assumed, but because the motion is described with a scalar velocity potential and the component $h_{00}$. The Lorentz invariance is eventually violated and the specific gauge (preferred frame) has to be chosen in order to express the acoustic quantities which are relevant for the observer and not because of the nature of perturbation (waves) or sources. As will be discussed later, acoustic observers violate the Lorentz invariance because their time is not dictated by the acoustic spacetime, but a decoupled dimension in the (external) Newtonian absolute space and time, in which they (as well as solid boundaries; see section~\ref{ch:energy_and_charge}) exist. However, this does not preclude exploiting the analogy for calculations of metric perturbation, while leaving the choice of an appropriate frame for the final step.

From eq.~(\ref{particle_acceleration_spatial_Newtonian_gauge}) we can determine other acoustic quantities. Since particle velocity and potential $\phi$ are related by the equality
\begin{equation}\label{acoustic_velocity_potential}
v^k = \frac{d x^k}{dt}= \phi^{,k},
\end{equation}
we can associate $h_{00}$ with $\phi$ (similarly as in Newtonian gravitation):
\begin{equation}\label{acoustic_potential_h00}
\frac{d \phi}{dt} = \frac{c_0^2}{2} h_{00}.
\end{equation}
In a quiescent fluid acoustic pressure and potential are related by the equality
\begin{equation}\label{acoustic_pressure_potential}
p = -\rho_0 \frac{d \phi}{dt} = - \frac{\rho_0 c_0^2}{2} h_{00}.
\end{equation}
Finally, for a plane longitudinal sound wave the only non-zero component of particle velocity is in the direction of wave propagation (for example, the $x^3$-axis) and it equals
\begin{equation}\label{acoustic_velocity_pressure}
v^3 = \frac{p}{\rho_0 c_0} = - \frac{c_0}{2} h_{00}.
\end{equation}
This also validates the supposed non-relativistic particle motion ($|v^3| \ll c_0$), since $|h_{00}| \ll 1$.

A note should be made on the terms in the last two equations which involve the mean fluid density $\rho_0$. While $c_0$ in the numerator of the last term in eq.~(\ref{acoustic_velocity_pressure}) represents the theoretical maximum particle velocity, $\rho_0 c_0^2$ in eq.~(\ref{acoustic_pressure_potential}) does not entirely represent the maximum (static) pressure $p_0$. The latter is for a perfect gas $p_0 = \rho_0 c_0^2/\gamma$, where $\gamma$ is the heat capacity ratio ($\gamma \approx 1.4$ for air under usual conditions). This suggests that we should transform
\begin{equation}\label{density_correction}
\rho_0 \rightarrow \frac{\rho_0}{\gamma}
\end{equation}
when we switch from general relativity (and other analogue theories below, in which $\gamma = 1$) to acoustics, for example, when we calculate sound pressure from the relativistic results. In order to avoid confusion or doubling the expressions, we will ignore this most of the time and keep $\gamma = 1$, especially since we normally switch to the dynamic acoustic quantities only at the end of the calculation, if at all. Thermodynamics of the background fluid, which causes this difference, will be discussed in section~\ref{ch:background_fluid_thermodynamics}. As already stated, the analogy with general relativity discussed here does not cover the background fluid properties.

Based on the analysis above we can conclude that sound waves do correspond to a small curvature of the background acoustic spacetime,  a purely geometric quantity (the metric perturbation tensor) suffices for their description, and no further dynamic analogy with general relativity is required. The linearized theory is common for both types of waves. However, sound waves in fluids are naturally described only by the metric perturbation component $h_{00}$ in the Newtonian gauge, and not in a Lorenz gauge which is appropriate for gravitational waves. Such violation of the Lorentz invariance and the existence of a preferred frame are dictated by the acoustic observer, whose clocks are ticking in Newtonian time, not the acoustic spacetime. A solution of the governing linearized Einstein field equations (the wave equation with $\bar h^{\alpha\beta}$ as the unknown) in a Lorenz (or any other) gauge can still be calculated first, for example, to utilize the existing results of general relativity, but a transformation to the Newtonian (rather than transverse-traceless) gauge must be performed eventually, in order to obtain classical acoustic quantities.

Lastly, it is important to note that the value of $h_{00}$ in the Newtonian gauge does not have to be of the same order of magnitude as $\bar{h}^{TT}_{\alpha\beta}$ (or $\bar{h}_{\alpha\beta}$ or the initial perturbation $h_{\alpha\beta}$ from eq.~(\ref{gauging_order_of_magnitude})). In fact, the following analysis of compact sources of waves will show that the acoustic perturbation is only a very weak, higher-order component of $\bar{h}_{\alpha\beta}$. As such, it does not satisfy the Lorenz gauge condition and vanishes in the transverse-traceless gauge, which captures only the transverse waves of the leading order. For further insights we need to investigate how the component $h_{00}$ is determined by different sources of sound, namely point monopole, dipole, and quadrupole. With the (apparent) exception of the discussion on the cosmological constant in section~\ref{ch:dark_energy}, the study of non-compact sources (for which the longitudinal component may become comparable to the transverse component) is outside the scope of this work.

\section{Generation of sound waves in acoustic spacetime}\label{ch:sound_wave_generation_in_spacetime}

In the previous section we saw how a metric perturbation of the acoustic spacetime carries the information on a propagating sound wave, regardless of the actual source of perturbation. In this section we treat the three types of sources of sound in free space -- monopole, dipole, and quadrupole -- using the same relativistic formalism and the linearized Einstein field equations as the governing equations. It turns out that, as in linearized general relativity, it is the unsteady or moving mass-energy\footnote{We postpone the discussion on the steady mass of the background fluid  until section~\ref{ch:mass_background_fluid}, because its density $\rho_0$ does not follow from the relativistic but thermodynamic analogy. However, the source of expansion of the (bounded) universe, which is associated with the cosmological constant, will be treated in section~\ref{ch:dark_energy} as a low-frequency, but unsteady source. While section~\ref{ch:sound_wave_generation_in_spacetime} deals with compact sources of sound, the source at a cosmological scale (or scale of the bounded acoustic spacetime) is obviously non-compact and deserves a separate treatment.} in the source region which curves the acoustic spacetime around it and generates fields.

Both mass-energy and momentum must be conserved. While the dynamics of general relativity (in particular the conservation equations~(\ref{conservation_laws_Einstein_tensor})) and fluid dynamics (the Navier-Stokes equations) are generally different, an important regime in which the analogy remains kinematic and valid is for small (non-relativistic) velocities (recall also eq.~(\ref{Einstein_field_equations_linear_non-relativistic_source}) and the discussion which followed). In (aero)acoustics this means that the fluid particles move with velocities much below the speed of sound, not only in the sound propagation region (as in section~\ref{ch:sound_wave_in_spacetime}), but also in the source region. This is the case with essentially incompressible low Mach number turbulent flows. In general relativity this condition is satisfied by matter moving much slower than the speed of light, when mass constitutes the dominant part of mass-energy. Hence, in this section we extend the acoustic analogy to the sources of sound, at least in low Mach number flows.

Type of a source of waves which is undoubtedly shared by general relativity and acoustics in fluids is quadrupole and therefore it will be considered first\footnote{This is rather unusual for acoustic literature, but physically well justified if unbounded fluids without external sources of sound are studied first. It is also interesting to notice that the aeroacoustic and relativistic theories of quadrupole wave radiation occurred in the mid-twentieth century and developed almost simultaneously, independent of each other.}. While typical relevant quadrupole sources of gravitational waves are binary stars and black holes, an aeroacoustic quadrupole is associated with pure instability of the unbounded fluid (the medium) in the form of vorticity or turbulence. The analogy between a co-rotating vortex pair and a black hole binary, being closest to a pure instability of the relativistic medium (the vacuum itself), will be examined in section~\ref{ch:acoustic_binary}. Before that we will derive Lighthill's general scaling low for power of an isentropic incompressible aeroacoustic quadrupole, starting with the linearized Einstein field equations with a non-relativistic source. This should already indicate equal nature of the quadrupole sources. For completeness, even though irrelevant for the theory of gravitational waves, we will show in sections~\ref{ch:pulsating_sphere} and \ref{ch:oscillating_sphere} that acoustic monopole and dipole sources can also be described (though inefficiently) by the same equations, as long as the fluid in the source region is essentially incompressible.

\subsection{Aeroacoustic quadrupole}\label{ch:aeroacoustic_sound_generation}

In order to study the mechanisms of wave generation, we refer back to the linear wave equation~(\ref{Einstein_field_equations_linear}) with the source term. Unsteady fluid in the source region has to satisfy the conservation laws, eq.~(\ref{conservation_laws_Einstein_tensor}):
\begin{equation}\label{conservation_laws}
	T^{\alpha\beta}{}_{;\beta} = 0.
\end{equation}
In general relativity the frame-independent stress-energy tensor of a perfect fluid equals
\begin{equation}\label{stress-energy_tensor_perfect_fluid}
	\boldsymbol T = (\rho c_0^2+p) \vec{U} \otimes \vec{U} + p \boldsymbol g^{-1},
\end{equation}
with $\rho$ denoting mass-energy density, $p$ pressure, $\vec{U}$ normalized four-velocity of the fluid particles ($\vec{U} \otimes \vec{U} = U^\alpha U^\beta$ in a given frame), and $\boldsymbol g$ metric tensor. If all particles move with non-relativistic three-dimensional velocity, $|\vec v| \ll c_0$, the approximation $\vec{U} \approx (1,\vec v/c_0) = [1,v^1/c_0, v^2/c_0, v^3/c_0]$ holds and the components of the stress-energy tensor in nearly flat spacetime (eq.~(\ref{eq:metric_tensor_eta})) are given by
\begin{equation}\label{eq:stress-energy_tensor_perfect_fluid_components}
	T^{\alpha\beta} = 
	\begin{bmatrix}
		\rho c_0^2 & (\rho + \frac{p}{c_0^2}) c_0 v^1 & (\rho+\frac{p}{c_0^2}) c_0 v^2 & (\rho + \frac{p}{c_0^2}) c_0 v^3\\
		 (\rho + \frac{p}{c_0^2}) c_0 v^1 & (\rho+\frac{p}{c_0^2})v^1 v^1 + p & (\rho+\frac{p}{c_0^2})v^1 v^2 & (\rho+\frac{p}{c_0^2})v^1 v^3\\
		 (\rho + \frac{p}{c_0^2}) c_0 v^2 & (\rho+\frac{p}{c_0^2})v^1 v^2 & (\rho+\frac{p}{c_0^2})v^2 v^2 + p & (\rho+\frac{p}{c_0^2})v^2 v^3\\
		 (\rho + \frac{p}{c_0^2}) c_0 v^3 & (\rho+\frac{p}{c_0^2})v^1 v^3 & (\rho+\frac{p}{c_0^2})v^2 v^3 & (\rho+\frac{p}{c_0^2})v^3 v^3 + p
	\end{bmatrix}.
\end{equation}
Like metric tensor, it is a symmetric second-order tensor.

We shall already notice the similarity between the spatial part of the stress-energy tensor, $T^{jk}$, and Lighthill's tensor\cite{Lighthill1952}, which is a pure aeroacoustic source of sound in unbounded fluids, as well as the monopole mass source, $\rho$, from eq.~(\ref{Einstein_field_equations_linear_non-relativistic_source}) in $T^{00}$, and the dipole momentum source, $\rho \vec v$, in the components $T^{0i} = T^{i0}$. Apart from a different constant $c_0$ (the speed of sound instead of the speed of light), the main differences are that in classical fluid dynamics $\rho$ denotes usual matter rather than energy-mass density and the appearance of pressure in the sum $\rho+p/c_0^2$ in the momentum and stress terms. However, when\footnote{\label{speed_of_sound_and_light}Relativistic fluids are not considered here. For example, the early universe consisted likely of massless particles. It is interesting to note that such a fluid still posses compressibility and the speed of sound can be defined~\cite{Partovi1994}. Moreover, the maximum speed of compressible waves is of the same order of magnitude as the speed of light, smaller only by the factor $\sqrt{3}$. This allows a more direct use of the analogy in the context of relativistic acoustics. Quantum and thermodynamic origins of mass and the background density $\rho_0$ in acoustic spacetime will be discussed in section~\ref{ch:acoustic_Lagrangians}.} $|\vec v| \ll c_0$, ordinary mass has a dominant contribution in the total mass-energy and $p/c_0^2 \ll \rho$~\cite{Schutz2017}. In aeroacoustics and fluid dynamics the same condition is satisfied in incompressible ($\rho = \rho_0$) subsonic flows with low Mach number values. Therefore, the non-relativistic stress-energy tensor reduces to
\begin{equation}\label{eq:stress-energy_tensor_perfect_fluid_components_non-relativistic}
T^{\alpha\beta} = 
\begin{bmatrix}
\rho_0 c_0^2 & \rho_0 c_0 v^1 & \rho_0 c_0 v^2 & \rho_0 c_0 v^3\\
\rho_0 c_0 v^1 & \rho_0 v^1 v^1 + p & \rho_0 v^1 v^2 & \rho_0 v^1 v^3\\
\rho_0 c_0 v^2 & \rho_0 v^1 v^2 & \rho_0 v^2 v^2 + p & \rho_0 v^2 v^3\\
\rho_0 c_0 v^3 & \rho_0 v^1 v^3 & \rho_0 v^2 v^3 & \rho_0 v^3 v^3 + p
\end{bmatrix}.
\end{equation}

Another difference between the spatial part of the stress-energy tensor and Lighthill's tensor is that the latter one follows from the conservation of mass and momentum after the weak acoustic parts (involving sound pressure or density) are shifted to the left-hand side of the wave equation to represent the radiated sound waves. For example, after subtracting divergence of the inviscid momentum equation from the time derivative of the mass equation and rearranging the terms:
\begin{equation}\label{Lighthill_analogy}
\begin{aligned}
&\frac{\partial^2 \rho}{\partial t^2} + \frac{\partial}{\partial t} \nabla \cdot (\rho \vec v) - \nabla \cdot \frac{\partial}{\partial t} (\rho \vec v) - \nabla \cdot \nabla \cdot (\rho \vec v \vec v) - \nabla \cdot \nabla p = 0\\
\Rightarrow& \frac{1}{c_0^2} \frac{\partial^2 \rho}{\partial t^2} - \nabla^2 \rho = \frac{1}{c_0^2} \nabla \cdot \nabla \cdot \left( \rho \vec v \vec v + (p - \rho c_0^2) \boldsymbol I \right),
\end{aligned}
\end{equation}
with the identity tensor $\boldsymbol I$. Therefore, even in its compressible form Lighthill's tensor as the source does not contain entirely conserved components. In contrast to this, the full stress-energy tensor, which we consider here as the source, satisfies the conservation laws in eq.~(\ref{conservation_laws}) in slightly perturbed spacetime. The entire stress-energy tensor is the source of (kinematic) perturbation of the spacetime itself, captured by the left-hand side of the wave equation~(\ref{Einstein_field_equations_linear}), so it need not be split into the source and propagation parts. This is arguably a more natural way of describing a pure instability of the background fluid. The difference between the two approaches becomes clearer if we recall that the single acoustically relevant component of the solution of eq.~(\ref{Einstein_field_equations_linear}) expressed in the Newtonian gauge is of sub-leading order and does not satisfy the Lorenz gauge condition, eq.~(\ref{Lorenz_gauge}). Rather than making the source non-conserved, it is only the component of the solution in a preferred frame of reference which is not conserved, but which happens to describe the longitudinal sound waves.

Finally, we should point out that the stress-energy tensor is inviscid. This will turn out to be one of the key differences in comparison to normally dissipated or damped sound fields in fluids. In section~\ref{source_of_entropy} it will even be argued that the lack of energy losses is the reason for the occurrence of relativistic time coupled with space, as well as for expansion of the universe. Time and boundedness of space will thus be interpreted as a counterpart of damping.

In the following treatment of aeroacoustic sound generation we follow Misner et al.~\cite{Misner2017} and consider a single isolated source of waves, far from which the spacetime is asymptotically flat towards the infinity, that is, the background metric converges to the one from eq.~(\ref{eq:metric_tensor_eta}). The weak metric perturbation of interest is defined with eq.~(\ref{metric_tensor_small_perturbation}) everywhere, including the source region. Linearized\footnote{It is interesting to note that even if the source is relativistic, that is, with high subsonic velocities in the source region, we can still define $\bar{h}_{\alpha\beta}$ in the same way and the derivation below holds. However, an appropriate stress-energy tensor of the compressible flow should be used, which will not be attempted here.} eq.~(\ref{Einstein_field_equations_linear}) holds under the condition in eq.~(\ref{Lorenz_gauge}). In addition to this, we expect that only a small fraction of the stress-energy tensor is responsible for the radiation of waves. In order to emphasize this, we formally split it into the dominant ``effective'' stress-energy tensor, $T^{\text{eff}}_{\alpha\beta}$, and the small component $t_{\alpha\beta}$: $T_{\alpha\beta} = T^{\text{eff}}_{\alpha\beta} + t_{\alpha\beta}$. This gives
\begin{equation}\label{Einstein_field_equations_linear_split}
	\Box \bar{h}_{\alpha\beta} = -\frac{2kG}{c_0^4} (T^{\text{eff}}_{\alpha\beta} + t_{\alpha\beta}).
\end{equation}
Still, no part of the stress-energy tensor, including the radiation-related part $t_{\alpha\beta}$, is shifted to the left-hand side of eq.~(\ref{Einstein_field_equations_linear_split}). With the usual free-space Green's function $\delta(t-\epsilon R/c_0)/(4\pi R)$ and $\delta$ denoting the Dirac delta function, a general solution for both ingoing ($\epsilon = -1$) and outgoing ($\epsilon = +1$) wave in the spacetime with signature $[-+++]$ is
\begin{equation}\label{Einstein_field_equations_linear_split_solution}
	\bar{h}_{\alpha\beta} = \frac{kG}{2\pi c_0^4} \int \frac{[T^{\text{eff}}_{\alpha\beta} + t_{\alpha\beta}]_{(t-\epsilon R/c_0)}}{R} d^3 \vec y,
\end{equation}
where the integral is over the entire three-dimensional space and $R = |x^i-y^i|$ is distance from the source. The values of $T^{\text{eff}}_{\alpha\beta}$ and $t_{\alpha\beta}$ are to be evaluated at the retarded time $t-\epsilon R/c_0$.

Next, we assume that the source is compact, so that its characteristic length scale satisfies the inequality $L \ll c_0/ \omega$, where $\omega$ is characteristic angular frequency of the oscillations. This is actually already satisfied by the non-relativistic motion, $|\vec v| \ll c_0$, since $|\vec v| \sim L \omega$. For this reason, the inequality is also referred to as slow-motion condition, which is fulfilled by non-relativistic sources. As a consequence, non-relativistic sources are necessarily compact and vice versa. We also consider far geometric field ($R \gg L$) and accordingly approximate
\begin{equation}\label{Einstein_field_equations_linear_split_solution_far_geometric_field}
	\bar{h}_{\alpha\beta} = \frac{kG}{2\pi r c_0^4} \int [T^{\text{eff}}_{\alpha\beta} + t_{\alpha\beta}]_{(t-\epsilon r/c_0)} d^3 \vec y,
\end{equation}
where $r$ is radial coordinate of the spherical coordinate system with the source in its origin.

From the conservation laws, eq.~(\ref{conservation_laws}), one can deduce the identity~\cite{Misner2017}
\begin{equation}\label{quadrupole_moment_tensor_identity}
	\frac{1}{c_0^2} \frac{d^2}{dt^2}\int (T^{\text{eff}}_{00} + t_{00}) x_j x_k d^3 \vec x = 2 \int (T^{\text{eff}}_{jk} + t_{jk}) d^3 \vec x.
\end{equation}
It relates spatial components of the stress-energy tensor, which are responsible for the quadrupole radiation, with the acoustically relevant component $T_{00}$. (The product with the spatial coordinates $x_j$ and $x_k$ gives the quadrupole radiation pattern.) Exactly due to this important relation stemming from the fact that the full stress-energy tensor satisfies the conservation laws, we can avoid dividing the source terms between the left-hand and right-hand side of the wave equation as in Lighthill's analogy. Its integral form is suitable for spatially limited source regions. The second-order time derivative on the left-hand side of eq.~(\ref{quadrupole_moment_tensor_identity}) takes over the role of the second-order derivatives of the source terms in Lighthill's analogy (such as the double divergence in eq.~(\ref{Lighthill_analogy})). These naturally appear when a second-order tensor reduces to a scalar and the reduction makes the longitudinal waves much weaker than the transverse waves generated by the same source, as we are about to see.

The integral on the left-hand side of eq.~(\ref{quadrupole_moment_tensor_identity}) together with the factor $1/c_0^2$ represents the second moment of the mass distribution and it is called quadrupole moment tensor of the mass distribution. It is commonly denoted with $I_{jk}$, which is, thus, by definition
\begin{equation}\label{quadrupole_moment_tensor}
	I_{jk} = \frac{1}{c_0^2} \int (T^{\text{eff}}_{00} + t_{00}) x_j x_k d^3 \vec x.
\end{equation}
The quantity which appears to be more convenient for mathematical description of gravitational wave generation is reduced quadrupole moment defined as
\begin{equation}\label{reduced_quadrupole_moment_tensor}
	\hcancel{I}_{jk} = I_{jk} - \frac{1}{3} \delta_{jk} I_l {}^l.
\end{equation}
From equations~(\ref{Einstein_field_equations_linear_split_solution_far_geometric_field})-(\ref{quadrupole_moment_tensor}):
\begin{equation}\label{Einstein_field_equations_linear_split_solution_far_geometric_field_quadrupole_moment_tensor_jk}
	\bar{h}_{jk} = \frac{kG}{4\pi r c_0^4} \frac{d^2}{dt^2} I_{jk}(t-\epsilon r/c_0).
\end{equation}

We now limit ourselves to the derivation of a scaling law for power of an aeroacoustic quadrupole. Classical acoustic approach excludes the irrelevant near-field effects by considering the far-field energy. However, this is not strictly necessary. It suffices to calculate reaction of the source to the far-field radiation, which can actually be performed in the acoustic near field (as a contribution of the far-field component in the near field). This approach as well quantifies the energy lost by the source due to the radiation of waves. Expanding $\bar{h}_{jk} (t-\epsilon r/c_0)$ in powers of $r$ for $\omega r/c_0 \ll 1$ (in the near field) gives
\begin{equation}\label{Einstein_field_equations_linear_split_solution_far_geometric_field_quadrupole_moment_tensor_jk_Taylor_expansion}
	\bar{h}_{jk} (t-\epsilon r/c_0) = \bar{h}_{jk} (t) - \frac{\epsilon r/c_0}{1!} \frac{d}{dt} \bar{h}_{jk} (t) + \frac{(\epsilon r/c_0)^2}{2!} \frac{d^2}{dt^2} \bar{h}_{jk} (t) - \frac{(\epsilon r/c_0)^3}{3!} \frac{d^3}{dt^3}  \bar{h}_{jk} (t) + ...
\end{equation}
where $...$ replaces the omitted higher-order terms. The terms involving even powers of $\epsilon r/c_0$ do not represent wave radiation, since they are invariant to the sign of $\epsilon$. In gravitation they add higher-order corrections to the leading-order Newtonian potential from eq.~(\ref{Einstein_field_equations_linear_non-relativistic_source_near_field}), causing the near-field effects such as perihelion shift. In the acoustic analogy, they contribute to the incompressible fluctuations which do not propagate into far field, but which are not already captured by the acoustic near field satisfying the scalar Laplace equation.

Leaving only the terms which correspond to the wave radiation with $\epsilon = 1$ for an outgoing wave,
\begin{equation}\label{Einstein_field_equations_linear_split_solution_far_geometric_field_quadrupole_moment_tensor_radiation_jk}
	\bar{h}^{\text{react}}_{jk} = -\frac{kG}{4\pi c_0^5} \frac{d^3}{dt^3} I_{jk}(t) -\frac{kG}{24\pi c_0^7} r^2 \frac{d^5}{dt^5} I_{jk}(t) + ...
\end{equation}
Using eq. (\ref{Lorenz_gauge}), from which
\begin{equation}
	\bar{h}_{0j,0} = \bar{h}_{j0,0} = -\bar{h}^{j0}{}_{,0} = \bar{h}^{jk}{}_{,k} = \bar{h}_{jk,k}
\end{equation}
and
\begin{equation}
	\bar{h}_{00,0} = \bar{h}^{00}{}_{,0} = -\bar{h}^{0j}{}_{,j} = \bar{h}_{0j,j},
\end{equation}
we can express the remaining components,
\begin{equation}\label{Einstein_field_equations_linear_split_solution_far_geometric_field_quadrupole_moment_tensor_radiation_0j}
	\bar{h}^{\text{react}}_{0j} = -\frac{kG}{12\pi c_0^6} x^k \frac{d^4}{dt^4} I_{jk}(t) -\frac{kG}{120\pi c_0^8} r^2 x^k \frac{d^6}{dt^6} I_{jk}(t) + ...
\end{equation}
and
\begin{equation}\label{Einstein_field_equations_linear_split_solution_far_geometric_field_quadrupole_moment_tensor_radiation_00}
	\bar{h}^{\text{react}}_{00} = -\frac{kG}{12\pi c_0^5} \frac{d^3}{dt^3} I_{jj}(t) -\frac{kG}{120\pi c_0^7} (r^2 \delta^{jk} + 2 x^j x^k) \frac{d^5}{dt^5} I_{jk}(t) + ...
\end{equation}
Components of the tensor $\bar h^\text{react}_{\alpha\beta}$ represent radiation reaction potentials in a Lorenz gauge\footnote{In electromagnetism the forces associated with analogue vector-potentials are often called radiation damping or Lorentz frictional forces. These will be investigated in section~\ref{ch:analogy_with_EMG}.}.

The components $\bar{h}^\text{react}_{0j}$ depend on $x^k$, which indicates transverse motion. However, equations~(\ref{Einstein_field_equations_linear_split_solution_far_geometric_field_quadrupole_moment_tensor_radiation_jk}), (\ref{Einstein_field_equations_linear_split_solution_far_geometric_field_quadrupole_moment_tensor_radiation_0j}), and (\ref{Einstein_field_equations_linear_split_solution_far_geometric_field_quadrupole_moment_tensor_radiation_00}) are written in a Lorenz gauge. In order to obtain a Newtonian form as in eq.~(\ref{particle_acceleration_spatial_Newtonian_gauge}), which is suitable for acoustic radiation, we first switch back to $h_{\alpha \beta} = \bar{h}_{\alpha \beta} - \bar{h}^\nu{}_\nu \eta_{\alpha \beta}/2$ from eq.~(\ref{metric_tensor_small_perturbation_Lorenz_gauge_inverted}) and change the coordinates to $x^\mu + \xi^\mu$, with
\begin{equation}\label{gauging_to_Newtonian_form1}
	\xi_0 = -\frac{kG}{12\pi c_0^4} \frac{d^2}{dt^2} I_{ll}(t) + \frac{kG}{48\pi c_0^6} x^j x^k \frac{d^4}{dt^4} I_{jk}(t) - \frac{kG}{48\pi c_0^6} r^2 \frac{d^4}{dt^4} I_{ll}(t)
\end{equation}
and
\begin{equation}\label{gauging_to_Newtonian_form2}
\xi_j = -\frac{kG}{8\pi c_0^5} x^k \frac{d^3}{dt^3} I_{jk}(t) + \frac{kG}{24\pi c_0^5} x^j \frac{d^3}{dt^3} I_{ll}(t).
\end{equation}
After applying eq.~(\ref{metric_tensor_small_perturbation_gauging}), the transformed component $h^\text{react}_{00}$ reads~\cite{Misner2017}
\begin{equation}\label{Einstein_field_equations_linear_split_solution_far_geometric_field_quadrupole_moment_tensor_radiation_00_gauging_to_Newtonian}
	h^{\text{react}}_{00} = -\frac{kG}{20\pi c_0^7}  x^j x^k \frac{d^5}{dt^5} \hcancel{I}_{jk}(t)
\end{equation}
in the leading order, while the remaining components from eq.~(\ref{particle_acceleration_spatial}) are of higher order: $h^{\text{react}}_{0j} \sim (\omega L/c_0) h^{\text{react}}_{00}$. Therefore, we have expressed the metric component $h_{00}$ in the Newtonian gauge. It was calculated in near field and in gravitation it corrects the Newtonian potential from eq.~(\ref{Einstein_field_equations_linear_non-relativistic_source_near_field}) by taking into account the radiation of gravitational waves. Analogously, it represents radiated longitudinal sound waves, which constitute a very small fraction of eq.~(\ref{Einstein_field_equations_linear_non-relativistic_source}) (see section~\ref{ch:pulsating_sphere}), in addition to the near field in eq.~(\ref{Einstein_field_equations_linear_non-relativistic_source_near_field}). Such longitudinal contributions have no relevance in general relativity, as they are negligible compared to the transverse waves.

Since it depends only on the spatial part of the reduced quadrupole moment tensor, we expect $h^{\text{react}}_{00}$ to capture fluid particle motion due to an incoming longitudinal sound wave generated by Lighthill's quadrupole source in an isentropic incompressible flow, $\rho_0 \vec v \vec v$. The geodesic equation~(\ref{particle_acceleration_spatial_Newtonian_gauge}) gives acceleration of a non-relativistic particle affected by the weak perturbation:
\begin{equation}\label{longitudinal_wave_acceleration}
	\frac{d^2 x^l}{dt^2} = \frac{c_0^2}{2} h^{\text{react},l}_{00} = -\frac{kG}{40\pi c_0^5}  \left( x^j x^k \frac{d^5}{dt^5} \hcancel{I}_{jk}(t) \right)^{,l}.
\end{equation}
In order to derive a scaling law for the source power, we suppose that $T_{jk}$ from eq.~(\ref{eq:stress-energy_tensor_perfect_fluid_components_non-relativistic}) scales as $\rho_0 |\vec v|^2$, as in free jet, where $\rho_0$ is density of the essentially incompressible fluid in the source region. From equations~(\ref{quadrupole_moment_tensor_identity})-(\ref{reduced_quadrupole_moment_tensor}), $|\hcancel{I}_{jk}| \sim \rho_0 L^5$, so the particle velocity from eq.~(\ref{longitudinal_wave_acceleration}) scales as
\begin{equation}\label{quadrupole_ac_velocity_scaling}
	|\vec v_{\text{ac}}| \sim \frac{kG \rho_0 L^2}{40\pi c_0} \left( \frac{\omega L}{c_0} \right)^4.
\end{equation}
The intensity (energy flux) scales as
\begin{equation}\label{quadrupole_ac_intensity_scaling}
	|\vec I| \sim \rho_0 c_0 |\vec v_\text{ac}|^2 \sim \frac{k^2 G^2 \rho_0^3 L^4}{(40\pi)^2c_0} \left( \frac{|\vec v|}{c_0} \right)^8,
\end{equation}
where we also replaced $\omega L \sim |\vec v|$. We have thus re-derived Lighthill's 8$^\text{th}$-power law, which for $L \sim r$ in the acoustic near field gives the scaling~\cite{Lighthill1952,Hirschberg2018} $|\vec I| \sim \rho_0 c_0^3 (|\vec v|/c_0)^8$. The result is a strong indication that linearized general relativity and Einstein field equations~(\ref{Einstein_field_equations_linear}) are able to capture not only sound waves, but also their quadrupole generation mechanism in unbounded fluids, at least in low Mach number flows.

Furthermore, neglecting the dimensionless multiplication constant involving $k$ (which will become relevant later, when a full analytical solution rather than a scaling law is obtained), the two expressions become equal if we set
\begin{equation}\label{gravitational_constant_and_Schwarzschild_radius}
	G = \frac{c_0^2 L}{2M} \sim \frac{c_0^2 }{\rho_0 L^2} ,
\end{equation}
where $M$ is total mass of the source. Indeed, in cosmology
\begin{equation}\label{Schwarzschild_radius_mass}
	L = \frac{2GM}{c_0^2}
\end{equation}
is Schwarzschild radius, which determines size of a black hole with mass $M$ and its event horizon. It is thus reasonable to introduce an analogue quantity, acoustic Schwarzschild radius, to represent the length scale of aeroacoustic sources in acoustic spacetime. These are typically acoustically compact vortices, which suggests a close relation between colliding vortices, as a source of sound, and merging black holes, as a source of gravitational waves, at least for non-relativistic velocities. A comparison of the acoustic and relativistic solutions obtained numerically will be presented in section~\ref{ch:acoustic_binary}. Here it is sufficient to note that eq.~(\ref{Schwarzschild_radius_mass}) can be understood as a definition of the gravitational constant. It relates mass of a source with length in the same manner as $c_0$ relates space and time, both in general relativity and the acoustic analogy. For constant $\rho_0$, its reciprocal value is proportional to the surface of the source. While this makes the constant appear as source-dependent from the point of view of classical Newtonian acoustics, it is not the case (by definition) in a consistent relativistic analogy which includes sources of sound.

Finally, we can compare the orders of magnitude of the obtained longitudinal component and the leading-order transverse component and prove that the former is much smaller. After replacing $G \sim c_0^2/(\rho_0 L^2)$, $r \sim L$, $d^2/dt^2 \sim \omega^2$, and $|I_{jk}| \sim \rho_0 L^5$ in eq.~(\ref{Einstein_field_equations_linear_split_solution_far_geometric_field_quadrupole_moment_tensor_jk}), we see that the metric perturbation and the amplitude of gravitational waves scale
\begin{equation}\label{gravitational_strain_scaling}
|\bar{h}^{TT}_{jk}| \sim |\bar{h}_{jk}| \sim \left( \frac{\omega L}{c_0} \right)^2,
\end{equation}
while inserting eq.~(\ref{quadrupole_ac_velocity_scaling}) into eq.~(\ref{acoustic_velocity_pressure}) gives scaling of the acoustic component \begin{equation}\label{acoustic_component_metric_perturbation_scaling}
|h_{00}| \sim \frac{|\vec v_{\text{ac}}|}{c_0} \sim \left( \frac{\omega L}{c_0} \right)^4.
\end{equation}
(The same far-field scaling holds for the near-field component in eq.~(\ref{Einstein_field_equations_linear_split_solution_far_geometric_field_quadrupole_moment_tensor_radiation_00}) multiplied with $\omega L/c_0$.) Orders of magnitude of the amplitudes of gravitational and sound waves differ by the factor $(\omega L/c_0)^2$, which is due to the longitudinal (scalar)  nature of sound waves and traces back to the second-order time derivative in eq.~(\ref{quadrupole_moment_tensor_identity}). Hence, for compact (non-relativistic) sources, longitudinal waves are much weaker than transverse waves\footnote{The same conclusion will follow from the analogy with transverse waves of the vector theory of electromagnetism in section~\ref{ch:aeroacoustic_sound_generation_EMG}.}. However, acoustic observers in fluids detect only the former.

The derivation of the scaling law presented here is entirely based on the relativistic formalism and results. It has certain advantages over Lighthill's original derivation based on classical fluid dynamics. The full stress-energy tensor is used as the source of waves due to pure instabilities of the background fluid (acoustic vacuum), without a delicate split into the source (flow) and propagation (sound) terms or selection of an appropriate quantity (sound pressure or density). The waves are a purely geometric perturbation of spacetime and thus decoupled from the dynamics in the bounded source region. Transition from the second-order tensor to the acoustic scalar is achieved by the choice of the specific Newtonian gauge for expressing the solution (metric perturbation), rather than by contracting the source term with double divergence.

Lighthill's analogy and the derivation above assume a quiescent background fluid outside the source region. Possible effects of a background flow on sound propagation, such as convection and refraction, can be taken into account by replacing the metric $\eta_{\alpha\beta}$ from eq.~(\ref{metric_tensor_small_perturbation}) with an appropriate background metric, for example, one from section~\ref{background_spacetime} (see also Ref.~\cite{Gregory2015a} and~\cite{Gregory2015b}), as long as the large length scales and time scales clearly distinguish the background flow from the acoustic perturbation. As already recognized in literature~\cite{Visser1998}, these background metrics cannot be calculated using an analogy with gravitation or the Einstein field equations, since geometry of the background acoustic spacetime depends on the externally conditioned properties of the background flow (mean flow velocity, temperature distribution, etc.), not the mass of the fluid particles itself. This limitation is explained by the ability of the acoustic analogy to capture kinematics, but not the dynamics of general relativity~\cite{Ewert2020}. However, the main reason is arguably the existence of external factors, not the fluid dynamics. The limitation does not necessarily extend to a perturbation of the background metric caused by the (moving mass of the) flow which is not externally dictated, such as the generated waves. As demonstrated above, the linearized Einstein field equations, including the source term, can be used for describing sound waves in the acoustic spacetime. This is, indeed, possible because sound generation in an incompressible flow is purely kinematic and the stress-energy tensor reduces to the same form in both theories for non-relativistic velocities\footnote{Non-relativistic velocities are also a prerequisite for the equivalence of general relativity and the theory of gravitoelectromagnetism~\cite{Mashhoon2008}. The acoustic analogy described here can be seen as a scalar-field complement. It is potentially even more relevant for the attempts to promote gravitoelectromagnetism to a full theory of gravitation alternative to general relativity~\cite{Hills2015}.}. Moreover, even the near-field Newtonian gravity complies with the analogy (eq.~(\ref{Einstein_field_equations_linear_non-relativistic_source_near_field})), with the same role of mass as the source. Hence, it is the external control of the background fluid and different governing equations for relativistic motion which separate the two theories, not the dynamics and role of mass as such. Finally, steady (and therefore unbounded) background fluid and external influences are even conceptually in contradiction with the theory of general relativity and cosmology, regardless of the dynamics. Not surprisingly, the background fluid density value and other properties of the flat acoustic spacetime will follow from the thermodynamic treatment in section~\ref{ch:background_fluid_thermodynamics}.

Unlike gravitational waves, which require a quadrupole source, sound waves in fluids can be generated also by monopole and dipole sources, represented by the components $T^{00}$ and $T^{0j} = T^{j0}$ of the stress-energy tensor, respectively (see eq.~(\ref{eq:stress-energy_tensor_perfect_fluid_components})). Although they will be considered in sections~\ref{ch:pulsating_sphere} and \ref{ch:oscillating_sphere}, some results can already be anticipated. When $T_{00} \sim \rho c_0^2$ (rather than $T_{jk} \sim \rho_0 |\vec v|^2$) dominates sound generation, as in jets with combustion or significant vapour condensation~\cite{Hirschberg2018}, power of the acoustic monopole source scales as $(\omega L/c_0)^4 \sim (|\vec v|/c_0)^4$ (rather than ($|\vec v|/c_0)^8$), which will agree with squared eq.~(\ref{acoustic_velocity_spherical_classic}). As discussed above, the factor $(\omega L/c_0)^2$ multiplying the amplitude is due to acoustic waves being longitudinal, characterized with a single scalar, and the second-order time derivative from eq.~(\ref{quadrupole_moment_tensor_identity}). A slightly different analysis in frequency domain will show in section~\ref{ch:pulsating_sphere} that the factor can also be explained by compressibility of sound waves, due to which the metric perturbation tensor has to satisfy the full Helmholtz equation, not the Laplacian equation satisfied by an incompressible perturbation (see eq.~(\ref{Einstein_field_equations_linear_low_indices_Helmhotlz}) and the discussion which follows it). Power of an acoustic dipole source with the dominant components $T_{0j} \sim \rho_0 c_0 |\vec v|$ scales accordingly as $(\omega L/c_0)^6 \sim (|\vec v|/c_0)^6$, which can be compared with squared eq.~(\ref{particle_acceleration_spatial_Newtonian_gauge_dipole}). In section~\ref{ch:aeroacoustic_sound_generation_EMG} we will see that the same scaling follows from the analogy with the vector theory of electromagnetic waves (which is also more natural for dipole sources), without referring to the more complicated second-order tensor theory of gravitation. Nevertheless, the amplitude of longitudinal waves will be shown to be lower than the amplitude of transverse waves by the same factor as in the case of gravitational waves, $(\omega L/c_0)^2$ (see eq.~(\ref{Maxwell_currents_Lorenz_gauge_split_solution_far_geometric_field_dipole_moment_j_scaling})).

\subsection{Acoustic binary}\label{ch:acoustic_binary}

Appearance of the eight power law for the Newtonian, longitudinal part of the relativistic quadrupole radiation indicates an analogy between rotating black holes, as sources of gravitational waves, and vorticity, an aeroacoustic source of sound due to a pure instability of the fluid, the acoustic vacuum. It is well known that a single isolated vortex does not produce sound~\cite{Hirschberg2018}. At least a vortex pair, typically a co-rotating vortex pair, is necessary to constitute a compact aeroacoustic quadrupole. With regard to this, it is not surprising that one of the strongest (and therefore most important) compact sources of gravitational waves are star and black hole binaries~\cite{Schutz2017,Baumgarte2010}. The previous analysis, thus, motivates comparison of wave radiation by a co-rotating vortex pair and a black hole binary moving with non-relativistic velocities. This will be done here with the aid of numerical relativistic solutions. In addition to the derived scaling law, the numerical solutions allow comparison of amplitudes, frequencies, and even waveforms of the generated waves. 

As explained in section~\ref{ch:sound_wave_in_spacetime}, metric perturbation due to a gravitational wave and associated gravitational strain are usually expressed in the transverse-traceless gauge. Since the longitudinal component is irrelevant, transformation to the Newtonian gauge is not performed in the far field. However, both acoustic and relativistic binary are compact sources of waves, so we can utilize the results of section~\ref{ch:aeroacoustic_sound_generation} to calculate the acoustically relevant component of the metric perturbation from the relativistic solution obtained numerically. This will then be compared with the direct numerical solution of sound radiation of a vortex pair, which is governed by the compressible Navier-Stokes equations~\cite{Mitchell1995}. For completeness, the two waveforms will be compared not only in the non-relativistic regime, but also as the two vortices (or black holes) approach each other with increasing velocities. In addition to this, the available analytical acoustic solution for low Mach number velocities~\cite{Mitchell1995} allows comparison of the amplitude and frequency of the generated wave in the non-relativistic regime\footnote{Since the acoustic solution from section~\ref{ch:aeroacoustic_sound_generation} is derived in the near field and the gravitational strain of the rotating source will be calculated at infinity, the initial phase of the wave will not be calculated, but adjusted manually.}.

For an outgoing wave in the geometric far field of a compact non-relativistic source, we can write from eq.~(\ref{Einstein_field_equations_linear_split_solution_far_geometric_field_quadrupole_moment_tensor_jk})
\begin{equation}\label{Einstein_field_equations_linear_split_solution_far_geometric_field_quadrupole_moment_tensor_jk_TT}
\bar{h}_{jk}^{TT} = \frac{kG}{4\pi r c_0^4} \frac{d^2}{dt^2} \hcancel{I}_{jk}^{TT}(t-r/c_0).
\end{equation}
We also used the fact that the quadrupole moment tensor ${I}_{jk}$ and reduced quadrupole moment tensor $\hcancel{I}_{jk}$ have equal transverse-traceless parts\cite{Misner2017}. By definition\cite{Baumgarte2010},
\begin{equation}\label{reduced_quadrupole_moment_tensor_TT}
\hcancel{I}_{jk}^{TT} = \left( P_j{}^l P_k{}^m - \frac{1}{2} P_{jk} P^{lm} \right) \hcancel{I}_{lm},
\end{equation}
where the projection operator is given by
\begin{equation}\label{projection_operator}
P_j{}^l = \eta_j{}^l - n_j n^l
\end{equation}
and $n^i = x^i/r$. After comparing equations~(\ref{Einstein_field_equations_linear_split_solution_far_geometric_field_quadrupole_moment_tensor_jk_TT})-(\ref{projection_operator}) with eq.~(\ref{Einstein_field_equations_linear_split_solution_far_geometric_field_quadrupole_moment_tensor_radiation_00_gauging_to_Newtonian}), we see that the two amplitudes are related by
\begin{equation}\label{radiation_reaction_00_Newtonian_gauuge}
|h^{\text{react}}_{00}| \frac{c_0}{\Omega r}= \frac{2}{5} \left( \frac{\Omega D}{2 c_0} \right)^2 |\bar{h}_{jk}^{TT}|,
\end{equation}
where $D$ is binary separation (distance between the two black holes), $\Omega$ is the angular frequency of their rotation (half of the frequency of the generated waves, $\omega$), and $|x^j x^k| = D/2$, since the black holes move on the boundary of the source region. Here $D$ is larger than the Schwarzschild radius of each black hole. As expected, eq.~(\ref{radiation_reaction_00_Newtonian_gauuge}) is in accordance with the result of the previous section, namely $|h_{00}| \sim (\omega L/c_0)^2 |\bar h^{TT}_{jk}|$ for $L \sim D/2$ and $\omega \sim \Omega$, since $h^\text{react}_{00}$ was calculated in the acoustic near field, so the far-field $|h_{00}| = |h^\text{react}_{00}|c_0/(\Omega r)$ to the lowest order (compare also equations~(\ref{Einstein_field_equations_linear_split_solution_far_geometric_field_quadrupole_moment_tensor_jk}) and (\ref{Einstein_field_equations_linear_split_solution_far_geometric_field_quadrupole_moment_tensor_radiation_jk})).
Therefore, the acoustically relevant component of the far-field metric perturbation equals
\begin{equation}\label{h_00_Newtonian_gauuge}
|h_{00}| = \frac{2}{5} \left( \frac{\Omega D}{2 c_0} \right)^2 |\bar{h}_{jk}^{TT}|
\end{equation}
and the acoustic component is, indeed, negligible compared to the transverse-traceless components for every compact (non-relativistic) source.

Gravitational wave strain is often expressed at future null infinity, which we denote as ${\mathfrak{R}} [h(t)]$. In doing so, the amplitude decay with the distance from the source, $1/r$ in eq.~(\ref{Einstein_field_equations_linear_split_solution_far_geometric_field_quadrupole_moment_tensor_jk_TT}), is replaced by $1/L$, so from eq.~(\ref{h_00_Newtonian_gauuge}):
\begin{equation}
|{\mathfrak{R}} [h(t)]| = \frac{r}{L} |\bar{h}_{jk}^{TT}| = \frac{5}{2} \left( \frac{2 c_0}{\Omega D} \right)^2 \frac{r}{L} |h_{00}|
\end{equation}
Finally, the component $h_{00}$ is related with the sound pressure via eq.~(\ref{acoustic_pressure_potential}), so
\begin{equation}\label{co-rotating_vortex_pair_acoustic_pressure_radiation_reaction_00_Newtonian_gauge}
|{\mathfrak{R}} [h(t)]| = \frac{5}{\rho_0 c_0^2} \left( \frac{2 c_0}{\Omega D} \right)^2 \frac{r}{L} |p| .
\end{equation}
Attaching the time dependence $\cos(2\Omega t + \Phi)$ with an undetermined initial phase $\Phi$,
\begin{equation}\label{co-rotating_vortex_pair_acoustic_pressure_radiation_reaction_00_Newtonian_gauge_waveform}
{\mathfrak{R}} [h(t)] = \frac{5}{\rho_0 c_0^2} \left( \frac{2 c_0}{\Omega D} \right)^2 \frac{r}{L} |p| \cos(2\Omega t + \Phi).
\end{equation}

For a non-relativistic rotation the angular frequency $\Omega$  can be estimated using the third law of Kepler:
\begin{equation}\label{angular_frequency_Kepler}
\Omega = \sqrt{\frac{GM_\text{tot}}{D^3}},
\end{equation}
where $M_\text{tot}$ is total mass of the binary. For two black holes of equal mass, $M = M_\text{tot}/2$, we can replace $Lc_0^2 = GM_\text{tot}$ from eq.~(\ref{Schwarzschild_radius_mass}) and obtain
\begin{equation}\label{angular_frequency_Kepler_equal_black_holes}
\Omega = c_0 \sqrt{\frac{L}{D^3}}.
\end{equation}
Notice, however, that in the acoustic analogy this frequency is dictated externally, by the flow, not by the masses of the vortices or the acoustic near field. Therefore, the rotation frequency cannot be determined only by the Schwarzschild radius and binary separation, but by the given flow (see eq.~(\ref{angular_frequency_co-rotating_vortex_pair}) below). This is due to the fact that the externally controlled flow is not covered by the analogy with general relativity (recall the final discussion in section~\ref{ch:aeroacoustic_sound_generation}). Still, the different causes of rotation do not imply different mechanisms of wave generation, once the rotation has occurred.

Equation~(\ref{co-rotating_vortex_pair_acoustic_pressure_radiation_reaction_00_Newtonian_gauge}) gives the scaling factor which should be used when comparing acoustic and relativistic far-field solutions, for example, results of numerical simulations. On the other hand, classical analytical solution for a co-rotating vortex pair~\cite{Mitchell1995} in three spatial dimensions gives the sound pressure amplitude
\begin{equation}\label{co-rotating_vortex_pair_acoustic_pressure_circulation}
|p| = \frac{\rho_0 \Gamma^4}{4 \pi^3 D^3 c_0^2 r} = \frac{\rho_0 \pi D^5 \Omega_\Gamma^4}{4 c_0^2 r},
\end{equation}
where $\Gamma = \pi D^2 \Omega_\Gamma = 2\pi L U_0$ is circulation of each vortex and $U_0$ is tangential speed at the radius $L$ from the centre of the vortex. Hence, the rotation rate of the vortices is
\begin{equation}\label{angular_frequency_co-rotating_vortex_pair}
\Omega_\Gamma = \frac{2 L U_0}{D^2}.
\end{equation}
The solution holds in the far field of acoustically compact vortices in an incompressible flow (at low Mach number $U_0/c_0$). As anticipated, the acoustic binary requires an additional parameter ($U_0$) to determine the rotation speed and frequency, in comparison to the gravitational analogue. However, this is not essential for the generation of waves by the rotating mass. Equations~(\ref{angular_frequency_Kepler_equal_black_holes}) and (\ref{angular_frequency_co-rotating_vortex_pair}) give a relation between $\Omega_\Gamma$ and $\Omega$,
\begin{equation}\label{angular_frequency_Keppler_vortex_pair}
\Omega_\Gamma = 2 \Omega \frac{U_0}{c_0} \sqrt{\frac{L}{D}}
\end{equation}
with which eq.~(\ref{co-rotating_vortex_pair_acoustic_pressure_circulation}) becomes
\begin{equation}\label{co-rotating_vortex_pair_acoustic_pressure_angular_frequency}
|p| = \frac{4 \rho_0 \pi D^3 \Omega^4 U_0^4 L^2}{c_0^6 r}.
\end{equation}
Therefore, the gravitational wave strain at future null infinity from eq.~(\ref{co-rotating_vortex_pair_acoustic_pressure_radiation_reaction_00_Newtonian_gauge_waveform}) expressed in terms of sound pressure equals
\begin{equation}\label{co-rotating_vortex_pair_acoustic_pressure_radiation_reaction_00_Newtonian_gauge_waveform_pressure}
{\mathfrak{R}} [h(t)] = 160\pi \left( \frac{\Omega D}{2 c_0} \right) \left( \frac{\Omega L}{c_0} \right) \left( \frac{U_0}{c_0} \right)^4 \cos(2\Omega t + \Phi).
\end{equation}
Its amplitude depends on compactness of the vortices and the entire binary as well as on the Mach number $U_0/c_0$. Moreover, after inserting eq.~(\ref{angular_frequency_Kepler_equal_black_holes}),
\begin{equation}\label{co-rotating_vortex_pair_acoustic_pressure_radiation_reaction_00_Newtonian_gauge_waveform_pressure_no_Omega}
{\mathfrak{R}} [h(t)] = 80\pi \left( \frac{L}{D} \right)^2 \left( \frac{U_0}{c_0} \right)^4 \cos(2\Omega t + \Phi).
\end{equation}
This relation can be used for a direct comparison of relativistic (left-hand side) and acoustic (right-hand side) solutions.

Gravitational metric perturbation of a black hole binary is obtained numerically using Einstein toolkit~\cite{EinsteinToolkit2019, Loeffler2012}. The simulation uses a modified setup of the black hole merger GW150914~\cite{GW150914} which produced the first directly observed gravitational waves~\cite{Abbot2016}. The initial values of the main physical parameters of the merger are listed in Table~\ref{tab:merger_simulation_parameters}. The black holes have equal mass, which is 32.5 solar mass ($M_\odot$), and spin. Total mass of the binary is thus $M_\text{tot} = 65M_\odot$. At the starting time $t=0$ both black holes are positioned on the $x$-axis of a local Cartesian coordinate system, centred around the origin (located at $x = 5L$ and $x=-5L$, where $L$ is the Schwarzschild radius of the black holes), and rotate clockwise in the plane $z=0$ around the origin, when observed from the half-space $z>0$, with $U_0/c_0 = 0.385$. Motion in radial and azimuthal directions is much slower and provides numerical stability of the calculations of the merger.

\begin{table}[h]
	\caption{Initial properties of the black hole binary.}
	\label{tab:merger_simulation_parameters}
	\begin{tabular}{ | l | l |}
		\hline
		\textbf{parameter} & \textbf{value} \\
		\hline
		binary separation & $10 L$ \\
		\hline
		black hole mass & $32.5 M_\odot$ \\
		\hline
		black hole spin & $-0.385 L^2 c_0^3/(4G)$ \\
		\hline
		radial linear momentum & $-0.0008454 L c_0^3/G$ \\
		\hline
		azimuthal linear momentum & $-0.0953 L c_0^3/G$ \\
		\hline
	\end{tabular}
\end{table}

\begin{figure}[h]
	\centering
	\begin{subfigure}{.48\textwidth}
		\centering
		\includegraphics[width=1\linewidth]{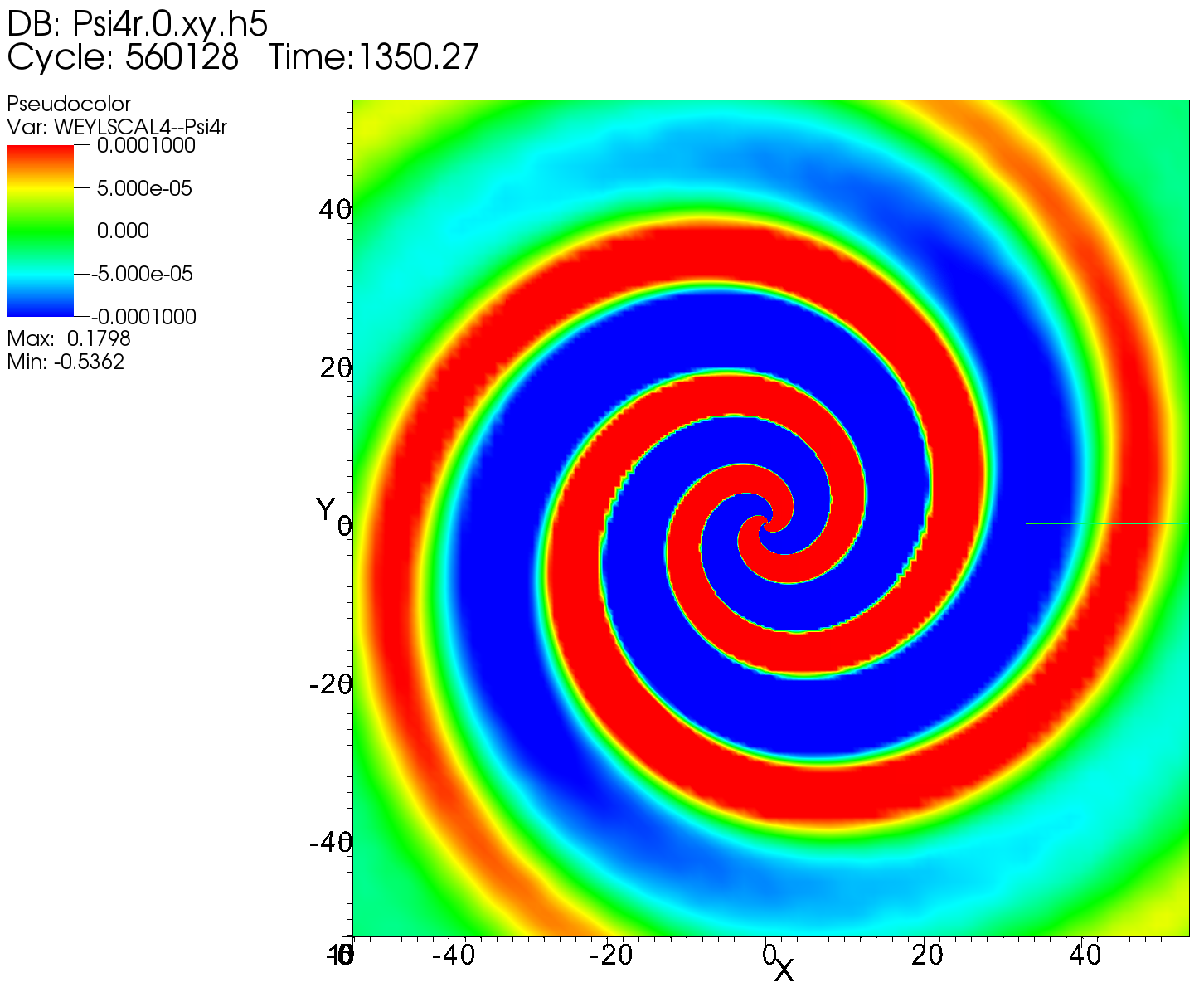}
		%\caption{A subfigure}
		\label{fig:source_function}
	\end{subfigure}%
	\begin{subfigure}{.52\textwidth}
		\centering
		\includegraphics[width=1\linewidth]{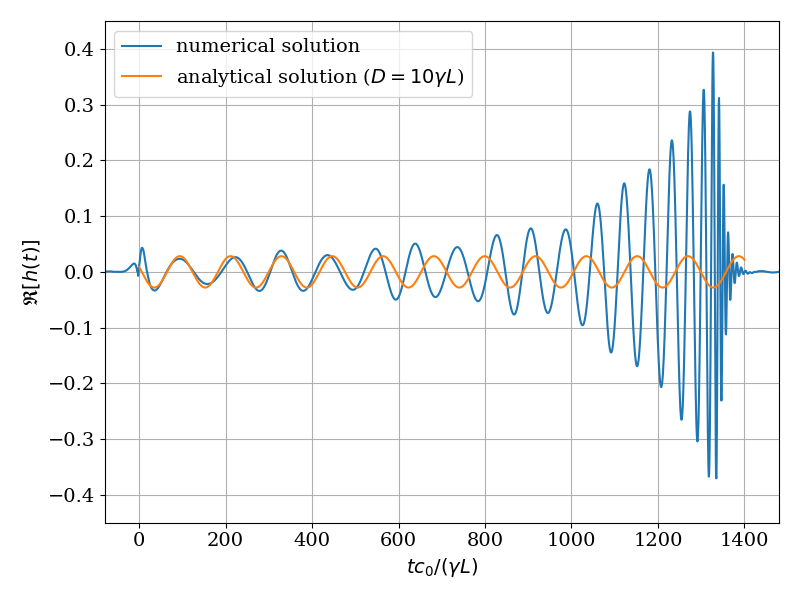}
		%\caption{A subfigure}
		\label{fig:wave}
	\end{subfigure}
	\caption{Scaled sound pressure of the acoustic binary: (left) close to the source and (right) at infinity. The numerical solution is obtained by means of numerical general relativity.}
	\label{fig:source_function_and_wave}
\end{figure}

The gravitational wave strain at future null infinity is extracted from the numerical results using the open source Python package POWER~\cite{Johnson2017}. It is given by the blue curve in Fig.~\ref{fig:source_function_and_wave} (right) as a function of normalized time. Neglecting the numerical artefact at $t=0$, as the distance between the black holes decreases, the wave amplitude and frequency increase as well as the rotation speed of the black holes. Shortly before the merge the speed approaches $c_0$ and the binary reaches the relativistic regime.

The orange line in Fig.~\ref{fig:source_function_and_wave} (right) represents the acoustic solution (right-hand side of eq.~(\ref{co-rotating_vortex_pair_acoustic_pressure_radiation_reaction_00_Newtonian_gauge_waveform_pressure_no_Omega})) at $t=0$. However, a small adjustment is necessary in order to take into account the effect of the heat capacity ratio $\gamma \neq 1$ from eq.~(\ref{density_correction}). Since mass of the black hole should be replaced with $M/\gamma$, we replace equivalently $L \rightarrow \gamma L$, according to eq.~(\ref{Schwarzschild_radius_mass}). Here $\gamma = 1.4$. Consequently, $D = 10 \gamma L$ at $t=0$, the time axis is scaled with\footnote{Indeed, the time intervals and distances should be scaled with $\gamma$ as above because of the thermodynamic origin of time in the background fluid with the mean density $\rho_0$, as will be discussed in section~\ref{source_of_entropy}.} $c_0/(\gamma L)$, while eq.~(\ref{angular_frequency_Kepler_equal_black_holes}) still holds. The amplitude at $t = 0$ is thus $|{\mathfrak{R}} [h(t)]| = 0.028$ for $U_0/c_0 = 0.385$. Lastly, the undetermined phase $\Phi$ is adjusted manually for easier visual comparison with the numerical solution.

Amplitudes and frequencies of the two waveforms match very well in the initial part of the merging process. For the observed compact quadrupole source and non-relativistic motion, the far-field analytical acoustic solution corresponds to the solution of the linearized Einstein field equations expressed in the Newtonian gauge. In addition to this, the entire waveform of the numerical solution is qualitatively comparable with Figure~5 in Ref.~\cite{Mitchell1995}, which represents the aeroacoustic source function calculated as a solution of the compressible Navier-Stokes equations, which also indicates a similar nature of wave radiation. Figure~\ref{fig:source_function_and_wave} (left) supports this further by showing the real part of the Weyl scalar $\Psi_4$ close to the source at $t = 1350.27 \gamma L/c_0$. The Weyl scalar measures the outgoing gravitational radiation~\cite{Baumgarte2010} and, like the derived acoustic quantities, already includes the second-order time derivative. At the chosen time $t$ the diagram can be directly compared with Figure~6 in Ref.~\cite{Mitchell1995} and both show the radiation pattern of a rotating quadrupole. Still, a more detailed comparison in the relativistic case is outside the scope of this text.

The obtained results prove that the analogy between acoustics and general relativity, in particular between sound radiation by a compact aeroacoustic quadrupole in an incompressible flow and gravitational radiation of compact sources governed by the linearized Einstein field equations, is not only capable of capturing accurately the power scaling low of the quadrupole source, but also the exact amplitude and waveform of the generated waves. The moving mass exhibits the same behaviour in the source region and, with appropriate parameter values, radiates waves of equal strength. Even though the analogy does not cover the externally controlled dynamics of the background fluid and, consequently, the background spacetime (which is reflected in the difference between $\Omega_\Gamma$ and $\Omega$ above) and certain correction of the mass-related parameters is necessary (scaling with the heat capacity ratio), it has been confirmed that the analogy holds for the kinematic source of perturbation. In other words, exclusion of the dynamics of the background fluid does not mean that the analogy is useless for studying sources of waves and that the (linearized) Einstein field equations cannot be used for acoustic calculations.

Due to the complexity of the relativistic problem, we had to refer to the numerical simulations for calculating amplitudes and waveforms of the weak perturbation. The initial phase remained undetermined. The complete analytical solutions are much simpler to obtain for monopole and dipole sources, which will be studied next. Although irrelevant for the theory of transverse gravitational waves, these solutions are important in the acoustic analogy for two reasons. First, in order to show that the relativistic framework (under the discussed conditions) covers not only the quadrupole sources, which are common for both theories, but also monopole and dipole sources of longitudinal sound waves, in the same sense and following the same procedures as above. Second, comparison with the well-known exact acoustic solutions allows fixing all remaining constants of the theory.

\subsection{Pulsating sphere}\label{ch:pulsating_sphere}

Acoustic monopole is the simplest source of sound in fluids, which has no counterpart in gravitation or electromagnetism. Unlike longitudinal sound waves, transverse gravitational and electromagnetic waves cannot be generated, for example, by pulsating spherical objects. This is because sound waves in fluids are compressible and characterized by the scalar velocity potential. Therefore, it is of interest to study how monopole radiation can be captured by the analogy with general relativity.

As a typical example of an ideal monopole source, we consider acoustically compact pulsating sphere, for which the exact solution is well-known~\cite{Dowling1983}. We start with the linearized Einstein field equations~(\ref{Einstein_field_equations_linear_split}),
\begin{equation}\label{Einstein_field_equations_linear_low_indices}
	\Box \bar{h}_{\alpha\beta} = -\frac{2kG}{c_0^4} T_{\alpha\beta} = -\frac{2kG}{c_0^4} (T^{\text{eff}}_{\alpha\beta} + t_{\alpha\beta}),
\end{equation}
and solve them similarly as in Ref.~\cite{Schutz2017}. Supposing simple oscillations of the radiation-related part of the stress-energy tensor, $t_{\alpha\beta}$, with angular frequency $\omega$ and complex amplitude $S_{\alpha\beta}$,
\begin{equation}\label{stress-energy_tensor_simple_ocsillations}
	t_{\alpha\beta} = S_{\alpha\beta} e^{-j \omega t}.
\end{equation}
Since the source is compact, $\omega L/c_0 \ll 1$, with $L$ denoting radius of the sphere. A solution of eq.~(\ref{Einstein_field_equations_linear_low_indices}) for an outgoing wave has the form
\begin{equation}\label{Einstein_field_equations_linear_low_indices_solution_outgoing_wave}
\bar{h}_{\alpha\beta} = \frac{A_{\alpha\beta}}{r}e^{-j \omega (t-r/c_0)},
\end{equation}
where $r$ represents distance from the source. Thereby we neglect all terms which decay with $r^{-2}$ (the near-field terms which do not contribute to the radiation). After replacing $\partial^2/\partial t^2 = (-j\omega)^2 = -\omega^2$ and removing the common time dependence $e^{-j\omega t}$ on the two sides of eq.~(\ref{Einstein_field_equations_linear_low_indices}), we obtain a second-order tensor form of the time-independent Helmholtz equation,
\begin{equation}\label{Einstein_field_equations_linear_low_indices_Helmhotlz}
	[ (\omega/c_0)^2 + \nabla^2 ] \left( \frac{A_{\alpha\beta}}{r}e^{j \omega r/c_0} \right) = -\frac{2kG}{c_0^4} S_{\alpha\beta}.
\end{equation}

Integrating the left-hand side over the compact source region with the volume $V = 4 \pi L^3/3$, bounded with the spherical surface $S = 4\pi L^2$ gives the following terms:
\begin{equation}\label{Einstein_field_equations_linear_low_indices_Helmhotlz_integral_L}
	\int_V{\frac{\omega^2}{c_0^2} \frac{A_{\alpha\beta}}{r}e^{j \omega r/c_0} d^3 \vec y} = \frac{4 \pi L^2}{3} \frac{\omega^2}{c_0^2} A_{\alpha\beta} e^{j \omega L/c_0} = \frac{4 \pi}{3} \left( \frac{\omega L}{c_0} \right)^2 A_{\alpha\beta},
\end{equation}
and
\begin{equation}\label{Einstein_field_equations_linear_low_indices_Helmhotlz_integral_T}
\begin{aligned}
	\int_V{\nabla^2 \left( \frac{A_{\alpha\beta}}{r}e^{j \omega r/c_0} \right) d^3 \vec y} &= \oint_S \vec n \cdot \nabla \left( \frac{A_{\alpha\beta}}{r}e^{j \omega r/c_0} \right) d^2 \vec{y} \\
	&= 4 \pi L^2 \frac{d}{dr} \left( \frac{A_{\alpha\beta}}{r}e^{j \omega r/c_0} \right)_{r=L} \\
	&= 4 \pi L^2 \left( -\frac{A_{\alpha\beta}}{r^2} e^{j \omega r/c_0} + \frac{j \omega}{c_0} \frac{A_{\alpha\beta}}{r}e^{j \omega r/c_0} \right)_{r=L} \\
	&= - 4 \pi A_{\alpha\beta} + j 4 \pi \left( \frac{\omega L}{c_0} \right) A_{\alpha\beta},
\end{aligned}
\end{equation}
with $\vec n$ the unit vector normal to the surface of the sphere pointing outwards. In the last equality in eq.~(\ref{Einstein_field_equations_linear_low_indices_Helmhotlz_integral_T}) we also approximated $e^{j\omega L/c_0} \approx 1$ for $\omega L/c_0 \ll 1$. Hence, the three radiation-related terms are:
\begin{equation}\label{Term1}
\text{T1} = - 4 \pi A_{\alpha\beta},
\end{equation}
\begin{equation}\label{Term2}
\text{T2} = j 4 \pi \left( \frac{\omega L}{c_0} \right) A_{\alpha\beta},
\end{equation}
\begin{equation}\label{Term3}
\text{T3} = \frac{4 \pi}{3} \left( \frac{\omega L}{c_0} \right)^2 A_{\alpha\beta}.
\end{equation}

For the compact source, T1 $\gg$ T2 $\gg$ T3. However, longitudinal sound waves must be associated with the weakest term T3, since the first term on the left-hand side of eq.~(\ref{Einstein_field_equations_linear_low_indices_Helmhotlz}) cannot be negligible compared to the second one, which gives the terms T1 and T2. The compressible waves satisfy the full Helmholtz equation not the Laplace equation and the term T3 is balanced by only a small fraction of the terms T1 and T2. In contrast to this, the leading terms of eq.~(\ref{Einstein_field_equations_linear_low_indices_Helmhotlz_integral_T}) can be associated with transverse (``incompressible'') waves. In particular, the strongest term T1 is associated with the second-order tensor gravitational waves~\cite{Schutz2017}. The scaling T3 $\sim (\omega L/c_0)^2$ T1 is entirely in accordance with equations~(\ref{gravitational_strain_scaling}) and (\ref{acoustic_component_metric_perturbation_scaling}) or eq.~(\ref{h_00_Newtonian_gauuge}). The much weaker (scalar) component of the metric perturbation vanishes in the transverse-traceless gauge. The compressible waves are thereby left out as higher-order gauge terms, which do not even satisfy the Lorenz gauge condition. Consequently, the spherically symmetric pulsations of the source can produce only longitudinal waves and not transverse gravitational waves\footnote{It is interesting to notice that the constant factor of T1 and T3 becomes $\pm 4\pi$ for both types of waves if $c_0$ in T3 is replaced with $c_0/\sqrt{3}$. As explained in footnote~\ref{speed_of_sound_and_light}, the maximum speed of compressible waves is exactly $\sqrt{3}$ smaller than the speed of light.}.

After integrating the right-hand side of eq.~(\ref{Einstein_field_equations_linear_low_indices_Helmhotlz}) over the source region, the leading term T1 gives the following expression for the polarization tensor of gravitational waves:
\begin{equation}\label{amplitude_gravitational_waves}
	A_{\alpha\beta} = \frac{kG}{2\pi c_0^4} \int_{V} S_{\alpha\beta} d^3 \vec y.
\end{equation}
Since the acoustic waves are weaker by the order $(\omega L/c_0)^2$, only a small fraction of $t_{\alpha\beta}$ (which is already a small fraction of $T_{\alpha\beta}$) actually generates the longitudinal waves. Besides multiplying with the factor T3/T1 = $-(\omega L/c_0)^2/3$ we should multiply with 2, in order to take into account the fraction of the Terms T1 and T2 which also contributes to the compressible waves\footnote{This corresponds to the factor of 1/2, which appeared by switching to the transverse-traceless gauge using the projection-operator term $P_j{}^l P_k{}^m - P_{jk} P^{lm}/2$ in eq.~(\ref{reduced_quadrupole_moment_tensor_TT}).}. Therefore, the polarization tensor which captures the sound waves reads
\begin{equation}\label{amplitude_acoustic_waves}
	A_{\alpha\beta} = -\frac{kG}{3\pi c_0^4} \left( \frac{\omega L}{c_0} \right)^2 \int_{V} S_{\alpha\beta} d^3 \vec y.
\end{equation}
In other words, longitudinal waves of the compact monopole source are described by the equation
\begin{equation}\label{Einstein_field_equations_linear_non-relativistic_source_acoustic}
\Box \bar{h}^{00} = \frac{4kG}{3 c_0^2} \left( \frac{\omega L}{c_0} \right)^2 \rho,
\end{equation}
which is only a small fraction of eq.~(\ref{Einstein_field_equations_linear_non-relativistic_source}). Their near-field contribution is accordingly much smaller than in eq.~(\ref{Einstein_field_equations_linear_non-relativistic_source_near_field}). After applying eq.~(\ref{amplitude_acoustic_waves}) in  eq.~(\ref{Einstein_field_equations_linear_low_indices_solution_outgoing_wave}),
\begin{equation}\label{Einstein_field_equations_linear_low_indices_solution_outgoing_wave_acoustic}
	\bar{h}_{\alpha\beta} = -\frac{kG}{3\pi c_0^4} \left( \frac{\omega L}{c_0} \right)^2 \frac{e^{j\omega r/c_0}}{r}  \int_{V} t_{\alpha\beta} d^3 \vec y.
\end{equation}
This metric perturbation describes the radiated longitudinal sound waves generated by a compact (not necessarily monopole) spherical source. We emphasize again that it does not have to satisfy the condition $\bar{h}^{\alpha\beta}{}_{,\beta} = 0$, even though the components are expressed in a Lorenz gauge, because it is of sub-leading order. (The Lorenz gauge condition is satisfied at the leading order of the term T1.)

The form of the stress-energy tensor in eq.~(\ref{eq:stress-energy_tensor_perfect_fluid_components}) suggested that the monopole source represented by the component $T^{00}$ is due to the unsteady density $\rho$ (as in eq.~(\ref{Einstein_field_equations_linear_non-relativistic_source})). However, the fluid in a compact source region is essentially incompressible, described by the stress-energy tensor from eq.~(\ref{eq:stress-energy_tensor_perfect_fluid_components_non-relativistic}), so the source of sound waves is in fact the unsteady volume, that is, radius of the pulsating sphere, $L$. This becomes clearer in the integral form of eq.~(\ref{Einstein_field_equations_linear_low_indices_solution_outgoing_wave_acoustic}) than in the differential eq.~(\ref{Einstein_field_equations_linear_non-relativistic_source}). Indeed, mass injection in a compact source region comes down to the displacement of a volume fraction of the fluid around it~\cite{Hirschberg2018}. Hence, even the compact acoustic monopole is a kinematic source. This explains why mass has the same role in non-relativistic gravitation and the acoustic analogy, even for a monopole source of sound and without an analogy with compressible fluid dynamics. The only necessary condition is that the source is non-relativistic and, therefore, compact.

With respect to the above, the single non-zero far-field component $\bar{h}_{00}$ equals\footnote{This solution also follows from the Schwarzschild metric, which is the only spherically symmetric and asymptotically (for $r \rightarrow \infty$) flat solution of the Einstein field equations in vacuum. Radius of the pulsating sphere represents an unsteady Schwarzschild radius from eq.~(\ref{Schwarzschild_radius_mass}). For linearized eq.~(\ref{Einstein_field_equations_linear_low_indices}) and large $r$ (in the far field), the solution in a Lorenz gauge reads~\cite{Misner2017} $\bar{h}_{00} = kMG/(2\pi r c_0^2)$, $\bar{h}_{0j} = \bar{h}_{jk} = 0$. The acoustic mass source, $m = M - M^{\text{eff}}$, is associated with the volume integral of $t_{00} = (\rho - \rho^\text{eff}) c_0^2$ over the source region and must be scaled with the factor $-2(\omega L/c_0)^2/3$, with the same reasoning which led to eq.~(\ref{amplitude_acoustic_waves}). This gives the contribution of $\bar h_{00}$ from eq.~(\ref{Einstein_field_equations_linear_low_indices_solution_outgoing_wave_acoustic_00}) in the near field ($e^{j\omega r/c_0} \approx 1$), where we expect the equivalence with Newtonian gravity. Otherwise, the obtained metric perturbation $\bar{h}_{\alpha\beta}$ (with the only non-zero and unsteady component $\bar{h}_{00}$) does not satisfy the Lorenz gauge condition or describe transverse waves, even at the leading order. It is purely acoustic.}
\begin{equation}\label{Einstein_field_equations_linear_low_indices_solution_outgoing_wave_acoustic_00}
	\bar{h}_{00} = -\frac{kG}{3\pi c_0^4} \left( \frac{\omega L}{c_0} \right)^2 \frac{e^{j\omega r/c_0}}{r}  \int_{V} t_{00} d^3 \vec y.
\end{equation}
We can now express the last integral in terms of the unsteady radius $L$, replace it with
\begin{equation}\label{T_00_integral}
	\int_{V} t_{00} d^3 \vec y = \rho_0 c_0^2 \frac{4}{3} (L+\bar{l} e^{-j\omega t})^3 \pi \approx 4 \rho_0 c_0^2 L^2 \pi \bar{l} e^{-j\omega t}
\end{equation}
at the first order, where $\bar{l} \ll L$ is amplitude of the oscillations around the average value $L$, and obtain
\begin{equation}\label{Einstein_field_equations_linear_low_indices_solution_outgoing_wave_acoustic_00_mass_source}
	\bar{h}_{00} = -\frac{4kG \rho_0 L^2}{3 c_0^2} \left( \frac{\omega L}{c_0} \right)^2 \bar{l} \frac{e^{-j\omega (t-r/c_0)}}{r}.
\end{equation}

From eq. (\ref{gravitational_constant_and_Schwarzschild_radius}) with $M = 4 \rho_0 L^3 \pi/3$, 
\begin{equation}\label{gravitational_constant_monopole}
	G = \frac{3 c_0^2}{8\rho_0 L^2 \pi}
\end{equation}
and therefore
\begin{equation}\label{Einstein_field_equations_linear_low_indices_solution_outgoing_wave_acoustic_00_mass_source_2}
	\bar{h}_{00} = -\frac{k}{2\pi} \left( \frac{\omega L}{c_0} \right)^2 \bar{l} \frac{e^{-j\omega (t-r/c_0)}}{r}.
\end{equation}
Following the procedure from sections~\ref{ch:sound_wave_in_spacetime} and \ref{ch:aeroacoustic_sound_generation}, we switch from $\bar{h}_{00}$ to $h_{00}$, using eq.~(\ref{metric_tensor_small_perturbation_Lorenz_gauge_inverted}):
\begin{equation}\label{h_00_and_h_00_bar_monopole}
	h_{00} = \bar h_{00} - \frac{1}{2} \eta_{00} \bar h^\nu{}_\nu = \bar h_{00} + \frac{1}{2} \bar h^\nu{}_\nu = \bar h_{00} + \frac{1}{2} \eta^{\mu\nu} \bar h_{\mu\nu} = \bar h_{00} + \frac{1}{2} \eta^{00} \bar h_{00} = \frac{1}{2} \bar h_{00}
\end{equation}
and $h_{0j} = 0$. As expected for the monopole source, the metric perturbation is already in the Newtonian gauge and from eq.~(\ref{particle_acceleration_spatial_Newtonian_gauge}) (or directly eq.~(\ref{acoustic_velocity_pressure})) radial component of the particle velocity due to the incoming spherical sound wave equals
\begin{equation}\label{longitudinal_wave_acceleration_pulsating_sphere}
	v_\text{ac}^r = \frac{k}{8\pi} c_0 \left( \frac{\omega L}{c_0} \right)^2 \bar{l} \frac{e^{-j\omega (t-r/c_0)}}{r}.
\end{equation}
The component proportional to $1/r^2$ has again been neglected and the purely acoustic metric perturbation does not produce transverse motion.

The classical acoustic solution for an acoustically compact ($\omega L/c_0 \ll 1$) pulsating sphere reads~\cite{Dowling1983}
\begin{equation}\label{acoustic_velocity_spherical_classic}
	v_\text{ac}^r = -c_0 \left( \frac{\omega L}{c_0} \right)^2 \bar{l} \frac{e^{-j \omega (t - r/c_0)}}{r}.
\end{equation}
In order to match the two exact results, we should adopt $k = -8 \pi$. In general relativity the dimensionless constant is~\cite{Schutz2017} $k = 8 \pi$. The opposite sign can be traced back to the opposite signs of equations~(\ref{amplitude_gravitational_waves}) and~(\ref{amplitude_acoustic_waves}) or the terms T1 and T3. A physical corollary is mass being an attracting source of gravity and a repelling acoustic source. While black hole is a typical example of the former, white hole is a more natural monopole source of sound.

The derivation above proves not only that the exact acoustic solutions can be obtained using the linearized Einstein field equations, including the stress-energy tensor as the source, but that the acoustic form of the equations differs only by the sign of the source term. We underline that it does not imply existence of longitudinal gravitational waves, which would satisfy eq.~(\ref{Einstein_field_equations_linear_non-relativistic_source}) at the leading order, rather than eq.~(\ref{Einstein_field_equations_linear_non-relativistic_source_acoustic}). Acoustic perturbation in the near field does not correspond to the Newtonian gravitation, but its relativistic correction due to radiation. As the final validation of the analogy, we examine sound radiation of a compact acoustic dipole relativistically.

\subsection{Oscillating sphere}\label{ch:oscillating_sphere}

Next we observe an acoustically compact sphere (with the fixed radius $L$) oscillating with velocity $\vec v \sim e^{-j\omega t}$ and magnitude $|\vec v| \ll c_0$. The derivation follows the one in section~\ref{ch:pulsating_sphere}, except that the integral in eq.~(\ref{T_00_integral}) is replaced by ($r$, $\phi$, and $z$ are in this case cylindrical coordinates with the sphere oscillating around the origin):
\begin{equation}\label{T_00_integral_dipole}
	\int_{V} t_{0j} d^3 \vec y = \rho_0 c_0 v_j \int_{0}^{L} r dr \int_{0}^{2 \pi} d\phi \int_{0}^{L} dz = \rho_0 c_0 L^3 \pi v_j.
\end{equation}
The remaining components, $t_{00}$ and $t_{jk}$, equal zero. In the last integral we took into account the fact that only half of the translating sphere acts on the surrounding fluid during its motion, with the scattering cross-sectional area $L^2 \pi$. Equation~(\ref{Einstein_field_equations_linear_low_indices_solution_outgoing_wave_acoustic}) gives then the only non-vanishing components of the metric perturbation:
\begin{equation}\label{Einstein_field_equations_linear_low_indices_solution_outgoing_wave_acoustic_0j}
	\bar{h}_{0j} = -\frac{kG \rho_0 L^3 }{3 c_0^3} \left( \frac{\omega L}{c_0} \right)^2 \frac{e^{j\omega r/c_0}}{r} v_j.
\end{equation}

Note that this metric does satisfy the Lorenz gauge condition ($\bar{h}^{\alpha\beta}{}_{,\beta} = \bar{h}^{0j}{}_{,j} = 0$) in the near field of the compact sphere ($e^{j\omega r/c_0} \approx 1$) to the lowest order of $1/r$, because the fluid around the sphere is incompressible ($v^j{}_{,j} = 0$). The motion is irrotational so we can write
\begin{equation}\label{Einstein_field_equations_linear_low_indices_solution_outgoing_wave_acoustic_0j_potential}
\bar{h}_{0j} = -\frac{kG \rho_0 L^3 }{3 c_0^3} \left( \frac{\omega L}{c_0} \right)^2 \frac{e^{j\omega r/c_0}}{r} \phi_{,j},
\end{equation}
with the scalar velocity potential $\phi$.
Since this metric perturbation has the same order of $\omega L/c_0$ as that of the monopole in eq.~(\ref{Einstein_field_equations_linear_low_indices_solution_outgoing_wave_acoustic_00_mass_source}), the acoustically relevant (compressible) part must be of even higher order in the case of the dipole.

The metric should be expressed in the Newtonian gauge, so we transform the coordinates, $x_\alpha \rightarrow x_\alpha + \xi_\alpha$, with
\begin{equation}\label{gauging_to_Newtonian_form_dipole1}
\xi_0 = -\frac{kG \rho_0 L^3 }{3 c_0^3} \left( \frac{\omega L}{c_0} \right)^2 \frac{e^{j\omega r/c_0}}{r} \phi
\end{equation}
and
\begin{equation}\label{gauging_to_Newtonian_form_dipole2}
\xi_j = -\frac{kG \rho_0 L^3 }{3 c_0^3} \left( \frac{\omega L}{c_0} \right)^2 \frac{e^{j\omega r/c_0}}{r} \phi r_{,j}.
\end{equation} Equations~(\ref{metric_tensor_small_perturbation_Lorenz_gauge_inverted}) and (\ref{metric_tensor_small_perturbation_gauging}) with $\bar h^\nu{}_\nu = 0$ give
\begin{equation}
h_{\alpha\beta} = \bar h_{\alpha\beta} - \xi_{\alpha,\beta} - \xi_{\beta,\alpha}
\end{equation}
and from equations~(\ref{gauging_to_Newtonian_form_dipole1}) and (\ref{gauging_to_Newtonian_form_dipole2}):
\begin{equation}\label{gauging_to_Newtonian_form_dipole1_derivative1}
\xi_{0,0} = \frac{j\omega kG \rho_0 L^3 }{3 c_0^4} \left( \frac{\omega L}{c_0} \right)^2 \frac{e^{j\omega r/c_0}}{r} \phi,
\end{equation}
\begin{equation}\label{gauging_to_Newtonian_form_dipole2_derivative}
	\xi_{j,0} = \frac{j\omega kG \rho_0 L^3 }{3 c_0^4} \left( \frac{\omega L}{c_0} \right)^2 \frac{e^{j\omega r/c_0}}{r} \phi r_{,j},
\end{equation}
\begin{equation}\label{gauging_to_Newtonian_form_dipole1_derivative2}
\begin{aligned}
\xi_{0,j} &= -\frac{kG \rho_0 L^3 }{3 c_0^3} \left( \frac{\omega L}{c_0} \right)^2 \frac{e^{j\omega r/c_0}}{r} \left( \phi_{,j} + \frac{j\omega}{c_0} \phi r_{,j} \right) \\
&= \bar h_{0j} -\frac{j\omega kG \rho_0 L^3 }{3 c_0^4} \left( \frac{\omega L}{c_0} \right)^2 \frac{e^{j\omega r/c_0}}{r} \phi r_{,j},
\end{aligned}
\end{equation}
to the lowest order of $1/r$. Therefore,
\begin{equation}\label{Einstein_field_equations_linear_low_indices_solution_outgoing_wave_acoustic_dipole_00}
	h_{00} = \bar{h}_{00} - \xi_{0,0} - \xi_{0,0} = -2 \xi_{0,0} = \frac{-2 j\omega kG \rho_0 L^3 }{3 c_0^4} \left( \frac{\omega L}{c_0} \right)^2 \frac{e^{j\omega r/c_0}}{r} \phi,
\end{equation}
while
\begin{equation}\label{Einstein_field_equations_linear_low_indices_solution_outgoing_wave_acoustic_dipole_0j}
h_{0j} = \bar{h}_{0j} - \xi_{0,j} - \xi_{j,0} = 0
\end{equation}
and $h_{jk} = 0$.

The metric perturbation is in the Newtonian form and we can again use eq.~(\ref{particle_acceleration_spatial_Newtonian_gauge}) to express the acoustic particle velocity:
\begin{equation}
\frac{d v_\text{ac}^k}{dt} = \frac{c_0^2}{2} h_{00}^{,k} = \frac{- j\omega kG \rho_0 L^3 }{3 c_0^2} \left( \frac{\omega L}{c_0} \right)^2 \frac{e^{j\omega r/c_0}}{r} \left( \phi^{,k} + \frac{j \omega}{c_0} \phi r^{,k} \right),
\end{equation}
neglecting the component $\sim 1/r^2$ again. Since $\phi^{,k} \sim \phi/L \gg j\omega \phi r^{,k}/c_0 \sim \omega \phi /c_0$ for the compact source, we can neglect the second term in favour of the first one. With $d v_\text{ac}^k/dt = -j\omega v_\text{ac}^k$ and $\phi^{,k} = v^k$,
\begin{equation}
v_\text{ac}^k = \frac{kG \rho_0 L^3 }{3 c_0^2} \left( \frac{\omega L}{c_0} \right)^2 \frac{e^{j\omega r/c_0}}{r} v^{k}.
\end{equation}
The constant $G$ can be expressed from eq.~(\ref{gravitational_constant_and_Schwarzschild_radius}). However, the effective Schwarzschild radius is $L/2$, since only one half of the sphere pushes the fluid. (Equivalently, the effective mass is $2M$, because the last integral in eq.~(\ref{T_00_integral_dipole}) covers only half of the mass between $-L$ and $L$.) Hence,
\begin{equation}\label{gravitational_constant_and_Schwarzschild_radius_dipole}
G = \frac{c_0^2 L}{4M} = \frac{3 c_0^2}{16\rho_0 L^2 \pi},
\end{equation}
and
\begin{equation}
v_\text{ac}^k = \frac{k c_0}{16 \pi \omega } \left( \frac{\omega L}{c_0} \right)^3 \frac{e^{j\omega r/c_0}}{r} v^{k}.
\end{equation}
Finally, the radial component equals
\begin{equation}\label{particle_acceleration_spatial_Newtonian_gauge_dipole}
	v_{\text{ac}}^r = \frac{k c_0}{16 \pi \omega} \left( \frac{\omega L}{c_0} \right)^3 |\vec{v}| \cos(\theta) \frac{e^{-j\omega (t-r/c_0)}}{r},
\end{equation}
where $\theta$ denotes the angle between $\vec v$ and the position vector $\vec r$. This matches the classical solution~\cite{Dowling1983} for $k = -8\pi$ (in contrast to the relativistic $k = 8\pi$) and thus confirms the value of $k$. Moreover, it shows that the analogy is capable of capturing acoustic dipole radiation, as well.

As a brief summary of the analogy described above, we can write the acoustic form of the linearized Einstein field equations,
\begin{equation}\label{Einstein_field_equations_linear_low_indices_acoustic}
\Box \bar{h}_{\alpha\beta} = \frac{16\pi G}{c_0^4} T_{\alpha\beta},
\end{equation}
in which the stress-energy tensor $T_{\alpha\beta}$ has the form as in eq.~(\ref{eq:stress-energy_tensor_perfect_fluid_components_non-relativistic}). (A monopole source determined by the component $T_{00}$ is due to the unsteady volume of the source region.) The equations hold under the Lorenz gauge condition in eq.~(\ref{Lorenz_gauge}). The analogy has been shown to be valid for low Mach number (that is, non-relativistic and therefore compact) sources and waves as a weak perturbation of the background spacetime. The quiescent background fluid represents the acoustic vacuum, which is not covered by the theory of unsteady perturbation. Particle motion due to the longitudinal sound waves is determined by the component $h_{00}$ of the metric perturbation expressed in the Newtonian gauge, according to the geodesic equation~(\ref{particle_acceleration_spatial_Newtonian_gauge}). This very weak component is of sub-leading order and alone does not satisfy the Lorenz gauge condition.

Although the analogy includes monopole and dipole sources, it is evident that the components $T_{jk}$ and $\bar h_{jk}$ of the second-order tensor theory are redundant in the absence of quadrupole sources of sound. Indeed, in section~\ref{ch:analogy_with_EMG} we will introduce conceptually very similar analogy to show that dipole radiation can be treated more easily using the vector theory of electromagnetism. Although electric charge takes over the role of mass, such an analogy should not be surprising, considering that well-established gravitoelectromagnetism~\cite{Mashhoon2008} also relates the two theories in the non-relativistic limit. Before that a note shall be made regarding the cosmological constant, which is related to the largest length and time scales of the theory (and the associated dark energy), but often mistakenly attributed to the (steady and unbounded) background spacetime. Indeed, it is finiteness and unsteadiness which allow the analogue interpretation of the cosmological constant as described in the next section. The same is not true for the steady background fluid, which is a subject of quantum thermodynamic analogies examined in section~\ref{ch:acoustic_Lagrangians}.
\section{Dark energy and energy of vacuum}\label{ch:dark_energy}

Although not necessary for classical acoustic calculations in unbounded steady fluids, it is interesting to note that the appearance of small but non-zero cosmological constant in eq.~(\ref{Einstein_field_equations}) can also be treated quasi-acoustically. This might be surprising if the cosmological constant were associated with the energy of the vacuum itself without any perturbation (that is $\rho_0$ above), since we have already stated that the acoustic analogy with general relativity does not cover dynamics of the background spacetime. Actually, the cosmological constant represents a purely geometric expansion of the bounded observable universe in the Big Bang cosmology, which is alternatively interpreted to be caused by the dark energy with density $\rho_\Lambda$~\cite{Rovelli2015}. The true cause of the expansion (as well as the boundedness of the observable universe) is arguably of quantum thermodynamic nature and will be discussed in section~\ref{source_of_entropy}. Here we only show how the cosmological constant represents the expansion and why the dark energy $\rho_\Lambda$ should not be confused with the energy of the (unobservable, steady, and unbounded) background spacetime $\rho_0$.

The key assumption is that the characteristic time scale of the radial expansion of the bounded universe is long enough that every observer is located in the near field of its source. The associated angular frequency $\omega$ is so low (but finite) that $\omega r/c_0 \ll 1$ holds for any distance $r$ smaller than the radius of the observable universe $R_U$. In other words, the frequency associated with the radius $R_U$,
\begin{equation}\label{lowest_frequency_universe}
\omega = 2 \pi c_0/R_U,
\end{equation}
can be considered as the lowest frequency of the theory, making $\rho_0$ (unlike usual unsteady $\rho$) unobservable.

At such low frequencies the wave operator in eq.~(\ref{Einstein_field_equations_linear}) reduces to the Laplacian. Since the observer is by definition located outside the source region, which has the characteristic length scale $L \lesssim r$, the condition $\omega L/c_0 \ll 1$ implies also slow motion of the source. The stress-energy tensor component $T^{00} = \rho_\Lambda c_0^2$ dominates and we can write as in eq.~(\ref{Einstein_field_equations_linear_non-relativistic_source_near_field})
\begin{equation}\label{Einstein_field_equations_linear_low_freq_very_slow_source}
\nabla^2 \bar{h}^{00} = -\frac{2kG}{c_0^2} \rho_\Lambda.
\end{equation}
As already discussed, the Newtonian gravity of a non-relativistic source corresponds to an acoustic near field. In the Newtonian limit of general relativity the analogy covers the effects of the mass, but not the mass of the background fluid, since $\rho_\Lambda \ll \rho_0$ must be satisfied in the linearized theory. We should also reiterate that the far-field extension of the leading-order solution of eq.~(\ref{Einstein_field_equations_linear_low_freq_very_slow_source}) (compare with eq.~(\ref{Einstein_field_equations_linear_non-relativistic_source})) does not match the sound fields calculated in the previous sections (recall eq.~(\ref{Einstein_field_equations_linear_non-relativistic_source_acoustic})). It rather corresponds to non-physical higher-order longitudinal gravitational waves.

Since $\omega$ is finite ($\rho_\Lambda \neq \rho_0$), so is the radius of the observable universe,
\begin{equation}\label{Laplacian_as_cosmological_constant}
|\nabla^2| \gtrsim \left( \frac{2\pi}{R_U} \right)^2 > 0
\end{equation}
is also finite, and every observable $\bar h^{00}$ must be unsteady, with the characteristic length scale smaller than or at most comparable to $R_U$. Hence, even in the absence of any other perturbation, there must be an always (and for every observer, regardless of their location) present source with density $\rho_\Lambda \neq 0$, which is the dark energy density. Notice that such defined $\rho_\Lambda$ is only a model for the expansion, which makes it possible that the source is omnipresent, including all possible observer locations. The actual cause of expansion is not a distinct source, but the macroscopic observer and the inherent thermodynamics, as will be discussed later.

The ``source'' of expansion stems from the second term on the left-hand side of eq.~(\ref{Einstein_field_equations}). In order to treat it as a usual source, we shift the term to the right-hand side of the equation and obtain for its contribution to the stress-energy tensor~\cite{Schutz2017}
\begin{equation}\label{stress-energy_tensor_dark_energy}
	T_\Lambda^{\alpha\beta} = -\frac{\Lambda c_0^4}{kG} \eta^{\alpha\beta} = 
	\begin{bmatrix}
	\rho_\Lambda c_0^2 & 0 & 0 & 0\\
	0 & p_\Lambda & 0 & 0\\
	0 & 0 & p_\Lambda & 0\\
	0 & 0 & 0 & p_\Lambda
	\end{bmatrix}.
\end{equation}
The unperturbed background metric is assumed to be the Minkowski metric and for the last equality we used accordingly the general form of the stress-energy tensor from eq.~(\ref{eq:stress-energy_tensor_perfect_fluid_components}) with $\vec v = 0$. The dark energy density $\rho_\Lambda$ and pressure $p_\Lambda$ equal
\begin{equation}\label{dark_energy_density}
	\rho_\Lambda = -\frac{p_\Lambda}{c_0^2} = \frac{\Lambda c_0^2}{kG}.
\end{equation}
Inserting this in eq.~(\ref{Einstein_field_equations_linear_low_freq_very_slow_source}) gives
\begin{equation}\label{cosmological_constant_as_source}
\nabla^2 \bar{h}^{00} = -2 \Lambda.
\end{equation}
Hence, the monopole source with which the radial expansion is modelled is determined equivalently by $\Lambda$, $\rho_\Lambda$, or $p_\Lambda$.

Poisson's eq.~(\ref{cosmological_constant_as_source}) has the solution
\begin{equation}
\bar h^{00} = \frac{1}{4\pi r} \int_V 2\Lambda d^3 \vec y = \frac{2}{3} \Lambda r^2,
\end{equation}
where we integrate over the entire sphere with the radius $r$ (recalling that the ``source'' term originates from the left-hand side of eq.~(\ref{Einstein_field_equations})) and we replace the spherically symmetric source with the radius $r$ with a point source in its centre. The solution is in the Newtonian form.
Equations~(\ref{h_00_and_h_00_bar_monopole}) and (\ref{particle_acceleration_spatial_Newtonian_gauge}) give
\begin{equation}
h_{00} = \frac{1}{2} \bar h_{00} = \frac{1}{3} \Lambda r^2
\end{equation}
and
\begin{equation}
\frac{d^2 r}{dt^2} = \frac{c_0^2}{2} \frac{\partial h_{00}}{\partial r} = \frac{1}{3} c_0^2 \Lambda r.
\end{equation}
The last equation leads to the exponentially increasing radius
\begin{equation}\label{expanding_radius}
r = r_0 e^{\sqrt{\Lambda/3} c_0 t}
\end{equation}
with $r_0 = r(t=0)$, which can also be written as
\begin{equation}\label{expanding_radius_imaginary_frequency}
r = r_0 e^{-j \omega t}
\end{equation}
in terms of the imaginary angular frequency $\omega = j \sqrt{\Lambda/3} c_0$. Therefore, the condition $\omega r/c_0 \ll 1$ becomes $\sqrt{\Lambda/3} r \ll 1$.

In cosmology $R_U \approx 4.4 \cdot 10^{26}$\,m and $\Lambda \approx 1.1 \cdot 10^{-52} \text{m}^{-2}$. The second value is obtained very closely from the first one if eq.~(\ref{Laplacian_as_cosmological_constant}) is used in eq.~(\ref{cosmological_constant_as_source}) with the maximal unit amplitude of the background metric ($(2\pi/R_U)^2/2 \approx 1.02 \cdot 10^{-52}\text{m}^{-2}$). The near-field condition becomes effectively $r \ll R_U$, so the model applies to the distances much smaller than the radius of the observable universe. Moreover, $\rho_\Lambda \approx 0.6 \cdot 10^{-26} \text{kg/m}^3$ from eq.~(\ref{dark_energy_density}) with $k=8\pi$, which is supposedly much smaller than $\rho_0$ (see section~\ref{ch:acoustic_Lagrangians}). Using the Hubble radius, $R_H \approx R_U/3.4$, as the Schwarzschild radius of the source (for no mass-energy can move with speed higher than $c_0$) and $M = 4 \rho_H R_H^3 \pi/3$ with $\rho_H$ the Hubble energy density, eq.~(\ref{Schwarzschild_radius_mass}) gives
\begin{equation}\label{gravitational_constant_and_Schwarzschild_radius_universe}
\rho_H = \frac{3 c_0^2}{8G R_H^2 \pi} \approx 0.96 \cdot 10^{-26} \text{kg/m}^3.
\end{equation}
In fact, the dark energy constitutes around 68\% of the total energy density. Finally, $(\sqrt{\Lambda/3}c_0)^{-1} \approx 5.5 \cdot 10^{17}$\,s, which is close to the Hubble time $t_H \approx 4.55 \cdot 10^{17}$\,s, for $r = r_H$ the Hubble radius in eq.~(\ref{expanding_radius}).

It is interesting to note that the source $\rho_\Lambda$ is a monopole source of increasing entropy (see section~\ref{source_of_entropy}). As such, it is entirely in alignment with Lighthill's analogy, as the only aeroacoustic source in the absence of boundaries of the acoustic spacetime, besides the quadrupole source from sections~\ref{ch:aeroacoustic_sound_generation} and \ref{ch:acoustic_binary}. Recalling the discussion preceding eq.~(\ref{Einstein_field_equations_linear_low_indices_solution_outgoing_wave_acoustic_00}), the compact (non-relativistic) monopole source is naturally characterized by a dimensionless injected volume fraction due to the expansion, $\beta$, which is according to eq.~(\ref{expanding_radius})
\begin{equation}\label{injected_volume_fraction_as_source}
\beta = e^{\sqrt{3 \Lambda} c_0 t}.
\end{equation}
Thus, the increasing entropy of the universe, its expansion, and boundedness appear to be closely related.

It is clear from above that $\rho_\Lambda$ does not follow in any way from the background value $\rho_0$, even though $\rho_\Lambda$ is also constant (but representing expansion in time which is observed only in the near field and sets the minimum frequency of the theory in eq.~(\ref{lowest_frequency_universe})).
\iffalse
Equation~(\ref{metric_tensor_small_perturbation_Lorenz_gauge_inverted}) gives further
\begin{equation}
h_{00} = \bar h_{00} - \frac{1}{2} \eta_{00} \bar h^\nu{}_\nu = \bar h_{00} + \frac{1}{2} \eta^{\mu\nu} \bar h_{\mu\nu} = \bar h_{00} + \frac{1}{2} (-\bar h_{00} + 3 (-\bar h_{00})) = -\bar h_{00},
\end{equation}
so
\begin{equation}
\nabla^2 h^{00} = \frac{2kG}{c_0^2} \rho,
\end{equation}
which is already in the Newtonian gauge.

In fact, the very low-frequency source can be modelled as
\begin{equation}
\rho_\Lambda = \frac{\rho_0 \Lambda}{c_0^2} \int_{V_U} \frac{\partial^2 r}{\partial t^2} d^3 \vec x,
\end{equation}
where $\beta$ denotes a dimensionless injected volume fraction due to the expansion.
\fi
The non-zero cosmological constant and dark energy are a direct consequence of the expansion of the finite spacetime, not the dynamics of any form of matter or energy. Consequently, they do not have counterparts in an unbounded acoustic spacetime, for which angular frequencies can be arbitrarily small and in which far field exists outside every source region. From practical point of view, the cosmological constant is usually set to zero in acoustically relevant calculations of gravitational radiation on all scales much smaller than the cosmological scale, so this difference between the two theories is rarely significant.

In contrast to general relativity, the background fluid density $\rho_0$ relevant for classical acoustics can be measured. The acoustic vacuum without boundaries consists of the same particles (fluid molecules exhibiting random motion at the microscopic scale) inside an aeroacoustic source region as well as outside it, in the propagation region or an undisturbed fluid. Boundaries of the source region are often fuzzy and cannot be determined exactly. On the other hand, the mass-energy of sources and pure vacuum appear to be more clearly separated in general relativity (the two sides of equations~(\ref{Einstein_field_equations}) and (\ref{Einstein_field_equations_linear})). However, the model of vacuum consisting of elementary particles of spacetime is entirely in accordance with the lattice theories of quantum gravity\cite{Rovelli2015}. These particles, which are not accessible to the classical field theories, give rise to the steady density $\rho_0$, as will be shown with the aid of the quantum thermodynamic analogy in section~\ref{ch:acoustic_Lagrangians}. But only a macroscopic disturbance involving a large number of these particles can be observed, either as a moving mass-energy of the sources or a metric perturbation.

\section{Analogy with covariant electromagnetism}\label{ch:analogy_with_EMG}

So far we have seen how acoustic scalar fields in fluids can be calculated based on the analogy with the second-order tensor fields of general relativity. The analogy appeared to be particularly useful for treating quadrupole sources of sound in incompressible (compact, non-relativistic) flows within the linearized theory, but unnecessarily complicated for monopole and dipole sources, which are typically associated with scalar and vector fields, respectively. Therefore, before referring to the micro-scale quantum theories it is reasonable to establish a classical acoustic analogy with the linearized vector theory of electromagnetism, in particular its Lorentz-invariant form -- covariant electromagnetism. The kinematic analogy is expected to be suitable for dipole sources in incompressible flows. Holding in the same non-relativistic limit as gravitoelectromagnetism, the analogy can be seen as the scalar-field extension to a more general gravito-electromagnetic-acoustic theory. 

Like the metric tensor in gravitation, the four-vector potential $\bar{A}^\mu = [V/c_0, \vec A]$ is the central quantity in the covariant formulation of classical electromagnetism. The components $V$ and $\vec A$ are called scalar and vector potential, respectively, and $c_0$ is the speed of light. Since classical electromagnetism is already a linear theory, $\bar A^\mu$ is to be compared with $\bar h^{\mu\nu}$ from the linearized Einstein field equations, rather than a general metric $g^{\mu\nu}$ (or $h^{\mu\nu}$; as we will see shortly, $\bar A^\mu$ satisfying the wave equation satisfies also the Lorenz gauge condition). This is emphasized here with the bar above the symbol $A$, which is usually omitted in literature.

An analogy between charge-governed electromagnetism and mass-governed fluid dynamics and classical acoustics can be effectively established by comparing the equations of motion. Charged particles in electromagnetic fields move according to the Lorentz force law~\cite{Griffiths2013},
\begin{equation}\label{Lorentz_force}
	\vec f = -q\nabla V - q\frac{\partial \vec A}{\partial t} - q (\nabla \times \vec A) \times \vec v,
\end{equation}
where $\vec f$ is force per unit volume and $q$ is charge density. This matches Crocco's form of the momentum equation in fluid dynamics \cite{Hirschberg2018} in the non-relativistic regime ($|\vec v| \ll c_0$ with $c_0$ the speed of sound) of an essentially incompressible fluid,
\begin{equation}\label{Crocco_momentum_equation}
	\vec f = \rho_0 \nabla B + \rho_0 \frac{\partial \vec v}{\partial t} + \rho_0 (\nabla \times \vec v) \times \vec v,
\end{equation}
where $B$ denotes the total enthalpy. A comparison of the two equations suggests the following substitution of electromagnetic with mechanical quantities\footnote{Of course, an equally valid analogy could be established with opposite signs: $V \rightarrow -B$, $\vec A \rightarrow -\vec v$, $\bar{A}^\mu = [V/c_0, \vec A] \rightarrow [-B/c_0, -\vec v]$, and $q \rightarrow \rho_0$. We opt for the convention above.}: $V \rightarrow B$, $\vec A \rightarrow \vec v$, and therefore $\bar{A}^\mu = [V/c_0, \vec A] \rightarrow [B/c_0, \vec v]$ and $-q \rightarrow \rho_0$ representing fluid rather than charge density. As will become evident later, occurrence of both positive and negative mass is physically justified. It only determines the sign of the source term (or direction of the velocity vector of a dipole source) in the incompressible flow. As with general relativity, the constant $\rho_0$ of the background fluid remains external to the analogy with electromagnetism. In the specific case of a motionless background fluid, small perturbation of enthalpy can be expressed in terms of pressure as $B = p/\rho_0$, while in a homentropic flow $p = \rho c_0^2$, so $\bar{A}^\mu = [V/c_0, \vec A] \rightarrow [\rho c_0/\rho_0, \vec v]$. This implies that the scalar potential $V$ corresponds to compressible sound waves in the propagation region of the fluid. Indeed, the four-vector potential $\bar A^\mu$ quantifies electromagnetic fields, which are caused by currents as sources.

Using the substitutions above, we can introduce the analogue electric field
\begin{equation}\label{electric_field}
	\vec E = -\nabla V - \frac{\partial \vec A}{\partial t} \rightarrow -\nabla B - \frac{\partial \vec v}{\partial t}
\end{equation}
and magnetic field
\begin{equation}\label{magnetic_field}
	\vec B = \nabla \times \vec A \rightarrow \nabla \times \vec v = \vec \omega,
\end{equation}
with $\vec \omega$ denoting the vorticity vector. Since particle motion due to sound waves in fluids is irrotational ($\vec \omega = 0 \Rightarrow \vec B = 0$), we see that sound fields are captured only by the analogue electric fields. In terms of the fields $\vec E$ and $\vec B$, eq.~(\ref{Lorentz_force}) reads
\begin{equation}\label{Lorentz_force_E_B}
\vec f = q \vec E - q \vec B \times \vec v.
\end{equation}

The two Maxwell's equations which do not include currents as sources are readily satisfied:
\begin{equation}\label{Maxwell_curl_E}
	\nabla \times \vec E + \frac{\partial \vec B}{\partial t} = 0
\end{equation}
and
\begin{equation}\label{Maxwell_div_B}
	\nabla \cdot \vec B = 0.
\end{equation}
The remaining two Maxwell's equations are obtained under the Lorenz gauge condition,
\begin{equation}\label{EMG_Lorenz_gauge}
\bar{A}^{\alpha}{}_{,\alpha} = 0
\end{equation}
or equivalently
\begin{equation}\label{EMG_Lorenz_gauge_classical}
\frac{1}{c_0^2} \frac{\partial V}{\partial t} + \nabla \cdot \vec A = 0.
\end{equation}
They read
\begin{equation}\label{Maxwell_div_E}
	\nabla \cdot \vec E = -\nabla^2 V + \frac{1}{c_0^2} \frac{\partial^2 V}{\partial t^2} = -\Box V = \frac{1}{\epsilon_0} \frac{J^0}{c_0}
\end{equation}
and
\begin{equation}\label{Maxwell_curl_B}
\begin{split}
	\nabla \times \vec B - \frac{1}{c_0^2} \frac{\partial \vec E}{\partial t} &= \nabla \times (\nabla \times \vec A) - \nabla (\nabla \cdot \vec A) + \frac{1}{c_0^2} \frac{\partial^2 \vec A}{\partial t^2} \\ &= -\nabla^2 \vec A + \frac{1}{c_0^2} \frac{\partial^2 \vec A}{\partial t^2} = -\Box \vec A = \mu_0 \vec J.
\end{split}
\end{equation}
Here $J^\alpha = [J^0, \vec J] = [q c_0, q \vec v] \rightarrow [-\rho_0 c_0, -\rho_0 \vec v]$ is four-current (the source of $A^\alpha$) satisfying the conservation of charge
\begin{equation}\label{conservation_of_charge}
	J^\alpha{}_{,\alpha} = 0
\end{equation}
with $c_0^2 \epsilon_0 \mu_0 = 1$. The vector $q \vec v$ corresponds to momentum as the acoustic dipole source. Equations~(\ref{Maxwell_div_E}) and (\ref{Maxwell_curl_B}) can be merged into a single four-vector wave equation
\begin{equation}\label{Maxwell_currents_Lorenz_gauge}
	\Box \bar{A}^\mu = -\mu_0 J^\mu,
\end{equation}
which governs the linear electromagnetic theory. Notice that the current has been introduced as the source of the field in which charged particles move much like the stress-energy tensor was introduced as the source of the field in which particles with mass-energy move or its acoustic analogue in an incompressible flow as the source of the sound field in which massive particles move. Massive (charged) particles are present in both the source and propagation regions. This holds in the relativistic analogy from previous sections, with $\rho_0$ the mass-energy density of the undisturbed vacuum.

For non-relativistic motion in the source region, $|\vec v| \ll c_0$ and the component $J^0$ dominates. Maxwell's equations become the scalar wave equation
\begin{equation}\label{Maxwell_currents_Lorenz_gauge_non_relativistic}
\Box \bar{A}^0 = -\mu_0 J^0.
\end{equation}
In the near field of the source ($\omega r/c_0 \ll 1$), they reduce to Poisson's equation for electrostatics,
\begin{equation}\label{Maxwell_currents_Lorenz_gauge_non_relativistic_near_field}
\nabla^2 \bar{A}^0 = -\mu_0 J^0,
\end{equation}
and the Lorentz invariance is violated. Hence, Coulomb’s electrostatics corresponds to acoustic near field in the same manner as Newtonian gravity in eq.~(\ref{Einstein_field_equations_linear_non-relativistic_source_near_field}). The charge $q$ can even vary over time, as long as the near-field condition is satisfied. As usual, an acoustic monopole is associated with an unsteady $\rho$ (as in equations~(\ref{Einstein_field_equations_linear_non-relativistic_source}) and (\ref{Einstein_field_equations_linear_non-relativistic_source_near_field})) or, in effect, volume injection, not with the positive constant $\rho_0$, so both positive and negative charge are physical in the analogy.

We can now compare equations~(\ref{Maxwell_currents_Lorenz_gauge}), (\ref{EMG_Lorenz_gauge}), and (\ref{conservation_of_charge}) against equations~(\ref{Einstein_field_equations_linear}), (\ref{Lorenz_gauge}), and \ref{conservation_laws}. If we substitute the constant on the right hand side of the wave equation as $2kG/c_0^2 \rightarrow K\mu_0$ (with $K$ a dimensionless constant analogue to $k$ which is to be determined later in section~\ref{ch:pulsating_and_oscillating_sphere_EMG}; analogously to $G$, $\mu_0$ relates charge and length), the non-relativistic stress-energy tensor from eq.~(\ref{eq:stress-energy_tensor_perfect_fluid_components_non-relativistic}) obtains the following analogue form in nearly flat spacetime:
\begin{equation}\label{eq:stress-energy_tensor_perfect_fluid_components_EMG}
T^{\mu\nu} \rightarrow 
\begin{bmatrix}
-c_0 J^0 & -c_0 J^1 & -c_0 J^2 & -c_0 J^3\\
-c_0 J^1 & T^{11} & T^{12} & T^{13}\\
-c_0 J^2 & T^{12} & T^{22} & T^{23}\\
-c_0 J^3 & T^{13} & T^{23} & T^{33}
\end{bmatrix}
\end{equation}
with $T^{\mu0} = T^{0\mu} \rightarrow -c_0 J^\mu$, while a weak metric perturbation in Lorenz gauge can be written consequently as
\begin{equation}\label{eq:metric_perturbation_components_EMG}
\bar{h}^{\mu\nu} \rightarrow 
\begin{bmatrix}
-\bar{A}^0/c_0 & -\bar{A}^1/c_0 & -\bar{A}^2/c_0 & -\bar{A}^3/c_0\\
-\bar{A}^1/c_0 & \bar{h}^{11} & \bar{h}^{12} & \bar{h}^{13}\\
-\bar{A}^2/c_0 & \bar{h}^{12} & \bar{h}^{22} & \bar{h}^{23}\\
-\bar{A}^3/c_0 & \bar{h}^{13} & \bar{h}^{23} & \bar{h}^{33}
\end{bmatrix}
\end{equation}
with $\bar{h}^{\mu 0} = \bar{h}^{0\mu} \rightarrow -\bar{A}^\mu/c_0$. In other words, $\bar A^\mu$ describes propagating waves and corresponds to the first row/column of $\bar h^{\mu\nu}$, while the source-current $J^\mu$ corresponds to the first row/column of $T^{\mu\nu}$. As in equations~(\ref{eq:stress-energy_tensor_perfect_fluid_components}) and (\ref{eq:stress-energy_tensor_perfect_fluid_components_non-relativistic}), we notice that $J^k$ corresponds to a dipole source in an inviscid fluid and $J^0$ (with unsteady charge density or volume) to an acoustic monopole. Again, both sides of the wave equation~(\ref{Maxwell_currents_Lorenz_gauge}) are conserved (equations~(\ref{EMG_Lorenz_gauge}) and (\ref{conservation_of_charge})) and we do not split the equations of fluid dynamics into acoustic source and propagation terms. The Newtonian gauge will be used again to extract the acoustically relevant component from the solution.

However, the substitutions in equations~(\ref{eq:stress-energy_tensor_perfect_fluid_components_EMG}) and (\ref{eq:metric_perturbation_components_EMG}) hold only in the contravariant vector space. In the framework of general relativity the analogue components $A^\mu$ and $J^\mu$ transform as rows and columns of the second-order tensors, not as four-vectors. With the adopted signature $[-+++]$ the difference becomes evident for the components $_0$ in the covariant vector space and it reflects in the sign of the charge. For example, in flat spacetime $\bar{h}_{0 0} = \eta_{0 0} \bar{h}^{0 0} \eta_{0 0} = \bar{h}^{0 0}$, while for a four-vector $A_0 = \eta_{00} A^0 = -A^0$, with the opposite sign of the charge (or the slowly varying source function). Thus, in the acoustic analogy the observer sets not only the preferred Newtonian frame of reference, but also the contravariant vector space in which the analogue charge is observed. This should be taken into account when using the gravitoelectromagnetic analogy for acoustic calculations but, like the choice of the preferred frame, it becomes relevant only at the end of the calculations. We also notice that a small change of coordinates ($x^\alpha \rightarrow x^\alpha + \xi^\alpha$) does not affect the order of magnitude, so $|{A}^\mu|/c_0 \sim |\bar{A}^\mu|/c_0 \ll 1$ holds for a weak acoustic perturbation in the propagation region, since $\bar{A}^\mu \rightarrow [\rho c_0/\rho_0, \vec v]$. This is equivalent to eq.~(\ref{gauging_order_of_magnitude}) of linearized general relativity. 

For the sake of completeness, we can also introduce the electromagnetic tensor
\begin{equation}\label{eq:EMG_tensor}
F^{\alpha\beta} = \bar{A}^{\beta,\alpha} - \bar{A}^{\alpha,\beta}
\end{equation}
with the components
\iffalse
\begin{equation}\label{eq:EMG_tensor_components}
F^{\alpha\beta} = 
\begin{bmatrix}
0 & E^1/c_0 & E^2/c_0 & E^3/c_0\\
-E^1/c_0 & 0 & B^3 & -B^2\\
-E^2/c_0 & -B^3 & 0 & B^1\\
-E^3/c_0 & B^2 & -B^1 & 0
\end{bmatrix},
\end{equation}
or after lowering the indices
\fi
\begin{equation}\label{eq:EMG_tensor_components_low_indices}
F^{\alpha\beta} =
\begin{bmatrix}
0 & E^1/c_0 & E^2/c_0 & E^3/c_0\\
-E^1/c_0 & 0 & B^3 & -B^2\\
-E^2/c_0 & -B^3 & 0 & B^1\\
-E^3/c_0 & B^2 & -B^1 & 0
\end{bmatrix}.
\end{equation}
Its relations with the electric and magnetic fields are
\begin{equation}\label{eq:electric_field_electromagnetic_tensor}
E^i = c_0 F^{0i}
\end{equation}
and
\begin{equation}\label{eq:magnetic_field_electromagnetic_tensor}
B^i = \frac{1}{2} \epsilon^i{}_{jk} F^{jk},
\end{equation}
where $\epsilon_{ijk}$ is the Levi-Civita tensor in three spatial dimensions. The electric field is determined by the first row or column of the electromagnetic tenor, while the remaining components determine the magnetic field. Taking the divergence of eq.~(\ref{eq:EMG_tensor}) and using equations~(\ref{EMG_Lorenz_gauge}) and (\ref{Maxwell_currents_Lorenz_gauge}) gives
\begin{equation}\label{Maxwell_equations_EMG_tensor}
F^{\alpha\beta}{}_{,\beta} = \bar{A}^{\beta,\alpha}{}_{,\beta} - \bar{A}^{\alpha,\beta}{}_{,\beta} = -\Box \bar{A}^{\alpha} = \mu_0 J^{\alpha}.
\end{equation}

Similarly as in general relativity, the Lorenz gauge condition from eq.~(\ref{EMG_Lorenz_gauge}) reduces the number of unknown components in eq.~(\ref{Maxwell_currents_Lorenz_gauge}) from four to three. The number can be reduced further to two (one electric and one magnetic component) by selecting a specific Lorenz gauge. However, rather than pursuing the description of transverse electromagnetic waves, we shall (as in section~\ref{ch:sound_wave_in_spacetime}) analyse kinematics of
particles affected by the field. To this end we notice that eq.~(\ref{Lorentz_force_E_B}) is the spatial part of the equation of motion in covariant form,
\begin{equation}\label{geodesic_equation_curved_spacetime_EMG}
\rho_0 \frac{dU^\alpha}{d\tau} = q \eta^{\alpha\beta} F_{\beta\mu} U^\mu,
\end{equation}
which, indeed, describes motion of non-relativistic charged particles in a flat background spacetime. With the analogy $q \rightarrow -\rho_0$, it can be compared with eq.~(\ref{geodesic_equation_curved_spacetime}). In the lowest-order approximation
\begin{equation}\label{particle_acceleration_EMG}
\frac{dU^\alpha}{d\tau} = -\eta^{\alpha\beta} F_{\beta 0} = \eta^{\alpha\beta} (\bar{A}_{\beta,0} - \bar{A}_{0,\beta}).
\end{equation}
As already anticipated, particle motion in the acoustic analogy depends only on the first row/column of $F_{\alpha\beta}$, that is, the electric field. The three-dimensional acceleration of a non-relativistic particle ($\tau \approx t$) is thus
\begin{equation}\label{particle_acceleration_spatial_Lorenz_gauge_EMG}
\frac{d^2 x^k}{dt^2} = c_0 \eta^{k l} (\bar{A}_{l,0} - \bar{A}_{0,l}).
\end{equation}
Switching to a general $A_\alpha$ with $A_l = \bar{A}_l$ and $A_0 = \bar{A}_0/2$, which follow from eq.~(\ref{metric_tensor_small_perturbation_Lorenz_gauge_inverted}) and $h^{\alpha 0} \rightarrow -A^\alpha/c_0$ (but not $h^{\alpha 0} = h_{\alpha 0} \rightarrow -A_\alpha/c_0$, according to the above) with the trace equal to $\bar{A}^0/c_0 = -\bar{A}_0/c_0$, we obtain
\begin{equation}\label{particle_acceleration_spatial_EMG}
\frac{d^2 x^k}{dt^2} = c_0 \eta^{k l} (A_{l,0} - 2A_{0,l}).
\end{equation}
This can be compared\footnote{Absence of the factor 1/2 on the right-hand side (or a term symmetric to $A_l$), the appearance of the factor 2 in the last term after switching to the Newtonian gauge, as well as our convention $q \rightarrow -\rho_0$ will lead to the value $K = -4$ later.} with eq.~(\ref{particle_acceleration_spatial}). Finally, the component relevant for an acoustic observer is expressed in the Newtonian gauge, in which vorticity and transverse motion are suppressed and the Lorentz invariance is broken. Supposing $|A_l| \ll |A_0|$,
\begin{equation}\label{particle_acceleration_spatial_Newtonian_gauge_EMG}
\frac{d^2 x^k}{dt^2} = - 2 c_0 A_0{}^{,k},
\end{equation}
and $A_0$ is not necessarily of the same leading order as $\bar A_\mu$ and $A_\mu$, since it alone does not satisfy the Lorenz gauge condition and does not capture the transverse waves. As usual, the particle motion due to a longitudinal sound wave is expressed by means of the gradient of a single scalar, $A_0$. Recalling eq.~(\ref{particle_acceleration_spatial_Newtonian_gauge}) and the discussion which follows it, this completes the description of sound waves within the analogy with covariant electromagnetism and we can switch to the analysis of sources.

\subsection{Aeroacoustic dipole}\label{ch:aeroacoustic_sound_generation_EMG}

First we derive the scaling law for power of an aeroacoustic dipole in an incompressible flow. The generated sound is typically due to a turbulent low Mach number flow acting on a foreign solid body in the source region. The derivation is very similar to that for an aeroacoustic quadrupole in section~\ref{ch:aeroacoustic_sound_generation}, but based on the established analogy with the vector theory of electromagnetism. The exact solutions for point monopole and dipole sources will be obtained in section~\ref{ch:pulsating_and_oscillating_sphere_EMG}. Unlike a relativistic quadrupole, an acoustic dipole can appear only at the physical boundaries of the fluid, in the absence of external forces. (As opposed to the term in eq.~(\ref{Lighthill_analogy}) involving the tensor $\rho \vec v \vec v$, dipole source terms do not follow from splitting the conservation laws of fluid dynamics\footnote{The last free-space term in eq.~(\ref{Lighthill_analogy}) involving the scalar $p - \rho c_0^2$ captures monopole sound generation due to entropy change. It will come into focus in the thermodynamic analogy in section~\ref{ch:acoustic_Lagrangians}.}.) Effects of the boundaries in the source region\footnote{Boundary conditions in the wave propagation region will not be considered in the electromagnetic analogy presented here.}, which are captured the most generally and efficiently by the Ffowcs Williams and Hawkings equation, will be examined in more detail in section~\ref{ch:energy_and_charge}.

As in section~\ref{ch:aeroacoustic_sound_generation}, we observe an isolated source of waves (captured by $\bar A_\alpha$) in free space (a quiescent background fluid) and we formally split the source ($J_\alpha$) into the effective ($J^\text{eff}_\alpha$) and weak part ($j_\alpha$). Only the latter part is responsible for the radiation of waves. Solution of the equation
\begin{equation}\label{Maxwell_currents_Lorenz_gauge_split}
\Box \bar{A}_\alpha = -\mu_0 (J^\text{eff}_\alpha + j_\alpha)
\end{equation}
reads
\begin{equation}\label{Maxwell_currents_Lorenz_gauge_split_solution}
\bar{A}_\alpha = \frac{\mu_0}{4\pi} \int \frac{[J^\text{eff}_\alpha + j_\alpha]_{(t-\epsilon R/c_0)}}{R} d^3 \vec y
\end{equation}
with $R$ and $\epsilon$ (and several other quantities in the following) defined as in section~\ref{ch:aeroacoustic_sound_generation}. For a compact source with the characteristic length scale $L \ll c_0/\omega$ (or, equivalently, for a non-relativistic motion in the source region or incompressible flows, $|\vec v| \ll c_0$) and far geometric field, $R \gg L$,
\begin{equation}\label{Maxwell_currents_Lorenz_gauge_split_solution_far_geometric_field}
\bar{A}_\alpha = \frac{\mu_0}{4\pi r} \int [J^\text{eff}_\alpha + j_\alpha]_{(t-\epsilon r/c_0)} d^3 \vec y.
\end{equation}
From eq.~(\ref{conservation_of_charge})
\begin{equation}\label{dipole_moment_identity}
	\frac{d}{dt} d_j = \frac{1}{c_0} \frac{d}{dt} \int (J^\text{eff}_0 + j_0) x_j d^3 \vec x = \int (J^\text{eff}_j + j_j) d^3 \vec x
\end{equation}
with the introduced dipole moment
\begin{equation}\label{dipole_moment}
	d_j = \frac{1}{c_0} \int (J^\text{eff}_0 + j_0) x_j d^3 \vec x.
\end{equation}
Equation~(\ref{dipole_moment_identity}) follows from the conservation of charge and relates the spatial components responsible for dipole radiation with the acoustically relevant temporal component. The product with $x_j$ gives the dipole directivity pattern. The time derivative replaces a divergence of the force vector in the source region, which is usual in aeroacoustic analogies and makes acoustically compact dipoles inefficient sources of longitudinal waves. Spatial part of the solution can then be expressed in terms of the dipole moment as
\begin{equation}\label{Maxwell_currents_Lorenz_gauge_split_solution_far_geometric_field_dipole_moment_j}
\bar{A}_j = \frac{\mu_0}{4\pi r} \frac{d}{dt} d_j(t-\epsilon r/c_0).
\end{equation}

In order to estimate the sound energy radiated into the far field, we expand the last result in powers of $r$ for $\omega r/c_0 \ll 1$ in the acoustic near field. This gives
\begin{equation}\label{Maxwell_currents_Lorenz_gauge_split_solution_far_geometric_field_dipole_moment_j_Taylor_expansion}
\bar{A}_j (t-\epsilon r/c_0) = \bar{A}_j (t) - \frac{\epsilon r/c_0}{1!} \frac{d}{dt} \bar{A}_j (t) + \frac{(\epsilon r/c_0)^2}{2!} \frac{d^2}{dt^2} \bar{A}_j (t) - \frac{(\epsilon r/c_0)^3}{3!} \frac{d^3}{dt^3}  \bar{A}_j (t) + ...
\end{equation}
Leaving only the terms with odd powers of $\epsilon r/c_0$ (associated with waves) and with $\epsilon = 1$ for outgoing waves, we obtain spatial components of the radiation reaction potential
%What do the omitted (even-powers) near-field terms correspond to in electromagnetism? Some very near field correction of particle charge?
\begin{equation}\label{Maxwell_currents_Lorenz_gauge_split_solution_far_geometric_field_dipole_moment_radiation_j_Taylor_expansion}
\bar{A}^\text{react}_j = - \frac{\mu_0}{4 \pi c_0} \frac{d^2}{dt^2} d_j (t) - \frac{\mu_0}{24 \pi c_0^3} r^2 \frac{d^4}{dt^4} d_j (t) + ...
\end{equation}
Using the Lorenz gauge condition, eq.~(\ref{EMG_Lorenz_gauge}), from which it follows
\begin{equation}
	\bar{A}_{0,0} = -\bar{A}^0{}_{,0} = \bar{A}^j{}_{,j} = \bar{A}_{j,j},
\end{equation}
we calculate the remaining component
\begin{equation}\label{Maxwell_currents_Lorenz_gauge_split_solution_far_geometric_field_dipole_moment_radiation_0_Taylor_expansion}
\bar{A}^\text{react}_0 = - \frac{\mu_0}{8 \pi c_0^2} x^j \frac{d^3}{dt^3} d_j (t) - \frac{\mu_0}{96 \pi c_0^4} r^2 x^j \frac{d^5}{dt^5} d_j (t) + ...
\end{equation}
In order to switch to the Newtonian gauge, we recall that $A_0 = \bar{A}_0/2$ and $A_l = \bar{A}_l$, so
\begin{equation}\label{Maxwell_currents_Lorenz_gauge_split_solution_far_geometric_field_dipole_moment_radiation_0_Taylor_expansion_transformed}
A^\text{react}_0 = - \frac{\mu_0}{16 \pi c_0^2} x^j \frac{d^3}{dt^3} d_j (t) - \frac{\mu_0}{192 \pi c_0^4} r^2 x^j \frac{d^5}{dt^5} d_j (t) + ...
\end{equation}
and $A^\text{react}_j = \bar{A}^\text{react}_j$.

Finally, we notice\cite{Landau2000} that (unlike $\bar{h}^\text{react}_{0j}$ in eq.~(\ref{Einstein_field_equations_linear_split_solution_far_geometric_field_quadrupole_moment_tensor_radiation_0j})) $\bar{A}^\text{react}_j = A^\text{react}_j$ does not contain the coordinates explicitly. Therefore, the magnetic field must be zero, according to eq.~(\ref{magnetic_field}), and the component $A^\text{react}_0$ already describes irrotational motion due to acoustic radiation, without any further change of coordinates. The gauge is Newtonian, even though the components $A^\text{react}_j$ are of lower order and dominant over $A^\text{react}_0$. After replacing $q \rightarrow -\rho_0$ and $\vec f = \rho_0 d\vec v/dt$, the equation of motion (eq.~(\ref{Lorentz_force_E_B})) of a particle due to the incoming wave reads
\begin{equation}\label{Lorentz_force_B_zero}
\frac{d\vec v}{dt} = - \vec E = \frac{\partial \vec A}{\partial t},
\end{equation}
where we also used eq.~(\ref{electric_field}) and the fact that $V/c_0 \ll |\vec A|$. The equality follows also from eq.~(\ref{particle_acceleration_spatial_EMG}) (but not eq.~(\ref{particle_acceleration_spatial_Newtonian_gauge_EMG}), since $|A_l| \gg |A_0|$). The force $\vec f$ which acts on the particle due to the radiation is called Abraham–Lorentz, radiation damping, or Lorentz frictional force~\cite{Landau2000}. As in the case of gravitational waves, far-field longitudinal waves associated with $A_0^\text{react}$ vanish from the theory of (stronger) transverse electromagnetic waves, which satisfy the Lorenz gauge condition, eq.~(\ref{EMG_Lorenz_gauge}).

Scaling law for the dipole source can now be readily derived. According to equations~(\ref{dipole_moment}) and (\ref{Maxwell_currents_Lorenz_gauge_split_solution_far_geometric_field_dipole_moment_radiation_0_Taylor_expansion_transformed}), $d_j \sim q L^4$ and $A^\text{react}_0 \sim \mu_0 q \omega^3 L^5 /c_0^2$, respectively, which together with eq.~(\ref{Lorentz_force_B_zero}) gives
\begin{equation}\label{dipole_ac_velocity_scaling}
|\vec v_{\text{ac}}| \sim \frac{k c_0}{K} \left( \frac{\omega L}{c_0} \right)^3,
\end{equation}
where we again replaced $q$ with $-\rho_0$, $\mu_0$ with $2kG/(K c_0^2)$, and for $G$ we used eq.~(\ref{gravitational_constant_and_Schwarzschild_radius_dipole}). Such scaling agrees with the result in eq.~(\ref{particle_acceleration_spatial_Newtonian_gauge_dipole}), since $\omega r \sim \omega L \sim |\vec v|$ for the compact oscillating sphere as an ideal acoustic dipole and $\vec v$ its velocity. Neglecting the factor $k/K$, the intensity scales as
\begin{equation}\label{dipole_ac_power_scaling}
|\vec I| \sim \rho_0 c_0 |\vec v_\text{ac}|^2 \sim \rho_0 c_0^3 \left( \frac{|\vec v|}{c_0} \right)^6
\end{equation}
and we have obtained the 6$^\text{th}$-power law for the aeroacoustic dipole, which was already anticipated at the end of section~\ref{ch:aeroacoustic_sound_generation}. In principle, the same result could be derived using the linearized Einstein field equations with the stress-energy tensor from eq.~(\ref{eq:stress-energy_tensor_perfect_fluid_components_EMG}) as the source, with the four-current in the first row/column, and following the procedure in section~\ref{ch:oscillating_sphere}. Here we utilized the simpler electromagnetic analogy for the vector fields. The only reference to the theory of relativity is in the final replacement of constants.

Similarly like the linearized Einstein field equations, Maxwell equations can be used (even more efficiently) for acoustic calculations involving free-space compact dipole sources in incompressible flows. As in section~\ref{ch:aeroacoustic_sound_generation}, we did not need to split the conserved current $J_\alpha$ into sound generation and propagation terms, since the generated waves are described by the four-vector potential on the left-hand side of eq.~(\ref{Maxwell_currents_Lorenz_gauge}) and the full vector current is treated as the source. Usual contraction of the source vector with divergence is replaced by the time derivative in eq.~(\ref{dipole_moment_identity}), which makes the longitudinal sound waves much weaker than the transverse electromagnetic waves for compact sources. Indeed, eq.~(\ref{Maxwell_currents_Lorenz_gauge_split_solution_far_geometric_field_dipole_moment_j}) implies that the amplitude of the transverse waves scales as (with  $\mu_0 \sim k/(K \rho_0 L^2)$, $d_j \sim \rho_0 L^4$, $r \sim L$, and $d/dt \sim \omega$)
\begin{equation}\label{Maxwell_currents_Lorenz_gauge_split_solution_far_geometric_field_dipole_moment_j_scaling}
|\vec v| \sim |\bar{A}_j| \sim \frac{k c_0}{K} \left( \frac{\omega L}{c_0} \right),
\end{equation}
which is larger than the amplitude in eq.~(\ref{dipole_ac_velocity_scaling}) by the factor $(\omega L/c_0)^{-2}$. The same factor was obtained in equations~(\ref{gravitational_strain_scaling}) and (\ref{acoustic_component_metric_perturbation_scaling}). Even though $|\bar{A}_j|/c_0$ is stronger than $|\bar{h}_{jk}|$ by $(\omega L/c_0)^{-1}$, the factor remained unchanged, because the component $A^\text{react}_0$ is not weakened by an additional change of coordinates like $h^\text{react}_{00}$. The cause of the factor is again to be found in the longitudinal character of sound waves in fluids, which satisfy the Helmholtz rather than the Laplace equation, as will be shown in the next section.

Like the linear relativistic analogy, the electromagnetic analogy allows calculations of sound fields, in particular when these are generated by dipole sources. As already mentioned, in contrast to quadrupole sources, acoustic dipoles do not appear in unbounded fluids alone nor the associated terms follow directly from the conservation laws of fluid dynamics. The dipoles typically form as a reduction of quadrupole sources in turbulent flows (section~\ref{ch:aeroacoustic_sound_generation}) at acoustically compact bodies with rigid surfaces. The constrained turbulent flow has to satisfy additional boundary conditions at the rigid surfaces, which turns the kinematic quadrupole mechanism of sound generation into more efficient dipole. The resulting forces acting on the foreign body introduce external forces which correspond to the current $q \vec v \rightarrow -\rho_0 \vec v$. The Ffowcs Williams and Hawkings aeroacoustic analogy provides an exact mathematical formulation of this process. But this also means that the analogies established above permit a relation between the two types of sources -- the stress-energy tensor and four-current -- in which the latter is a reduction of the former at a boundary in the background medium. This relation is the topic of section~\ref{ch:energy_and_charge}. Next we refer once again to the analytical solutions for pulsating and oscillating compact spheres, which shall provide further insights, beyond the power scaling law, as well as the exact value of the constant $K$.

\subsection{Pulsating and oscillating sphere}\label{ch:pulsating_and_oscillating_sphere_EMG}

The exact solutions for compact pulsating and oscillating spheres can be obtained with the same procedure as in sections~\ref{ch:pulsating_sphere} and \ref{ch:oscillating_sphere}, after replacing $\bar{h}_{\alpha 0}$ and $T^\text{eff}_{\alpha 0} + t_{\alpha 0}$ with appropriate components of the four-vector potential and four-current, according to equations~(\ref{eq:stress-energy_tensor_perfect_fluid_components_EMG}) and (\ref{eq:metric_perturbation_components_EMG}), while $\bar{h}_{jk} = 0$, $T^\text{eff}_{j k} + t_{j k} = 0$, and $2kG/c_0^2$ is replaced by $K\mu_0$. However, it is easier to mimic the procedure in the electromagnetic analogy, without referring to the second-order tensors. To this end we solve eq.~(\ref{Maxwell_currents_Lorenz_gauge_split}) in frequency domain by supposing a compact source of waves, four-current of the form
\begin{equation}\label{current_simple_ocsillations}
j^{\alpha} = S^{\alpha} e^{-j \omega t},
\end{equation}
and an outgoing spherical wave in the far field
\begin{equation}\label{Maxwell_equations_linear_low_indices_solution_outgoing_wave}
\bar{A}^{\alpha} = \frac{C^{\alpha}}{r}e^{-j \omega (t-r/c_0)}.
\end{equation}
After cancelling the time-dependence term on both sides of the wave equation~(\ref{Maxwell_currents_Lorenz_gauge_split}), we obtain a four-vector form of the Helmholtz equation,
\begin{equation}\label{Maxwell_equations_linear_low_indices_Helmhotlz}
[ (\omega/c_0)^2 + \nabla^2 ] \left( \frac{C^{\alpha}}{r}e^{j \omega r/c_0} \right) = -\mu_0 S^{\alpha}.
\end{equation}

Integration over the compact spherical source region with radius $L$ ($\omega L/c_0 \ll 1$) leaves three terms on the left-hand side of the equation:
\begin{equation}\label{Maxwell_equations_linear_low_indices_Helmhotlz_integral_L}
\int_V{\frac{\omega^2}{c_0^2} \frac{C^{\alpha}}{r}e^{j \omega r/c_0} d^3 \vec y} = \frac{4 \pi}{3} \left( \frac{\omega L}{c_0} \right)^2 C^{\alpha}
\end{equation}
and
\begin{equation}\label{Maxwell_equations_linear_low_indices_Helmhotlz_integral_T}
\begin{aligned}
\int_V{\nabla^2 \left( \frac{C^{\alpha}}{r}e^{j \omega r/c_0} \right) d^3 \vec y} = - 4 \pi C^{\alpha} + j 4 \pi \left( \frac{\omega L}{c_0} \right) C^{\alpha},
\end{aligned}
\end{equation}
where $C^{\alpha}$ is constant over the compact source region. We again associate the term in eq.~(\ref{Maxwell_equations_linear_low_indices_Helmhotlz_integral_L}) with acoustic waves. Therefore, the leading-order solution (which satisfies the Laplace rather than Helmholtz equation and describes transverse electromagnetic waves)
\begin{equation}\label{amplitude_electromagnetic_waves}
C^{\alpha} = \frac{\mu_0}{4\pi} \int_{V} S^{\alpha} d^3 \vec y
\end{equation}
should be multiplied (recalling the discussion on the terms T1, T2, and T3 in section~\ref{ch:pulsating_sphere}) with the factor $-(2/3) (\omega L/c_0)^2$ in order to obtain the acoustically relevant
\begin{equation}\label{amplitude_acoustic_waves_EMG}
C^{\alpha} = -\frac{\mu_0}{6\pi} \left( \frac{\omega L}{c_0} \right)^2 \int_{V} S^{\alpha} d^3 \vec y
\end{equation}
and from eq.~(\ref{Maxwell_equations_linear_low_indices_solution_outgoing_wave})
\begin{equation}\label{Maxwell_equations_linear_low_indices_solution_outgoing_wave_acoustic}
\bar{A}^{\alpha} = -\frac{\mu_0}{6\pi} \left( \frac{\omega L}{c_0} \right)^2 \frac{e^{j\omega r/c_0}}{r}  \int_{V} j^{\alpha} d^3 \vec y.
\end{equation}
This four-vector potential is of sub-leading order and does not have to satisfy the Lorenz gauge condition, eq.~(\ref{EMG_Lorenz_gauge}). The same scaling factor $(\omega L/c_0)^2$ appears between transverse and longitudinal waves as in the relativistic analogy, which was also concluded in section~\ref{ch:aeroacoustic_sound_generation_EMG}. Consequently, the acoustically relevant contribution is negligible in the near field and electrostatic eq.~(\ref{Maxwell_currents_Lorenz_gauge_non_relativistic_near_field}) and vanishes in a Lorenz gauge. For example, sound field of an acoustic monopole satisfies the wave equation
\begin{equation}\label{electromagnetic_analogy_acoustic_monopole}
	\Box \bar{A}^0 = \frac{2}{3} \left( \frac{\omega L}{c_0} \right)^2 \mu_0 c_0 q,
\end{equation}
not eq.~(\ref{Maxwell_currents_Lorenz_gauge_non_relativistic}). It is the leading-order term in eq.~(\ref{Maxwell_equations_linear_low_indices_Helmhotlz_integral_T}) which satisfies the Lorenz gauge condition.

In the case of a compact pulsating sphere, the essential kinematic mechanism of monopole radiation is not unsteady charge/mass density in it, but its volume. Therefore, we can use $q \rightarrow -\rho_0$ even for a monopole source and replace
\begin{equation}\label{J_0_integral}
\int_{V} j^{0} d^3 \vec y = -\rho_0 c_0 \frac{4}{3} (L+\bar{l} e^{-j\omega t})^3 \pi \approx -4 \rho_0 c_0 L^2 \pi \bar{l} e^{-j\omega t}
\end{equation}
at the leading order. The only non-zero component of $\bar{A}^{\alpha}$ is
\begin{equation}\label{Maxwell_equations_linear_low_indices_solution_outgoing_wave_acoustic_0_mass_source}
\bar{A}^{0} = \frac{2 \mu_0 \rho_0 c_0 L^2}{3} \left( \frac{\omega L}{c_0} \right)^2 \bar{l} \frac{e^{-j\omega (t-r/c_0)}}{r}.
\end{equation}
Substituting $\mu_0$ with $2kG/(K c_0^2)$ and $G$ with $3 c_0^2/(8\rho_0 L^2 \pi)$ from eq.~(\ref{gravitational_constant_monopole}) gives $\mu_0 = 3k/(4K\rho_0 L^2 \pi)$ and
\begin{equation}\label{Maxwell_equations_linear_low_indices_solution_outgoing_wave_acoustic_0_mass_source_2}
\bar{A}^{0} = \frac{k c_0}{2\pi K} \left( \frac{\omega L}{c_0} \right)^2 \bar{l} \frac{e^{-j\omega (t-r/c_0)}}{r}.
\end{equation}
Switching to
\begin{equation}\label{Maxwell_equations_linear_low_indices_solution_outgoing_wave_acoustic_0_mass_source_3}
A^{0} = \frac{\bar{A}^0}{2} = \frac{k c_0}{4\pi K} \left( \frac{\omega L}{c_0} \right)^2 \bar{l} \frac{e^{-j\omega (t-r/c_0)}}{r}
\end{equation}
(which is already in the Newtonian gauge) and using eq.~(\ref{particle_acceleration_spatial_Newtonian_gauge_EMG}) gives
\begin{equation}\label{particle_acceleration_spatial_Newtonian_gauge_EMG_pulsating_sphere}
v_\text{ac}^r = -\frac{k}{2\pi K} c_0 \left( \frac{\omega L}{c_0} \right)^2 \bar{l} \frac{e^{-j\omega (t-r/c_0)}}{r}
\end{equation}
at the leading order of $1/r$. This matches the result in eq.~(\ref{longitudinal_wave_acceleration_pulsating_sphere}) for $K = -4$. The factor follows from the different factors on the right-hand sides of equations~(\ref{particle_acceleration_spatial_Newtonian_gauge_EMG}) and (\ref{particle_acceleration_spatial_Newtonian_gauge}) (recall also the comparison of equations~(\ref{particle_acceleration_spatial_EMG}) and (\ref{particle_acceleration_spatial})).

In the case of a compact sphere oscillating with velocity $\vec v$ ($|\vec v| \ll c_0$),
\begin{equation}\label{J_00_integral_dipole}
\int_{V} j^{j} d^3 \vec y = -\rho_0 v^j \int_{0}^{L} r dr \int_{0}^{2 \pi} d\phi \int_{0}^{L} dz = -\rho_0 L^3 \pi v^j
\end{equation}
and from eq.~(\ref{Maxwell_equations_linear_low_indices_solution_outgoing_wave_acoustic})
\begin{equation}\label{Maxwell_equations_linear_low_indices_solution_outgoing_wave_acoustic_j}
\bar{A}^{j} = \frac{\mu_0 \rho_0 L^3 }{6} \left( \frac{\omega L}{c_0} \right)^2 \frac{e^{j\omega r/c_0}}{r} v^j,
\end{equation}
while $j^0 = 0$ and $\bar{A}^0 = 0$. (See section~\ref{ch:oscillating_sphere} for clarification of symbols.) The Lorenz gauge condition is satisfied for the essentially incompressible flow around the sphere ($v^j{}_{,j} = 0$) and the acoustic component is of higher order. The sphere oscillates irrotationally, so after expressing
\begin{equation}\label{Maxwell_equations_linear_low_indices_solution_outgoing_wave_acoustic_j_scalar_potential}
	\bar{A}^{j} = \frac{\mu_0 \rho_0 L^3 }{6} \left( \frac{\omega L}{c_0} \right)^2 \frac{e^{j\omega r/c_0}}{r} \phi^{,j}
\end{equation}
with the scalar velocity potential $\phi$, we can introduce a suitable change of coordinates $x^\mu \rightarrow x^\mu + \xi^\mu$ with
\begin{equation}\label{gauging_to_Newtonian_form_dipole1_EMG}
\xi^0 = -\frac{\mu_0 \rho_0 L^3 }{6 c_0} \left( \frac{\omega L}{c_0} \right)^2 \frac{e^{j\omega r/c_0}}{r} \phi
\end{equation}
and
\begin{equation}\label{gauging_to_Newtonian_form_dipole2_EMG}
\xi^j = \frac{\mu_0 \rho_0 L^3 }{6 c_0} \left( \frac{\omega L}{c_0} \right)^2 \frac{e^{j\omega r/c_0}}{r} \phi r^{,j}.
\end{equation}
Equation~(\ref{metric_tensor_small_perturbation_gauging}) determines the transformation (with $h^{\alpha 0} \rightarrow -A^\alpha/c_0$):
\begin{equation}\label{four_vector_potential_small_perturbation_gauging}
\frac{A^{\alpha}}{c_0} \rightarrow \frac{A^{\alpha}}{c_0} + \xi^{\alpha,0} + \xi^{0,\alpha}.
\end{equation}
Here
\begin{equation}
	\xi^{0,0} = -j \omega \frac{\mu_0 \rho_0 L^3 }{6 c_0^2} \left( \frac{\omega L}{c_0} \right)^2 \frac{e^{j\omega r/c_0}}{r} \phi,
\end{equation}
\begin{equation}
	\xi^{j,0} = j \omega \frac{\mu_0 \rho_0 L^3 }{6 c_0^2} \left( \frac{\omega L}{c_0} \right)^2 \frac{e^{j\omega r/c_0}}{r} \phi r^{,j},
\end{equation}
and
\begin{equation}
	\xi^{0,j} = -\frac{\bar A^j}{c_0} - j\omega \frac{\mu_0 \rho_0 L^3 }{6 c_0^2} \left( \frac{\omega L}{c_0} \right)^2 \frac{e^{j\omega r/c_0}}{r} \phi r^{,j}
\end{equation}
to the lowest order of $1/r$. Hence, in this Newtonian gauge ($A^\mu = \bar A^\mu$, according to eq.~(\ref{metric_tensor_small_perturbation_Lorenz_gauge_inverted}), since $\bar{A}^0$ determines the trace $\bar{h}^\mu {}_\mu$ and equals zero)
\begin{equation}\label{Maxwell_equations_linear_low_indices_solution_outgoing_wave_acoustic_dipole_0}
A^{0} = 2 c_0\xi^{0,0} = j \frac{k }{8 \pi K} \left( \frac{\omega L}{c_0} \right)^3 \frac{e^{j\omega r/c_0}}{r} \phi,
\end{equation}
and $A^j = 0$. We also replaced $\mu_0 \rightarrow 2kG/(K c_0^2) = -3k/(8K \rho_0 L^2 \pi)$ since, as in section~\ref{ch:oscillating_sphere}, $G$ takes half of the value for a pulsating sphere (eq.~(\ref{gravitational_constant_and_Schwarzschild_radius_dipole})) as well as the opposite sign (or direction of motion of the adopted negative charge). Finally, we use eq.~(\ref{particle_acceleration_spatial_Newtonian_gauge_EMG}) to obtain at the leading order of $1/r$ and $\omega L/c_0$
\begin{equation}\label{particle_acceleration_spatial_Newtonian_gauge_dipole_EMG}
v_{\text{ac}}^k = -\frac{k c_0}{4 \pi K \omega} \left( \frac{\omega L}{c_0} \right)^3 \frac{e^{j\omega r/c_0}}{r} v^k,
\end{equation}
with $\phi^{,k} = v^k$, or
\begin{equation}
	v_{\text{ac}}^r = -\frac{k c_0}{4 \pi K \omega} \left( \frac{\omega L}{c_0} \right)^3 |\vec{v}| \cos(\theta) \frac{e^{-j\omega (t-r/c_0)}}{r}.
\end{equation}
This is equal to eq.~(\ref{particle_acceleration_spatial_Newtonian_gauge_dipole}) for $K = -4$.

We can conclude that the analogy between acoustics and covariant theory of electromagnetism is governed by Maxwell's equations in the form
\begin{equation}\label{Maxwell_currents_Lorenz_gauge_acoustic_analogy}
\Box \bar{A}^\mu = -\frac{4\pi G}{c_0^2} J^\mu,
\end{equation}
with $q \rightarrow -\rho_0$ in the four-current, and the Lorenz gauge condition in eq.~(\ref{EMG_Lorenz_gauge}). It holds for a low Mach number flow in the source region or, equivalently, acoustically compact sources. Particle motion due to an incoming longitudinal sound wave is described by the component $A^0$ in the (Newtonian) frame in which the magnetic field vanishes, according to eq.~(\ref{particle_acceleration_spatial_Newtonian_gauge_EMG}). Importantly, the acoustic quantities are eventually calculated over the relativistic analogy. Components of the four-vector potential in Lorenz gauge, $\bar A^\mu$, should be expressed in contravariant vector space and transformed to the Newtonian form as a row or column of a metric perturbation tensor ($-\bar A^\mu/c_0 \rightarrow \bar h^{\mu 0} = \bar h^{0\mu}$). Both the preferred Newtonian frame and the contravariant vector space are dictated by the acoustic observer. Still, we do not need to work with full second-order tensors in calculations of monopole and dipole radiation, which can be performed more efficiently using the vector theory of electromagnetism. Furthermore, if a monopole source is dominant, the components $\bar A^j$ and $J^j$ become redundant and the usual scalar acoustic theory suffices.

Both analogies considered so far, relativistic and electromagnetic, are kinematic. In contrast to mass-energy represented by $\rho_0$, charge can be both positive or negative, signifying two opposite directions of the dipole current (vector $\rho_0 \vec v$). Finally, like the relativistic analogy, the electromagnetic analogy does not imply existence of longitudinal electromagnetic waves at the leading order. For example, near field of the considered scalar component from eq.~(\ref{electromagnetic_analogy_acoustic_monopole}) is only a very small correction of electrostatic eq.~(\ref{Maxwell_currents_Lorenz_gauge_non_relativistic_near_field}). Nevertheless, the vector analogy and Maxwell equations can be used for obtaining exact acoustic solutions, as demonstrated above.

\subsection{Relation between analogue mass-energy and charge}\label{ch:energy_and_charge}

In section~\ref{ch:sound_wave_generation_in_spacetime} we established an analogy between acoustic quadrupole source in an unbounded incompressible fluid and non-relativistic stress-energy tensor, the source of gravitational waves. In this section an analogy is found between free-space acoustic dipole in an incompressible flow and non-relativistic four-current, the source of electromagnetic waves. Since aeroacoustic dipoles are typically caused by the quadrupole sound generation mechanism in combination with fluid boundaries in the source region, the two analogies invoke a relation between the analogue sources of waves. In particular, the well-known mechanism of reduction of an aeroacoustic quadrupole (vorticity or turbulence) to dipole (unsteady loading force) at an acoustically compact foreign body in the flow~\cite{FfowcsWilliams1969} implies a common nature of mass-energy and charge in the analogue sources. It is also the only mechanism of dipole sound generation in a fluid in the absence of external forces.

In order to describe the process of reduction, we follow Ffowcs Williams and Hawkings~\cite{FfowcsWilliams1969} and multiply eq.~(\ref{conservation_laws}) in a flat background acoustic spacetime with the Heaviside step function $H(f)$:
\begin{equation}\label{conservation_laws_H}
	H T^{\alpha\beta}{}_{,\beta} = (H T^{\alpha\beta}{})_{,\beta} - T^{\alpha\beta} H_{,\beta} = 0,
\end{equation}
with $f(t,\vec x)$ an arbitrary smooth function of both space and time such that $f(t,\vec x) = 0$ at the supposed boundary of spacetime\footnote{In general, a control surface $f(t,\vec x) = 0$ can be defined independently of actual foreign bodies which bound the medium, but it usually matches the physical surfaces at all times $t$.}, $f(t,\vec x) < 0$ outside spacetime, and $f(t,\vec x) > 0$ inside spacetime. In this way we formally extend the validity of the conservation laws beyond the boundary of spacetime, where eq.~(\ref{conservation_laws_H}) still applies. If the boundary moves with velocity $\vec u$, then $\partial H/\partial t + \vec u \cdot \nabla H = 0$. However, $|\vec u| \ll c_0$ must hold for non-relativistic sources and the expression is not Lorentz invariant. Like acoustic receivers, spacelike boundaries of the acoustic spacetime require breaking the Lorentz invariance and dictate a preferred frame. In this frame the covariant four-vector associated with the boundary, which is normal to all vectors tangent to the boundary, has the components $[1,0,0,0]$. Moreover, $\nabla H = \nabla f \delta(f) = |\nabla f| \delta(f) \vec n$ in this frame, where $\vec n$ is unit vector normal to the boundary pointing into the space and $\delta$ denotes Dirac delta function.

The term $(H T^{\alpha\beta}{})_{,\beta}$ in eq.~(\ref{conservation_laws_H}) captures the free-space quadrupole located outside the boundary, where $H = 1$. This source was treated in section~\ref{ch:sound_wave_generation_in_spacetime} and can be neglected here, particularly in the presence of a more efficient dipole. In contrast to it, the remaining term involves $H_{,\beta} = [-\vec u \cdot \nabla H/c_0, \nabla H]$ and is non-zero only at the boundary. Integrating it over space (extended beyond the boundary) gives
\begin{equation}\label{surface_dipole_source1}
\begin{aligned}
- \int T^{\alpha\beta} H_{,\beta} d^3 \vec x &= - \int (T^{\alpha j} H_{,j} + T^{\alpha 0} H_{,0}) d^3 \vec x \\
&= - \int (T^{\alpha j} - T^{\alpha 0} u^j/c_0) |\nabla f| \delta(f) n_j d^3 \vec x \\
&= - \int_S (T^{\alpha j} - T^{\alpha 0} u^j/c_0) n_j d^2 \vec x,
\end{aligned}
\end{equation}
where the last integral is over the two-dimensional surface only. Therefore, a source attached to the boundary is added to the free-space quadrupole and it has the form of the four-vector current (recall eq.~(\ref{eq:stress-energy_tensor_perfect_fluid_components_EMG})), a dipole source,
\begin{equation}\label{surface_dipole_source_current}
J^\alpha = T^{\alpha j} n_j/c_0 - T^{\alpha 0} u^j n_j/c_0^2.
\end{equation}

After inserting the values $T^{00} = \rho_0 c_0^2$, $T^{0j} = T^{j0} = \rho_0 c_0 v^j$, and $T^{kj} = \rho_0 v^k v^j + p \delta^{kj}$ from eq.~(\ref{eq:stress-energy_tensor_perfect_fluid_components_non-relativistic}), which apply for the incompressible flow at the surface of an acoustically compact body,
\begin{equation}\label{surface_dipole_source_current_FW-H}
\begin{aligned}
J^\alpha &= [\rho_0 (v^j - u^j) n_j, \rho_0 (v^j-u^j) v^k n_j/c_0 + p \delta^{jk} n_j/c_0] \\
& = [\rho_0 (\vec v - \vec u) \cdot \vec n, \rho_0 (\vec v - \vec u)\vec v \cdot \vec n/c_0 + p \vec n/c_0].
\end{aligned}
\end{equation}
The four components correspond to the boundary source terms in the Ffowcs Williams and Hawkings equation~\cite{Hirschberg2018,FfowcsWilliams1969}. This becomes more evident after calculating the four-dimensional divergence\footnote{As already discussed, the divergence is actually replaced by the time derivative in eq.~(\ref{dipole_moment_identity}) in the considered electromagnetic analogy, in the same way as the double divergence of Lighthill's analogy is replaced by the second-order time derivative in eq.~(\ref{quadrupole_moment_tensor_identity}).}
\begin{equation}\label{surface_dipole_source_current_FW-H_divergence}
\begin{aligned}
J^\alpha{}_{,\alpha} &= (\rho_0 v^j n_j - \rho_0 u^j n_j)_{,0} + (\rho_0 (v^j n_j - u^j n_j)v^k/c_0 + p \delta^{kj} n_j/c_0)_{,k}
\\&= \frac{\partial}{\partial t} \left( \rho_0 (\vec v - \vec u) \cdot \vec n/c_0 \right) + \nabla \cdot \left( \rho_0 \vec v (\vec v - \vec u) \cdot \vec n/c_0 + p \vec n/c_0 \right).
\end{aligned}
\end{equation}
The first term represents the monopole source of thickness noise, while the second term is the dipole due to unsteady lifting force acting on the boundary, responsible for loading noise.
\iffalse
If the surface is motionless ($\vec u \cdot \vec n= 0$) and essentially flat ($\vec n$ changes much slower over $\vec x$ compared to $p$), the first-order approximation is
\begin{equation}
\begin{aligned}
J^\alpha{}_{,\alpha} = \frac{\rho_0}{c_0} \frac{\partial \vec v}{\partial t} \cdot \vec n + \frac{1}{c_0} \nabla p \cdot \vec n.
\end{aligned}
\end{equation}
The two terms are components of $(\rho_0/c_0) \partial \vec v/\partial t$ and $(\rho_0/c_0) \nabla B$ normal to the surface, with $B = p/\rho_0$ for small perturbations of enthalpy in the absence of background flow. They are analogous to the normal component of the electric field from eq.~(\ref{electric_field}) multiplied with $q/c_0 \rightarrow -\rho_0/c_0$ and after the replacements $V \rightarrow B$ and $\vec A \rightarrow \vec v$.
\fi

However, in order for the electromagnetic analogy to hold, the current in equations~(\ref{surface_dipole_source_current}) and (\ref{surface_dipole_source_current_FW-H}) must be conserved, that is, the divergence in eq.~(\ref{surface_dipole_source_current_FW-H_divergence}) must equal zero in the source region. This is the case when the boundary is rigid (impenetrable, so $\vec u \cdot \vec n = \vec v \cdot \vec n$), motionless in the selected frame ($\nabla p \cdot \vec n = 0$), and compact (so $\int_V p \nabla \cdot \vec n d^3 \vec x = p \int_V \nabla \cdot \vec n d^3 \vec x = p \int_S \vec n d^2 \vec x = 0$, since $p$ is constant over the surface of the body $S$ enclosing the volume $V$). The first and second conditions are fulfilled by a motionless solid body located in fluid, which thus bounds the acoustic medium. In terms of the analogy, they set a boundary of spacetime which, as already pointed out, violates Lorentz invariance. An aeroacoustic dipole is typically formed by sharp edges in turbulent flows. Limit of the observable universe at the Planck scale may present their relativistic analogue, as will be discussed in section~\ref{ch:mass_background_fluid}. The last condition does not introduce any new constraints, since the source was already assumed to be non-relativistic. Under these conditions the current reduces to
\begin{equation}\label{conserved_induced_current}
J^\alpha = [0,p \vec n /c_0],
\end{equation}
which is the most commonly dominant source of loading noise in bounded turbulent flows~\cite{Hirschberg2018}. The remaining part of the current in equations~(\ref{surface_dipole_source_current}) and (\ref{surface_dipole_source_current_FW-H}), which is not conserved, may still be acoustically relevant. It is of higher order for $|\vec v| \ll c_0$ (the term $\rho_0 (\vec v - \vec u) \vec v \cdot \vec n/c_0$) or responsible for monopole radiation (the component $J^0 = \rho_0 (\vec v - \vec u) \cdot \vec n$, which is considered next) and vanishes from the theory of electromagnetic waves.
\iffalse
A less rigorous alternative procedure for obtaining dipole radiation directly from the perturbation of acoustic spacetime close to the boundary (without referring to the stress-energy tensor and four-vector current as sources) is described in Appendix~A. As such, it is more compatible with the theory of loop quantum gravity.
\fi

A quadrupole source in a bounded medium can reduce to a dipole. Similarly, a dipole source can reduce to a monopole at a boundary. Multiplying eq.~(\ref{conservation_of_charge}) with the Heaviside function $H(f)$ and expanding it as above gives
\begin{equation}\label{conservation_laws_H_EMG}
H J^{\alpha}{}_{,\alpha} = (H J^{\alpha}{})_{,\alpha} - J^{\alpha} H_{,\alpha}= 0.
\end{equation}
In addition to the free-space dipole term with $HJ^\alpha$, the source term which is non-zero only at the boundary integrated over space gives
\begin{equation}\label{conservation_laws_H_surface_term_EMG}
\begin{aligned}
-\int J^{\alpha} H_{,\alpha} d^3 \vec x &= -\int (J^{0} H_{,0} + J^{j} H_{,j}) d^3 \vec x \\&= -\int (-J^{0} u^j/c_0 + J^{j}) |\nabla f| \delta(f)n_j d^3 \vec x \\
&= -\int_S (-J^{0} u^j/c_0 + J^{j}) n_j d^2 \vec x.
\end{aligned}
\end{equation}
The monopole source at the boundary is therefore expressed by the term
\begin{equation}\label{induced_monopole}
-J^{0} u^jn_j/c_0 + J^{j} n_j = q (v^j - u^j) n_j = q (\vec v - \vec u) \cdot \vec n \rightarrow -\rho_0 (\vec v - \vec u) \cdot \vec n
\end{equation}
and it coincides with the source of thickness noise. Since the boundary is stationary in the considered frame, monopole radiation does not take place if the boundary is also rigid (true boundary of the acoustic spacetime, violating the Lorentz invariance). Otherwise, the source term has the same form as $J^0$ in eq.~(\ref{surface_dipole_source_current_FW-H}), which alone does not satisfy the conservation equation $J^\alpha{}_{,\alpha} = 0$. In acoustics it is usually due to a moving body in the fluid~\cite{Hirschberg2018}. It has no relativistic counterpart, since motion of the body would require an action of external forces. The only possible analogue (free-space) monopole source is of thermodynamic nature, captured by the last term in eq.~(\ref{Lighthill_analogy}), and will be studied in section~\ref{source_of_entropy}.

The short analysis based on the Ffowcs Williams and Hawkings aeroacoustic analogy provides an effective mathematical description for reduction of quadrupole to dipole and dipole to monopole sources at the boundaries of acoustic spacetime. In the former case, analogue mass-energy acts on a boundary of the medium introduced by a compact foreign body and induces analogue charge. Surface of the body appears as charged, which makes the electromagnetic analogy suitable for calculations of fields generated by dipole sources even when they originate from quadrupole sources. In contrast to free-space currents, which are related to the stress-energy tensor via $J^\alpha = -T^{\alpha 0}/c_0$, the boundary currents take the form $J^\alpha = T^{\alpha j} n_j/c_0 - T^{\alpha 0} u^j n_j /c_0^2$. 

Both reduction mechanisms take place in incompressible flows at compact boundaries. A reduction to a conserved dipole current requires that the boundary is rigid and motionless (no external forces act to move the body) in the specific frame of reference dictated by both acoustic receivers and the boundaries. The resulting current is $J^\alpha = [0,p \vec n /c_0]$. If the boundary is not rigid, a quadrupole can reduce further to monopole, which is represented by the (non-conserved) unsteady mass $\rho_0 (\vec v - \vec u) \cdot \vec n/c_0$, rather than the free-space component $T^{00}/c_0^2$. Alternatively, an external force is required to move the body. Either way the scalar acoustic theory applies.

Adding to the discussion in section~\ref{ch:dark_energy}, we can also conclude that, if external forces are excluded, the elementary particles which build the acoustic spacetime (the fluid molecules) also make acoustic charge in the presence of rigid and compact bodies, as well as quadrupole sources. Like Newtonian mass, Coulomb’s charge is near field of the sources. However, the acoustically relevant scalar component is of higher order (recall equations (\ref{Maxwell_currents_Lorenz_gauge_non_relativistic}) and (\ref{electromagnetic_analogy_acoustic_monopole})) and longitudinal electromagnetic (and gravitational) waves are not physical. Micro-boundaries of the observable spacetime (for example at the Planck scale) are thus responsible for the mass-induced charge (as explained in this section) as well as the thermodynamic monopole source of time, expansion of the observable universe (since they introduce thermodynamic uncertainty of all observations, as will be discussed in section~\ref{source_of_entropy}), and the omnipresent source of dark energy $\rho_\Lambda$ due to increasing entropy (recall the last term of eq.~\ref{Lighthill_analogy}). On the other hand, thermodynamics of the elementary particles determines properties of the background spacetime. Value of the analogue mass-energy $\rho_0$ can be derived only at the smallest length scale of a complete theory, after a quantum analogy has been established.

\section{Summary of the classical analogies}\label{ch:unified_analogy}

Before we switch to the acoustic analogy with quantum fields it is worthwhile to summarize the essentials of the two analogies with classical field theories developed so far. They are both kinematic, valid for non-relativistic velocities of particles (with the speed of sound replacing the speed of light) in essentially incompressible and inviscid flows and for acoustically compact sources. For simplicity, we also assume that the background spacetime is flat, that is, we ignore wave propagation effects in moving or inhomogeneous fluids, such as convection and refraction. Even more so since externally controlled flows cannot be derived using the analogies, but have to be given.

Free-space aeroacoustic quadrupole radiation in fluids can be described by a weak metric perturbation tensor $\bar{h}^{\alpha\beta}$ of general relativity. At the leading order it satisfies the linearized Einstein field equations,
\begin{equation}\label{Einstein_field_equations_linear2}
\Box \bar{h}^{\alpha\beta} = -\frac{2kG}{c_0^4} T^{\alpha\beta},
\end{equation}
under the Lorenz gauge condition,
\begin{equation}\label{Lorenz_gauge2}
\bar{h}^{\alpha\beta}{}_{,\beta} = 0.
\end{equation}
In the acoustic analogy $k=-8\pi$ and the speed of sound $c_0$ couples the space and time of the acoustic spacetime. As in general relativity, the constant $G$ additionally relates mass (or $\rho_0$, the background fluid density) and length scale of the source. The symmetric stress-energy tensor $T^{\alpha\beta}$ acts as the source and satisfies the conservation laws, which in the flat background spacetime read
\begin{equation}\label{conservation_laws_Einstein_tensor2}
T^{\alpha\beta}{}_{,\beta} = 0.
\end{equation}
In incompressible flows it obtains the form from eq.~(\ref{eq:stress-energy_tensor_perfect_fluid_components_non-relativistic}). It is only the spatial components which are responsible for quadrupole radiation and in section~\ref{ch:aeroacoustic_sound_generation} we learned that a compact quadrupole source can be written in the integral form as
\begin{equation}\label{quadrupole_source}
	\int T_{jk} d^3 \vec x = \frac{1}{2c_0^2} \frac{d^2}{dt^2} \int T_{00} x_j x_k d^3 \vec x = \frac{1}{2} \frac{d^2}{dt^2} I_{jk},
\end{equation}
where $I_{jk}$ is quadrupole moment tensor.

Equation of particle motion due to a metric perturbation $h^{\alpha \beta}$ reads
\begin{equation}\label{particle_acceleration_2}
\frac{dU^\alpha}{d\tau} = -c_0 \Gamma^\alpha {}_{00} = -\frac{c_0}{2} \eta^{\alpha\beta} (h_{\beta 0, 0} + h_{0 \beta, 0} - h_{0 0, \beta}).
\end{equation}
Classical acoustic acceleration is expressed by its three-dimensional form
\begin{equation}\label{particle_acceleration_spatial_Newtonian_gauge2}
\frac{d^2 x^k}{dt^2} = \frac{c_0^2}{2} h_{00}^{,k},
\end{equation}
where
\begin{equation}\label{metric_tensor_small_perturbation_Lorenz_gauge_inverted2}
h_{\alpha\beta} = \bar h_{\alpha\beta} - \frac{1}{2} \eta_{\alpha\beta} \bar h^\nu{}_\nu - \xi_{\alpha,\beta} - \xi_{\beta,\alpha}
\end{equation}
and the transformation of coordinates $x^\alpha \rightarrow x^\alpha + \xi^\alpha$ is performed such that the Newtonian gauge is achieved, in which the particle motion is longitudinal. The preferred frame of reference is dictated by the acoustic observer. Acoustically relevant part of $h_{\alpha\beta}$ is of sub-leading order ($h_{00} \ll |\bar h_{\alpha \beta}| \sim |h_{\alpha \beta}|$) and the associated fraction of $\bar h_{\alpha\beta}$ does not satisfy eq.~(\ref{Lorenz_gauge2}). In the context of gravitation it represents a small correction of Newtonian gravity (the near field of eq.~(\ref{Einstein_field_equations_linear2})) due to wave radiation, not longitudinal gravitational waves.

Radiation of a free-space dipole is described more naturally with the electromagnetic four-vector potential $\bar{A}^\alpha$ satisfying (linear) Maxwell's equations,
\begin{equation}\label{Maxwell_currents_Lorenz_gauge2}
\Box \bar{A}^\alpha = -\mu_0 J^\alpha,
\end{equation}
and the Lorenz gauge condition,
\begin{equation}\label{EMG_Lorenz_gauge2}
\bar{A}^{\alpha}{}_{,\alpha} = 0.
\end{equation}
The analogy between (density of) charge and mass $q \rightarrow -\rho_0$ is adopted and, like $G$ above, the constant $\mu_0 \rightarrow -kG/(2 c_0^2)$ couples the mass/charge with the length scale of the source. The source current satisfies the conservation of charge in the flat spacetime,
\begin{equation}\label{conservation_of_charge2}
J^\alpha{}_{,\alpha} = 0.
\end{equation}
In incompressible flows $J^\alpha = [q c_0, q \vec v] \rightarrow [-\rho_0 c_0, -\rho_0 \vec v]$ and only the three spatial components lead to dipole radiation. Thus, opposite charges can still be represented by the strictly positive value of $\rho_0$ by changing the direction of the velocity vector $\vec v$. In section~\ref{ch:aeroacoustic_sound_generation_EMG} we saw that a compact dipole is given by
\begin{equation}\label{dipole_source}
	\int J^j d^3 \vec x = \frac{1}{c_0} \frac{d}{dt} \int J_0 x^j d^3 \vec x = \frac{d}{dt} d^j, 
\end{equation}
where $d^{j}$ is dipole moment. 

Equation of motion of a particle in the field is
\begin{equation}\label{particle_acceleration_EMG_2}
\frac{dU^\alpha}{d\tau} = -\eta^{\alpha\beta} F_{\beta 0} = \eta^{\alpha\beta} (\bar{A}_{\beta,0} - \bar{A}_{0,\beta})
\end{equation}
and the acoustic acceleration equals
\begin{equation}\label{particle_acceleration_spatial_Newtonian_gauge_EMG2}
\frac{d^2 x^k}{dt^2} = - 2 c_0 A_0^{,k},
\end{equation}
where
\begin{equation}
	\frac{A^{\alpha}}{c_0} = \frac{\bar A^{\alpha}}{c_0} + \xi^{\alpha,0} + \xi^{0,\alpha}.
\end{equation}
The last equality follows from eq.~(\ref{metric_tensor_small_perturbation_Lorenz_gauge_inverted2}), the analogy $h^{\alpha 0} \rightarrow -A^\alpha/c_0$ (similarly $T^{\alpha 0} \rightarrow -c_0 J^\alpha$), and the trace $\bar{h}^\nu {}_\nu = \bar A^0/c_0 = 0$. The transformation of coordinates ensures a vorticity-free gauge (with vanishing magnetic field), which is thus analogue to the Newtonian gauge. In this gauge the acoustically relevant component $A^0$ is of sub-leading order ($A^{0} \ll |\bar A^{\alpha}| \sim |A^{\alpha}|$) and the associated part of $\bar A^\alpha$ does not satisfy eq.~(\ref{EMG_Lorenz_gauge2}). In electromagnetism it quantifies weak forces which are due to radiation and add to electrostatic Coulomb's force (the near field of eq.~(\ref{Maxwell_currents_Lorenz_gauge2})), rather than longitudinal electromagnetic waves.

Referring to relativistic eq.~(\ref{metric_tensor_small_perturbation_Lorenz_gauge_inverted2}) for obtaining the Newtonian form means that the components of the four-vector potential and current transform as the first row or column of the second-order tensors of general relativity, not as four-vectors. This is necessary for eventually obtaining the quantities which are relevant for an acoustic observer, who also sets the preferred contravariant vector space in addition to the Newtonian frame.

In the absence of external forces, a free-space dipole can form only at boundaries of foreign bodies inside the source region of a quadrupole. Like acoustic observers, the boundaries necessary break the Lorentz invariance and confine the acoustic spacetime. In order to preserve incompressibility of the flow around it, the body has to be acoustically compact. We saw in section~\ref{ch:energy_and_charge} how a quadrupole source given by $T^{\alpha\beta}$ reduces to a dipole, given by the conserved current $J^\alpha \rightarrow [0, p\vec n/c_0]$, at a rigid boundary of a compact and motionless body (in the preferred frame with no external forces acting on the body). The compact dipole source can thus be modelled as the free-space source
\begin{equation}\label{surface_dipole_source}
\frac{1}{c_0} \int p n^j d^3 \vec x = \frac{d}{dt} d^j
\end{equation}
and the analogue charge is mass-induced. In the acoustic analogy the fluid molecules are elementary particles which build not only the acoustic spacetime and quadrupole sources, but charges, as well.

\iffalse
Since we would like to combine it with the second-order tensor from above, we can use alternatively
(see equations~(\ref{particle_acceleration_EMG_2}) and (\ref{eq:EMG_tensor})) the antisymmetric tensor $J^{\beta,\alpha} - J^{\alpha,\beta}$. From eq.~(\ref{Maxwell_currents_Lorenz_gauge2}),
\begin{equation}
\Box F^{\alpha\beta} = \Box (\bar{A}^{\beta,\alpha}-\bar{A}^{\alpha,\beta}) = -\mu_0 (J^{\beta,\alpha}-J^{\alpha,\beta})
\end{equation}
and for a compact dipole source
\begin{equation}\label{dipole_source_tensor}
\int (J_{k,j}-J_{j,k}) d^3 \vec x = \frac{d}{dt} (d_{k,j} - d_{j,k}).
\end{equation}
\fi

Finally, acoustic monopole radiation can be described simply with a dimensionless scalar $\bar \chi \ll 1$ satisfying the scalar wave equation of the linearized theory,
\begin{equation}\label{scalar_wave_equation2}
\Box \bar \chi = -\Psi \rho,
\end{equation}
where $\rho$ represents the source of unsteady mass density and $\Psi$ is a constant with the unit m/kg. Comparison with eq.~(\ref{Einstein_field_equations_linear2}) with $T^{\alpha\beta}$ from eq.~(\ref{eq:stress-energy_tensor_perfect_fluid_components}) and $\bar \chi = \bar h^{00}$ gives $\Psi = 2kG/c_0^2$. A compact monopole source is expressed by the unsteady mass
\begin{equation}\label{monopole_source}
	\int \rho d^3 \vec x = m.
\end{equation}
Alternatively, it can be modelled with constant mass density $\rho_0 = T^{00}/c_0^2$ from eq.~(\ref{eq:stress-energy_tensor_perfect_fluid_components_non-relativistic}) and unsteady volume of the source region, as in section~\ref{ch:pulsating_sphere}.

From eq.~(\ref{particle_acceleration_spatial_Newtonian_gauge2}), velocity of a particle in the field equals 
\begin{equation}\label{acoustic_velocity}
	\frac{d^2 x^k}{dt^2} = \frac{c_0^2}{2} \chi^{,k}.
\end{equation}
For a non-relativistic particle, this can be seen as the spatial part of
\begin{equation}\label{particle_acceleration_acoustics}
	\frac{dU^\alpha}{d\tau} = \frac{c_0}{2} \eta^{\alpha\beta} \chi_{, \beta},
\end{equation}
which agrees with eq.~(\ref{particle_acceleration_2}). Moreover, from eq.~(\ref{metric_tensor_small_perturbation_Lorenz_gauge_inverted}) or eq.~(\ref{h_00_and_h_00_bar_monopole}), $\chi = \bar \chi/2$ in the Newtonian gauge. 

In section~\ref{ch:energy_and_charge} we also saw how an aeroacoustic quadrupole can reduce to a free-space monopole given by the scalar $\rho = \rho_0 (\vec v - \vec u) \cdot \vec n/c_0$ at a non-rigid boundary or an externally moved compact body in the source region. Such a compact monopole can thus be modelled as the free-space source
\begin{equation}\label{surface_monopole_source}
\int \rho_0 (v^j - u^j) n_j/c_0 d^3 \vec x = m,
\end{equation}
which is not conserved. Since they require action of external forces or a penetrable boundary of spacetime, such sources of mass do not appear in general relativity.

Bringing equations~(\ref{quadrupole_source}), (\ref{surface_dipole_source}), and (\ref{surface_monopole_source}) together after replacing $\bar A^\alpha$ with $-c_0 \bar h^{\alpha 0}$, $\bar \chi$ with $\bar h^{00}$, $\mu_0$ with $-kG/(2c_0^2)$, and $\Psi$ with $2kG/c_0^2$, we can list the integral forms of the aeroacoustic sources of metric perturbation of the acoustic spacetime in unsteady incompressible flows:
\begin{equation}
-\frac{2kG}{c_0^2} m = - \frac{2kG}{c_0^4} \int \rho_0 c_0 (v^j - u^j) n_j d^3 \vec x,
\end{equation}
\begin{equation}
-\frac{kG}{2c_0^3} \frac{d}{dt} d^j = -\frac{kG}{2c_0^4} \int p n^j d^3 \vec x,
\end{equation}
and
\begin{equation}
- \frac{2kG}{c_0^4} \frac{1}{2} \frac{d^2}{dt^2} I^{jk} = - \frac{2kG}{c_0^4} \int T^{jk} d^3 \vec x,
\end{equation}
with $T^{jk}$ given by eq.~(\ref{eq:stress-energy_tensor_perfect_fluid_components_non-relativistic}). The corresponding far-field solutions for outgoing waves of isolated compact sources read
\begin{equation}
\bar h^{00} = \bar \chi = \frac{kG}{2 \pi r c_0^4} \int \left( \rho_0 c_0 (v^j - u^j) n_j \right)_{(t-r/c_0)} d^3 \vec x,
\end{equation}
\begin{equation}
\bar h^{j 0} = -\frac{\bar A^j}{c_0} = \frac{kG}{8 \pi r c_0^4} \int \left( p n^j \right)_{(t-r/c_0)} d^3 \vec x,
\end{equation}
and
\begin{equation}
\bar h^{jk} = \frac{kG}{2\pi r c_0^4} \int \left( T^{jk} \right)_{(t-r/c_0)} d^3 \vec x.
\end{equation}
The first is monopole contribution due to unsteady fluid displacement caused by a moving body in the flow or a non-rigid boundary, the second, dipole contribution is due to the loading force acting on a rigid motionless body, and the third is Lighthill's inviscid quadrupole. All can be associated with the source terms of the inviscid and adiabatic Ffowcs Williams and Hawkings equation.

The only remaining free-space aeroacoustic source, the monopole source of entropy as in eq.~(\ref{Lighthill_analogy}), will turn out to be responsible for the expansion of the observable universe and, moreover, for its finiteness and time itself. Its true quantum thermodynamic nature shall be studied in the next section. In general relativity it is only modelled by means of the small cosmological constant or dark energy (not to be confused with the much larger energy of the background spacetime). In the classical acoustic analogy above it is a low-frequency monopole source, the maximal wavelength of which corresponds closely to the radius of the observable universe. Every observer is thus located in its near field. The fact that the limitation of observers on the micro-scale (the ability to interact only with a large number of elementary particles) gives rise to its thermodynamic nature makes the source not a distinct monopole source (as the one in eq.~(\ref{surface_monopole_source}), for example), but an apparent source, which is equally present at all times and locations in the observable universe. Somewhat paradoxically, the microscopic limitation of the observers determines the macroscopic size of the entire observable universe as well as its expansion (the increase of entropy in time).

Since it is also determined by an external thermodynamic theory of the elementary particles which constitute the spacetime (and sources in it), the constant $\rho_0$ remained unobservable and given in the classical analogies above. The only formal difference is the occurrence of the heat capacity ratio $\gamma$, which equals 1.4 in perfect diatomic gases that are typical in classical acoustics, and which should be replaced by 1 when switching to the analogue relativistic and electromagnetic theories.
\section{Analogue quantum fields and thermodynamics}\label{ch:acoustic_Lagrangians}

Classical thermodynamics is relevant for acoustics in fluids mainly for deriving equation of state of the fluid and expressing the background fluid quantities, such as the mean density $\rho_0$. With the exception of certain specific sources of sound, thermodynamics is largely an independent theory decoupled from the theory of sound waves and the common macroscopic acoustic quantities are calculated from the wave equation and used without reference to the laws of thermodynamics. However, the two theories become much closer in the realm of quantum fields, when microscopic quantum particles are treated as fields as well as thermodynamically. It is in this sense of the analogy between classical acoustic and quantum fields that we will associate the two theories in this section.

Acoustic analogies with quantum field theory and quantum thermodynamics shall be studied for several reasons:
\begin{itemize}
	\item Similarly as the classical analogies above, they provide alternative formalisms for acoustic problems and new insights into the analogue theory, especially regarding the nature of mass-energy and the background density $\rho_0$.
	\item They point to a possible connection between the indeterminism of observations at the micro-scale, associated thermodynamics of unobservable elementary particles, and the monopole source of expansion of the observable universe, which could only be modelled in the classical theories.
	\item They reveal parallels between many important concepts of quantum physics and their (less abstract) counterparts in classical acoustics, which are rarely acknowledged. This places acoustics closer to the remaining two major theories of modern physics from Fig.~\ref{fig:physics_and_acoustics}.
	\item The obtained theory at the microscopic (quantum) scale is complementary to the macroscopic (classical) theory developed so far and allows formulation of a unified acoustic theory, which is the topic of section~\ref{ch:unified_theory}.
\end{itemize}
With regard to this, we establish first acoustic analogies with quantum mechanics and quantum field theory, before treating the analogue fields thermodynamically.

Before starting, a note should be made on the following analogue quantization of acoustic fields. Although necessary for appropriate interpretation of the measurements of quantum particles, promotion of quantum fields to operators acting in a Hilbert or Fock space is an extension of the theory which is not essential for the underlying physical phenomena. All the relevant information on particles is assumed to be given by their fields and therefore we will not need to treat them further as operators. In other words, we will refer very little to the corpuscular appearance of virtual acoustic quantum particles and focus on the fields, especially since we are eventually interested only in macroscopic properties of sound fields, not in the unobservable motion of isolated acoustic particles. This does not make the theory any less quantum.

\subsection{Lagrangian formalism}\label{ch:Lagrangian_formalism}

Both classical~\cite{Landau2000} and quantum field theories~\cite{Schwartz2014} can be formulated in terms of Lorentz-invariant Lagrangians (or Hamiltonians) and the principle of least action. While such formulations might not be essential for the former theories ruled by certain governing equations, they make a starting point for the mathematical treatment of quantum fields. Accordingly, we shall first show that classical acoustics, considered in the relativistic framework of the preceding sections, can also be based on Lorentz-invariant Lagrangians.

For a given Lagrangian (or, more precisely, Lagrangian density) $\mathcal{L}$, the action is defined as its integral over the spacetime\cite{Landau2000}:
\begin{equation}\label{action}
S = \int \mathcal{L}(x) d^4 x,
\end{equation}
where $x^\mu = [c_0 t,\vec x]$ is denoted here with $x$ for brevity, and, as before, $c_0$ is the reference speed of sound in acoustic problems. The equation of motion of the field theory (such as wave equation) follows from the principle of least action under variation of the field. For example, for a scalar field $\phi$, which can describe sound fields in fluids, the Lagrangian is a functional only of $\phi$ and its first derivatives, $\phi_{,\mu}$. The variation $\phi \rightarrow \phi + \delta \phi$ leads to the change of action
\begin{equation}\label{action_variation}
\begin{aligned}
\delta S &= \int \delta \mathcal{L}(\phi,\phi_{,\mu}) d^4 x = \int \left[ \frac{\partial \mathcal{L}}{\partial \phi} \delta \phi + \frac{\partial \mathcal{L}}{\partial (\phi_{,\mu})} \delta (\phi_{,\mu}) \right] d^4 x \\
&= \int \left[ \frac{\partial \mathcal{L}}{\partial \phi} \delta \phi + \left( \frac{\partial \mathcal{L}}{\partial (\phi_{,\mu})} \delta \phi \right)_{,\mu} - \left( \frac{\partial \mathcal{L}}{\partial (\phi_{,\mu})} \right)_{,\mu} \delta \phi \right] d^4 x.
\end{aligned}
\end{equation}
The middle term in the last equality reduces to an integral over the boundary of spacetime. It vanishes in an unbounded spacetime, assuming that the field decays to zero at infinity (which is typical in quantum field theory), or, more generally, if $\phi$ is fixed ($\delta \phi = 0$) at the boundary, even if the spacetime is bounded\footnote{\label{Heaviside_function_Lagrangian_action}If $\phi$ is not fixed at the boundary, we can multiply the Lagrangian with the Heaviside function $H(f)$, as in section~\ref{ch:energy_and_charge}, and thus formally extend the spacetime to infinity, where $\delta \phi = 0$. This will be done later in eq.~(\ref{Neother_current_complex_scalar_field2_boundary}). Furthermore, if we are interested in action, not Lagrangian (which is the case with the derivation of stress-energy tensor; see equations~(\ref{Lagrangian_global_symmetry_derivative})-(\ref{Neother_current_stress_energy_tensor_boundary})), we could multiply the entire variation in eq.~(\ref{action_variation}) with $H(f)$.}. Minimizing the action, $\delta S/\delta \phi = 0$, leads to the Euler-Lagrange equation
\begin{equation}\label{action_variation_zero}
\frac{\partial \mathcal{L}}{\partial \phi} - \left( \frac{\partial \mathcal{L}}{\partial (\phi_{,\mu})} \right)_{,\mu} = 0.
\end{equation}

The most general Lorentz-invariant Lagrangian of a real scalar field, up to the second order\footnote{In section~\ref{ch:mass_background_fluid} we will see that higher-order terms can be used for modelling the background fluid as a massive field as well as its interaction with the massless acoustic field.} of $\phi$ is
\begin{equation}\label{Lagrangian_real_scalar_field}
\mathcal{L} = \frac{1}{2} \phi^{,\mu} \phi_{,\mu} + \frac{1}{2} \left( \frac{m c_0}{\hbar} \right)^2 \phi^2,
\end{equation}
where $m$ is particle\footnote{We already emphasize that a single particle may actually be a system of elementary particles, much like every particle in continuum mechanics and classical acoustics involves many elementary particles, such as fluid molecules.} mass, $\hbar$ is the reduced Planck constant ($\hbar = 1.05 \cdot 10^{-34}$\,kg\,m$^2$/s), and the normalization with 1/2 is conventional. The Lagrangian is also invariant under the discrete symmetry $\phi \rightarrow -\phi$. The mass term can be seen as a particular form of the potential functional $\mathcal{V}(\phi)$ which depends on the particle's environment (and may explicitly break any symmetry of the Lagrangian), that is
\begin{equation}\label{Lagrangian_real_scalar_field_potential}
\mathcal{L} = \frac{1}{2} \phi^{,\mu} \phi_{,\mu} + \mathcal{V}(\phi),
\end{equation}
where the first term is the kinetic term. Indeed, we will see that a particle obtains its mass only in interaction with the environment and will be able to model different environments with different mass terms. For now we observe ``free" particles, even if they are massive. Inserting eq.~(\ref{Lagrangian_real_scalar_field}) into eq.~(\ref{action_variation_zero}) gives the Klein-Gordon equation in flat Minkowski spacetime:
\begin{equation}\label{action_variation_zero_real_scalar_field}
- \frac{1}{2}\left( \frac{\partial (\phi^{,\mu} \phi_{,\mu})}{\partial (\phi_{,\mu})} \right)_{,\mu} + \left( \frac{m c_0}{\hbar} \right)^2 \phi  = - \left( \phi^{,\mu} \right)_{,\mu} +\left( \frac{m c_0}{\hbar} \right)^2 \phi = - \Box \phi + \left( \frac{m c_0}{\hbar} \right)^2 \phi = 0.
\end{equation}
For $m = 0$, the equation reduces to eq.~(\ref{scalar_wave_equation2}) without the source term,
\begin{equation}\label{scalar_wave_equation}
\Box \phi = 0.
\end{equation}
Hence, acoustic fields in fluids are naturally specified by the Lorentz-invariant Lagrangian of a real scalar field
\begin{equation}\label{Lagrangian_real_scalar_field_massless}
\mathcal{L} = \frac{1}{2} \phi^{,\mu} \phi_{,\mu} = \frac{1}{2} \eta^{\mu\nu} \phi_{,\mu} \phi_{,\nu}
\end{equation}
and free (of any environment) and massless particles. In a curved background spacetime, which will not be considered here, the Lagrangian can be generalized to
\begin{equation}\label{Lagrangian_real_scalar_field_massless_curved_background_spacetime}
\mathcal{L} = \frac{1}{2} g^{\mu\nu} \phi_{,\mu} \phi_{,\nu},
\end{equation}
which results in the d'Alembertian from eq.~(\ref{eq:d'Alembertian}) in the equation of motion.

\subsection{Mass in acoustic near field}\label{ch:mass_near_field}

Although sound field is massless, the non-acoustic massive field $\phi$ from eq.~(\ref{action_variation_zero_real_scalar_field}) is still of great interest. In fact, massless acoustic field necessarily becomes massive in the acoustic near field of a compact (non-relativistic) source or a foreign object interacting with the field. In order to show this, we use the Planck–Einstein relation between energy of a massless field (which should not be confused with usual energy of sound waves) and frequency,
\begin{equation}\label{Planck_Einstein}
E = \hbar \omega,
\end{equation}
and Einstein's expression for the rest mass of a non-relativistic particle,
\begin{equation}\label{Einstein_energy_mass}
E = m c_0^2.
\end{equation}
Combined together, these two equations give
\begin{equation}\label{reduced_Planck_constant}
\hbar = \frac{mc_0^2}{\omega}.
\end{equation}
which can be taken as a definition of the constant $\hbar$. It couples mass and frequency in the quantum theory in the same way as the constants $G$ and $\mu_0$ connect mass and length in the classical analogies above (recalling eq.~(\ref{gravitational_constant_and_Schwarzschild_radius}) and $\mu_0 \rightarrow -kG/(2 c_0^2)$). Equating the two energies is also analogous to the Compton scattering of a photon (massless particle) by a massive charged particle (electron) with the Compton wavelength $2\pi\hbar/(mc_0)$.

Equation~(\ref{action_variation_zero_real_scalar_field}) in frequency domain (with the time dependence $e^{-j\omega t}$) reads
\begin{equation}\label{action_variation_zero_real_scalar_field_freq_domain}
\left( \frac{\omega}{c_0} \right)^2 \phi + \nabla^2 \phi - \left( \frac{m c_0}{\hbar} \right)^2 \phi = 0
\end{equation}
\iffalse
This is equal to the time-independent Schrödinger equation,
\begin{equation}\label{Schroedinger_equation_time_independent}
\frac{2mE}{\hbar^2} \phi + \nabla^2 \phi - \frac{2mV}{\hbar^2} \phi = 0
\end{equation}
for $E = \hbar^2 \omega^2/(2m c_0^2)$ and $V = mc_0^2/2$, which is satisfied for a free particle in the uniform potential in the absence of dispersion. Indeed, the massive Klein-Gordon equation reduces to the Schrödinger equation in the non-relativistic limit of high mass\cite{Schwartz2014}.
\fi
and after inserting eq.~(\ref{reduced_Planck_constant}) it becomes Laplace's equation
\begin{equation}\label{action_variation_zero_real_scalar_field_freq_domain_spontaneously_broken_symmetry_non-relativistic_particle}
\nabla^2 \phi = 0.
\end{equation}
Like Newtonian gravity and Coulomb's electrostatics, it is not Lorentz invariant and has the same operator as equations~(\ref{Einstein_field_equations_linear_non-relativistic_source_near_field}) and (\ref{Maxwell_currents_Lorenz_gauge_non_relativistic_near_field}). In classical acoustics it describes incompressible ($\rho = \rho_0$ and $c_0 \rightarrow \infty$) fluctuations in the acoustic near field at distances $l \ll c_0/\omega$. Hence, in the acoustic analogy a massive (or charged) particle is actually near field of a compact source (or body) which breaks the Lorentz invariance\footnote{Once we introduce phonons as the elementary particles of sound fields, we will see that they are described by the Lagrangian in eq.~(\ref{Lagrangian_real_scalar_field_massless}) in the far field and by the Lagrangian in eq.~(\ref{Lagrangian_real_scalar_field}) in the near field. Notice, however, that even such ``elementary" particles involve many fluid molecules.}, but not the source (or body) itself. Incompressibility of the fluid in the near field makes the region appear as a stiff object with density $\rho_0$ and the background fluid density receives its ``manifestation'' which it otherwise lacks in free space. However, an additional mechanism is necessary in order to associate the unsteady field with the background medium, which would point to the common elementary particles building both. As will be demonstrated next, the mechanism can be modelled as a spontaneously broken symmetry. The length scale $L$ at which the symmetry breaking takes place is given roughly by
\begin{equation}\label{spontaneous_symmetry_breaking_length_scale}
\frac{\omega L}{c_0} = \frac{m c_0 L}{\hbar} = 1.
\end{equation}
Notice that it depends on frequency and therefore mass-energy of the particle. Nevertheless, the value of $\rho_0$ is still dictated externally and not given by quantum field theory. 

\subsection{Massive fluid field}\label{ch:mass_background_fluid}

The previous analysis indicates an interaction between the acoustic quantum field and the background fluid. The former obtains mass of the latter in the near field. In the following we consider how this process can be modelled in quantum field theory, bearing in mind that the origin of the background fluid remains outside the scope of the theory.

\subsubsection{Background fluid as a quantum field and acoustic particles}\label{background_fluid_and_acoustic_particles}

Another important occurrence of a massive field is in the model of the background fluid, that is the acoustic medium, which is also treated as a quantum field. Next we consider $\phi$ to be the field of fluid particles in general, regardless of the acoustic field. Each particle actually contains many fluid molecules, which cannot be observed. The Lorentz-invariant Lagrangian contains the next allowed higher order of $\phi$ in the Taylor expansion of a general potential~\cite{Schwartz2014}:
\begin{equation}\label{Lagrangian_real_scalar_field_spontaneous_symmetry_breaking}
\mathcal{L} = \frac{1}{2} \phi^{,\mu} \phi_{,\mu} + \frac{1}{2} \left( \frac{m c_0}{\hbar} \right)^2 \phi^2 + \frac{\lambda}{4!} \phi^4,
\end{equation}
where $\lambda$ is a constant coefficient and the division with 4! is conventional. As before, the Lagrangian exhibits the symmetry $\phi \rightarrow -\phi$. Furthermore, the squared mass appears essentially as the coefficient of the lowest, second-order term of an arbitrary potential in which a particle can be found (the first-order term is zero close to equilibrium, when the expansion holds, and the arbitrary constant part of the potential is set to zero).

In the mechanism of spontaneous symmetry breaking both $\lambda$ and $m$ can depend on certain parameter $l$. In particular, $m^2 > 0$ when $l > L$, where we suppose that $L$ is the point of symmetry breaking from eq.~(\ref{spontaneous_symmetry_breaking_length_scale}), and $m^2 < 0$ when $l < L$. In the latter case, $m^2$ in the Lagrangian has to be replaced with positive $-m^2$, so
\begin{equation}\label{Lagrangian_real_scalar_field_spontaneous_symmetry_breaking_strong_perturbations}
\mathcal{L} = \frac{1}{2} \phi^{,\mu} \phi_{,\mu} - \frac{1}{2} \left( \frac{m c_0}{\hbar} \right)^2 \phi^2 + \frac{\lambda}{4!} \phi^4.
\end{equation}
The mean, background value of $\phi$, which we denote $\phi_0$, corresponds to the minimum of the potential in equilibrium and can be calculated from
\begin{equation}
\frac{\partial}{\partial \phi} \left( \frac{1}{2} \left( \frac{m c_0}{\hbar} \right)^2 \phi^2 - \frac{\lambda}{4!} \phi^4 \right) = \left( \frac{m c_0}{\hbar} \right)^2 \phi - \frac{\lambda}{6} \phi^3 = 0.
\end{equation}
It is $\phi_0 = \pm \sqrt{6m^2 c_0^2/(\lambda \hbar^2)}$ and both non-zero values break the symmetry $\phi \rightarrow -\phi$. We are free to choose one of the two vacua, say
\begin{equation}\label{real_scalar_field_vacuum}
\phi_0 = \sqrt{\frac{6m^2 c_0^2}{\lambda \hbar^2}}.
\end{equation}
Now supposing a perturbation of the background fluid, we can expand $\phi = \phi_0 + \tilde{\phi}$ around the vacuum, which gives
\begin{equation}\label{Lagrangian_real_scalar_field_spontaneous_symmetry_breaking_strong_perturbations_perturbed}
\begin{aligned}
\mathcal{L} &= \frac{1}{2} \tilde{\phi}^{,\mu} \tilde{\phi}_{,\mu} - \frac{1}{2} \left( \frac{m c_0}{\hbar} \right)^2 (\phi_0^2 + 2\phi_0 \tilde{\phi} + \tilde{\phi}^2) + \frac{\lambda}{4!} (\phi_0^4 + 4\phi_0^3 \tilde{\phi} + 6 \phi_0^2 \tilde{\phi}^2 \\
&+4\phi_0 \tilde{\phi}^3 + \tilde{\phi}^4) = \frac{1}{2} \tilde{\phi}^{,\mu} \tilde{\phi}_{,\mu} - \left( \frac{1}{2} \left( \frac{m c_0}{\hbar} \right)^2 \phi_0^2 - \frac{\lambda}{4!} \phi_0^4 \right) - \left( \left( \frac{m c_0}{\hbar} \right)^2 \phi_0 - \frac{\lambda}{6} \phi_0^3 \right) \tilde{\phi} \\
&- \left( \frac{1}{2} \left( \frac{m c_0}{\hbar} \right)^2 - \frac{\lambda}{4} \phi_0^2 \right) \tilde{\phi}^2 + \frac{\lambda}{6} \phi_0 \tilde{\phi}^3 + \frac{\lambda}{4!} \tilde{\phi}^4 \\
&= \frac{1}{2} \tilde{\phi}^{,\mu} \tilde{\phi}_{,\mu} - \frac{3}{2\lambda} \left( \frac{m c_0}{\hbar} \right)^4 + \left( \frac{m c_0}{\hbar} \right)^2 \tilde{\phi}^2 + \sqrt{\frac{\lambda m^2 c_0^2}{6 \hbar^2}} \tilde{\phi}^3 + \frac{\lambda}{4!} \tilde{\phi}^4.
\end{aligned}
\end{equation}
This Lagrangian is not invariant under the change of sign $\tilde{\phi} \rightarrow -\tilde{\phi}$ like the one in eq.~(\ref{Lagrangian_real_scalar_field}), but it is symmetric under $\tilde{\phi} \rightarrow -\tilde{\phi} -2 \phi_0$, as we would expect from a perturbation added to the steady background fluid. Hence, the field of fluid particles $\phi$ with its Lagrangian in eq.~(\ref{Lagrangian_real_scalar_field_spontaneous_symmetry_breaking}) captures both the background fluid (the non-zero constant $\phi_0$) and its perturbation (the field $\tilde \phi$, which alone could be represented by the Lagrangian in eq.~(\ref{Lagrangian_real_scalar_field}), without the third term). For example, for compressible sound waves and scalar acoustic fields it is natural that $\phi$ represents the fluid density, so $\phi_0 = \rho_0$ in eq.~(\ref{real_scalar_field_vacuum}) and $\lambda = 6 (mc_0/\hbar)^2/\rho_0^2$. The field $\tilde \phi$ is massive in the acoustic near field, according to the mechanism of spontaneous symmetry breaking, while in the far field acoustic $\tilde \phi = \rho$ is given by the Lagrangian from eq.~(\ref{Lagrangian_real_scalar_field_massless}). Since both the medium and the perturbation are contained in a single field $\phi$, the elementary particles must be common (they are, indeed, both built by the fluid molecules). However, the value of $\rho_0$ must follow from the thermodynamics of the background field.
%check this!!

Although massive, the field $\tilde \phi$ is the near-field extension of purely acoustic waves and carries the complete information on the sound field (recall section~\ref{ch:mass_near_field} and eq.~(\ref{reduced_Planck_constant}), but also the approach which lead to eq.~(\ref{Einstein_field_equations_linear_split_solution_far_geometric_field_quadrupole_moment_tensor_jk_Taylor_expansion})). Moreover, because of its mass it also obeys the laws of quantum mechanics and analogue acoustic quantum particles can be defined. These massive particles are simply materialization of massless acoustic particles (which we will actually call phonons) in the acoustic near field. 

In order to show this, we first notice that $\phi_0 = \rho_0$ is real and positive, so we can express it as the product
\begin{equation}\label{background_fluid_density_QFT}
\phi_0 = \psi_0^* \psi_0 = \rho_0,
\end{equation}
where $\psi_0^*$ is complex conjugate of $\psi_0$ and $\psi_0$ is in general a complex field (called state or wave function in quantum mechanics) with an arbitrary phase. Then
\begin{equation}\label{background_fluid_mass_QFT}
\int_V \psi_0^*\psi_0 d^3 \vec{x} = N_m M_0,
\end{equation}
where $M_0$ is the fluid molecular mass (in standard quantum mechanics it is normalized to 1 and $\psi_0^* \psi_0$ is interpreted as probability density of finding a particle in some region of space) and $N_m$ is the number of molecules (elementary particles) inside the volume $V$. Adding a weak perturbation $\tilde{\psi} \ll \psi_0$ to it, we can write $\psi = \psi_0 + \tilde{\psi}$ and
\begin{equation}\label{quantized_density_field}
\begin{aligned}
\int_V \psi^*\psi d^3 \vec{x} &= \int_V \psi_0^*\psi_0 d^3 \vec{x} + \int_V \tilde{\psi}^*\psi_0 d^3 \vec{x} + \int_V \psi_0^*\tilde{\psi} d^3 \vec{x} + \int_V \tilde{\psi}^*\tilde{\psi} d^3 \vec{x} \\
&= N_m M_0 + N_{ph} M.
\end{aligned}
\end{equation}
The key assumption here is that $\psi_0$ and $\tilde{\psi}$ are mutually uncorrelated (random motion of molecules is not correlated with the motion due to sound waves), so the middle two integrals vanish. The perturbation appears to introduce new ``elementary" particles~\cite{Brennan2016}, the phonons, while both $\psi_0$ and $\tilde \psi$ describe systems of true elementary particles, the fluid molecules. The total number of elementary particles is thus
\begin{equation}
N_\text{tot} = N_m + N_{ph},
\end{equation}
with, by definition,
\begin{equation}\label{acoustic_field_mass_QFT}
\int_V \tilde{\psi}^*\tilde{\psi} d^3 \vec{x} = N_{ph} M
\end{equation}
and $\tilde{\psi}^*\tilde{\psi} = |\rho_\text{ac}|$ is amplitude of the acoustic density (in standard quantum mechanics -- probability density of finding the phonon). Since the massive particles entirely correspond to the sound field, we will call them massive phonons~\cite{Brennan2016}, each with the mass $M$, and $N_{ph}$ is their number in $V$.

The mass $M$ follows from eq.~(\ref{reduced_Planck_constant}). The total quantum energy is given by
\begin{equation}\label{total_quaantum_energy}
\begin{aligned}
E & = \frac{1}{M_0} \int_V \psi_0^* j\hbar \frac{\partial}{\partial t} \psi_0 d^3 \vec{x} + \frac{1}{M} \int_V \tilde{\psi}^* j\hbar \frac{\partial}{\partial t} \tilde{\psi} d^3 \vec{x} \\
& = \frac{1}{M_0} \int_V \psi_0^* j\hbar \frac{\partial}{\partial t} \psi_0 d^3 \vec{x} + N_{ph} \hbar \omega.
\end{aligned}
\end{equation}
We normalized with the masses $M_0$ and $M$, according to the definitions in equations~(\ref{background_fluid_mass_QFT}) and (\ref{acoustic_field_mass_QFT}), and supposed $\tilde{\psi}(\vec x, t) = \tilde{\psi}(\vec x) e^{-j\omega t}$. This confirms that phonons can be treated both as massless (with energy $\hbar \omega$, as in eq.~(\ref{Planck_Einstein})) and massive particles\footnote{This duality can be compared with vacuum polarization in quantum field theory -- alternate appearance and disappearance of massive particles in pure vacuum in the presence of a massless field. In the context here, it may represent apparent alternation between massive and massless phonons in the near field of omnipresent sources (the monopole source of entropy) or compact boundaries (dipole radiation at the microscopic boundary of the observable universe). Both occurrences are apparent (due to the observer), taking place at the smallest, Planck scale.} (with mass $M = \hbar \omega/c_0^2$ from equations~(\ref{Einstein_energy_mass}) and (\ref{reduced_Planck_constant})). The first term in eq.~(\ref{total_quaantum_energy}) can be treated similarly, if random molecular motion is modelled as uncorrelated high-frequency oscillations of $\psi_0 (\vec x, t) = \psi_0(\vec x) e^{-j\omega_{0} t}$ with $\omega_0 \gg \omega$. The frequency $\omega_0$ is high enough that the fluctuations of $\psi_0$ and its phase cannot be observed. The integrated vacuum energy of the dual massless particles is then
\begin{equation}\label{background_frequency}
\begin{aligned}
E & = \frac{1}{M_0} \int_V \psi_0^* j\hbar \frac{\partial}{\partial t} \psi_0 d^3 \vec{x} = N_m \hbar \omega_0
\end{aligned}
\end{equation}
and from eq.~(\ref{reduced_Planck_constant})
\begin{equation}\label{the_highest_frequency}
	\omega_0 = \frac{M_0 c_0^2}{\hbar}.
\end{equation}

Since the fluid particles are massive and moving at speeds much below the speed of sound, both wave functions, $\psi_0$ and $\tilde \psi$ (we will write only $\psi$, for brevity) associated with the massive background field and phonons, must obey the Schrödinger equation,
\begin{equation}\label{Schroedinger_equation}
j\hbar \frac{\partial}{\partial t} \psi = \hat H \psi = -\frac{\hbar^2}{2 M_\psi} \nabla^2 \psi + V \psi.
\end{equation}
Here $V(\vec{x},t)$ is a real potential, $\hat H$ is the Hamilton operator (operators are indicated by the hat in their symbols) which is defined by the second equality, and $\psi$ is time-dependent (in the so-called Schrödinger picture of quantum states) pure state of the system with mass $M_\psi$. The state function is normalized to satisfy eq.~(\ref{background_fluid_mass_QFT}) for eq.~(\ref{acoustic_field_mass_QFT}) and the system consists of $N_m$ molecules or $N_{ph}$ phonons. Like the more general Dirac equation, the Schrödinger equation is the square root of the Klein-Gordon equation in terms of their operators, but, unlike the former, only in the limit of large mass, which is relevant here. The other two equations present a Lorentz-invariant and ultraviolet completion (or as eq.~(\ref{Helmholtz_eq_from_Schroedinger}) will show, a far-field completion, when the symmetry is not spontaneously broken to give rise to the massive particles) of the Schrödinger equation~\cite{Schwartz2014}. In order to show this, we calculate
\begin{equation}
\left(j\hbar \frac{\partial}{\partial t} \right)^2 = -\hbar^2 \frac{\partial^2 }{\partial t^2}
\end{equation}
and
\begin{equation}\label{Schroedinger_to_Klein-Gordon_Taylor}
\begin{aligned}
\left( -\frac{\hbar^2}{2 M_\psi} \nabla^2 \right)^2 &\approx \left( M_\psi c_0^2 - \frac{\hbar^2}{2 M_\psi} \nabla^2 - \frac{\hbar^4}{8 c_0^2 M_\psi^3} \nabla^4 - ... \right)^2 \\&= \left( \sqrt{M_\psi^2 c_0^4 - \hbar^2 c_0^2 \nabla^2} \right)^2 = M_\psi^2 c_0^4 - \hbar^2 c_0^2 \nabla^2.
\end{aligned}
\end{equation}
The approximation with the Taylor series holds for $\hbar^2 \nabla^2 \sim \hbar^2 \omega^2/c_0^2 \ll M_\psi^2 c_0^2$, that is, $M_\psi c_0^2 \gg \hbar \omega$ (which is satisfied by large collections of molecules or phonons, when $\hbar \omega = M_0 c_0^2$ or $\hbar \omega = M c_0^2$, respectively) and because the term $M_\psi c_0^2$ does not affect appreciably dynamics of the non-relativistic theory. Therefore, squared operator of eq.~(\ref{Schroedinger_equation}) with $V = 0$ (free particle) divided with $(\hbar c_0)^2$ matches the operator in eq.~(\ref{action_variation_zero_real_scalar_field}), for $M_\psi \gg m$. The analogue quantum mechanics applies only for large collections of elementary particles and this fact will provide a basis for the thermodynamic treatment later. While elementary particles satisfy eq.~(\ref{action_variation_zero_real_scalar_field}), the Schrödinger equation is satisfied only by a system of such particles, the large mass of which violates the Lorentz invariance, as before.

In the absence of external sources of mass, total mass of the analogue quantum particles is conserved. From eq.~(\ref{Schroedinger_equation}) it follows~\cite{Brennan2016}:
\begin{equation}
\begin{aligned}
&\psi^* \left( -\frac{\hbar^2}{2 M_\psi} \nabla^2 \psi + V \psi \right) - \psi \left( -\frac{\hbar^2}{2 M_\psi} \nabla^2 \psi + V \psi \right)^* = -\frac{\hbar^2}{2M_\psi} (\psi^* \nabla^2 \psi - \psi \nabla^2 \psi^*)\\
&= \psi^* j\hbar \frac{\partial}{\partial t} \psi - \psi \left( j\hbar \frac{\partial}{\partial t} \psi \right)^* = j\hbar \left( \psi^* \frac{\partial}{\partial t} \psi + \psi \frac{\partial}{\partial t} \psi^* \right)
\end{aligned}
\end{equation}
and with
\begin{equation}
\begin{aligned}
\nabla \cdot (\psi^* \nabla \psi - \psi \nabla \psi^*) = \psi^* \nabla^2 \psi - \psi \nabla^2 \psi^*
\end{aligned}
\end{equation}
we obtain
\begin{equation}\label{Shroedinger_conservation_of_mass}
\frac{\partial}{\partial t} (\psi^* \psi) - \frac{j\hbar}{2M_\psi} \nabla \cdot (\psi^* \nabla \psi - \psi \nabla \psi^*) = 0.
\end{equation}
This matches the continuity equation of fluid dynamics for $\psi^* \psi = \phi = \rho = \rho_0 + \rho_\text{ac} > 0$ and $- \frac{j\hbar}{2M_\psi} (\psi^* \nabla \psi - \psi \nabla \psi^*) = \rho \vec v$. Furthermore, in section~\ref{ch:currents_as_sources} we will see that the four-vector $J^\mu = [-\rho_0 c_0, -\rho_0 \vec v]$ based on such mass and momentum in an incompressible flow is a conserved Noether current ($J^\mu{}_{,\mu} = 0$) and it matches the current from section~\ref{ch:analogy_with_EMG}. A quick comparison of eq.~(\ref{Shroedinger_conservation_of_mass}) and the divergence of eq.~(\ref{Neother_current_complex_scalar_field2}) shows that
\begin{equation}\label{mass_phonon_Schroedinger}
2M_\psi = \hbar \omega/c_0^2 = m
\end{equation}
is the mass of the field $\phi$ in section~\ref{ch:currents_as_sources} and it is much larger than the mass of each elementary particle, $M_0$ or $M$. Hence, the current $J^\mu$ (as well as the stress energy tensor $T^{\mu\nu}$), analogue charged particles and mass in the source regions always involve many fluid molecules (that is, systems of elementary particles) in classical acoustics and classical analogies. The analogue non-relativistic, massive particles obey the Schrödinger equation. Incidentally, the factor of 2 in eq.~(\ref{mass_phonon_Schroedinger}) stems from the Taylor expansion in eq.~(\ref{Schroedinger_to_Klein-Gordon_Taylor}) and leads to dispersion of waves described by the Schrödinger equation~\cite{Brennan2016,Griffiths2004}.

Without the second quantization of quantum field theory, $\psi$ is in general a vector of the pure state of the small fluid portion observed as a single entity, rather than a collection of its constitutive elementary particles. Any vector of a pure state (which does not have to be a determinate eigenstate) can be expressed as a weighted sum of orthonormal eigenvectors $\ket{i}$:
\begin{equation}\label{state_vector_expansion}
\ket{\psi} = \sum_i \psi_i \ket{i}
\end{equation}
with
\begin{equation}
\psi_i = \bra{i} \hat I \ket{\psi}.
\end{equation}
Here we introduced Dirac's bra-ket notation, according to which $\ket \psi = \psi$ is a ket-vector and $\bra \psi = \int \psi^* (...) d^3 \vec x$ is a bra-vector in the covariant vector space, where $(...)$ is to be replaced by a ket-vector. $\hat I$ denotes the identity operator. The orthonormal vectors satisfy
\begin{equation}\label{orhonormal_modes}
\sum_i \ket{i} \bra{i} = \hat I.
\end{equation}

The previous analysis showed that the quantum mechanical treatment is appropriate for studying large analogue quantum systems, while quantum field theory applies to the analogue elementary particles\footnote{The latter can be used, for example, for studying interactions at the level of elementary particles of different types, such as the mechanisms of aeroacoustic sound generation, which can be modelled as an interaction of the massive fluid field and radiated phonons (see section~\ref{ch:background_field_dipole_quadrupole}).} However, it should be noted that the two analogue theories reduce to the same basis in frequency domain and near field of massive particles, when the symmetry is spontaneously broken (see above), as well as in the far field of massless particles. For the usual time dependence of $\psi(\vec x,t) = \psi(\vec x) e^{-jEt/\hbar}$, the time-independent Schrödinger equation reads
\begin{equation}\label{time-independent_Schroedinger_equation}
E\psi = -\frac{\hbar^2}{2M_\psi} \nabla^2 \psi + V\psi
\end{equation}
and for $V=0$ (free particle)
\begin{equation}\label{time-independent_Schroedinger_equation_zero_potential}
E\psi + \frac{\hbar^2}{2M_\psi} \nabla^2 \psi = 0.
\end{equation}
After replacing $2 M_\psi = m$ and multiplying both sides of the equation with $m/\hbar^2$, a form of the Helmholtz equation is obtained:
\begin{equation}\label{Helmholtz_eq_from_Schroedinger}
\frac{Em}{\hbar^2} \psi + \nabla^2 \psi = 0.
\end{equation}
The same form follows from the Klein-Gordon equation in frequency domain (eq.~(\ref{action_variation_zero_real_scalar_field_freq_domain})) for massless particles (with $E = \omega \hbar$ and $m = E/c_0^2$), or in acoustic near field of massive particles (with $E = m c_0^2$), when the symmetry is spontaneously broken (so the first term in eq.~(\ref{action_variation_zero_real_scalar_field_freq_domain}) vanishes in favour of the second one and $m^2 \rightarrow -m^2$). As already anticipated, the Klein-Gordon equation is a far-field completion of the Schrödinger equation. As an important consequence, the determinate states of energy (eigenfunctions of the Hamiltonian~\cite{Griffiths2004}) correspond entirely to usual acoustic modes, solutions of the homogeneous Helmholtz equation, in the regions of space where $V=0$. The associated eigenvalues are $E = \omega^2\hbar^2/(m c_0^2)$. This holds in spite of the fact that the time-dependent Schrödinger equation does not describe massless fields and that its solutions $\psi$ have the time dependence $e^{-jEt/\hbar}$, rather than $e^{-j\omega t}$. It is only its first-order time derivative which makes a difference between massive quantum systems and Lorentz-invariant fields (massive in the near field or massless in the far field). Otherwise, small portions of fluid and phonons share the time-independent eigenmodes (determinate states), because the first quantization is common to quantum mechanics and quantum field theory.

The eigenmodes representing determinate states are mutually orthogonal (focusing on the perturbation and denoting it with $\psi$ rather than $\tilde \psi$ in the following):
\begin{equation}\label{modes_orthogonality}
\int_{V_S} \psi_m^* (\vec x) \psi_n (\vec x) d^3 \vec x = \bra{\psi_m} \hat I \ket{\psi_n} = \begin{cases} N_{ph,n} M_n &\text{for $m = n$}\\
0 &\text{otherwise}
\end{cases}
\end{equation}
with $N_{ph,n}$ the number of phonons in determinate state $\psi_n$ ($\sum_n N_{ph,n} = N_{ph}  \gg 1$, so $m \gg M$; the same relations hold for $\psi_0$ with $N_{m,n} M_0$ on the right-hand side of eq.~(\ref{modes_orthogonality}), $\sum_n N_{m,n} = N_{m}  \gg 1$, and $m \gg M_0$). Since we supposed that the modes are discrete (the indices $m$ and $n$ are positive integers), $N_{ph,n} M_n$ is finite, the vector space of $\vec x$ where the particle can be found is bounded, its volume $V_S$ is finite, and $\hat I$ is the identity matrix. Indeed, equation~(\ref{modes_orthogonality}) is also satisfied by eigenmodes of bounded acoustic systems, for example, a closed room or cavity with the interior volume $V_S$, when $N_{ph,n} M_n = \mathrm{P}_nV_S$. The constant $\mathrm{P}_n$ represents the positive amplitude of acoustic density for the mode $n$ averaged over $V_S$ (recall that ${\psi}^*{\psi} = |\rho_\text{ac}|$).

In general, acoustic modes are determined down to an arbitrary multiplication constant which can be absorbed in $\mathrm{P}_n$. In standard quantum mechanics, the modes are orthonormal (as the vectors in eq.~(\ref{orhonormal_modes})), that is, normalized such that in total $N_{ph} M = 1$ and $|\psi|^2$ expresses the probability distribution of location of the single particle, which has to integrate to one in the entire volume $V_S$. In the acoustic analogy, the value on the right-hand side of eq.~(\ref{modes_orthogonality}) is not only dictated by the definitions in equations~(\ref{background_fluid_mass_QFT}) and (\ref{acoustic_field_mass_QFT}), but also more appropriate for physical interpretations. Namely, it allows the particle mass-energy to appear not only in the exponent of the time-dependent wave function $\psi \sim e^{-jEt/\hbar}$, but in the amplitude, as well (the value of $M$ or $M_0$). This makes the summation of energies of particles meaningful in the acoustic context. In addition to this, $\psi_n$ and acoustic density amplitude are determined by the number of elementary particles (phonons) $N_{ph,n}$, which can be interpreted as the second quantization of quantum field theory. In acoustics, the arbitrary multiplication constant is fixed by the source of sound.

In an unbounded space the eigenmodes receive the well-known form $e^{j \vec k \cdot \vec x}$ of plane waves, where $\vec k$ is wave vector in continuous vector space and the sum in eq.~(\ref{state_vector_expansion}) should be replaced by a Fourier integral over $\vec k$, since the modes are not discrete. The wave function $\psi$ is not normalizable (the integral in eq.~(\ref{modes_orthogonality}) becomes a scaled delta function which is non-zero for equal modes) and, consequently, a free particle in an unbounded space cannot have a definite energy~\cite{Griffiths2004}. In closed spaces (particle in a potential well), a state vector is analogous to the modes of the room, the boundaries of which correspond to the potential. The boundaries can be imposed by other particles in the environment, as in the case of $\psi_0$ in an unbounded fluid, or some additional constrains.

With regard to the above, we can conclude that both the background fluid field and its perturbation follow the rules of quantum mechanics. They involve a large number of elementary particles (fluid molecules or phonons), the determinate states of which correspond to the classical acoustic modes with appropriate normalization. These depend on the environment. However, before continuing with the thermodynamic treatment in the next subsection, we should point to several differences between classical acoustic and quantum modes. While $|\rho_\text{ac} (\vec x)| = {\psi}^*(\vec x) {\psi}(\vec x)$ in the quantum analogy, contribution of a mode with eigenfrequency $\omega_n$ to the acoustic density amplitude at the receiver location $\vec x$ is in room acoustics~\cite{Masovic2021}
\begin{equation}\label{eq:solution_tailored_Green_Helmholtz_modes}
	|\rho_\text{ac} (\vec x)| = \frac{|p_\text{ac} (\vec x)|}{c_0^2}  = \frac{1}{c_0^2} |Q(\vec y)| |G_\text{tail}(\vec x | \vec y)| = |Q(\vec y)| \left| \frac{\psi (\vec x) \psi^* (\vec y)}{ C_n(\omega_n^2+2j \zeta_n \omega_n -\omega^2)} \right|,
\end{equation}
with $p_\text{ac}$ acoustic pressure, $\zeta_n$ (in 1/s) real and positive damping constant of the mode (which quantifies energy losses and exponential decay of the amplitude in time, $|\rho_\text{ac} (\vec x)| \sim e^{-\zeta t}$), and $C_n$ value of the integral in eq.~(\ref{modes_orthogonality}). The equality holds for a compact monopole source located at $\vec y$ with the source function $Q$ (in kg/s$^2$) and the tailored Green's function (frequency response of the room) $G_\text{tail}(\vec x | \vec y)$, which describes wave propagation from $\vec y$ to $\vec x$. The contribution is maximal at the eigenfrequency of the mode ($\omega = \omega_n$),
\begin{equation}\label{eq:solution_tailored_Green_Helmholtz_modes_at_eigenfrequency}
	|\rho_\text{ac} (\vec x)| = |Q(\vec y)| \left| \frac{\psi (\vec x) \psi^* (\vec y)}{ 2j \zeta_n \omega_n C_n} \right|.
\end{equation}
The following deviations from the analogy with quantum mechanics should be noticed:
\begin{itemize}
	\item Normalization with $C_n$ diminishes the effect of an arbitrary factor with which modal functions can be multiplied. On the other hand, it is the source function (strength) $Q$ which determines the resulting amplitude $|\rho_\text{ac}|$, and which should be associated with the number of analogue particles in the determinate state.
	\item The modes of real rooms are always damped, due to energy losses inside the rooms (for example, dissipation in air) or at the (sound absorbing) boundaries. This is not the case with isolated systems in standard quantum mechanics. However, a change of entropy can be associated with the damping and included within the framework of quantum thermodynamics (see section~\ref{ch:acoustic_modes_thermodynamics}).
	\item The appearance of $\vec y = \vec x$ in the quantum analogy\footnote{In the usual case in room acoustics, when $\vec y \neq \vec x$, the point monopole source and receiver can be treated as two identical interacting particles in the same potential of the environment (room), which exchange massless phonons. The associated interacting fields fit $\phi$ from section~\ref{ch:Lagrangian_formalism} rather than $\psi$ of quantum mechanics and, indeed, the Feynman propagator of the quantum field theory for the scalar fields $\phi$ matches the tailored Green's function.} can be justified only by a uniformly distributed source, existing at every point of observation, such as an apparent source of entropy due to the observer (sections~\ref{ch:dark_energy} and \ref{ch:acoustic_modes_thermodynamics}).
\end{itemize}
Taking these into account, eq.~(\ref{eq:solution_tailored_Green_Helmholtz_modes_at_eigenfrequency}) becomes
\begin{equation}\label{eq:solution_tailored_Green_Helmholtz_modes_at_eigenfrequency_quantum_mechanics}
	|\rho_\text{ac} (\vec x)| = |Q(\vec x)| \frac{\psi (\vec x) \psi^* (\vec x)}{ 2 \zeta_n \omega_n C_n},
\end{equation}
which equals $\psi (\vec x) \psi^* (\vec x)$ for
\begin{equation}\label{entropy_source_function_amplitude}
	|Q(\vec x)| = 2 \zeta_n \omega_n C_n  = 2 \zeta_n \omega_n \mathrm{P}_n V_S = 2 \zeta_n \omega_n N_{ph,n} M_n
\end{equation}
in the quantum analogy. The source must take into account not only the energy of phonons, but the damping, as well. In fact, it is driven by the damping and vanishes if $\zeta_n = 0$.

\subsubsection{Thermodynamics of the steady background fluid}\label{ch:background_fluid_thermodynamics}

Next we consider only the background fluid and suppose its simplest form -- uniform and steady fluid, such as quiescent air. Its thermodynamics can, of course, be treated classically, by modelling the fluid particles much like microscopic solid bodies with stochastic motion and bouncing off each other in space. However, it is of interest here to approach the problem within the framework of quantum thermodynamics, based on the quantum analogy established so far. With regard to this, we observe a quantum system which is coupled (entangled) to its environment (the so-called heat bath) E. The two together build a complete, isolated system. The total Hamiltonian can be expressed in terms of the Hamiltonians of the system ($\hat H$) and the environment ($\hat H_\text{E}$) as~\cite{Deffner2019}
\begin{equation}
\hat H_\text{tot} = \hat H \otimes \hat I_\text{E} + \hat I \otimes \hat H_\text{E},
\end{equation}
where we suppose an ultraweak coupling, that is, we neglect any energy exchange between the system and environment in the total energy, which is $E_\text{tot} = E + E_\text{E}$ and conserved. Hamiltonian is defined in eq.~(\ref{Schroedinger_equation}), $\otimes$ denotes a tensor product, and $\hat I_\text{E}$ and $\hat I$ are identity matrices of the environment and system, respectively. We should already note that boundedness of the observable universe allows expressing every state of each system as a sum of discrete modes from eq.~(\ref{modes_orthogonality}). The minimum frequency and frequency resolution of the observer are limited. This will be elaborated further in section~\ref{source_of_entropy}.

If the heat bath consists of $N_{m,\text{E}}$ fluid molecules as identical ultraweakly coupled sub-systems in different states, then
\begin{equation}\label{energy_system_environment}
E_\text{E} = \sum_{i=1}^{n} N_{m,\text{E},i} E_{\text{E},i}
\end{equation}
and
\begin{equation}\label{number_of_states_environment}
\sum_{i=1}^{n} N_{m,\text{E},i} = N_{m,\text{E}},
\end{equation}
with $n$ denoting the total number of energy/frequency bins (each characterized with energy $E_{\text{E},i}$) and frequency $\omega_i$, while $N_{m,\text{E},i}$ is the number of molecules in the state with energy $E_{\text{E},i}$. $E_{\text{E},i}$ and $\omega_i$ can represent centre values of the finite energy and frequency bins. If $n = 2$, the states are also called qubits~\cite{Deffner2019}. In view of the previous subsection, $E_{\text{E},i} = \omega_i^2\hbar^2/(M_0 c_0^2)$ and $N_{m,\text{E}} \gg 1$. The summation of energy/frequency in eq.~(\ref{energy_system_environment}) may at first seem unphysical in the acoustic context. However, as already discussed, it is allowed by the presence of elementary particles on the right-hand side of eq.~(\ref{modes_orthogonality}), so that the (very small) amplitude associated with a single elementary particle is also directly related to the frequency and the total energy is obtained after multiplying with the number of particles.

For the given energy of the environment, $E_\text{E}$, the number of states in which the environment consisting of distinguishable molecules can be found (its degeneracy) is~\cite{Griffiths2004}
\begin{equation}
\mathcal N_{m,\text{E}} (E_\text{E}) = \frac{N_{m,\text{E}}!}{\prod_{i=1}^{n} (N_{m,\text{E},i}!/d_i^{N_{m,\text{E},i}})},
\end{equation}
where $d_i$ is degeneracy (or modal density -- the number of modes sharing the same frequency or frequency bin) for energy $E_{\text{E},i}$. For large numbers $N_{m,\text{E},i} \gg 1$, Stirling's approximation gives
\begin{equation}
\begin{aligned}
\ln(\mathcal N_{m,\text{E}} (E_\text{E})) &= \ln(N_{m,\text{E}}!) + \sum_{i = 1}^{n} \left( N_{m,\text{E},i} \ln(d_i) - \ln(N_{m,\text{E},i}!) \right) \\
&\approx N_{m,\text{E}} \ln(N_{m,\text{E}}) + \sum_{i = 1}^{n} \left( N_{m,\text{E},i} \ln(d_i) - N_{m,\text{E},i} \ln(N_{m,\text{E},i}) \right).
\end{aligned}
\end{equation}
Canonical equilibrium is by definition a configuration of the heat bath in which the maximum number of energy eigenvalues $E_{\text{E},i}$ are occupied. Furthermore, the method of Lagrange multipliers applied to maximize $\ln(\mathcal N_{m,\text{E}} (E_\text{E}))$ under the conditions in equations~(\ref{energy_system_environment}) and (\ref{number_of_states_environment}) gives
\begin{equation}
\begin{aligned}
\frac{\partial}{\partial N_{m,\text{E},i}}& \left( \ln(\mathcal N_{m,\text{E}} (E_\text{E})) + \alpha \left( N_{m,\text{E}} - \sum_{i=1}^{n} N_{m,\text{E},i} \right) + \beta \left( E_\text{E} - \sum_{i=1}^{n} N_{m,\text{E},i} E_{\text{E},i} \right) \right) \\
&\approx \ln(d_i) - \ln(N_{m,\text{E},i}) - \alpha - \beta E_{\text{E},i} = 0
\end{aligned}
\end{equation}
and the Boltzmann-Gibbs distribution as the solution~\cite{Griffiths2004}:
\begin{equation}\label{Boltzmann-Gibbs_distribution}
N_{m,\text{E},i} = d_i e^{-\alpha - \beta E_{\text{E},i}} = d_i e^{-\alpha - E_{\text{E},i}/(k_B T_0)},
\end{equation}
where $\alpha$ and $\beta$ are the Lagrange multipliers and the equilibrium temperature is defined as $T_0 = 1/(\beta k_B)$ with $k_B$ the Boltzmann constant ($k_B = 1.38 \cdot 10^{-23}$\,J/K).

The exact values of $d_i$ and $E_{\text{E},i}$ depend on the actual potential in $\hat H_\text{E}$ in the same way as modal density and eigenfrequencies of acoustic modes depend on the room. For example, if the molecules are located in a three-dimensional infinite square well (which is analogous to a rectangular room\footnote{Compare with eq.~(\ref{wave_equation_and_rigid_boundary}) and the problem discussed there. Micro-cavity at the molecular (acoustic Planck) scale (acoustic Planck length $L_P$ will be defined in eq.~(\ref{Planck_length})) is formed by the potential of the fluid particles in the environment and relates to the high frequency $\omega_0$. Acoustic waves with frequencies much smaller than $\omega_0$ require larger bounded spaces. Still, the quantum analogy holds for both length scales, as already indicated in the context of eq.~(\ref{Helmholtz_eq_from_Schroedinger}). For example, in section~\ref{acoustic_modes_and_equilibrium} focusing on acoustic fields we will see that a diffuse sound field entirely corresponds to the thermodynamic equilibrium defined here.} with volume $V_S$ and we let large $\omega_i$ formally take continuous values $\omega>0$, the value of $d$ within a small frequency range $d\omega$ equals\cite{Griffiths2004,Kuttruff2009}
\begin{equation}\label{modal_density_Schroeder}
d = \frac{V_S}{2 \pi^2 c_0^3} \omega^2 d\omega.
\end{equation}
Moreover, it can be shown that this equality holds in acoustically large rooms regardless of their shapes, when $\omega L/c_0 \gg 1$, with $L \sim \sqrt[3]{V_S}$ characteristic length scale of the room. Thus, the environment does not strictly have to build a rectangular lattice. Notice also that $\omega_0 L/c_0 \gg 1$ for all observable length scales $L$.

After replacing $E_{\text{E},i}$ with continuous $\omega^2\hbar^2/(M_0 c_0^2)$, the eigenvalues of eq.~(\ref{Helmholtz_eq_from_Schroedinger}) for a single molecule, and letting the highest frequency approach the infinity, we obtain
\begin{equation}\label{number_of_states_environment_rectangular_room}
N_{m,\text{E}} = \frac{V_S e^{-\alpha}}{2 \pi^2 c_0^3} \int_{0}^{\infty} e^{- \omega^2 \hbar^2/(k_B T_0 M_0 c_0^2)} \omega^2 d\omega = V_S e^{-\alpha} \left( \frac{k_B T_0 M_0}{4 \pi \hbar^2} \right)^{3/2}
\end{equation}
and, multiplying additionally with the molecular energy $\omega^2\hbar^2/(M_0 c_0^2)$,
\begin{equation}\label{energy_system_environment_rectangular_room}
\begin{aligned}
E_\text{E} &= \frac{V_S e^{-\alpha} \hbar^2}{2 \pi^2 M_0 c_0^5} \int_{0}^{\infty} e^{- \omega^2 \hbar^2/(k_B T_0 M_0 c_0^2)} \omega^4 d\omega = \frac{V_S e^{-\alpha} \hbar^2}{2 \pi^2 M_0 c_0^5} \frac{3 \sqrt{\pi}}{8} \left( \frac{k_B T_0 M_0 c_0^2}{\hbar^2} \right)^{5/2} \\
&= \frac{3 V_S e^{-\alpha} k_B T_0}{2} \left( \frac{k_B T_0 M_0}{4 \pi \hbar^2} \right)^{3/2} = \frac{3N_{m,\text{E}} k_B T_0}{2}.
\end{aligned}
\end{equation}
These values of the integrals hold for any $\omega$ in the upper limit of the integral which satisfies $\omega^2 \hbar^2/(k_B T_0 M_0 c_0^2) \gtrapprox 4$. Together with $E_{\text{E},i} = \omega_i^2\hbar^2/(M_0 c_0^2)$ and the equation of state~(\ref{speed_of_sound_temperature}) which we are about to derive, this condition is automatically satisfied when $N_{m,\text{E},i} \gg 1$, which was already assumed in Stirling's approximation. This result shows that for the universal modal density in eq.~(\ref{modal_density_Schroeder}) at high enough frequencies, all eigenvalues $E_{\text{E},i}$ are equally probable ($N_{m,\text{E}}$ and $E_\text{E}$ do not depend on $\omega$, so all $\omega_i$ are equally probable). Not only is the maximal number of energy eigenvalues (frequency bins) occupied, but they are also all equally likely from an observer's point of view (similarly as in an unbounded free space, in which angular frequency of plane waves can take any real positive value, or in a diffuse sound field in a room). The entropy is thus maximal. It takes only acoustically large system with a large number of elementary particles in thermodynamic equilibrium.

The last equality in eq.~(\ref{energy_system_environment_rectangular_room}) represents also the expression for average kinetic energy of $N_{m,\text{E}}$ atoms of an ideal monoatomic gas. The factor of 3 is due to the three degrees of freedom per atom~\cite{Gemmer2010}, which actually belong to the same sub-system (here a fluid molecule) and the factor of 1/2 is due to eq.~(\ref{mass_phonon_Schroedinger}) ($E_\text{E} = 3 M_{\psi,\text{E}} c_0^2 = 3N_{m,\text{E}} M_0 c_0^2/2$). Accordingly, with $M_0 c_0^2 = 2 E_\text{E}/(3N_{m,\text{E}}) = k_B T_0$ multiplied with the heat capacity ratio $\gamma$ according to eq.~(\ref{density_correction}), we obtain the equation of state
\begin{equation}\label{speed_of_sound_temperature}
M_0 c_0^2 = \gamma k_B T_0
\end{equation}
and $T_0$ is the reference temperature of the theory, a consequence of random motion of the gas molecules. Similarly as $c_0$ relating space and time, $G$ and $\mu_0$ relating mass/charge and space, or $\hbar$ relating mass and frequency, $k_B$ relates mass-energy and temperature and eq.~(\ref{speed_of_sound_temperature}) can be seen as its definition for the molecular mass and background temperature. The latter can also be expressed in terms of $\omega_0$ from eq.~(\ref{the_highest_frequency}) as
\begin{equation}\label{equilibrium_frequency}
T_0 = \frac{\hbar \omega_0}{\gamma k_B }.
\end{equation}
The associated background fluid density (in the context of quantum thermodynamics also referred to as typical or average state~\cite{Gemmer2010}) is
\begin{equation}\label{equation_of_state_pressure}
\phi_0 = \rho_0 = \gamma p_0/c_0^2 = \mathrm{P}_0
\end{equation}
($p_0$ denotes the static pressure and $\gamma p_0 = \rho_0 c_0^2$ is the bulk modulus of the ideal gas), following from equations~(\ref{background_fluid_density_QFT}) and (\ref{modes_orthogonality}).
%prove the second equality

Thus, the key properties of the steady background fluid are completely determined by the high-frequency eigenmodes ($\omega_0 \gg \omega$ with $\omega$ frequency of any acoustic perturbation), as long as the observed region of space is acoustically large, with the minimum length scale $L \gg c_0/\omega_0$, and with a large number of modes and molecules per mode. The mean background density follows then from quantum thermodynamics of the massive system of distinguishable fluid molecules in canonical equilibrium, interacting only ultraweakly with its environment. More specifically, $d_i$ appearing in eq.~(\ref{Boltzmann-Gibbs_distribution}) was expressed in eq.~(\ref{modal_density_Schroeder}) using modal analysis of the background fluid field, while the exponential distribution was a result of the statistical treatment. It is interesting to note that in quantum theory such canonical equilibrium at high frequencies would also prohibit a possible collapse of the massive system to the lowest energy configuration~\cite{Schwartz2014}, which would otherwise result in a low entropy state and the violation of the second law of thermodynamics. The non-zero background density is thus permitted by the laws of thermodynamics and maximization of entropy, not the Fermi exclusion principle or quantum field theory alone. Furthermore, massless particles obtain mass from the background field in the near field (section~\ref{ch:mass_near_field}), so potential of any large environment which is measurable induces the mass-energy of particles~\cite{Hills2015}.

Finally, a few remarks on the required size of the system and the large number of elementary particles in it can be given. The condition $\omega_0 L/c_0 \gg 1$ gives
\begin{equation}
\frac{M_0 c_0}{\hbar} \frac{2GM_\text{E,tot}}{c_0^2} \gg 1,
\end{equation}
where the first term follows from eq.~(\ref{reduced_Planck_constant}) and the second term from eq.~(\ref{Schwarzschild_radius_mass}) with the mass of the system $M_\text{E,tot} = N_{m,\text{E}} M_0$. As expected,
\begin{equation}
N_{m,\text{E}} \gg \frac{\hbar c_0}{2GM_0^2} = \frac{2 \hbar G}{ L_0^2 c_0^3} = \frac{\hbar G}{ L_P^2 c_0^3} = 1,
\end{equation}
where we replaced $M_0 = L_0 c_0^2/(2G)$, again according to eq.~(\ref{Schwarzschild_radius_mass}). Acoustic Planck length $L_P$ is the length scale of a fluid molecule and the last two equalities will follow from eq.~(\ref{Planck_length}). Therefore, thermodynamic behaviour of the system does not take an extremely large number of constituent particles~\cite{Gemmer2010}. On the other hand, a large number of molecules, $N_{m,\text{E}}$, ensures a sufficiently high energy $E_\text{E}$ to provide an acoustically large space at frequency $\omega_0$, with many frequency bins populated with modal eigenfrequencies. Hence, the conditions of an acoustically large space and large total number of particles and modes are essentially equivalent. However, they still do not imply a large number of particles in each frequency bin, which is necessary for Stirling's approximation (for example, if majority of the particles belong to only a few possible modes). This is guaranteed by the high entropy of the canonical equilibrium at high frequencies. In contrast to these conditions, a small number of possible modes can be expected in acoustically small bounded spaces with fewer elementary particles (when the analogy with quantum field theory is more natural), while a non-uniform occupancy of frequency bins (low entropy) can also occur in acoustically large rooms at low frequencies.

Besides the steady background fluid in equilibrium, the quantum thermodynamic analogy can be applied to the acoustic fields, which are not necessarily in the state of canonical equilibrium and maximal entropy. This will be the subject of section~\ref{ch:acoustic_modes_thermodynamics}. Before that we give a brief overview of the Lagrangian formalism and treatment of the background fluid for the non-scalar field theory of dipole and quadrupole radiation.

\subsubsection{Analogue quantum fields for dipole and quadrupole radiation}\label{ch:background_field_dipole_quadrupole}

So far we have treated only real scalar quantum fields, which are sufficient for compressible sound waves in fluids and simple monopole sources. As in the classical analogies, a generalization to vector ($\bar A^\mu$) and second-order ($\bar h^{\mu\nu}$) fields is necessary in order to include dipole and quadrupole radiation. These fields should also be described by Lorentz-invariant Lagrangians and interact with the background field in a similar way as the scalar fields. However, they are not essential for the thermodynamic analysis, so we will give only a brief overview of their treatment in quantum field theory.

Modelling the background fluid turns out to be quite versatile and adaptable to the field in question. The mean density $\rho_0$ is an appropriate quantity for compressible waves. However, it is not the most natural choice for the vector and second-order tensor fields, which are generated by incompressible flows, describe incompressible waves at the leading order, and capture kinematics of particles located in the fields. Their coupling with the massive background fluid must also be kinematic and we should expect the background field to be expressed in terms of $c_0$ rather than $\rho_0$ or $p_0$. Still, the mechanism of spontaneous symmetry breaking in the near field must still hold.

Motion of particles due to an incompressible perturbation cannot be generally captured with a single real scalar field and the Lagrangians above. The next simplest Lagrangian is for a complex scalar field $\phi$. Since it has two degrees of freedom, the real and imaginary parts, we can impose an additional constraint when describing a physical fluid. It is usual to force the symmetry $\phi \rightarrow e^{j\alpha(x)} \phi$, where $\alpha(x)$ is an arbitrary phase ($x$ denotes a point in spacetime), which makes amplitude the only physically relevant part of the field. Notice that such a convention differs from the one which is common in classical acoustics and which takes the real part of a complex field as physically relevant.

At first glance, the most general Lorentz-invariant Lagrangian up to the second order, with the given symmetry $\phi \rightarrow e^{j\alpha} \phi$ of a massive field, may seem to be
\begin{equation}\label{Lagrangian_complex_scalar_field}
\mathcal{L} = \phi^{*,\mu} \phi_{,\mu} + \left( \frac{m c_0}{\hbar} \right)^2 \phi^*\phi
\end{equation}
(compare with eq.~(\ref{Lagrangian_real_scalar_field}); the Lagrangian for complex $\phi$ is conventionally without the multiplication factor 1/2). The two equations of motion (for $\phi$ and $\phi^*$, which are often referred to as particle and antiparticle) are calculated from the principle of least action similarly as in section~\ref{ch:Lagrangian_formalism} and read
\begin{equation}\label{equations_of_motion_scalar_field}
\Box \phi - \left( \frac{m c_0}{\hbar} \right)^2 \phi = 0 \text{ and } \Box \phi^* - \left( \frac{m c_0}{\hbar} \right)^2 \phi^* = 0.
\end{equation}
However, the values of $\phi$ at different points $x$  (and, therefore, its derivatives and other related mathematical expressions) still depend on the arbitrary choice of $\alpha(x)$ which can be made at each particular $x$. In order to fix this, a connection has to be introduced~\cite{Schwartz2014}, which is a vector gauge field $\bar A^\mu$ coupled to the field $\phi$. The simplest Lagrangian necessarily becomes
\begin{equation}\label{Lagrangian_complex_scalar_field_gauge_field}
\mathcal{L} = -\frac{1}{4} F_{\mu\nu} F^{\mu\nu} + (D_\mu \phi)^* (D^\mu \phi) + \left( \frac{m c_0}{\hbar} \right)^2 \phi^*\phi,
\end{equation}
which is, indeed, the Lagrangian of scalar quantum electrodynamics. As before, $F_{\mu\nu} = \bar{A}_{\nu,\mu} - \bar{A}_{\mu,\nu}$, while $D_\mu \phi = \phi_{,\mu} + je\bar{A}_\mu \phi/\hbar$ and $(D_\mu \phi)^* = \phi^*_{,\mu} - je\bar{A}_\mu \phi^*/\hbar$ are covariant derivatives, the second terms of which couple the fields $\bar A^\mu$ and $\phi$, and $e$ is a constant (elementary positive charge).

From section~\ref{ch:analogy_with_EMG} we know that $\bar{A}^\mu$ captures sound waves of dipole sources. The complex field $\phi$ in the Lagrangian in eq.~(\ref{Lagrangian_complex_scalar_field_gauge_field}) thus turns out to represent the fluid particles moving in the field, which are responsible for the propagation of sound waves and act like compact dipole sources in free space\footnote{The dipole source of waves, current $J^\mu$, will be derived in section~\ref{ch:currents_as_sources} which deals with the sources of quantum fields (see eq.~(\ref{Neother_current_complex_scalar_field2})). A similar distinction between the fluid particles carrying the waves and the field did not appear in the case of the real scalar fields, because the compressible waves already represent the unsteady mass of the particles.}. In order to model the steady background fluid, as well, we observe spontaneous breaking of now continuous symmetry $\phi \rightarrow e^{j\alpha(x)} \phi$. Canonically normalized Lagrangian of the next allowed higher order of $\phi$ is
\begin{equation}\label{Abelian_Higgs_model}
\mathcal{L} = -\frac{1}{4} F_{\mu\nu} F^{\mu\nu} + (D_\mu \phi)^* (D^\mu \phi) - \left( \frac{m c_0}{\hbar} \right)^2 |\phi|^2 + \frac{\lambda}{4} |\phi|^4,
\end{equation}
where $\phi$ now represents fluid particles in general, with or without a sound field. Sign of the mass term has been changed in the near field for the same reason as in eq~(\ref{Lagrangian_real_scalar_field_spontaneous_symmetry_breaking_strong_perturbations}). The potential has a minimum for
\begin{equation}
\frac{\partial}{\partial |\phi|} \left( \left( \frac{m c_0}{\hbar} \right)^2 |\phi|^2 - \frac{\lambda}{4} |\phi|^4 \right) = 2 \left( \frac{m c_0}{\hbar} \right)^2 |\phi| - \lambda |\phi|^3 = 0,
\end{equation}
which is for non-zero $|\phi_0| = \sqrt{2m^2c_0^2/(\lambda \hbar^2)}$, or $\phi_0 = \sqrt{2m^2c_0^2/(\lambda \hbar^2)} e^{j\theta}$ with arbitrary $\theta(x)$. We can choose $\phi_0 = \sqrt{2m^2 c_0^2/(\lambda \hbar^2)}$, which matches the expression in eq.~(\ref{real_scalar_field_vacuum}). The multiplication factors in the square roots become equal when $\lambda$ here is replaced by $\lambda/3$, in order to fit the canonical normalization in eq~(\ref{Lagrangian_real_scalar_field_spontaneous_symmetry_breaking_strong_perturbations}). However, the field $\phi$ considered here is purely kinematic and we should not expect $\phi_0 = \rho_0$ as in section~\ref{background_fluid_and_acoustic_particles}.

Next we could expand $\phi = \phi_0 + \tilde{\phi}$ with a weak perturbation $\tilde \phi$, as in section~\ref{background_fluid_and_acoustic_particles}. However, a more elegant approach for the symmetry $\phi \rightarrow e^{j\alpha(x)} \phi$ is to use the two real degrees of freedom of $\tilde{\phi}(x)$, denoted with $\sigma(x)$ and $\pi(x)$, to write
\begin{equation}
\phi = \left( \phi_0 + \frac{\sigma}{\sqrt{2}} \right) e^{j \pi/ (\sqrt{2} \phi_0)}.
\end{equation}
Only $\pi(x)$ is associated with the gauge field $\bar A^\mu$ and sound waves. Inserting this into eq.~(\ref{Abelian_Higgs_model}) gives~\cite{Schwartz2014}
\begin{equation}\label{Abelian_Higgs_model_expanded}
\begin{aligned}
\mathcal{L} &= -\frac{1}{4} F_{\mu\nu} F^{\mu\nu} \\
&- \left( \phi_0 + \frac{\sigma}{\sqrt{2}} \right)^2 \left( \frac{\sigma_{,\mu}}{\sqrt{2} \phi_0 + \sigma} -j \frac{\pi_{,\mu}}{\sqrt{2} \phi_0} - j\frac{e\bar{A}_\mu}{\hbar} \right) \left( \frac{\sigma^{,\mu}}{\sqrt{2} \phi_0 + \sigma} +j \frac{\pi^{,\mu}}{\sqrt{2} \phi_0} + j\frac{e\bar{A}^\mu}{\hbar} \right) \\
&- \left( \frac{1}{\lambda}\left( \frac{m c_0}{\hbar} \right)^4 - \left( \frac{m c_0}{\hbar} \right)^2 \sigma^2 - \frac{m c_0 \sigma^3}{2 \hbar} \sqrt{\lambda} - \frac{\lambda \sigma^4}{16} \right).
\end{aligned}
\end{equation}
%check this!!!!!!!!!!
In order to decouple and remove $\sigma(x)$, we can let $m \rightarrow \infty$ and $\lambda \rightarrow \infty$ (keeping $\phi_0$ unaffected), so that the Lagrangian becomes
\begin{equation}\label{Abelian_Higgs_model_expanded_decoupled_sigma}
\mathcal{L} = -\frac{1}{4} F_{\mu\nu} F^{\mu\nu} - \left( \frac{e \phi_0}{\hbar} \right)^2 \left( \bar{A}_\mu + \frac{\hbar \pi_{,\mu}}{\sqrt{2} e \phi_0} \right) \left( \bar{A}^\mu + \frac{\hbar \pi^{,\mu}}{\sqrt{2} e \phi_0} \right).
\end{equation}
We can also choose $\alpha(x)$ such that $\pi(x) = 0$ (the so-called unitary gauge). The obtained
\begin{equation}\label{Abelian_Higgs_model_expanded_decoupled_sigma_unitary_gauge}
\begin{aligned}
\mathcal{L} &= -\frac{1}{4} F_{\mu\nu} F^{\mu\nu} - \left( \frac{e \phi_0}{\hbar} \right)^2 \bar{A}_\mu \bar{A}^\mu
\end{aligned}
\end{equation}
is the Lorentz-invariant Proca Lagrangian for the massive gauge field $\bar A^\mu$ with mass $m = \sqrt{2}e\phi_0/c_0$. Indeed, the equation of motion derived with this Lagrangian (section~\ref{ch:Lagrangian_formalism}) is~\cite{Schwartz2014}
\begin{equation}\label{action_variation_zero_vector_field}
\Box \bar{A}^\mu - \left( \frac{m c_0}{\hbar} \right)^2 \bar{A}^\mu = 0,
\end{equation}
which is the four-vector form of eq.~(\ref{action_variation_zero_real_scalar_field}), and the Lorenz gauge condition is satisfied, $\bar{A}^\mu{}_{,\mu} = 0$. Laplace's equation is obtained in frequency domain when eq.~(\ref{reduced_Planck_constant}) holds,
\begin{equation}\label{action_variation_zero_vector_field_near_field}
	\nabla^2 \bar{A}^\mu = 0,
\end{equation}
and massive $\bar A^\mu$, indeed, represents massless (acoustic) $\bar A^\mu$ in the acoustic near field, when the symmetry is spontaneously broken. In quantum field theory the process of obtaining mass is called Higgs mechanism. The field $\pi(x)$ (often referred to as pion) represents a Goldstone boson and the field $\sigma(x)$ is known as Higgs boson. Since the positive charge $e$ should be analogous to the particle mass $m$, we find $\phi_0 = m c_0/(\sqrt{2} e) = c_0/\sqrt{2}$ and $\lambda = 4 m^2/\hbar^2$. The background value $\phi_0$ is essentially the speed of sound.
%...p565,p575

A complex scalar field describes kinematically, in terms of velocity, dipole radiation and the background fluid in the same manner as a real scalar field describes monopole radiation and the background fluid in terms of density. Extending the analysis to  quadrupole radiation, a symmetry of a (kinematic) field of fluid particles with additional degrees of freedom should give rise to the second-order tensor gauge field $\bar h^{\mu\nu}$. This essentially massless field should obtain mass in the acoustic near field by means of spontaneous symmetry breaking. In addition to this, it should be noted that in standard quantum electrodynamics the complex scalar field $\phi$ from above is replaced by a field of spin 1/2. A generalization to higher spins based on the tetrad formalism~\cite{Rovelli2015} allows inclusion of quadrupole (and other higher multipole) radiation. Accordingly, we should expect the most general field of fluid particles to be a spinor field\footnote{Such spinor field in loop quantum gravity does not only couple with the background metric (as the field $\phi$ with the gauge field above), but constitutes it. At the price of mathematical complexity, this concept seems to be more suitable for the acoustic spacetime, as well, which is built by the same particles as the sources and fields in it. The fluid particles are not immersed in the acoustic spacetime (like, for example, fermions in quantum field theory in curved spacetime), they constitute it, in the same way as the spin foam makes the spacetime in loop quantum gravity. In other words, we should deal with a pure (and possibly perturbed) vacuum only, rather than vacuum-field interactions. Nevertheless, we take the simpler approach, which is closer to the classical theories, as sufficient and leave the characterization of the background spacetime to a separate thermodynamic theory.}. However, this mathematically more involved and abstract approach is not necessary for the basic considerations here.

%not an incompressible near field, which is in eq.~(\ref{Maxwell_currents_Lorenz_gauge_split_solution_far_geometric_field_dipole_moment_j_Taylor_expansion}) (eq.(5.31)) and hence in the first term in the last equation!! Only negative mass term gives Laplace's equation. So what is this in acoustics, acoustic short circuit?

Although we will not generalize the Lagrangians in equations~(\ref{Lagrangian_complex_scalar_field_gauge_field}) and (\ref{Abelian_Higgs_model}) further to quadrupole radiation\footnote{In section~\ref{ch:currents_as_sources} we will actually argue that the Lagrangian in eq.~(\ref{Lagrangian_complex_scalar_field}) suffices for the description of fluid particles (without the gauge field) in this case, as well, since the field $\phi$ already captures their kinematics.} nor provide a detailed analysis of spontaneous symmetry breaking, several important concepts can be established. Interestingly, many of them appear to have analogues in the theory of quantum gravity. For example, in the absence of other sources or boundaries, the only ``foreign" objects which appear in vacuum are apparent microscopic boundaries at the smallest observable length scale. In classical acoustics, this is at least the scale of random molecular motion in the fluid. As in the case of other sources and compact bodies, spontaneous symmetry breaking takes place in the near field of such boundaries. According to eq.~(\ref{spontaneous_symmetry_breaking_length_scale}), radius of the near-field region can be estimated as $L_0 = \hbar/(M_0 c_0)$, where $M_0$ is molecular mass of the fluid, and it can also be associated with the maximum frequency of the theory, $\omega_0 = c_0/L_0$.

In the acoustic near field the massless field $\bar h^{\mu\nu}$ obtains mass of the background fluid, the thermodynamics of which was already discussed in section~\ref{ch:background_fluid_thermodynamics}. On the other hand, the indeterminism of particle motion in such small regions means that the particles can be considered to move stochastically\footnote{Thermodynamics of the fields and occurrence of monopole sources of entropy is the subject of the next subsection.} and, accordingly, observed as a micro-turbulence, the aeroacoustic quadrupole source from section~\ref{ch:aeroacoustic_sound_generation}. Therefore, the microscopic limitation of the observers causes not only boundaries of the observable universe, but also apparent quadrupole sources. Equation~(\ref{Schwarzschild_radius_mass}) relates then the mass of such a source with its acoustic Schwarzschild radius $L_0$: $M_0 = L_0 c_0^2/(2G)$. Equating the Schwarzschild radius with the radius of the near field and the molecular mass with the mass of the quadrupole source, we find
\begin{equation}\label{Planck_length}
L_0 = \sqrt{\frac{2\hbar G}{c_0^3}} = \sqrt{2} L_P,
\end{equation}
where $L_P = \sqrt{\hbar G/c_0^3}$ is the acoustic Planck length, and $M_0 = m_P/\sqrt{2}$, with the Planck mass by definition $m_P = \sqrt{\hbar c_0/G}$. A very small fluid element with the volume $V_0 \sim L_0^3$ behaves as an acoustic Planck particle. Within this volume, all fields are massive (massless sound fields and phonons lose meaning) and the mass originates from the unobservable fluid molecules, the elementary particles, and thermodynamics of the background fluid. In fact, this is the smallest meaningful source region, in which the massive perturbation of the acoustic spacetime cannot be distinguished from the generated fields and the separation of the left- and right-hand sides of the Einstein field equations becomes inadequate. Supposing that the background fluid is an ideal gas, we can relate the acoustic Planck particle directly to the reference temperature, as well. According to eq.~(\ref{speed_of_sound_temperature}),
\begin{equation}\label{Planck_temperature}
T_0 = \frac{M_0 c_0^2}{\gamma k_B} = \frac{m_P c_0^2}{\gamma k_B \sqrt{2}} = \frac{T_P}{\gamma \sqrt{2}},
\end{equation}
where $T_P = m_P c_0^2/k_B$ is Planck temperature.

The duality of the compact source-boundary has another important consequence. Section~\ref{ch:energy_and_charge} suggests that the aeroacoustic quadrupole source (micro-turbulence) reduces to a more efficient dipole source at the compact boundary, inducing charge at it. (Since the boundary set by the observer is motionless and rigid, no further reduction to monopole takes place.) The source strength is amplified by the factor~\cite{Howe2003} $c_0/(\omega_0 L_0)$ (recall also the discussion at the end of section~\ref{ch:aeroacoustic_sound_generation} and the $8^\text{th}$- and $6^\text{th}$-power laws of aeroacoustic quadrupoles and dipoles from equations~(\ref{quadrupole_ac_intensity_scaling}) and (\ref{dipole_ac_power_scaling})). Including this in eq.~(\ref{Planck_temperature}) and using again $L_0 = 2GM_0/c_0^2$ and $M_0c_0^2 = \hbar \omega_0$, we obtain
\begin{equation}\label{Hawking_radiation}
T_H = \frac{c_0 T_0}{\omega_0 L_0} = 4\pi\frac{\hbar c_0^3}{8 \pi \gamma G k_B M_0}.
\end{equation}
Apart from the factor $4 \pi$, this is the expression for Hawking radiation of a black hole with mass $M_0$. The more efficient dipole radiation at the boundary represents analogue electromagnetic radiation at the event horizon of the microscopic acoustic black hole. Since every macroscopic black hole (turbulence) is built by such elementary black holes, the same expression holds in general, with $M_0 \rightarrow M$ denoting mass of the black hole. On the other hand, the microscopic quadrupoles generate a very weak random noise in the rest of the space, as well. Reduced to dipoles at the micro-boundaries, the noise they generate may be amplified and analogous to the cosmic microwave background.

Finally, thermodynamics of fields at the Planck scale also gives rise to a (monopole) source of entropy, as will be shown shortly.

\subsection{Thermodynamics of acoustic modes}\label{ch:acoustic_modes_thermodynamics}

Unlike a steady background fluid, acoustic fields are not necessarily in the state of thermodynamic equilibrium and for their treatment we will need a more general quantum thermodynamic approach than in section~\ref{ch:background_fluid_thermodynamics}. The extension is also necessary for characterization of the sources of entropy, because the entropy is not maximal in general. Since we will use the analogy with sound fields in bounded spaces more extensively in the following considerations, we will denote quantum states with $\psi$ and refer to the elementary particles as phonons, although it is clear from section~\ref{background_fluid_and_acoustic_particles} that the approach applies to the background fluid field ($\psi_0$), as well.

\subsubsection{Acoustic entropy and equilibrium}\label{acoustic_modes_and_equilibrium}

Quantum states are often neither pure (a weighted sum of determinate states) nor well determined. For example, this is the case with both acoustic and background fields at the acoustic Planck scale or when an acoustic system consists of many interacting sub-systems which can be observed only in their integrity. With regard to that, we introduce a (probability) density matrix $\hat \rho$ in eq.~(\ref{modes_orthogonality}), such that
\begin{equation}\label{density_matrix_states}
\bra{\psi} \hat \rho \ket{\psi} = \sum_{i,j = 1}^n \psi_i^* \hat \rho_{ij} \psi_j
\end{equation}
is not necessarily zero, even for different modes ($i \neq j$), and $n$ is the total number of possible states or frequency bins (due to the finite frequency
resolution of the observer). The probability of finding the quantum system in a pure state $\ket{\psi}$ is
\begin{equation}\label{probability_pure_state}
P(\ket{\psi}) = \frac{1}{\mathrm{P}_\text{tot}} \bra{\psi} \hat \rho \ket{\psi} = \frac{1}{\mathrm{P}_\text{tot}} \sum_{i=1}^{n} \psi_i^* \hat \rho_{ii} \psi_i,
\end{equation}
where $\mathrm{P}_\text{tot} = \sum_{i=1}^{n} \mathrm{P}_i = N_{ph} M/V_S$. The trace of $\hat \rho$ equals
\begin{equation}\label{density_matrix_trace}
\Tr{\hat \rho} = \sum_{i=1}^{n} \hat \rho_{ii} = 1
\end{equation}
and the expectation value of some observable $A$ with the operator $\hat A$ equals
\begin{equation}\label{observable_expectation}
A = \langle \hat A \rangle = \Tr{\hat A \hat \rho} = \sum_{i,j=1}^{n} \hat A_{ij} \hat \rho_{ij}.
\end{equation}
For example, internal energy of the system is
\begin{equation}\label{energy_density_matrix_hamiltonian}
E = \langle \hat H \rangle = \Tr{\hat H \hat \rho}.
\end{equation}
While pure states satisfy the Schrödinger equation, the density operator satisfies the Liouville-von Neumann equation~\cite{Gemmer2010}
\begin{equation}\label{Liouville-von Neumann equation}
j\hbar \frac{\partial \hat \rho}{\partial t} = [\hat H, \hat \rho] = \hat H \hat \rho - \hat \rho \hat H,
\end{equation}
with the commutator $[\text{ }]$ defined by the second equality.

The non-negative von Neumann entropy is by definition
\begin{equation}\label{von_Neumann_entropy}
S = -k_B \Tr{\hat \rho \ln \hat \rho}.
\end{equation}
It is minimal ($S_\text{min} = 0$) for any single determinate state (when one component $\hat \rho_{ii}$ equals one) and maximal ($S_\text{max} = k_B \ln (n)$) when all determinate states are equally probable, that is when
\begin{equation}
\hat \rho_{ij} = \begin{cases} 1/n &\text{for $i = j$}\\
0 &\text{otherwise.}
\end{cases}
\end{equation}
Hence, in the acoustic analogy the entropy quantifies ``tonality'' of the power spectral density (or sound pressure/density amplitude per hertz) averaged over all source and receiver locations inside a room with the volume $V_S$.

Recalling the concluding comments of section~\ref{background_fluid_and_acoustic_particles}, certain care is necessary when using power spectra to obtain acoustically meaningful entropy. First, the acoustic power spectral density depends not only on the system and environment (room), but also on the excitation (source strength). Accordingly, we will usually assume that the source function takes the same value in all frequency bins (a uniform excitation), in order to suppress its effects on entropy. Second, each quantum mode (determinate state) is associated only with the frequency bin containing its eigenfrequency. In contrast to this, damped acoustic modes have finite widths which can extend over several adjacent narrow bins. On the other hand, if the bins are too wide, the frequency resolution may be unsatisfactory and fine variations of the power spectra over frequency cannot be captured well enough. As a reasonable compromise, we first suppose that the damping constant $\zeta_i = \zeta$ is equal for all modes and bins $i$ in the considered frequency band $\Delta f_\text{band} = \Delta \omega_\text{band}/(2\pi)$ (typically an octave or third-octave band), which is segmented in $n$ equally wide and non-overlapping frequency bins $\Delta f_i = \Delta f = \Delta \omega/(2\pi) = \Delta f_\text{band}/n$. This implies choosing the frequency bands such that the damping (or reverberation time) in the room does not vary significantly within each band, although it may differ between bands. Then we specify the bin width equal to the half-width of the modes, which is\cite{Kuttruff2009} $\Delta \omega = 2\zeta$. In contrast to this, the modal width in quantum thermodynamics can be seen as normalized, similarly as the function of state $\psi$, so the frequency resolution (dictated here by the damping constant) disappears from the considerations.

At the boundaries of a frequency bin with the width $\Delta \omega$, centred at a mode's eigenfrequency, sound energy of the mode drops to 50\% of its peak value and the amplitude drops by the factor $\sqrt{2}$. The thermodynamic analogy thus becomes inadequate if modes overlap and interfere significantly, so their energies outside the associated bins cannot be neglected. This is the case with high Helmholtz numbers, $\omega L/c_0 \gg 1$, with $L$ characteristic length scale of the room. In this regime, which matches the equilibrium from section~\ref{ch:background_fluid_thermodynamics}, the modal density is high and statistical energy analysis (SEA) is sufficiently accurate and preferred. The analysis is based on the assumption of a diffuse field, which turns out to be analogues to the canonical thermodynamic equilibrium. The quantum thermodynamic analogy is also invalid if the number of frequency bins is very small, so $\Delta \omega$ is comparable to $\Delta \omega_\text{band}$, which may be the case with (fractional) octave bands at very low frequencies, especially in strongly damped rooms. The analogy is thus most accurate around and below the Schroeder frequency~\cite{Kuttruff2009},
\begin{equation}\label{Schroeder_frequency}
	f_\text{Schroed} = \sqrt{\frac{3 c_0^3}{4 \zeta V_S}},
\end{equation}
as long as $\Delta \omega_\text{band} \gg \Delta \omega$. Hence, it provides a mid-frequency extension of SEA and the equilibrium thermodynamics of section~\ref{ch:background_fluid_thermodynamics}, when the modal density is not high enough for the diffuse-field/equilibrium assumption.

Continuing with the analogy, acoustic entropy can be defined simply as
\begin{equation}\label{acoustic_entropy}
S = -k_B \sum_{i=1}^{n} \frac{\mathrm{p}_i^2}{\sum_{j=1}^{n} \mathrm{p}_j^2} \ln \frac{\mathrm{p}_i^2}{\sum_{j=1}^{n} \mathrm{p}_j^2},
\end{equation}
where $\mathrm{p}_i^2$ is the mean squared sound pressure amplitude per hertz in the bin $i$. Due to the normalization with $\sum_{j=1}^{n} \mathrm{p}_j^2$, the two fractions correspond to the diagonal elements of $\hat \rho$ and can be seen as the probability of a determinate state $i$ (fraction of the total energy in the frequency band $\Delta \omega_\text{band}$ due to the mode $i$). Moreover, any classical first-order acoustic scalar can be used in place of sound pressure, or a second-order quantity instead of its squared value. If $\mathrm{p}_i^2 = \mathrm{p}_\text{diff}^2$ is equal in every bin, the entropy is maximal, $S_\text{diff} = k_B \ln (n)$. This is a property of a diffuse field (with uniform excitation, as defined above) and, as already noticed in literature\cite{LeBot2017}, diffuse field is the analogue state of canonical thermodynamic equilibrium at high frequencies.  If $\mathrm{p}_i^2 \neq 0$ in a single frequency bin, the entropy is minimal, $S_\text{min} = 0$, since in this bin $\mathrm{p}_i^2 = \sum_{j=1}^{n} \mathrm{p}_j^2$.

The entropy defined in eq.~(\ref{acoustic_entropy}) can in principle be calculated from any concrete power spectrum available. However, an analytically obtained value in a certain simplified scenario can bring additional practical value and provide more information on the physical meaning of acoustic entropy, especially with regard to the equilibrium and diffuse field value $S_\text{diff}$. In accordance with the above, we first suppose that all modes are excited equally and all $m$ ($m \leq n$) non-zero values of $\mathrm{p}_i^2$ are equal, $\mathrm{p}_i^2 = \mathrm{p}^2$. The entropy simplifies to
\begin{equation}\label{acoustic_entropy_constant_damping}
S = -k_B \ln \frac{\mathrm{p}^2}{m \mathrm{p}^2} = k_B \ln (m).
\end{equation}
Under the assumption that only one mode can dominate in a single frequency bin, $m$ is the number of modes in $\Delta \omega_\text{band}$. At high frequencies, when many modes strongly overlap (and the analogy becomes inaccurate), the number of modes can actually surpass the number of bins, leading to non-physical $S > S_\text{diff}$. At these frequencies $S = S_\text{diff}$ must be forced and the analogy should reduce to SEA.

Now we can compare the entropy from eq.~(\ref{acoustic_entropy_constant_damping}) with $S_\text{diff}$ at low-to-middle frequencies. For equal total energies in $\Delta \omega_\text{band}$, $m \mathrm{p}^2 = n \mathrm{p}_\text{diff}^2$ and we obtain
\begin{equation}\label{entropy_SPL_correction}
S = -k_B \ln \frac{\mathrm{p}^2}{n \mathrm{p}_\text{diff}^2} = S_\text{diff} - k_B \ln \frac{\mathrm{p}^2}{\mathrm{p}_\text{diff}^2}.
\end{equation}
Hence, the energy ratio ${\mathrm{p}^2}/{\mathrm{p}_\text{diff}^2}$ quantifies the deviation of $S$ from its maximal equilibrium value. For $m \leq n$, $\mathrm{p}^2 \geq \mathrm{p}_\text{diff}^2$ and $S_\text{diff} - S \geq 0$. On the other hand, we can also express the non-negative sound pressure level difference
\begin{equation}
\Delta L = 10\log_{10} \frac{\mathrm{p}^2}{\mathrm{p}_\text{diff}^2},
\end{equation}
which can be used as a correction of the diffuse-field value (which can be calculated, for example, by means of SEA). From eq.~(\ref{entropy_SPL_correction}) it is evident that this correction amounts to the entropy difference between diffuse and (possibly) non-diffuse field:
\begin{equation}\label{SEA_SPL_correction}
\Delta L = (10 \log_{10} e) \frac{S_\text{diff} - S}{k_B} = (10 \log_{10} e) \left( \ln n - \frac{S}{k_B} \right).
\end{equation}
For $m = n$, $\Delta L = 0$ and the model, indeed, reduces to the diffuse field. For $m < n$, the entropy $S < S_\text{diff}$ and correction $\Delta L$ quantify the presence of narrow-band peaks in the spectrum, which deviate from the diffuse-field model. Furthermore, the constant $k_B$ from the definition of entropy cancels out and does not affect the value of $\Delta L$.

In order to illustrate the meaning of acoustic entropy, the readily available analytical solution for sound fields in rectangular rooms can be used. The number of modes $m$ in eq.~(\ref{acoustic_entropy_constant_damping}) can be estimated using an extended version of eq.~(\ref{modal_density_Schroeder})\cite{Kuttruff2009} as
\begin{equation}\label{rect_room_number_of_modes_extended}
b = \sum_{i=1}^{n} \left( \frac{V_S \omega_i^2}{2 \pi^2 c_0^3} + \frac{S_S \omega_i}{8 \pi c_0^2} + \frac{L_S}{16 \pi c_0} \right) \Delta \omega_i
\end{equation}
with $V_S = L_1 L_2 L_3$, $S_S = 2(L_1 L_2 + L_2 L_3 + L_1 L_3)$, $L_S = 4(L_1 + L_2 + L_3)$, $L_1$, $L_2$, and $L_3$ dimensions of the rectangular room, and $\omega_i$ the centre frequency of $\Delta \omega_i$. The additional two terms make the estimation of the fraction of axial and tangential modes more accurate at low and middle frequencies. Since $b/\Delta f_\text{band} = m/(n \cdot 1\text{\,Hz})$ is the modal density in the entire frequency band (number of modes per hertz, recalling that at most one mode is expected to dominate in a single frequency bin),
\begin{equation}\label{modal_density_per_hertz_and_bin}
m = nb \frac{1\text{\,Hz}}{\Delta f_\text{band}} = b \frac{1\text{\,Hz}}{\Delta f}.
\end{equation}
This gives a very simple and crude, real-valued approximation of $m$, rather than an integer number of bins in eq.~(\ref{acoustic_entropy_constant_damping}). However, the rounding has no significant effect if the number of modes in $\Delta f_\text{band}$ is large. At the lowest frequencies, when $m$ is comparable to or even smaller than one, the thermodynamic analogy is not valid anyway. (In practice, the Helmholtz equation can then be solved directly, typically using numerical schemes, since the number of modes is small and the computational burden is low.). Formally, we can set $\Delta L = 0$ for $b < 1/3$, that is, assume a flat power spectrum well below the lowest eigenfrequency. As already discussed, we also set $S = S_\text{diff} = k_B \ln n$ and $\Delta L = 0$ for $m > n$.

\begin{figure}[p]
	\centering
	\begin{subfigure}{.68\textwidth}
		%\centering
		\includegraphics[width=1\linewidth]{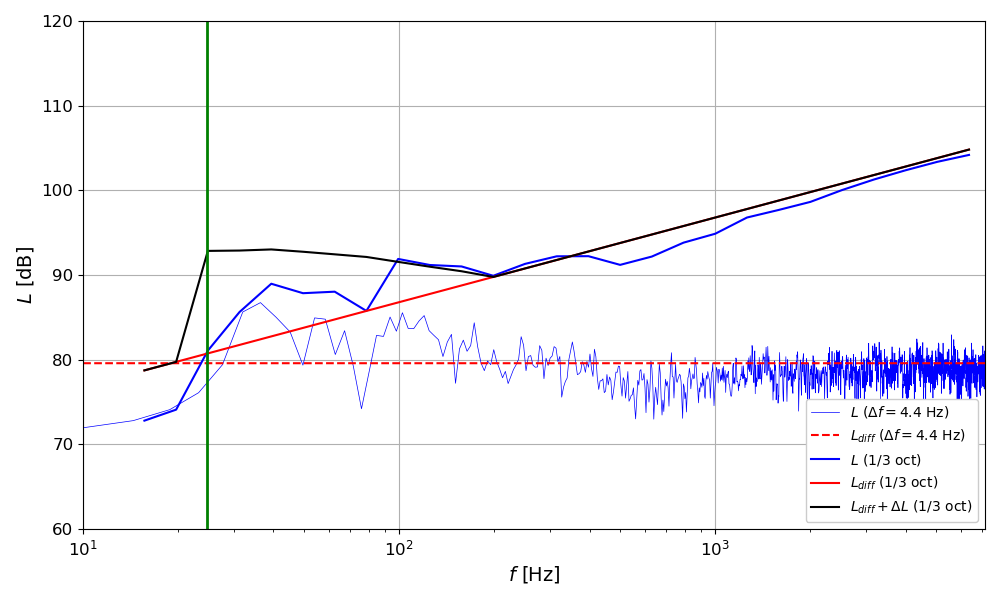}
		%\caption{A subfigure}
		\label{fig:L_room_T0-5s}
	\end{subfigure}
	\begin{subfigure}{.68\textwidth}
		%\centering
		\includegraphics[width=1\linewidth]{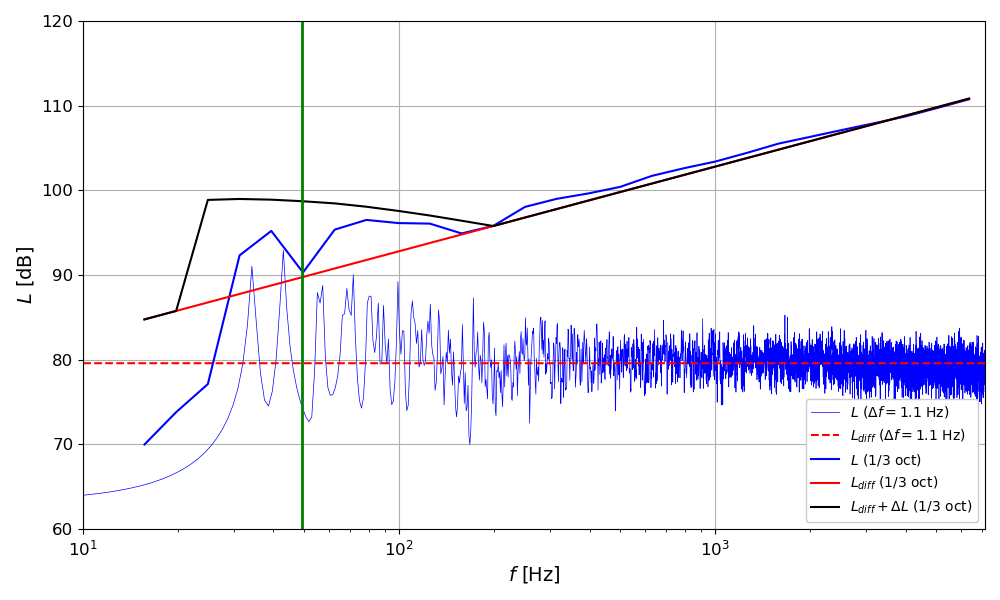}
		%\caption{A subfigure}
		\label{fig:L_room_T2s}
	\end{subfigure}
	\begin{subfigure}{.68\textwidth}
		%\centering
		\includegraphics[width=1\linewidth]{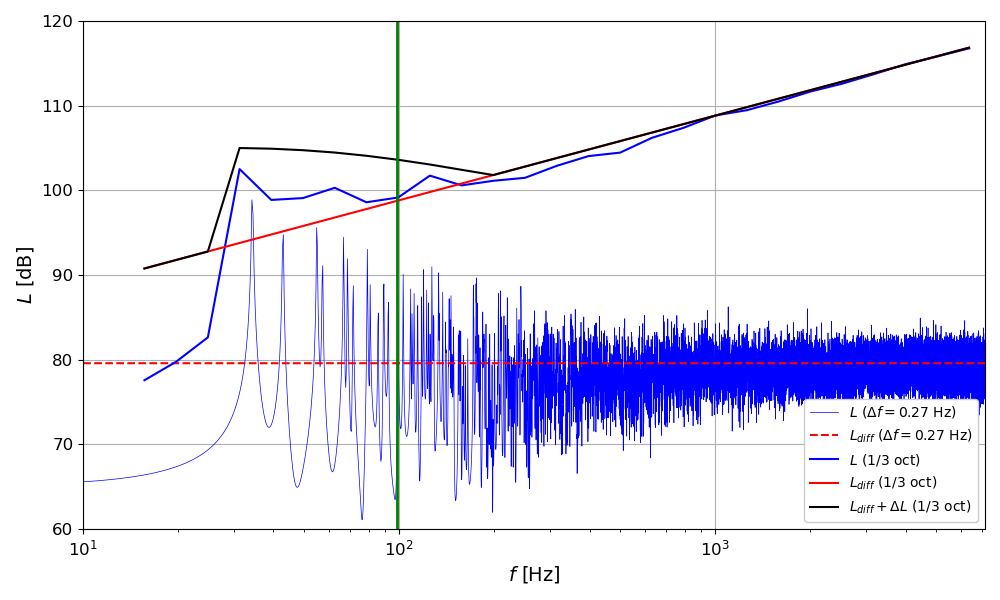}
		%\caption{A subfigure}
		\label{fig:L_room_T8s}
	\end{subfigure}
	\caption{Sound pressure level in the rectangular room (5m x 4m x 3m) with rigid walls and reverberation time (top) 0.5\,s, (middle) 2\,s, and (bottom) 8\,s. (Refer to the text for further details.)}
	\label{fig:mean_sound_level_room}
\end{figure}

As an example, Fig.~\ref{fig:mean_sound_level_room} shows the calculated spatially averaged sound pressure level in a rectangular room with rigid walls and dimensions ($L_1,L_2,L_3$) = (5\,m, 4\,m, 3\,m), filled with air at room temperature ($c_0 = 343$\,m/s, $\rho_0 = 1.2$\,kg/m$^3$). The entire considered frequency range ($f_\text{min},f_\text{max}$) = (10\,Hz, 7130\,Hz) is divided into frequency bins with $\Delta f = \zeta/\pi$ and~\cite{Kuttruff2009}
\begin{equation}
\zeta = \frac{3\ln (10)}{T_{60}},
\end{equation}
as well as the standard third-octave bands (each of which represents one band $\Delta f_\text{band}$ from above). For simplicity, the reverberation time $T_{60}$ is frequency-independent and three values are considered: 0.5\,s, 2\,s, and 8\,s. Sound energy density per hertz and unit volume is estimated for a diffuse field (the red lines) using the analytical solution~\cite{Kuttruff2009}
\begin{equation}\label{diffuse_field_energy_source_absorption}
E_{\text{diff},i} = \frac{4 P_W}{c_0 A}
\end{equation}
with acoustic power of the source per hertz 
\begin{equation}\label{source_power_strength}
P_W = \frac{Q^2/1\text{Hz}}{8\pi \rho_0 c_0},
\end{equation}
unit source strength $Q = 1$\,kg/s$^2$, and the equivalent absorption area estimated using Sabine's formula,
\begin{equation}\label{equivalent_absorption_area_Sabine}
A = \frac{8 \zeta V_S}{c_0}.
\end{equation}

%\afterpage{%
%}

The exact mean squared sound pressure amplitude per hertz (the blue lines) is calculated in each bin $i$ as~\cite{Masovic2021}
\begin{equation}\label{mean_squared_sound_pressure_amplitude_per_hertz_bin}
\mathrm{p}_i^2 = \frac{Q^2}{1\text{Hz}} \left| \frac{8 c_0^2}{V_S} \sum_{s} \frac{\prod_{j=1}^{3} \cos(k_{sj} x_j) \prod_{l=1}^{3} \cos(k_{sl} y_l)}{ C_s(\omega_s^2 + 2j \zeta_s \omega_s -\omega_i^2)} \right|^2,
\end{equation}
with $\zeta_s = \zeta$ for each mode $s$. The analytical solution holds for a rectangular room with hard walls at frequencies $\omega \gg \zeta$ (a weakly damped room). The eigenfrequencies $\omega_s = k_s c_0$ are estimated using the wave vector components $k_{s1}$, $k_{s2}$, and $k_{s3}$:
\begin{equation}\label{modes_wavenumbers_rectangular_room}
k_s = \sqrt{k_{s1}^2 + k_{s2}^2 + k_{s3}^2} = \sqrt{\left( \frac{s_1 \pi}{L_1} \right)^2 + \left( \frac{s_2 \pi}{L_2} \right)^2 + \left( \frac{s_3 \pi}{L_3} \right)^2},
\end{equation}
where $s_1$, $s_2$, and $s_3$ are non-negative integers and
\begin{equation}
C_s =
\begin{cases}
4 &\text{for single $s_i \neq 0$ (axial modes)}\\
2 &\text{for two $s_i \neq 0$ (tangential modes)}\\
1 &\text{for all three $s_i \neq 0$ (oblique modes)}.
\end{cases} 
\end{equation}
With the exception of the non-acoustic mode ($s_1$, $s_2$, $s_3$) = (0, 0, 0), all modes with $\omega_s/(2\pi) \leq 1.1 f_\text{max}$ are included in the sum in eq.~(\ref{mean_squared_sound_pressure_amplitude_per_hertz_bin}). The spatial averaging over all source ($\vec y$) and receiver ($\vec x$) locations in $V_S$ is approximated by averaging the values of $\mathrm{p}_i^2$ for ten different pairs of source  and receiver locations. The components ($y_1, y_2, y_3$) and ($x_1, x_2, x_3$) are selected randomly using uniform distributions in the ranges (0, $L_1$), (0, $L_2$), and (0, $L_3$). The average sound energy density per hertz and unit volume is then calculated with
\begin{equation}\label{acoustic_energy_pressure}
E_i = \frac{\mathrm{p}_i^2}{2 \rho_0 c_0^2}.
\end{equation}

For both the diffuse-field estimation from eq.~(\ref{diffuse_field_energy_source_absorption}) and exact spectra obtained using eq.~(\ref{acoustic_energy_pressure}), sound pressure level ($L_\text{diff}$ and $L$, respectively) in the bin $i$ is calculated as
\begin{equation}
L_i = 10\log_{10} \frac{E_i \Delta \omega c_0}{2\pi p_\text{ref}^2/(\rho_0 c_0)},
\end{equation}
with $p_\text{ref} = 2 \cdot 10^{-5}$\,Pa. Notice that $\Delta \omega/2$ cancels with $\zeta = \Delta \omega/2$ from eq.~(\ref{equivalent_absorption_area_Sabine}) in the case of a diffuse field. Since the width of the frequency bins adapts to the damping, the values of $L_{\text{diff},i}$ do not change with reverberation time. The third-octave values are obtained by means of energy summation:
\begin{equation}
L_\text{1/3 oct} = 10 \log_{10} \left( \sum_{i} 10^{L_i/10} \right),
\end{equation}
where the sum is over all bins which are part of a third-octave band. In addition to the analytical solutions, the correction of the diffuse-field values, $\Delta L$, is calculated in third-octave bands (with the black lines in the figure showing the corrected values $L_\text{diff} + \Delta L$) using eq.~(\ref{SEA_SPL_correction}) and the entropy from equations~(\ref{acoustic_entropy_constant_damping}), (\ref{rect_room_number_of_modes_extended}), and (\ref{modal_density_per_hertz_and_bin}). As already stated, the constant $k_B$ does not affect the correction and it is set to 1\,J/K. The corrected values capture the trends of the exact spectra in non-diffuse (``lower-frequency", lower-entropy) fields, within the range of validity of the analogy. The lower entropy leads in general to higher values of the spectra than estimated with the diffuse-field assumption. Of course, the simple analytical expressions from equations (\ref{rect_room_number_of_modes_extended}) and (\ref{modal_density_per_hertz_and_bin}) cannot replicate details of the spectra and the thermodynamic analogy breaks at very low frequencies, far from the equilibrium. The vertical green lines in Fig.~\ref{fig:mean_sound_level_room} indicate the Schroeder frequencies calculated using eq.~(\ref{Schroeder_frequency}).

In order to understand better the relation between the diffuse field or canonical equilibrium, which was supposed in section~\ref{ch:background_fluid_thermodynamics}, and acoustic modes and fields in general, which are considered here, it is instructive to derive the diffuse-field expression
\begin{equation}\label{mean_square_pressure_per_hertz_diffuse_field}
\mathrm{p}_{\text{diff},i}^2 = \frac{8 \rho_0 c_0 P_W}{A} = \frac{Q^2/1\text{Hz}}{\pi A} = \frac{Q^2 c_0 /1\text{Hz}}{8 \pi \zeta V_S},
\end{equation}
which follows from equations~(\ref{diffuse_field_energy_source_absorption})-(\ref{equivalent_absorption_area_Sabine}) and (\ref{acoustic_energy_pressure}), from the exact solution in eq.~(\ref{mean_squared_sound_pressure_amplitude_per_hertz_bin}). The latter holds for a weakly damped rectangular room with hard walls. We first average eq.~(\ref{mean_squared_sound_pressure_amplitude_per_hertz_bin}) over all source and receiver locations in the volume $V_S$:
\begin{equation}
\begin{aligned}
\frac{1}{V_S^2} \int_{V_S} &\int_{V_S} \mathrm{p}_i^2 d^3 \vec x d^3 \vec y \\
&= \frac{64 c_0^4 Q^2/1\text{Hz}}{V_S^4} \int_{V_S} \int_{V_S} \left| \sum_{s} \frac{\prod_{j=1}^{3} \cos(k_{sj} x_j) \prod_{l=1}^{3} \cos(k_{sl} y_l)}{ C_s(\omega_s^2 + 2j \zeta_s \omega_s -\omega_i^2)} \right|^2 d^3 \vec x d^3 \vec y.
\end{aligned}
\end{equation}
At high enough frequencies ($\omega L/c_0 \gg 1$, with $L$ characteristic length of the room; the regime of eq.~(\ref{modal_density_Schroeder})), much above the Schroeder frequency, the modal density is high and many modes overlap with different phases, in every finite frequency range of interest. Therefore, they sum energetically and we can approximate
\begin{equation}
\begin{aligned}
\frac{1}{V_S^2} \int_{V_S} &\int_{V_S} \mathrm{p}_i^2 d^3 \vec x d^3 \vec y \\
&= \frac{64 c_0^4 Q^2/1\text{Hz}}{V_S^4} \int_{V_S} \int_{V_S} \sum_{s} \left| \frac{\prod_{j=1}^{3} \cos(k_{sj} x_j) \prod_{l=1}^{3} \cos(k_{sl} y_l)}{ C_s(\omega_s^2 + 2j \zeta_s \omega_s -\omega_i^2)} \right|^2 d^3 \vec x d^3 \vec y \\
&= \frac{64 c_0^4 Q^2/1\text{Hz}}{V_S^4} \sum_{s} \frac{\int_{V_S} \int_{V_S} \prod_{j=1}^{3} \cos^2(k_{sj} x_j) \prod_{l=1}^{3} \cos^2(k_{sl} y_l) d^3 \vec x d^3 \vec y}{ C_s^2 \left| \omega_s^2 + 2j \zeta_s \omega_s -\omega_i^2  \right|^2},
\end{aligned}
\end{equation}
after changing the order of the sum and integrals. Furthermore, we can consider only the prevailing oblique modes (eq.~(\ref{modal_density_Schroeder}) and the first term in eq.~(\ref{rect_room_number_of_modes_extended})) with $C_s = 1$.

In a rectangular room with dimensions $L_1$, $L_2$, and $L_3$ the integrals split:
\begin{equation}
\begin{aligned}
\int_{V_S} \int_{V_S} \prod_{j=1}^{3} &\cos^2(k_{sj} x_j) \prod_{l=1}^{3} \cos^2(k_{sl} y_l) d^3 \vec x d^3 \vec y \\
&= \prod_{j=1}^{3} \int_{0}^{L_j} \cos^2(k_{sj} x_j) dx_j \prod_{l=1}^{3} \int_{0}^{L_l} \cos^2(k_{sl} y_l) dy_l.
\end{aligned}
\end{equation}
Next we notice that for $k_{sj} = s_j \pi/L_j \neq 0$ (from eq.~(\ref{modes_wavenumbers_rectangular_room}), for oblique modes)
\begin{equation}
\begin{aligned}
\int_{0}^{L_j} \cos^2(k_{sj} x_j) dx_j &= \frac{1}{k_{sj}}\int_{0}^{k_{sj} L_j} \cos^2(k_{sj} x_j) d(k_{sj}x_j) &\\
&= \frac{1}{2 k_{sj}} [ (k_{sj} L_j) + \sin(k_{sj} L_j) \cos(k_{sj} L_j)] = \frac{L_j}{2}&
\end{aligned}
\end{equation}
and similar holds for the integrals over $y_l$. Hence,
\begin{equation}
\frac{1}{V_S^2} \int_{V_S} \int_{V_S} \mathrm{p}_i^2 d^3 \vec x d^3 \vec y = \frac{c_0^4 Q^2/1\text{Hz}}{V_S^2} \sum_{s} \frac{1}{\left| \omega_s^2 + 2j \zeta_s \omega_s -\omega_i^2  \right|^2}.
\end{equation}
Since the modal density is high, we can replace the sum over modes $s$ with an integral over continuous frequency. We also consider only oblique modes, with the continuous modal density $V_S \omega^2/(2 \pi^2 c_0^3)$ from eq.~(\ref{modal_density_Schroeder}). Accordingly,
\begin{equation}\label{modes_superposition_to_diffuse_field}
\begin{aligned}
\frac{1}{V_S^2} \int_{V_S} \int_{V_S} \mathrm{p}_i^2 d^3 \vec x d^3 \vec y &= \frac{c_0 Q^2/1\text{Hz}}{2\pi^2 V_S} \int_{\omega_i - \Delta \omega/2}^{\omega_i + \Delta \omega/2} \frac{\omega^2}{\left| \omega^2 + 2j \zeta \omega -\omega_i^2  \right|^2} d\omega \\
&= \frac{c_0 Q^2/1\text{Hz}}{2\pi^2 V_S} \int_{\omega_i - \Delta \omega/2}^{\omega_i + \Delta \omega/2} \frac{\omega^2}{ (\omega^2 - \omega_i^2)^2 + 4 \zeta^2 \omega^2 } d\omega.
\end{aligned}
\end{equation}
Here, $\Delta \omega$ is a small frequency band centred at $\omega_i$. We also assumed that the low damping is constant in $\Delta \omega$ ($\zeta_s = \zeta \ll \omega_s$), so most of the energy of the mode $s$ is distributed at frequencies $\omega \approx \omega_s$. Consequently, only the modes with eigenfrequencies $\omega_s$ within $\Delta \omega$ contribute significantly to the energy and practically their entire energy is contained\footnote{This evidently does not hold if $\Delta \omega$ is close to or smaller than half-width of the modes, which was actually the case in the definition of acoustic entropy above. However, this does not prohibit defining the diffuse field independently. It just implies that the state of equilibrium is meaningful only for frequency ranges which include many frequency bins.} in $\Delta \omega$. Therefore, we can formally let $\Delta \omega \rightarrow \infty$.

The integral can be solved by introducing the variable $u = (\omega^2-\omega_i^2)/(2\omega)$. After noticing that $u \rightarrow 0$, because $\omega \approx \omega_i$ for a small $\Delta \omega$, and therefore $du = (1-u/\omega)d\omega \approx d\omega$, the integral becomes
\begin{equation}
\begin{aligned}
\frac{1}{4} \int \frac{1}{ u^2 + \zeta^2 } du = \frac{1}{4 \zeta} \tan^{-1} \left( \frac{u}{\zeta} \right) = \frac{\pi}{4 \zeta}.
\end{aligned}
\end{equation}
Substituting this in eq.~(\ref{modes_superposition_to_diffuse_field}), we finally obtain
\begin{equation}
\begin{aligned}
\frac{1}{V_S^2} \int_{V_S} \int_{V_S} \mathrm{p}_i^2 d^3 \vec x d^3 \vec y = \frac{c_0 Q^2/1\text{Hz}}{8\pi V_S \zeta},
\end{aligned}
\end{equation}
which is equal to eq.~(\ref{mean_square_pressure_per_hertz_diffuse_field}). It follows that the canonical equilibrium corresponds to a weakly damped sound field in a room, the energy of which is averaged over all possible source/receiver locations, at high frequencies, when the modal density is high enough that many (predominantly oblique) modes can be energetically summed within the observed frequency range. The range should be much larger than the half-width of the modes. Although this has been demonstrated for the specific case of a rectangular room, the same holds for acoustically large rooms of arbitrary shapes, similarly as eq.~(\ref{modal_density_Schroeder}). Section~\ref{ch:background_fluid_thermodynamics} showed that this condition is satisfied by the background fluid field in every observed region of space which is much larger than the acoustic Planck length and therefore contains many fluid molecules.

\subsubsection{The source of entropy}\label{source_of_entropy}

The previous analysis substantiated the analogy between canonical equilibrium and diffuse sound fields for large enough observed spaces (that is, Helmholtz number $\omega L/c_0 \gg 1$, with $L \gg L_0 = c_0/\omega_0$ in the case of the background fluid) and frequency ranges much larger than the modal half-width. Focusing on the background fluid, both conditions are fulfilled due to the limitations of the observers (the finite spatial and frequency resolution). In general, the lower the Helmholtz number value, the lower the entropy of the state which deviates from the equilibrium, assuming a fixed $\Delta \omega$.

In addition to this, section~\ref{background_fluid_and_acoustic_particles} pointed to an important difference between acoustic and quantum modes, which is the treatment of the modal width (and therefore minimal $\Delta \omega_0$), which is associated with damping and increasing entropy. In quantum thermodynamics it is normalized such that $\Delta \omega_0 = 2 \zeta_0 = 1\text{\,s}^{-1}$ and the frequency resolution of the observer drops out from the theory (the integrals over frequency, such as the one in eq.~(\ref{modes_superposition_to_diffuse_field}), reduce to sums over modal numbers, such as the one in eq.~(\ref{probability_pure_state})). In classical acoustics, however, a finite width of a mode is always a consequence of damping, together with the decay of sound energy density of the room mode in time, $E = E_0 e^{-2 \zeta_0 t}$, where $E_0 = E(t=0)$. The half-width of the modes and time are thus tightly related over energy. In contrast to this, energy in the bounded observable universe is conserved and constant. Therefore, the energy density can decay only by means of the increase of volume of the observable universe, that is the Hubble volume $V_H  = V_{H0} e^{2 \zeta_0 t} = 4\pi R_H^3/3 = 4\pi R_{H0}^3 e^{3t/t_H}/3$ and $R_H$ denotes the Hubble length. Thus, the analogue Hubble time is 
\begin{equation}\label{Hubble_time_damping_constant}
t_H = \frac{3}{2\zeta_0} = \frac{3}{\Delta \omega_0}.
\end{equation}

On the other hand, the frequency resolution $\Delta \omega_0$ is directly related to the maximal length scale. For example, the spatial Fourier transform of a rectangular pulse with the length $2 R_H$ gives
\begin{equation}
\frac{1}{2R_H} \int_{-R_H}^{R_H} e^{jkx} dx = \frac{\sin(k R_H)}{k R_H} = \text{sinc}(k R_H),
\end{equation}
with the wave number $k = \omega/c_0$. The sinc function takes the maximum value of 1 for $kR_H = 0$, while its squared value drops by 50\% (in alignment with the definition of the modal half-width from above) for $k R_H \approx \pm 1.39$. Hence,
\begin{equation}\label{delta_omega_i_and_R_H}
\Delta \omega_0 = c_0 \Delta k \sim \frac{2.78 c_0}{R_H} = \frac{2.78}{t_H},
\end{equation}
which matches closely the result in eq.~(\ref{Hubble_time_damping_constant}) and half of its value the minimum frequency from eq.~(\ref{lowest_frequency_universe}) (with $R_U \approx 3.4 R_H$). Thus, the finite modal width, which represents the lower limit of frequency resolution of an observer (or quantization of energy $E = \hbar \omega$), is the direct cause of the finite size of the observable universe with conserved energy. In quantum thermodynamics it is normalized to one, allowing eq.~(\ref{von_Neumann_entropy}), while the associated damping and increase of entropy are interpreted in cosmology as the (low-frequency) expansion of the universe and modelled with the weak and uniformly distributed dark energy and the associated cosmological constant (section~\ref{ch:dark_energy}).

Since $V_H = V_{H0} e^{2\zeta_0 t}$, frequency of the injected volume fraction from eq.~(\ref{injected_volume_fraction_as_source}) is imaginary, $\omega = j\Omega = j 2 \zeta_0 = j\Delta \omega_0$, and
\begin{equation}\label{injected_volume_fraction_damping_omega}
\beta = e^{-j\omega t} = e^{2\zeta_0 t} = e^{\Delta \omega_0 t}.
\end{equation}
The weak-damping condition is satisfied, since $\zeta_0 = \Delta \omega_0/2 = 1.39/t_H$ is much smaller than any measurable angular frequency, and $\Omega = \Delta \omega_0 = 2.78 c_0/R_H \approx 9.45 c_0/R_U$ approaches the minimum frequency from eq.~(\ref{lowest_frequency_universe}). Furthermore, by comparing equations~(\ref{injected_volume_fraction_damping_omega}) and (\ref{injected_volume_fraction_as_source}),
\begin{equation}\label{cosmological_constant_damping}
\Lambda = \frac{4\zeta_0^2}{3 c_0^2} = \frac{(\Delta \omega_0)^2}{3 c_0^2},
\end{equation}
which again makes the relation between the minimum observable frequency difference and expansion of the observable universe evident. Function of the weak source of the expansion equals
\begin{equation}\label{entropy_source_strength}
Q = \rho_0 \int_{V_H} \frac{\partial^2 \beta}{\partial t^2} d^3 \vec x = \rho_0 V_H \frac{\partial^2 \beta}{\partial t^2} = 4 \rho_0 V_H \zeta_0^2 \beta
\end{equation}
at the leading order. It is the apparent free-space compact monopole source due to increasing entropy, which was anticipated earlier (in eq.~(\ref{Lighthill_analogy}) and section~\ref{ch:dark_energy}), the location of which is distributed across the observable universe and in the near field of which every observer is found. Amplitude of the source function, $|Q| = 4\rho_0 V_H \zeta_0^2$ is also in full agreement with eq.~(\ref{entropy_source_function_amplitude}) of the quantum analogy, with fluid molecules as the elementary particles, their total mass $N_m M = \rho_0 V_H = \mathrm{P}_0 V_H$, and $\zeta_n \omega_n$ of the lowest mode replaced by $\zeta_0 \Omega = 2 \zeta_0^2$, which gives rise to the quantum mechanical interpretation of the source, besides the cosmological one.

Although all observers and, therefore, boundaries of the observable universe are located in the near field of the source of expansion and $\Omega L/c_0 \gg 1$ is not satisfied for any $L \ll R_U$, it is still of interest to calculate the high-frequency diffuse field generated by the source, since both the source and the universe boundaries are only apparent. This should lead to the characterization of the unobservable background fluid in the state of equilibrium, as in section~\ref{ch:background_fluid_thermodynamics}. The squared sound pressure in a diffuse field is from eq.~(\ref{mean_square_pressure_per_hertz_diffuse_field})
\begin{equation}
p_{\text{diff}}^2 = \mathrm{p}_{\text{diff}}^2 1\text{Hz} = \frac{Q^2 c_0}{8 \pi \zeta_0 V_H} = \frac{2}{\pi} \rho_0^2 V_H \zeta_0^3 c_0 \beta^2.
\end{equation}
Using $V_H = 4 R_H^3 \pi/3$ and $\zeta_0 = 3c_0/(2R_H)$,
\begin{equation}
p_{\text{diff}}^2 = 9 \rho_0^2 c_0^4 \beta^2
\end{equation}
or
\begin{equation}\label{pressure_universe_t_0}
|p_{\text{diff}}| = 3 \rho_0 c_0^2.
\end{equation}
Removing again the factor of 3 due to the three degrees of freedom per atom, as in eq.~(\ref{speed_of_sound_temperature}), $|p_{\text{diff}}| = \rho_0 c_0^2 = p_0$, with $p_0$ the maximum (static) pressure from eq.~(\ref{equation_of_state_pressure}) with $\gamma = 1$.

Besides the equilibrium value, this is also the value of $p_{\text{diff}}$ at the initial time $t=0$, when $\beta = 1$. Hence, the observed universe at $t=0$ matches the background fluid in the canonical equilibrium, with maximum values of pressure, density, and temperature. While the unobservable background fluid remains in the same state of maximal entropy for $t>0$, for an observer of quantized and conserved energy, the universe appears to expand, with decreasing energy density and increasing entropy (as the Helmholtz number $\omega_0 R_U/c_0$ increases). On the other hand, $\omega_0$ is the constant highest frequency of the theory associated with the Planck particles with density $\rho_0$ ($\omega_0 L_0/c_0 = \sqrt{2} \omega_0 L_P/c_0 = 1$; section~\ref{ch:background_field_dipole_quadrupole}). Therefore, the most distant, earliest state of the observable universe (with the size of a Planck particle, $V_{H0} \sim 4 L_P^3 \pi/3$) paradoxically provides a picture of the unobservable background fluid, the constitutive elementary particles at the Planck scale, and the observable universe as a whole, altogether. Finally, the source in eq.~(\ref{entropy_source_strength}) with unsteady $\beta$ from eq.~(\ref{injected_volume_fraction_damping_omega}) thus becomes the source of time in the observable universe. It also sets the ``arrow" of time, because the oscillatory function reduces to the exponential.

\subsubsection{Analogue thermodynamics}

Using the methods of stochastic thermodynamics, we can derive the first two laws of thermodynamics and thus formally complete the analogue thermodynamic theory. Following Deffner and Campbell~\cite{Deffner2019}, we observe again a system which is ultraweakly coupled with its environment and initially in the state of canonical equilibrium $\hat \rho_0$. Here, $\hat \rho_0$ represents simply the probability matrix of the system with the characteristic length $L \gg c_0/\omega$, consisting of many sub-systems, at time $t=0$. Since the system is in equilibrium, it has the Boltzmann-Gibbs distribution, that is
\begin{equation}\label{Boltzmann-Gibbs_distribution_density_matrix}
\hat \rho_0 = \frac{1}{Z_0} e^{-\hat H(\hat \rho_0, \lambda_0)/(k_B T_0)}.
\end{equation}
This can be compared with eq.~(\ref{Boltzmann-Gibbs_distribution}) with the probability density matrix containing the normalized numbers of elementary particles in different states. However, all details of the modal density and eigenfrequencies, which are dictated by the potential of the environment, are now contained in the Hamiltonian $\bar H$. The initial Hamiltonian is $\hat H_0 = \hat H(\hat \rho_0, \lambda_0)$ and the distribution is normalized with the partition function $Z_0$, while $\lambda_0$ denotes an externally controlled work parameter with the time scale of variation much larger than the time scale of the system.

The partition function is defined generally as
\begin{equation}\label{partition_function}
Z = \int e^{-\hat H(\hat \rho, \lambda)/(k_B T_0)} d\hat \rho = e^{-F/(k_B T_0)},
\end{equation}
where $F$ is the so-called Helmholtz free energy. For $n$ discrete determinate states,
\begin{equation}
Z = \Tr{e^{-\hat H/(k_B T_0)}} = \sum_{i=1}^{n} e^{-E_i/(k_B T_0)}.
\end{equation}
For the equilibrium state, the Liouville-von Neumann equation (\ref{Liouville-von Neumann equation}) gives $[\hat H_0,  \hat \rho_0] = 0$, while from eq.~(\ref{energy_density_matrix_hamiltonian})
\begin{equation}\label{energy_density_matrix_hamiltonian_Boltzmann-Gibs_state}
E_0 = \Tr{\hat \rho_0 \hat H_0}.
\end{equation}
Using eq.~(\ref{partition_function}) with $Z_0$ in eq.~(\ref{Boltzmann-Gibbs_distribution_density_matrix}),
\begin{equation}
\hat \rho_0 = e^{-(\hat H_0 - F_0)/(k_B T_0)}
\end{equation}
and we can express the entropy from eq.~(\ref{von_Neumann_entropy}) as
\begin{equation}\label{entropy_energy_free_energy_canonical_equilibrium}
S_0 = \frac{k_B}{k_B T_0} \Tr{\hat \rho_0 (\hat H_0 - F_0)} = \frac{1}{T_0} (E_0 - F_0).
\end{equation}
For the last equality we used equations (\ref{energy_density_matrix_hamiltonian_Boltzmann-Gibs_state}) and (\ref{density_matrix_trace}). The entropy contains the remaining information, after the normalization free-energy term from equations~(\ref{Boltzmann-Gibbs_distribution_density_matrix} and (\ref{partition_function})) is subtracted from the total energy. This is in agreement with the acoustic entropy of eq.~(\ref{acoustic_entropy}), a quantifier of tonality, which is independent of the absolute values of sound energy of a spectrum. For an isothermal quasi-static process, we obtain
\begin{equation}\label{first_law_thermodynamics}
dE = T_0 dS + dF,
\end{equation}
\iffalse
\begin{equation}
dS = \frac{1}{T_0} \left( \Tr{(d \hat \rho_0) \hat H_0} + \Tr{\hat \rho_0 d \hat H_0} - dF \right) = \frac{1}{T_0} \Tr{(d \hat \rho_0) \hat H_0},
\end{equation}
where we used~\cite{Deffner2019}
\begin{equation}
dF = \Tr{\hat \rho_0 d \hat H_0}.
\end{equation}
On the other hand, from eq.~(\ref{energy_density_matrix_hamiltonian}),
\begin{equation}
dE = \Tr{d(\hat \rho_0) \hat H_0 + \hat \rho_0 d \hat H_0} = \Tr{d(\hat \rho_0) \hat H_0} + d F.
\end{equation}
Hence,
\begin{equation}
dE = \Tr{d(\hat \rho_0) \hat H_0} + \Tr{\hat \rho_0 d \hat H_0} = T_0 dS + dF,
\end{equation}
\fi
which is a formulation of the first law of thermodynamics. The two terms represent the change of entropy (density matrix) or heat and the useful work (change of Hamiltonian as a function of the external parameter), respectively. Both contribute to the change of energy. In the acoustic analogy they represent change of the spectrum shape and overall level, respectively.

The work at time $t$ which is required to displace the system out of the canonical equilibrium depends on the initial state and is by definition
\begin{equation}
W(\hat \rho_0) = \hat H(\hat \rho(\hat \rho_0), \lambda_t) - \hat H(\hat \rho_0, \lambda_0).
\end{equation}
Due to inherent indeterminism, quantum work is a random variable and we calculate the average over all trajectories of $W$, using $\hat \rho_0$ as the probability density:
\begin{equation}\label{Jarzynski equality}
\begin{aligned}
\overline {e^{-W/(k_B T_0)}} &= \int \hat \rho_0 e^{-W(\hat \rho_0)/(k_B T_0)} d\hat \rho_0 = \frac{1}{Z_0} \int e^{-(W(\hat \rho_0) + \hat H(\hat \rho_0, \lambda_0))/(k_B T_0)} d\hat \rho_0 \\&= \frac{1}{Z_0} \int e^{-\hat H(\hat \rho(\hat \rho_0), \lambda_t)/(k_B T_0)} d\hat \rho_0 = \frac{1}{Z_0} \int e^{-\hat H(\hat \rho, \lambda_t)/(k_B T_0)} \left| \frac{d \hat \rho}{d \hat \rho_0} \right|^{-1} d\hat \rho \\&= \frac{1}{Z_0} \int e^{-\hat H(\hat \rho, \lambda_t)/(k_B T_0)} d\hat \rho = \frac{Z_t}{Z_0} = e^{-(F_t - F_0)/(k_B T_0)} = e^{-\Delta F/(k_B T_0)}.
\end{aligned}
\end{equation}
According to Liouville’s theorem, volume of the phase space of states is conserved, which gives~\cite{Deffner2019} $|d\hat \rho/d \hat \rho_0|^{-1} = 1$. The result in eq.~(\ref{Jarzynski equality}) is known as Jarzynski equality. According to Jensen's inequality, $\overline {e^{-W/(k_B T_0)}} \geq {e^{-\overline{W}/(k_B T_0)}}$ and, therefore,
\begin{equation}
\overline{W} \geq \Delta F.
\end{equation}
The dissipated work is~\cite{Gemmer2010}
\begin{equation}
\overline W_\text{diss} = \overline W - \Delta F = T \Delta S - \Delta \overline Q \geq 0,
\end{equation}
with $T$ denoting temperature, $S$ entropy, and $Q$ heat. This is the Clausius inequality expressing the second law of thermodynamics, which is arguably responsible for the arrow of time. The latter appears then due to unobservable microscopic degrees of freedom and time appears as relevant only in the macroscopic theory, but not for the state of equilibrium. The dissipated work represents the entropic costs which penalize the excursion from the canonical equilibrium. In room acoustics the excursion takes place when the diffuse-field condition $\omega L/c_0 \gg 1$ is not satisfied, while decreasing the frequency or size of the room and, therefore, modal density (see, for example, the pronounced tonality at low frequencies in Fig.~\ref{fig:mean_sound_level_room}) and entropy. The directionality of time is thus equivalent to the monotonic decrease of the exponential function in eq.~(\ref{injected_volume_fraction_damping_omega}).%, while the radius of the observable universe plays the role of the external parameter $\lambda$. 

One should also notice that in the entire quantum thermodynamic treatment, starting with the time-independent Schrödinger equation~(\ref{time-independent_Schroedinger_equation}), time never appears explicitly. Its appearance is only indirect, over the finite modal width in the source term of the injected volume fraction in eq.~(\ref{injected_volume_fraction_damping_omega}). Moreover, unlike mass, which is proportional to length in eq.~(\ref{Schwarzschild_radius_mass}), energy in eq.~(\ref{Planck_Einstein}) is not related to time, but angular frequency. Similarly to this, time is an emergent macroscopic phenomenon in loop quantum gravity, which, orders the states with different values of entropy according to the second law of thermodynamics. In contrast to this, it is an independent variable in classical acoustics, much like the spatial coordinates. Consequently, no additional source of entropy is necessary to give rise to time, even for a steady background fluid, and the acoustic spacetime is unbounded.

In a more general case, when the initial equilibrium state, say $\hat \rho_1$, is not canonical (with the Boltzmann-Gibbs distribution), for example, when the system is not only ultraweakly coupled to the environment (the sound field in a room is not diffuse), and $\lambda_1 \neq \lambda_0$,
\iffalse we still have
\begin{equation}
E_1 = \langle \hat H_1 \rangle = \Tr{\hat \rho_1 \hat H_1 },
\end{equation}
as in eq.~(\ref{energy_density_matrix_hamiltonian_Boltzmann-Gibs_state}), so
\begin{equation}
d E_1 = \Tr{(d\hat \rho_1) \hat H_1 + \hat \rho_1 d \hat H_1}.
\end{equation}
\fi
the entropy from eq.~(\ref{von_Neumann_entropy}) reads~\cite{Deffner2019}
\begin{equation}\label{non_canonical_equilibrium_entropy}
\begin{aligned}
S_1 &= -k_B \Tr{\hat \rho_1 \ln \hat \rho_1} = -k_B \Tr{\hat \rho_1 \ln \hat \rho_0 + \hat \rho_1 \ln (\hat \rho_1/\hat \rho_0)} \\
&= \frac{1}{T_1} (E_1 - F_1) - S(\hat \rho_1 || \hat \rho_0) = \frac{1}{T_1} (E_1 - \mathcal{F}_1).
\end{aligned}
\end{equation}
We have defined the quantum relative entropy $S(\hat \rho_1 || \hat \rho_0) = -k_B \Tr{\hat \rho_1 \ln (\hat \rho_0/\hat \rho_1)}$,
\iffalse and we used
\begin{equation}
\frac{1}{T_1} (E_1 - F_1) = -k_B \Tr{\hat \rho_1 \ln \hat \rho_0}.
\end{equation}
%%%%%% why?
\fi
as well as the information free energy,
\begin{equation}
\mathcal{F}_1 = F_1 + T_1 S(\hat \rho_1 || \hat \rho_0).
\end{equation}
In this way the same form as in eq.~(\ref{entropy_energy_free_energy_canonical_equilibrium}) is obtained, with the information free energy generalizing the Helmholtz free energy by including the information on the difference between $\hat \rho_1$ and the Boltzmann-Gibbs state $\hat \rho_0$. The generalized first law of thermodynamics from eq.~(\ref{first_law_thermodynamics}) reads
\begin{equation}
dE = T_1 d S + d \mathcal{F}.
\end{equation}

Lastly, it is worth mentioning that the second law of thermodynamics can also follow from an extended model of the (acoustic) quantum field which includes damping~\cite{Deffner2019}. We consider massless phonons which satisfy eq.~(\ref{scalar_wave_equation}). The simplest solution of the wave equation is a plane wave ${\phi} = a(t) e^{j\vec k \cdot \vec x}$, where $\vec k$ is the wave vector. It turns the wave equation into the equation of a harmonic oscillator,
\begin{equation}
\frac{1}{c_0^2} \frac{\partial^2}{\partial t^2} a(t) + k^2 a(t) = 0,
\end{equation}
with $k = |\vec k|$. Switching to the particle location $\vec x$ as the main variable and adding a damping term with the coefficient $D$ as well as an external stochastic force $\vec F (t)$ due to thermal fluctuations, we can write
\begin{equation}
M_0 \frac{\partial^2}{\partial t^2} \vec x(t) + D \frac{\partial}{\partial t} \vec x(t) + M_0 c_0^2 k^2 \vec x(t) = \vec F(t).
\end{equation}
This is the equation of a damped oscillator, but also a form of the Langevin equation with quadratic potential. Finite systems exhibiting Langevin dynamics can be shown to satisfy the fluctuation theorem of stochastic thermodynamics\cite{Kurchan1998}, which is a generalization of the second law of thermodynamics to microscopic systems far from equilibrium. Each acoustic mode also behaves as a damped oscillator, which dominate in the frequency response of a room at low frequencies (in the low-entropy regime). Hence, Langevin dynamics offers an alternative approach for capturing microscopic fluctuations in time domain, which were described by the density matrix above. However, the mass $M_0$ (or temperature $T_0$) has to be given. Its value cannot be deduced from the quantum field theory.

Complementing the discussion from section~\ref{ch:dark_energy}, we can summarize the main findings of the quantum thermodynamic analogy as follows. Basic properties of the background fluid, such as mean density, pressure, and temperature, are determined by the thermodynamic behaviour of the modes (determinate states) of the background fluid field in the canonical equilibrium (the state of maximal entropy analogue to a high-Helmholtz-number diffuse field). Thus, the elementary particles (fluid molecules) posses the mass only in interaction with their environment -- other systems and particles. In the near field of its source (such as a microscopic turbulence at the acoustic Planck scale) or a compact boundary (of the observable universe at the Planck scale), every massless particle obtains the mean density of the background fluid owing to the mechanism of spontaneous symmetry breaking. Excursions from the canonical equilibrium, that is diffuse field, are characterized by acoustic entropy, which quantifies tonality of room frequency responses. It is maximal in a diffuse field and decreases with frequency or room size, that is, with the Helmholtz number value. It is independent of the absolute energy of the response, which corresponds to the Helmholtz free energy.

Every piece of the fluid which can be observed (with which an observer can interact) acts as a quantum system of elementary particles (fluid molecules) as sub-systems. The system is much larger than the acoustic Planck length and ultraweakly coupled with its environment. In addition to this, inability of the observer to measure (conserved) energy (or eigenfrequency) with infinite accuracy (the quantization of energy with resolution much larger than the half-width of the modes) makes the observable universe bounded, the steady background fluid unobservable, and gives rise to the source of the universe expansion (or dark energy), entropy, and time in which the exponential expansion takes place\footnote{Using a more rigorous statistical approach it can be shown that the probability of negative entropy production diminishes when large systems of particles are observed and the asymmetry (``arrow'') of (thermal) time appears.}. The initial state of the observable universe matches the states of the unobservable background fluid and elementary particles at the Planck scale. Since the source is apparent, it is distributed throughout the observable universe and every observer is located in its near field. In contrast to this, finite widths of classical acoustic modes lead to weak damping (exponential decay of energy in time), while time is an independent variable.

\subsection{Mass/potential as a boundary}\label{ch:mass_boundary}

So far we have used the quantum acoustic analogy mainly for treating the background fluid field with fluid molecules as the elementary particles. However, the analogy applies to sound fields and massless phonons, as well. As discussed in section~\ref{background_fluid_and_acoustic_particles}, this is due to the fact that quantum determinate states, like acoustic modes, satisfy a form of Helmholtz equation, eq.~(\ref{Helmholtz_eq_from_Schroedinger}), in vacuum with zero-potential. The additional potential term of eq.~(\ref{time-independent_Schroedinger_equation}) or the mass term\footnote{As mentioned in the context of eq.~(\ref{Lagrangian_real_scalar_field_potential}), both mass and potential are due to the particle's environment and the potential term can be seen as a generalized mass term.} of the Klein-Gordon equation~(\ref{action_variation_zero_real_scalar_field}) can then be used to model the fluid/vacuum boundaries, such as due to solid bodies or room boundaries, rather than an inherent mass of the field. These necessarily violate the Lorentz invariance, as in eq.~(\ref{conservation_laws_H}). Here, however, we focus on their interaction with the massless acoustic field outside the source regions.

As the first example, we replace $m^2$ with $m^2 H(-f)$ in the Klein-Gordon equation, which thus becomes
\begin{equation}\label{wave_equation_and_rigid_boundary}
\Box \phi = \left( \frac{m c_0}{\hbar} \right)^2 H(-f) \phi.
\end{equation}
As before, $H$ denotes the Heaviside step function, $f(\vec x) = 0$ at the stationary boundary, $f(\vec x) > 0$ outside the body, in the propagation region of the massless field, and $f(\vec x) < 0$ inside the massive body. For $m \rightarrow \infty$, the only physically meaningful solution must satisfy
\begin{equation}\label{wave_equation_free_space}
\Box \phi = 0
\end{equation}
outside the body, with $\phi = 0$ at the boundary and inside the body, in order to preserve the continuity of $\phi(\vec x)$ at the boundary. The problem reduces to solving the wave equation with the boundary condition $\phi = 0$ introduced by the spatially confined mass. If (real) $\phi$ represents sound pressure or density and the body encloses a cavity, the time-independent version of the problem comes down to calculating eigenmodes of the cavity with acoustically soft boundaries. If (complex) $\phi$ represents particle displacement, velocity, or acceleration in one-dimensional space,  the boundary is acoustically hard. For instance, $f(x<0) > 0$, $f(x>0) < 0$, and $f(x=0) = 0$ for a rigid boundary at $x = 0$ and sound waves propagating in the half-space $x<0$. In both cases the boundaries are modelled by means of the infinite mass of the field when $f(\vec x) < 0$, which replaces additional boundary conditions. Their explicit inclusion in the wave equation using the Heaviside function is in accordance with section~\ref{ch:energy_and_charge}, but also with the suppression of the second term in eq.~(\ref{action_variation}), since the bounded vacuum is formally extended to infinity.

In the second example we limit ourselves to plane waves propagating in one-dimensional space along the $x$-axis and replace $m^2$ with $m^2 C \delta(x)$, where $C$ is a constant. Equation~(\ref{action_variation_zero_real_scalar_field}) in frequency domain becomes
\begin{equation}\label{Helmholtz_equation_one_dimensional_rigid_wall}
\left( \frac{\omega}{c_0} \right)^2 \phi + \frac{d^2 \phi}{d x^2} - \left( \frac{m c_0}{\hbar} \right)^2 C \delta(x) \phi = 0.
\end{equation}
The general solution for $x<0$ is $\phi(x<0) = I e^{jkx} - IR e^{-jkx}$, while for $x>0$ it reads $\phi(x>0) = I T e^{jkx}$, where $k = \omega/c_0$ is wave number, $I$ is complex amplitude of the incident plane wave, and $R$ and $T$ are reflection and transmission coefficients, respectively. Thereby, we exclude all energy losses as well as waves arriving from $x = +\infty$ and associate $\phi$ with particle velocity. The latter leads to the minus sign in the solution for $x<0$, due to the opposite directions of the incoming and reflected waves. The two solutions can be related by integrating eq.~(\ref{Helmholtz_equation_one_dimensional_rigid_wall}) around $x = 0$\cite{Griffiths2004}:
\begin{equation}\label{Helmholtz_equation_one_dimensional_rigid_wall_interal}
\int_{-\epsilon}^{\epsilon} \left[ \left( \frac{\omega}{c_0} \right)^2 \phi + \frac{d^2 \phi}{d x^2} - \left( \frac{ m c_0}{\hbar} \right)^2 C \delta(x) \phi \right] d x = 0.
\end{equation}
For $\epsilon \rightarrow 0$, the first term vanishes in favour of the second one (acoustic near field), while the second term becomes difference of the derivatives $d\phi/dx$ at $x = \epsilon$ and $x = -\epsilon$. Therefore,
\begin{equation}\label{Helmholtz_equation_one_dimensional_rigid_wall_interal2}
\left( \frac{d \phi}{d x} \right)_{x = \epsilon} - \left( \frac{d \phi}{d x} \right)_{x = -\epsilon} - \left( \frac{m c_0}{\hbar} \right)^2 C \phi(0)  = 0.
\end{equation}

After inserting $(d\phi/d x)_{x<0} = jk (Ie^{jkx} + IR e^{-jkx})$ and $(d\phi/d x)_{x>0} = jk IT e^{jkx}$, as well as $\phi(0) = I - IR$, we obtain
\begin{equation}\label{Helmholtz_equation_one_dimensional_rigid_wall_interal3}
jk (IT - I - IR) - \left( \frac{m c_0}{\hbar} \right)^2 C (I - IR)  = 0.
\end{equation}
Continuity of the solution at $x = 0$ forces $I - IR = IT$ and consequently
\begin{equation}\label{Helmholtz_equation_one_dimensional_rigid_wall_interal4}
2 jk (IT - I) - \left( \frac{m c_0}{\hbar} \right)^2 C IT  = 0,
\end{equation}
from which it follows
\begin{equation}\label{Helmholtz_equation_one_dimensional_rigid_wall_interal5}
T = \frac{1}{1 +j C m^2 c_0^2/(2k\hbar^2)}
\end{equation}
and
\begin{equation}\label{Helmholtz_equation_one_dimensional_rigid_wall_interal6}
|T|^2 = \frac{1}{1 + [C m^2 c_0^2/(2k\hbar^2)]^2}.
\end{equation}
This result is very similar to the expression for transmission loss of an acoustically thin homogeneous wall\cite{Kuttruff2007},
\begin{equation}\label{thin_homogeneous_wall}
|T|^2 = \frac{1}{1 + [\omega m_\text{wall}/(2 \rho_0 c_0)]^2},
\end{equation}
where $m_\text{wall}$ is mass of the wall per unit area and $\rho_0$ and $c_0$ are density and speed of sound in the surrounding air, respectively. In eq.~(\ref{Helmholtz_equation_one_dimensional_rigid_wall_interal6}) we need to apply again the equality $\hbar \omega = m c_0^2$, which holds in the near field of the acoustically compact body, to arrive at
\begin{equation}\label{Helmholtz_equation_one_dimensional_rigid_wall_interal7}
|T|^2 = \frac{1}{1 + [C \omega /(2 c_0)]^2}.
\end{equation}
Hence, the mass term in eq.~(\ref{Helmholtz_equation_one_dimensional_rigid_wall}) with $C = m_\text{wall}/\rho_0$ models an acoustically thin homogeneous rigid wall hit by a plane sound wave at normal incidence and the equation captures sound transmission through the wall. All boundary conditions are again replaced by the infinite but spatially confined mass of the field at $x=0$, in an acoustically compact region.

In quantum mechanics $|T|^2$ represents the probability of a quantum particle passing through the infinite delta-function potential barrier (the so-called tunnelling effect). In the acoustic analogy, the barrier appears as a local infinite increase of the background fluid density. The solutions are equal because of the analogy between quantum determinate states (in this case scattering, rather than bound states, with continuous $k$ and $\omega$) are analogous to acoustic modes (plane waves in free space satisfying the Helmholtz equation). Both examples above prove that the quantum analogy can be used for sound fields at usual length scales and that spatially confined and appropriately scaled mass of the essentially massless field can be used to model solid boundaries, besides the background fluid and acoustic near field of the previous sections. 

\subsection{Currents as sources}\label{ch:currents_as_sources}

In order to complete the analogue quantum field theory, we have to include the sources of fields. The classical theories from sections~\ref{ch:motivation}-\ref{ch:unified_analogy} make a clear distinction between massless fields propagating in vacuum (the left-hand side of the wave equation) and their sources (mass-energy or charge terms on the right-hand side). In contrast to this, we have seen that the (far-field) massless and (near-field or background) massive quantum particles are much more closely related and interacting. As already stated, this leads to arguably a more natural model for (aero)acoustic sound generation, since the same massive particles, fluid molecules, are found both in the source regions and acoustic vacuum. This is another reason to study the analogue sources in quantum theories. With the exception of monopole sources (with thermodynamic origin), these appear as Noether currents associated with the massive field of fluid particles, as will be shown next.

Lagrangian of the fluid particles in the case of dipole radiation was presented in eq.~(\ref{Lagrangian_complex_scalar_field}). The associated current follows from its symmetry ($\phi \rightarrow e^{j\alpha} \phi$) and Noether's theorem in the following way~\cite{Schwartz2014}. We express the derivative of the Lagrangian with respect to $\alpha$:
\begin{equation}\label{Lagrangian_continuous_symmetry_derivative}
\begin{aligned}
\frac{\partial \mathcal{L}}{\partial \alpha} &= \sum_{n=1}^{2} \left( \frac{\partial \mathcal{L}}{\partial \phi_n} \frac{\partial \phi_n}{\partial \alpha} + \frac{\partial \mathcal{L}}{\partial (\phi_{n,\mu})} \frac{\partial (\phi_{n,\mu})}{\partial \alpha} \right) \\
&= \sum_{n=1}^{2} \left[ \frac{\partial \mathcal{L}}{\partial \phi_n} \frac{\partial \phi_n}{\partial \alpha} + \left( \frac{\partial \mathcal{L}}{\partial (\phi_{n,\mu})} \frac{\partial \phi_{n}}{\partial \alpha} \right)_{,\mu} - \left( \frac{\partial \mathcal{L}}{\partial (\phi_{n,\mu})} \right)_{,\mu} \frac{\partial \phi_{n}}{\partial \alpha} \right] = 0,
\end{aligned}
\end{equation}
where $\phi_1 = \phi$ and $\phi_2 = \phi^*$. Since $\alpha$ is the arbitrary phase, the derivative has to vanish. The first and the last terms cancel according to eq.~(\ref{action_variation_zero}) and the remaining part of the equation gives the conservation law, $J^\mu{}_{,\mu}$, for the Noether current
\begin{equation}\label{Neother_current_complex_scalar_field1}
J^\mu = \sum_{n=1}^{2} \frac{\partial \mathcal{L}}{\partial (\phi_{n,\mu})} \frac{\partial \phi_{n}}{\partial \alpha}.
\end{equation}
According to section~\ref{ch:energy_and_charge}, the conserved current can correspond to a free-space aeroacoustic dipole in an incompressible flow around a stationary acoustically compact rigid body. After inserting $\partial \phi_1/\partial \alpha = j\phi$, $\partial \phi_2/\partial \alpha = -j\phi^*$, $\partial \mathcal{L}/\partial (\phi_{1,\mu}) = \phi^{*,\mu}$, and $\partial \mathcal{L}/\partial (\phi_{2,\mu}) = \phi^{,\mu}$,
\begin{equation}\label{Neother_current_complex_scalar_field2}
J^\mu = j\phi \phi^{*,\mu} - j\phi^* \phi^{,\mu}.
\end{equation}
We can now express the terms of the Lagrangian in eq.~(\ref{Lagrangian_complex_scalar_field_gauge_field}) which capture the interaction of the fluid and acoustic gauge fields in terms of the current:
\begin{equation}\label{Lagrangian_complex_scalar_field_gauge_field_current}
\begin{aligned}
\mathcal{L} &= -\frac{1}{4} F_{\mu\nu} F^{\mu\nu} + (\phi^*_{,\mu} - je\bar{A}_\mu \phi^*/\hbar) (\phi^{,\mu} + je\bar{A}^\mu \phi/\hbar) + \left( \frac{m c_0}{\hbar} \right)^2 \phi^*\phi \\
&=-\frac{1}{4} F_{\mu\nu} F^{\mu\nu} + je\bar{A}^\mu \phi \phi^*_{,\mu}/\hbar - j e\bar{A}_\mu \phi^* \phi^{,\mu}/\hbar + e^2\bar{A}_\mu \bar{A}^\mu \phi^*\phi/\hbar^2 + \phi_{,\mu}^*\phi^{,\mu} + \left( \frac{m c_0}{\hbar} \right)^2 \phi^*\phi \\
&=-\frac{1}{4} F_{\mu\nu} F^{\mu\nu} + j e\bar{A}^\mu \phi \phi^*_{,\mu}/\hbar - j e\bar{A}^\mu \phi^* \phi_{,\mu}/\hbar + e^2\bar{A}_\mu \bar{A}^\mu \phi^*\phi/\hbar^2 + \phi_{,\mu}^*\phi^{,\mu} + \left( \frac{m c_0}{\hbar} \right)^2 \phi^*\phi \\
&=-\frac{1}{4} F_{\mu\nu} F^{\mu\nu} + e\bar{A}^\mu J_\mu/\hbar + e^2\bar{A}_\mu \bar{A}^\mu \phi^*\phi/\hbar^2 + \phi_{,\mu}^*\phi^{,\mu} + \left( \frac{m c_0}{\hbar} \right)^2 \phi^*\phi.
\end{aligned}
\end{equation}
Hence, the massless field $\bar{A}^\mu$ is at the leading order described by the Lagrangian
\begin{equation}\label{Lagrangian_vector_field_current}
\begin{aligned}
\mathcal{L} =-\frac{1}{4} F_{\mu\nu} F^{\mu\nu} + e\bar{A}^\mu J_\mu/\hbar.
\end{aligned}
\end{equation}
This is the massless Proca (electromagnetic) Lagrangian with added interaction with the current $J_\mu$ and the interaction describes the mechanism of dipole generation. For comparison, in section~\ref{ch:analogy_with_EMG} we defined $J^\mu = [q c_0, q \vec v] \rightarrow [-\rho_0 c_0, -\rho_0 \vec v]$ and this Lagrangian indeed gives the equation of motion~(\ref{Maxwell_equations_EMG_tensor}) after multiplying the interaction term with $\mu_0/\mu_0$ and absorbing the constant $e/(\hbar \mu_0)$ into $J_\mu$. Hence, the Lagrangian in eq.~(\ref{Lagrangian_vector_field_current}) matches the classical analogy with electromagnetism from section~\ref{ch:analogy_with_EMG}. In addition to this, the quantum theory is extended by the current in eq.~(\ref{Neother_current_complex_scalar_field2}), which can be associated with massive constitutive particles in both source and propagation region, as well as the Lagrangian of the fluid particles in eq.~(\ref{Lagrangian_complex_scalar_field}) and the Lagrangian in eq.~(\ref{Abelian_Higgs_model}), if spontaneous symmetry breaking is relevant. In any case, the thermodynamic origin of charge/mass at the microscopic limit of the observable universe remains inaccessible. 

As a summary, the Lagrangians which give the classical equations of motion for monopole (eq.~(\ref{scalar_wave_equation2})), dipole (eq.~(\ref{Maxwell_currents_Lorenz_gauge2})), and quadrupole radiation (eq.~(\ref{Einstein_field_equations_linear2})) are, respectively,
\begin{equation}\label{Lagrangian_monopole}
\mathcal{L} = \frac{1}{2} \bar{\chi}^{,\mu} \bar{\chi}_{,\mu} - \Psi \rho \bar{\chi},
\end{equation}
which is eq.~(\ref{Lagrangian_real_scalar_field_massless}) with the source term,
\begin{equation}\label{Lagrangian_dipole}
\mathcal{L} = -\frac{1}{4} F_{\mu\nu} F^{\mu\nu} + \mu_0 J_\mu \bar{A}^\mu = \frac{1}{2} \bar{A}^\mu \Box \bar{A}_\mu - \frac{1}{2} \bar{A}^\mu \bar{A}^\nu{}_{,\mu\nu}+ \mu_0 J_\mu \bar{A}^\mu,
\end{equation}
and\cite{Schwartz2014}
\begin{equation}\label{Lagrangian_quadrupole}
\mathcal{L} = \frac{1}{2} \bar{h}^{\mu\nu} \Box \bar{h}_{\mu\nu} -\bar{h}^{\mu\nu} \bar{h}_{\nu\alpha,\mu}{}^{,\alpha} + \bar{h}^\alpha{}_\alpha \bar{h}_{\mu\nu}{}^{,\mu\nu} - \frac{1}{2} \bar{h}^\alpha{}_\alpha \Box \bar{h}^\alpha{}_\alpha - \frac{2kG}{c_0^4} T_{\mu\nu} \bar{h}^{\mu\nu}.
\end{equation}
The last one is the leading-order approximation of the Einstein-Hilbert Lagrangian of general relativity with current. The three source-currents are, respectively, $\rho$, $J_\mu$, and $T_{\mu\nu}$ and the interaction constants are $\Psi \rightarrow 2kG/c_0^2$, $\mu_0 \rightarrow -kG/(2 c_0^2)$, and $2kG/c_0^4$.

In the case of monopole radiation, the current is not conserved, because it is the source of unsteady mass of the fluid. The Lagrangian of the fluid particles is given in eq.~(\ref{Lagrangian_real_scalar_field}) or eq.~(\ref{Lagrangian_real_scalar_field_spontaneous_symmetry_breaking_strong_perturbations}) with spontaneous symmetry breaking and the mean density dictated by the thermodynamics of the steady background fluid. Lastly, the stress-energy tensor $T_{\mu\nu}$ is the Noether current obtained when the parameter $\alpha$ above is replaced by the coordinates $x^\nu$, which reflects the symmetry of action (not Lagrangian) under global spacetime translations. Equation~(\ref{Lagrangian_continuous_symmetry_derivative}) takes the form
\begin{equation}\label{Lagrangian_global_symmetry_derivative}
\begin{aligned}
\mathcal{L}^{,\nu} = \sum_{n=1}^{2} \left[ \frac{\partial \mathcal{L}}{\partial \phi_n} \phi_n{}^{,\nu} + \left( \frac{\partial \mathcal{L}}{\partial (\phi_{n,\mu})} \phi_n{}^{,\nu} \right)_{,\mu} - \left( \frac{\partial \mathcal{L}}{\partial (\phi_{n,\mu})} \right)_{,\mu} \phi_n{}^{,\nu} \right],
\end{aligned}
\end{equation}
which does not equal zero in general. However, eq.~(\ref{action_variation_zero}) applies again and
\begin{equation}\label{canonical_stress_energy_tensor}
\tau^{\mu\nu}{}_{,\mu} = \sum_{n=1}^{2} \left( \frac{\partial \mathcal{L}}{\partial (\phi_{n,\mu})} \phi_n{}^{,\nu} \right)_{,\mu} - \eta^{\mu\nu} \mathcal{L}_{,\mu} = 0
\end{equation}
with $\tau^{\mu\nu}$ the conserved canonical stress-energy tensor. This tensor is not necessarily symmetric, but it can be turned into the symmetric form of $T^{\mu\nu}$ after transformation
\begin{equation}\label{stress_energy_tensor_symmetrization}
\tau^{\mu\nu} \rightarrow \tau^{\mu\nu} + \Theta^{\alpha\mu\nu}{}_{,\alpha}
\end{equation}
with an appropriate $\Theta^{\alpha\mu\nu}$ satisfying the antisymmetry $\Theta^{\alpha\mu\nu} = -\Theta^{\mu\alpha\nu}$, so that $\Theta^{\alpha\mu\nu}{}_{,\alpha\mu} = 0$ and the conservation law $\tau^{\mu\nu}{}_{,\mu}$ still holds after the transformation.

Even without an explicit relation between $T^{\mu\nu}$ and $\phi$ we see that the fluid field is still sufficiently described by the complex scalar field $\phi$, as in the case of dipole radiation. This suggest that the complex field does not need any generalization in the analogue unified theory, similarly as the spinor fields of quantum electrodynamics suffice in loop quantum gravity, as well. The Lagrangian from eq.~(\ref{Lagrangian_complex_scalar_field}) (or eq.~(\ref{Abelian_Higgs_model}) with spontaneous symmetry breaking) is sufficient for capturing even non-translational and stochastic motion of the fluid particles. This was also anticipated in section~\ref{ch:background_field_dipole_quadrupole}. Nevertheless, the quantum thermodynamic treatment of the field performed in the previous sections is necessary in order to describe the steady background fluid.

Finally, for the consistency with section~\ref{ch:energy_and_charge}, the quadrupole current should reduce to dipole and dipole to monopole at compact boundaries in the quantum analogy, as well, recalling that the Lorentz invariance is violated. In order to include the effect of boundaries in the dipole-to-monopole reduction, we multiply the Lagrangian in eq.~(\ref{Lagrangian_complex_scalar_field}) with $H(f)$ from section~\ref{ch:energy_and_charge}. Since $f$ is not a function of $\alpha$, we can again use eq.~(\ref{action_variation_zero}) in eq.~(\ref{Lagrangian_continuous_symmetry_derivative}) (both multiplied with $H(f)$) and remove the first and the third term in the latter equation. This gives
\begin{equation}\label{Neother_current_complex_scalar_field2_boundary}
H(f) J^\mu{}_{,\mu} = 0,
\end{equation}
which is equal to eq.~(\ref{conservation_laws_H_EMG}) and the same considerations apply.
\iffalse
we replace $\phi$ with $H(f)\phi$ and the current in eq.~(\ref{Neother_current_complex_scalar_field2}) receives two additional terms which cancel:
\begin{equation}\label{Neother_current_complex_scalar_field2_boundary2}
\begin{aligned}
J^\mu &= jH(f)\phi (H(f) \phi^*)^{,\mu} - jH(f)\phi^* (H(f)\phi)^{,\mu} \\
&= j H^2(f)(\phi \phi^{*,\mu} - \phi^* \phi^{,\mu}) + j H(f) \phi \phi^* H(f)^{,\mu} - j H(f) \phi^* \phi H(f)^{,\mu} \\
&= j H^2(f)(\phi \phi^{*,\mu} - \phi^* \phi^{,\mu}).
\end{aligned}
\end{equation}
This confirms the finding in section~\ref{ch:energy_and_charge} that a rigid boundary does not induce new (monopole) sources.
\fi
In the case of quadrupole reduction, the function $f$ does depend on the coordinates $x^\nu$ which replace $\alpha$, so $H(f)$ in the Lagrangian would not separate readily in eq.~(\ref{Lagrangian_continuous_symmetry_derivative}). However, according to the above we should consider the action (integral of the Lagrangian over the spacetime from eq.~(\ref{action})), not the Lagrangian, so we can multiply entire eq.~(\ref{Lagrangian_global_symmetry_derivative}) with $H(f)$ (recall footnote~\ref{Heaviside_function_Lagrangian_action}). This leads to
\begin{equation}\label{Neother_current_stress_energy_tensor_boundary}
H(f) T^{\mu\nu}{}_{,\mu} = 0,
\end{equation}
which matches eq.~(\ref{conservation_laws_H}). Therefore, both classical reduction mechanisms captured by the Ffowcs Williams and Hawkings aeroacoustic analogy can be reproduced with Noether currents in the framework of quantum field theory. %An alternative model may be established by introducing Lagrangians involving interaction terms of the quadrupole current and vector gauge field or the dipole current and real scalar field, but this will not be attempted here.

\section{Unified acoustic theory}\label{ch:unified_theory}

Based on the established classical, quantum, and thermodynamic analogies it is not very difficult to define an analogue unified theory. It should bring together the major concepts and results of all treated analogies. In fact, two unified theories are possible -- one which matches usual classical acoustics in conjunction with fluid dynamics and one which agrees with modern Lorentz-invariant theories and uses their formalisms. Both shall be briefly discussed in this concluding section.

The basis (medium or vacuum) for all relevant physical phenomena is provided by the background fluid which consists of elementary particles, the fluid molecules, with mass $M_0$ or, equivalently, energy $M_0 c_0^2$. These particles are distributed uniformly over the three-dimensional medium which they build, but cannot be observed individually by any macroscopic observer. The observers can interact with them only integrally, that is, with small but finite collections (systems) of $N_m \gg 1$ molecules, within which motion of the constitutive particles can be considered to be stochastic. The minimum characteristic length scale of such systems corresponds closely to the acoustic Planck length $L_P$ from eq.~(\ref{Planck_length}).

In a uniform motionless background fluid, the systems are in the state of maximal entropy, the canonical equilibrium characterized by eq.~(\ref{speed_of_sound_temperature}) and the constant mean density $\rho_0$, pressure $p_0$, and temperature $T_0$ of the fluid. Hence, even though the elementary particles are in permanent motion, the time of macroscopic observers is of no relevance at their microscopic scales and the background fluid is steady. The mean values also define the maximum values for any macroscopic perturbation of the background fluid, which are sources or fields. For example, the stochastic motion of molecules can be treated as an undetectable microscopic turbulence, a very weak aeroacoustic quadrupole source of the fluid perturbation (section~\ref{ch:aeroacoustic_sound_generation}). However, the minimum Planck length also implies microscopic boundaries of the observable universe, much like event horizons of black holes. At such boundaries the quadrupole radiation mechanism reduces to the more efficient dipole radiation of analogue electromagnetic waves (section~\ref{ch:energy_and_charge}), which appears to be analogue to cosmic microwave background or (at larger black holes) Hawking radiation.

All observable fields and associated quantum particles in source or propagation regions are, thus, macroscopic perturbations of the background fluid (involving many fluid molecules) in, at these length scales, adjoined time. In classical acoustics the Newtonian time represents the fourth independent variable and the observable universe is therefore, like the background fluid, stationary and infinite. Moreover, time and space are continuous. More accurately, the space is discretized due to the Planck limit, but the independent time remains continuous. This is the first major deviation from the relativistic theories, in which space and time are coupled by the reference speed of waves in the vacuum, the constant $c_0$. Its value is also determined by the equilibrium state of the background fluid and sets the maximum speed of particles. The coupling of space and time into spacetime means that in these theories time is also discrete, with the minimum time scale corresponding to the Planck time $t_P = L_P/c_0$ and the associated maximum frequency $\omega_0 \sim c_0/L_P$.

Whenever an observer interacts with another macroscopic system, the latter is found in a state which is a superposition of weighted determinate states (modes) with associated eigenfrequencies or, equivalently, energies $E = \hbar \omega$. While the continuous frequency can be measured arbitrarily accurate in classical acoustics, it is, like space, discretized in quantum theories. The minimum difference between two eigenfrequencies, $\Delta \omega_0$, is thus finite. This is the second and deciding difference between the theories. In quantum thermodynamics $\Delta \omega_0$ is  normalized (to 1\,s$^{-1}$), so its value does not affect the thermodynamic behaviour of quantum systems and disappears from the theory (modal frequencies are replaced by modal numbers). This is particularly relevant for the fluid molecules as elementary particles which are also characterized by modes and quantum states. In fact, the canonical equilibrium is analogues to a diffuse sound field (see sections~\ref{ch:background_fluid_thermodynamics} and \ref{ch:acoustic_modes_thermodynamics}) which occurs for high Helmholtz number values ($\omega_0 L/c_0 \gg 1$ for any macroscopic length scale $L \gg L_P$), when the modal density is high.

The third and last major difference is in the physical meaning of the modes. In quantum theories they are also normalized and associated with the probability distribution of, for example, location of the (single) particle (as discussed with regard to eq.~(\ref{modes_orthogonality})).  In contrast to this, normalization of modes does not play any significant role in classical acoustics, since amplitude and energy of sound fields (and therefore the number of particles represented by a single mode) are ultimately determined by the sources of sound (recall eq.~(\ref{eq:solution_tailored_Green_Helmholtz_modes}) and the related discussion).

The discretization of frequency has two important implications. First, it sets the minimum frequency of the theory, $\Delta \omega_0/2$, and, consequently, the maximum distance, which is the radius of the observable universe, $R_U$ (see eq.~(\ref{delta_omega_i_and_R_H}) and the comments which follow it). This is not the case with the continuous frequency and unbounded medium of classical acoustics. Importantly, it makes the stationary background fluid in the state of maximal entropy unobservable. Second, the finite $\Delta \omega_0$ implies a finite minimum half-width of the modes on the frequency axis, which we approximate with $\Delta \omega_0$. In classical acoustics this is a property of damped modes with the damping constant $\zeta_0 = \Delta \omega_0/2$. The damping is associated with the increase of entropy and causes an exponential decrease of the associated energy density in time by the factor $e^{-2\zeta_0 t} = e^{-\Delta \omega_0 t}$. Conversely, unattenuated plane waves in unbounded free space can take any value of the continuous frequency. As opposed to this, the observable universe is bounded and the energy in it is conserved, so the exponential decay of energy density can be attributed only to an expansion of the observable universe in time. The expansion appears to be caused by a compact monopole source of entropy with the strength given in eq.~(\ref{entropy_source_strength}). Frequency of the source (eq.~(\ref{lowest_frequency_universe})) represents the minimum frequency of the theory, so every observer is located in its near field ($\omega r/c_0 \gg 1$ cannot be satisfied by any $r<R_U$). Moreover, the source appears to be steady, giving rise to the cosmological constant (equations~(\ref{cosmological_constant_damping}) and (\ref{injected_volume_fraction_as_source})) and associated dark energy (eq.~(\ref{dark_energy_density})). The apparent nature of the source makes it also possible that the compact source is found at every location in the observable universe. Finally, the source defines the macroscopic time with its direction (arrow) of energy decay. At the initial time $t=0$, $R_U \sim L_P$ and the observable universe has all properties of the smallest system of particles of the background fluid and the canonical equilibrium (recall eq.~(\ref{pressure_universe_t_0}) and the remarks below it). All this completes the characterization of the background medium and sets the stage for the (macroscopic) quantum and classical field theories.

In analogue quantum field theory the massive fluid particles (collections of molecules) are represented by a real scalar quantum field $\phi$ associated with density and the Lagrangian in eq.~(\ref{Lagrangian_real_scalar_field}), in the case of monopole radiation. In the case of dipole and quadrupole radiation, the particles are represented by a complex scalar field $\phi$ associated with velocity and the Lagrangian in eq.~(\ref{Lagrangian_complex_scalar_field}). The massless fields associated with sound waves are the real scalar field $\phi$ from eq.~(\ref{Lagrangian_real_scalar_field_massless}), vector gauge field $\bar A^\mu$, and second-order tensor field $\bar h^{\mu\nu}$ (section~\ref{ch:background_field_dipole_quadrupole}).

The background fluid field couples with a massless acoustic field and gives it mass in the acoustic near field of a compact source or boundary (compact solid body, including the microscopic boundaries of the observable universe) by means of spontaneous symmetry breaking. The process for monopole radiation is captured by the Lagrangian in eq.~(\ref{Lagrangian_real_scalar_field_spontaneous_symmetry_breaking_strong_perturbations}) with $\lambda = 6 (mc_0/\hbar)^2/\rho_0^2$ and $\phi_0 = \sqrt{6m^2 c_0^2/(\lambda \hbar^2)} = \rho_0$ and for dipole and quadrupole radiation by the Lagrangian in eq.~(\ref{Abelian_Higgs_model}) with $\lambda = 4 m^2/\hbar^2$ and $\phi_0 = \sqrt{2m^2 c_0^2/(\lambda \hbar^2)} = c_0/\sqrt{2}$. It supports the relation between mass, energy, and frequency according to equations~(\ref{Planck_Einstein})-(\ref{reduced_Planck_constant}) and reveals the dual nature of every massless far-field particle becoming massive in the near field.

The massive fluid field is also responsible for the occurrence of conserved currents in source regions, according to Noether's theorem. For dipole radiation it is the charge current of eq.~(\ref{Neother_current_complex_scalar_field2}) and for quadrupole it is the stress-energy tensor from equations~\ref{canonical_stress_energy_tensor} and~\ref{stress_energy_tensor_symmetrization}. Monopole source is an unsteady density of the fluid $\rho$, which is not conserved. All sources connect the generated massless fields with massive fluid particles, which is captured by the Lagrangians in equations~(\ref{Lagrangian_monopole})-(\ref{Lagrangian_quadrupole}). These Lagrangians correspond to the classical theories for the three types of radiation. In the case of microscopic boundaries in the propagation region and dipole radiation, the occurrence of currents is analogous to vacuum polarization.

Equations of motion derived using the Lagrangians in equations~(\ref{Lagrangian_monopole})-(\ref{Lagrangian_quadrupole}) and the principle of least action represent governing equations of the analogue classical theories. In the case of dipole (analogy with electromagnetism) and quadrupole radiation (analogy with linearized general relativity) the analogies are valid for low Mach number (incompressible) and inviscid flows in acoustically compact source regions (or regions around compact bodies and boundaries of the observable universe). Since the characteristic speeds are much lower than $c_0$, the mechanisms of sound generation are kinematic. Governing equations in the first case are Maxwell's equations~(\ref{Maxwell_currents_Lorenz_gauge}) with $\mu_0 \rightarrow 4\pi G/c_0^2$ and the current $J^\alpha = [q c_0, q \vec v] \rightarrow [-\rho_0 c_0, -\rho_0 \vec v]$ (since only the three spatial components are relevant for wave generation, opposite charge is equivalent to opposite direction of $\vec v$). In the second case these are the linearized Einstein field equations~(\ref{Einstein_field_equations_linear}) with $k=-8\pi$ and the stress-energy tensor from eq.~(\ref{eq:stress-energy_tensor_perfect_fluid_components_non-relativistic}). In the case of monopole radiation the governing scalar wave equation is eq.~(\ref{scalar_wave_equation2}) with $\Psi = -16 \pi G/c_0^2$,

Motion of a fluid particle located in the field of a dipole or quadrupole source in free space is given by eq.~(\ref{particle_acceleration_spatial_Newtonian_gauge_EMG}) or eq.~(\ref{particle_acceleration_spatial_Newtonian_gauge}), respectively, where the component $A_0$ is expressed in the gauge in which vorticity (magnetic field) is zero and $h_{00}$ is expressed in the Newtonian gauge. In both cases the preferred frame is due to the fact that the classical acoustic observer does not exist in the analogue acoustic spacetime, but in the Newtonian space and time. In the electromagnetic analogy the observer also sets the preferred contravariant vector space space, because the components of the four-vector potential $A^\mu \rightarrow -c_0 h^{\mu 0} = -c_0 h^{0 \mu}$ and current $J^\mu \rightarrow -T^{\mu 0}/c_0 = -T^{0 \mu}/c_0$ must transform as the first rows and columns of the metric perturbation and stress-energy tensor, respectively, not as four-vectors, in order to calculate usual acoustic quantities. Particle motion in the field of a monopole source is given by eq.~(\ref{acoustic_velocity}) in the Newtonian gauge. The following analogy applies: $\chi \rightarrow h^{00}$ and $\rho \rightarrow T^{00}$. In all cases the background acoustic spacetime formed by the steady quiescent background fluid is flat. If the fluid is not quiescent, the background Minkowski metric must be replaced with another appropriate metric, as in section~\ref{background_spacetime}. Such a metric does not follow from the analogy with general relativity, because the background fluid is externally controlled.

Acoustically compact foreign bodies in the fluid as well as the limits of the observable universe at the acoustic Planck scale introduce boundaries which violate the Lorentz invariance. An aeroacoustic quadrupole source in their vicinity reduces to dipole and, if the boundary is not rigid (that is, not a true, impenetrable boundary), a dipole source reduces to monopole, in the same manner as in the Ffowcs Williams and Hawkings analogy. If the boundary is stationary and rigid, the second-order stress-energy tensor reduces to the conserved four-vector current in eq.~(\ref{conserved_induced_current}), while if the stationary boundary is non-rigid, the four-vector current reduces to the scalar in eq.~(\ref{induced_monopole}). Therefore, if external forces are excluded, every analogue charge is induced by massive fluid particles acting on a compact boundary.

\section*{}\label{ch:references}

%\nocite{gruber}
%\nocite{laerm2}
%\nocite{din45657}
%\nocite{dinen60804}
%\nocite{dineniso266}
%\nocite{din45645}
%\nocite{vdi2565}
%\nocite{vdi2573}

%\bibliographystyle{plain}
%\bibliography{TaBib}

\end{document}